\documentclass[a4paper,12pt,times,custombib,
hidelinks,%for printing
]{PhDThesisPSnPDF}

\input{Preamble/preamble}
\newcommand{\unit}[1]{%
    \ensuremath{\,\mathrm{#1}}%
}

\newcommand{\E}[1]{%
    \cdot 10^{#1}%
}
\newcommand{\rarrow}{%
    \rightarrow%
}
\newcommand{\srarrow}{%
    \quad \rightarrow \quad%
}

\newcommand{\avg}[1]{%
    \ensuremath{\langle #1 \rangle}%
}
\newcommand{\fsl}[1]{%
    \ooalign{\(#1\)\cr\hidewidth\(/\)\hidewidth\cr}
}

\newcommand{\eqtext}[1]{%
    \quad\text{#1}\quad
}

\renewcommand{\nomgroup}[1]{%
\ifthenelse{\equal{#1}{Z}}{\item[\textbf{Acronyms / Abbreviations}]}{%
\ifthenelse{\equal{#1}{X}}{\item[\textbf{Symbols}]}
{}}}

\newcommand{\abbdef}[2]{%
    \phantomsection\label{abb:#1}%
    \nomenclature[Z]{#1}{#2 (Pag.~\pageref{abb:#1})%
    \phantomsection\label{abb:#1b}}%
    #2 (\hyperref[abb:#1b]{#1})%
}

\newcommand{\abb}[1]{\hyperref[abb:#1b]{#1}}
\newcommand{\sym}[1]{\hyperref[sym:#1b]{#1}}
\newcommand{\abbdefs}[3]{%
    \phantomsection\label{abb:#3}%
    \nomenclature[Z]{#1}{#2 (Pag.~\pageref{abb:#3})%
    \phantomsection\label{abb:#3b}}%
    #2 (\hyperref[abb:#3b]{#1})%
}
\newcommand{\abbs}[2]{\hyperref[abb:#2b]{#1}}

\newcommand{\be}{\begin{equation}\begin{gathered}}
\newcommand{\ee}{\end{gathered}\end{equation}}

\newcommand{\bea}{\begin{equation}\begin{aligned}}
\newcommand{\eea}{\end{aligned}\end{equation}}

\newcounter{feynmancounter}
\newenvironment{feynman}[1]
  {
    \addtocounter{feynmancounter}{1}
    \begin{fmffile}{feynm\thefeynmancounter}
	\begin{fmfgraph*}(#1)
  }
  { 
    \end{fmfgraph*}
    \end{fmffile}
  }

\newenvironment{preambletitle}[1]{
\cleardoublepage
\setsinglecolumn
\chapter*{\centering \Large #1}
\thispagestyle{empty}
}
  
\usepackage{amssymb}
\usepackage{mathtools}
\usepackage[italicdiff]{physics}
\usepackage{comment}
\usepackage{cancel}

\usepackage{feynmp-auto}

\usepackage{algorithm2e}
\usepackage[frozencache]{minted}
\usepackage{mmacells}
\usepackage{bm}

\usepackage{pdfpages}

\title{Exploration of possible signals beyond special relativity using high-energy astroparticle physics}
\author{Maykoll A. Reyes Hung}

\dept{Departamento de Física Teórica}

\university{Facultad de Ciencias, Universidad de Zaragoza}
\crest{
    \centering
    \includegraphics[width=7.5cm]{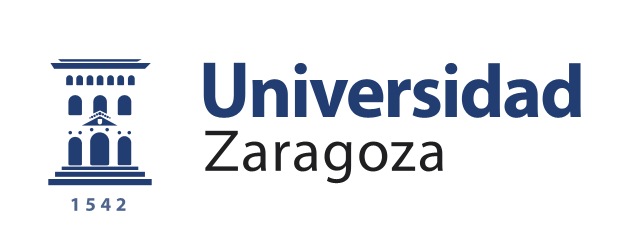}
    \hfill
    \includegraphics[width=7.5cm]{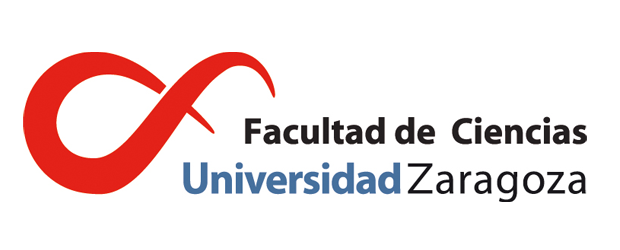}
}
\collegeshield{\includegraphics[width=0.3\textwidth]{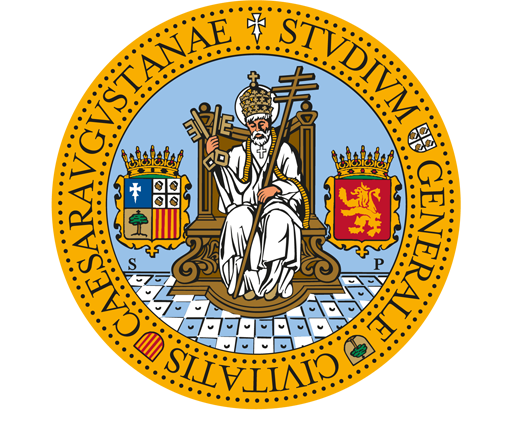}}

\supervisor{
José Manuel Carmona\newline
José Luis Cortés
}

 \supervisorrole{\textbf{Supervisors: }}
\renewcommand{\submissiontext}{Dissertation submitted for the degree of}

\degreetitle{Doctor of Philosophy}

\degreedate{\monthname[\the\month]\space\the\year} 

\ifdefineAbstract
\fi

\ifdefineChapter
\fi

\begin{document}
\includepdf[pages=-]{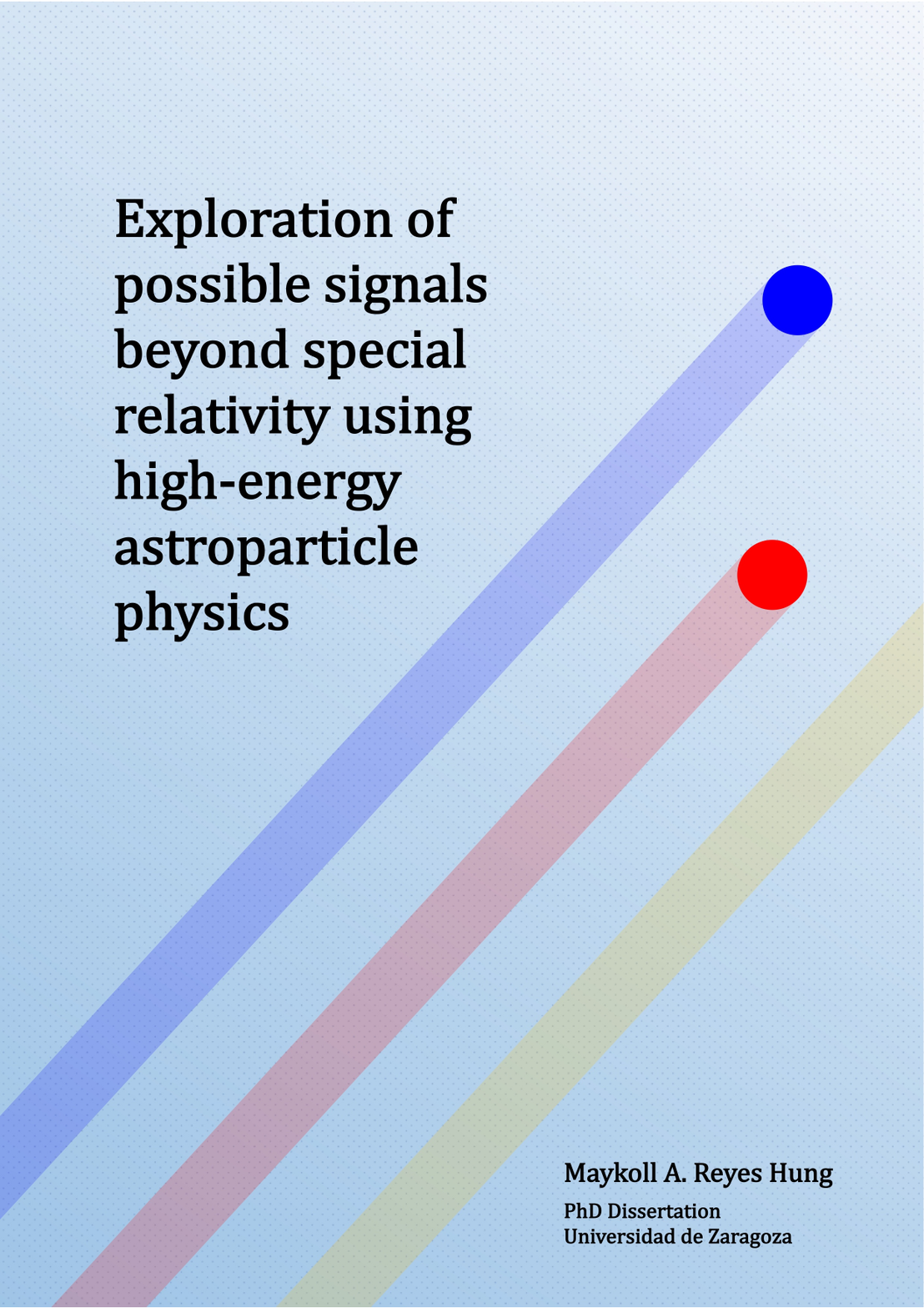}

%Fixing broken hbar (?)
\renewcommand{\hbar}{\hslash}

\frontmatter

\maketitle

% ******************************* Thesis Dedidcation ********************************

\begin{dedication} 

\begin{minipage}[l]{0.9\textwidth}\it
Dedicado a quien me inspira\\
a saber más cada día.\\
Para tí, papá.\\
\vspace{1em}

Dedicado a quien me inspira\\
a ser mejor persona.\\
Para tí, mamá.\\
\vspace{1em}

Dedicado a quien me inspira\\
a mirar al futuro.\\
Para tí, hermano.

\end{minipage}

\end{dedication}

%\include{Declaration/declaration}
% ************************** Thesis Acknowledgements **************************

\begin{preambletitle}{Agradecimientos/Acknowledgements}

Quiero agradecer al Ministerio de Ciencia e Innovación (MCI) y a la Agencia Estatal de Investigación (AEI) por la financiación de esta tesis a través de la ayuda predoctoral FPI PRE2019-089024 y al Fondo Social Europeo (FSE). Este trabajo también ha sido parcialmente financiado por las ayudas PGC2018-095328-B-I00 del MCI/AEI y la ayuda DGIID- DGA No. 2020-E21-17R de la Diputación General de Aragón (DGA).

Quiero agradecer a mis directores de tesis José Manuel Carmona y José Luis Cortés, por su eterna paciencia e invaluable ayuda durante la elaboración de esta tesis. No podría haber deseado unos mejores guías para mi iniciación a la investigación. También quiero agradecer a Javier Relancio por su valiosa contribución y orientación a esta investigación.

Quiero agradecer a los investigadores principales de mi ayuda FPI, Eduado Follana y Javier Redondo. Así como a Manuel Asorey, Jesús Clemente y José Esteve por un lado, y a Siannah Peñaranda e Inés Cavero por otro, por permitirme impartir docencia en sus respectivas asignaturas. También quiero agradecer en general al Departamento de Física Teórica de la Facultad de Ciencias de la Universidad de Zaragoza, por acogerme cálidamente desde mis años de estudiante hasta ahora.

I want to express my gratitude to the COST Action 18108 "Quantum gravity phenomenology in the multi-messenger approach", thanks to which I have been able to attend several training schools, conferences and other activities. I also want to thank the members therein, with many of them I have not only initiated interesting collaborations, but sometimes even a friendship.

I also want to express my gratitude to Denise Boncioli, for welcoming me during my three-month stay in L'Aquila. I appreciate your understanding and support during the final stage of preparation of my dissertation.

Quiero agradecer también a mis compañeros doctorandos de las áreas de física teórica y física nuclear por su gran esfuerzo en la divulgación a través de la Asociación $\hbar$. También quiero agradecer a todos mis amigos, con especial mención a Mario Tovar y Lucía Pereira, por los divertidos momentos vividos y por vivir.

Finalmente quiero agradecer toda mi familia, por su cariño y apoyo incondicional. En particular a mi padre Marcos Reyes, a mi madre Carmen Hung y a mi hermano César A. Reyes, a quienes amo con todo mi corazón, y por ello les dedico esta tesis.

\end{preambletitle}

% ************************** Thesis Abstract *****************************
% Use `abstract' as an option in the document class to print only the titlepage and the abstract.
\begin{abstract}
The Standard Model of particle physics and General Relativity are the most successful fundamental physics models up today. However, attempts to develop a new model that unifies the knowledge of both and extends their range of validity have been unsuccessful, due to the different role the spacetime plays in each one.

In order to construct this model of new physics, experimental feedback that could guide the theoretical developments would be very useful. However, the energies at which such effects clearly manifest are supposed to be of the order of the Planck scale, energies which are beyond the ones we can reach today in our experiments and in the highest energy astrophysical observations. This shows the importance of the research trying to find traces of this new physics at lower energies. The aim of this dissertation is to study possible signs of physics beyond Special Relativity in one of the most favorable scenarios, the physics of very high-energy astroparticles.

The fundamental ingredients of this study are presented in Chapters~\ref{chap:introduction} and \ref{chap:LIV-DSR}: the different astromessengers (neutrinos, gamma rays and cosmic rays), and the different ways to go beyond Special Relativity (a Violation of the Lorentz Invariance and Doubly Special Relativity). In Chapter~\ref{chap:LIV-DSR}, we also develop the use of the collinear approximation, a property of very high-energy interactions which allows us to simplify the calculation of decay widths and cross sections.

In Chapter~\ref{chap:neutrinos} we consider a model of Lorentz Invariance Violation with superluminal neutrinos. This model makes neutrinos unstable particles, which are able to decay through electron-positron and neutrino-antineutrino pair emission. We find that considering this model of new physics has relevant consequences on the expected neutrino flux at Earth, namely, the appearance of a cutoff in the spectrum, preceded by a small excess of flux. 

In Chapter~\ref{chap:cosmicrays} we repeat this analysis considering neutrinos produced during the propagation of very high-energy cosmic rays. We find that the existence of a cutoff in the spectrum can make disappear the cosmogenic neutrino peak that one expects to see as a consequence of cosmic ray interactions with the Cosmic Microwave Background. Moreover, the accumulation previous to the cutoff can coincide with the peak produced by cosmic ray interactions with the Extragalactic Background Light, enhancing its probability of detection.

In Chapter~\ref{chap:gammas}, we study the effects of Doubly Special Relativity on the propagation of gamma rays. We distinguish between two different phenomenological consequences: effects on the expected gamma ray flux and anomalies in the time of flights. Concerning the first one, we have focused in a particular model (DCL1) and studied its consequences on the transparency of the universe. We find, in comparison with the Special Relativity case, a (asymptotically constant) decrease of the transparency at high energies, preceded by an increase. In the same chapter we also present a general analysis, compatible with the Relative Locality of interactions, for the difference in times of flights between high- and low-energy photons in a flat spacetime.

Finally, in Chapter~\ref{chap:conclusions} we gather and summarize the most important results, limitations, and possible future extensions of this work.
\end{abstract}

\begin{preambletitle}{Resumen}
El Modelo Estándar de la física de partículas y la Relatividad General son los modelos de física fundamental más exitosos de los que disponemos hoy en día. Sin embargo, los intentos de construir un modelo que unifique el conocimiento de ambas y amplíe su rango de validez han sido infructuosos, debido al diferente papel que juega el espacio-tiempo en cada uno.

Para poder construir esta teoría de nueva física, sería muy útil información experimental que guíe los desarrollos teóricos. Sin embargo, se espera que las energías a las que dichos efectos son claramente manifiestos sean del orden de la escala de Planck, energías que quedan muy por encima de aquellas que somos capaces de alcanzar a día de hoy en nuestros experimentos y en las observaciones astrofísicas de más alta energía. De ahí la importancia de trabajos que se centren en la búsqueda de huellas de esta nueva física a energías menores. El objetivo de esta tesis es estudiar posibles señales de física más allá de Relatividad Especial en uno de los escenarios más favorables para ello, la física de astropartículas de muy alta energía.

En los Capítulos~\ref{chap:introduction} y \ref{chap:LIV-DSR} se presentan los ingredientes fundamentales de este estudio: los diferentes astromensajeros (neutrinos, rayos gamma y rayos cósmicos), y las diferentes formas de ir más allá de Relatividad Especial (Violación de la Invariancia Lorentz y Relatividad Doblemente Especial). En el Capítulo~\ref{chap:LIV-DSR}, también se desarrolla el uso de la aproximación colineal, una propiedad de las interacciones de muy alta energía que nos permitirá simplificar el cálculo de anchuras de desintegración y anchuras eficaces.

En el Capítulo~\ref{chap:neutrinos} consideramos un modelo de Violación de la Invariancia Lorentz con neutrinos superlumínicos. Dicho modelo convierte a los neutrinos en partículas inestables, permitiendo que decaigan mediante la emisión de pares electrón-positrón y neutrino-antineutrino. Encontramos que considerar este modelo nueva física tiene importantes consecuencias en el flujo de neutrinos esperado en la Tierra, concretamente, la aparición de un corte en el espectro, precedido de una pequeña acumulación. 

En el Capítulo~\ref{chap:cosmicrays} repetimos este análisis considerando neutrinos producidos durante la propagación de rayos cósmicos de muy alta energía. Encontramos que la existencia de un corte en el espectro puede hacer desaparecer el pico de neutrinos cosmogénicos que uno esperaría ver como consecuencia de las interacciones de los rayos cósmicos con el Fondo Cósmico de Microondas. Además de ello, la acumulación previa al cutoff puede coincidir con el pico producido por las interacciones de los rayos cósmicos con el Fondo de Luz Extragaláctico, incrementando la probabilidad de detectarlo.

En el Capítulo~\ref{chap:gammas}, estudiamos los efectos de la Relatividad Doblemente Especial en la propagación de rayos gamma. Podemos diferenciar dos implicaciones fenomenológicas diferentes: los efectos sobre el flujo y las anomalías en la diferencia de tiempos vuelo. Referente al primero, hemos considerado el modelo concreto (DCL1) y estudiado sus consecuencias en la transparencia del universo. Como resultado encontramos, respecto del caso de Relatividad Especial, un decremento (asintóticamente constante) de la transparencia a altas energías, precedido de un aumento. En el mismo capítulo también presentamos un modelo complemente general, y compatible con la Localidad Relativa de las interacciones, para la diferencia en tiempos de vuelo entre fotones de alta y baja energías en un espacio-tiempo llano.

Finalmente, en el Capítulo~\ref{chap:conclusions} recopilamos y resumimos los resultados más importantes, limitaciones, y posibles futuras extensiones de este trabajo.
\end{preambletitle}

\begin{preambletitle}{Publications}
The different parts of the research conducted during this dissertation have been published in the following articles,
\begin{refsection}
\nocite{Carmona:2019xxp,Carmona:2021lxr,Carmona:2022pro,Carmona:2023luz,Carmona:2022dtp,Carmona:2022jtc,ReyesHung:2022xut}
\printbibliography[heading=none]
\end{refsection}
and have contributed partially to the following reviews,
\begin{refsection}
\nocite{Addazi:2021xuf,Carmona:2022lyg}
\printbibliography[heading=none]
\end{refsection}

\end{preambletitle}

% *********************** Adding TOC and List of Figures ***********************

\tableofcontents

%\listoffigures

%\listoftables

% \printnomenclature[space] space can be set as 2em between symbol and description
%\printnomenclature[3em]

\printnomenclature

% ******************************** Main Matter *********************************
\mainmatter

\begin{refsection}
\chapter{Introduction}
\graphicspath{{Chapter1/Figs/}}
\label{chap:introduction}

\section{The Standard Model of particle physics and the \texorpdfstring{$\Lambda$}{Λ}CDM Model}
\label{sec:sm}

The ultimate goal of physics is to describe the fundamental constituents of the universe and the rules that govern the interactions among them. As of today, we have identified and classified many different elementary particles and four different interactions (gravitational, electromagnetic, weak and strong). This knowledge has been unified in two different models: the \abbdef{SM}{Standard Model} of particle physics, which uses the mathematical framework of \abbdef{QFT}{Quantum Field Theory} to describe the elementary particles, and the electromagnetic, weak, and strong interactions; and \abbdef{GR}{General Relativity}, which uses a geometrical framework to explain gravity, and is at the basis of the Concordance Cosmological Model or \abbdefs{$\Lambda$CDM}{Dark Energy and Cold Dark Matter Model}{LCDM}, which presently describes the evolution of the universe.

\subsection{The Standard Model of particle physics}

The \abb{SM} describes all known elementary particles and three (of the four known) interactions: electromagnetic, weak and strong interactions. The elementary particles of the \abb{SM} can be classified into two main groups: the particles that describe the interactions (bosons), and the particles that compose matter (fermions). 

\begin{figure}[tb]
    \centering
    \includegraphics[width=0.6\textwidth]{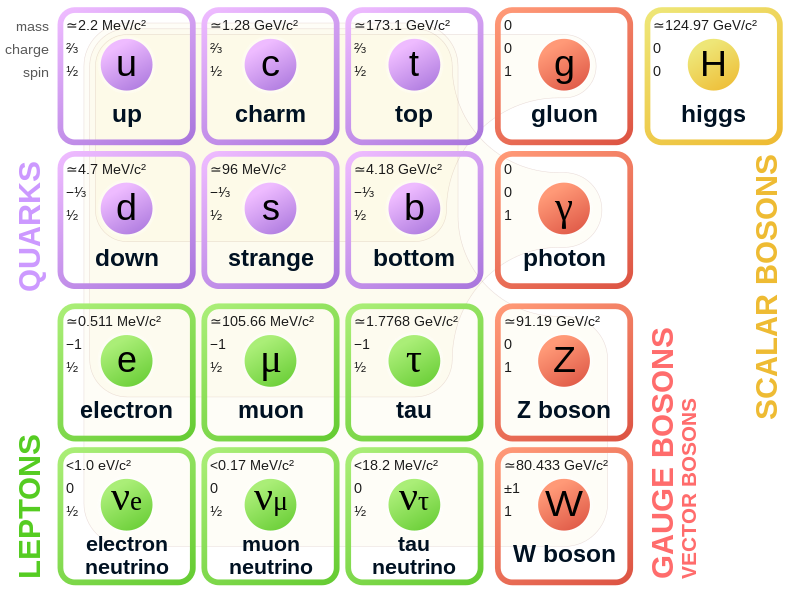}
    \caption{Elementary particles of the Standard Model of particle physics. From Wikipedia.}
    \label{fig:sm_particles}
\end{figure}

Bosons are characterized by an integer spin. The majority of bosons of the \abb{SM} have spin one: this includes the eight gluons, which carry the strong interaction; and the $W^{\pm}$, the $Z$, and the photon, which carry the electroweak (unification of electromagnetic and weak) interaction. Additionally, the \abb{SM} includes a boson of spin zero, the Higgs, which describes a mechanism to provide rest mass to some of the elementary particles.

Fermions, the particles that compose matter, are characterized by half-integer spin. They can be classified into two groups: quarks, the elementary particles that are affected by the strong interaction (up, down, charm, strange, top, bottom); and leptons, which are not (electron, muon, tau and their associated neutrinos). Quarks are always confined in groups (hadrons) of three (baryons, like the proton or neutron) or two (mesons, like the pion). Except for the proton, all hadrons are known to be unstable and disintegrate, but the protons and neutrons are able to form stable larger structures (nuclei).

All the previous ingredients are incorporated in the mathematical framework of \abb{QFT}~\cite{Lancaster:2014,Herrero:1998eq,Weinberg:1995}, in which elementary particles are described as excitations of quantum fields. In order to represent the propagation of a free (i.e. non-interacting) particle, one should construct a free Lagrangian (density) and minimize it in order to find the equations of motion. In order to describe (local) interactions between two or more elementary particles one should add new terms to the Lagrangian which involve products of two or more different fields at the same space-time point. Additionally, one can compute the probability amplitude, or transition amplitude, of the evolution of a certain initial state of elementary particles to a certain final state using the Fermi golden rule and perturbation theory; this allows us to compute decay widths and cross sections of different processes.

\subsection{Concordance Cosmological Model or \texorpdfstring{$\Lambda$}{Λ}CDM Model}
\label{sec:cosmology}

If the \abb{SM} and \abb{GR} were the ultimate theories to explain all the phenomena of our universe, one would expect that using the information provided by both models one could explain the evolution of the universe itself. A cosmological model should explain not only its current state, but also the past \textit{history} of the universe, to which we have access thanks to cosmological observations and measurements~\cite{Cherepashchuk:2013}. At present, the model which best fits our empirical cosmological observations is the \abbs{$\Lambda$CDM}{LCDM} model.

This model is able to explain the current expanding universe, in which the wavelengths of photons, or the de Broglie wavelengths of massive particles, expand when they travel through the space. This phenomenon causes energy loss, and it is encoded in a variable called redshift,
\be
    z_\text{\;at emission} \coloneqq \frac{\lambda_\text{detected}-\lambda_\text{emitted}} {\lambda_\text{emitted}} \,.
\ee
One can use the redshift as a large scale temporal coordinate to tag events of the history of the universe, as well as to characterize the location of astrophysical objects or particles travelling cosmic distances.

Another important feature of the universe is that it is homogeneous and isotropic at very large scales. In order to describe a universe like that in the framework of \abb{GR}, one should solve the Einstein field equations using a homogeneous, isotropic, and expanding metric, the \abbdef{FLRW}{Friedmann–Lemaître–Robertson–Walker} metric,
\be 
    ds^2 = dt^2 - a^2(t) \qty(\frac{dr^2}{1-k r^2}+r^2 d\Omega) \,,
    \label{eq:FLRW}
\ee
where $k$ is the curvature and $a(t)$ the scale factor of the universe, which is closely related to the redshift, $a(t)=1/(1+z(t))$. We can use this relation to also connect the temporal coordinate to the redshift variable,
\be 
    a(t)=\frac{1}{1+z(t)} \srarrow \frac{\dot a(t)}{a(t)} dt = -\frac{dz}{1+z(t)} \srarrow dt = -\frac{dz}{H(z)(1+z)} \,,
    \label{eq:dtdz}
\ee
where we have defined the Hubble parameter as $H=\dot a/a$.

Using the \abb{FLRW} metric to solve the Einstein equations leads to the Friedmann equations, which are the basis of the standard cosmology. The first Friedmann equation can be written in terms of the cosmic density of radiation ($\Omega_r$), matter ($\Omega_m$), curvature ($\Omega_k$), and dark energy ($\Omega_\Lambda$),
\be
    H(z) = H_0 \sqrt{\Omega_r (1+z)^4 + \Omega_m (1+z)^3 + \Omega_k (1+z)^2 + \Omega_\Lambda} \,.
    \label{eq:H(z)_completo}
\ee
The model does not predict the values of these cosmic densities, so they should be adjusted using cosmological observations and measurements. However, one finds that the numerical values obtained~\cite{ParticleDataGroup:2020} cannot be explained using the ingredients of the \abb{SM} of particle physics.

For instance, one would expect the matter contribution ($\Omega_m \approx 0.31$) to be dominated by the baryonic matter; however, one finds that the baryonic contribution is only $\Omega_b \approx 0.05$, so a new kind of matter, beyond the \abb{SM}, has to be assumed: (cold) dark matter ($\Omega_c \approx 0.26$).

Additionally, one finds the radiation and curvature contributions to be very small ($\Omega_r\approx 5.38\times 10^{-5}$ and $\Omega_k \approx 7.0\times 10^{-4}$). Then, as matter dominates, one would expect a shrinking universe due to the gravitational interaction. This way, in order to explain the observed accelerated expansion of the universe, one needs a very large dark energy contribution ($\Omega_\Lambda \approx 0.69$), which has not current explanation within the framework of the \abb{SM} of particle physics.

Let us note that, as the measured values of $\Omega_m$ and $\Omega_\Lambda$ are much larger than $\Omega_r$ and $\Omega_\kappa$, we can make the approximation
\be
    H(z) \approx H_0 \sqrt{\Omega_m (1+z)^3 +\Omega_\Lambda} \,,
    \label{eq:H(z)}
\ee
for small values of the redshift.

Finally, in order to explain the small value of the curvature contribution, one can introduce the inflaton field as a precursor of every particle. At the beginning of the universe, this very energetic field, in continuum quantum fluctuations, swiftly expanded erasing any previous large scale curvature, but preserving the inhomogeneities inherited from the quantum fluctuations at smaller scales. These inhomogeneities helped the gravitational interaction to form structures like galaxies and clusters afterwards.

The inflaton is assumed to decay in all the particles we know nowadays, forming a hot plasma of elementary particles. As the universe started cooling, the different particles decoupled from the plasma, escaping through the space and forming the primordial backgrounds, which still reach us today. First the neutrino decoupling, which leads to the \abbdefs{C$\nu$B}{Cosmic Neutrino Background}{CnB}. Then, after nucleosynthesis and recombination, light is able to decouple too, leading to the \abbdef{CMB}{Cosmic Microwave Background}.

Measuring the primordial backgrounds is a way to look at the state of the universe at the moment these particles decoupled, so it is a way to look at the first stages of the universe. Thanks to the \abb{CMB} we can go back up to redshifts around $z_\text{CMB}\approx 1100$ (when the universe was just 400.000 years old). If we were able to make direct measurements of the \abbs{C$\nu$B}{CnB}, we could go up to $z_\text{C$\nu$B}\approx 10^{10}$ (when the universe was just 1 second old)~\cite{Michney:2006mk}. Unfortunately, neutrinos are very poorly interacting particles, so the direct measurement of \abbs{C$\nu$B}{CnB} is quite challenging~\cite{Li:2015koa}. But that also means that, when we consider the current propagation of particles in the intergalactic medium, we can disregard the interactions of the particles with the \abbs{C$\nu$B}{CnB}. In contrast, the interactions of the extragalactic particles with the photon \abb{CMB} must be taken into account. In fact, it is not the only photon background we should consider.

After the formation of the first starts (redshifts between 1 and 3~\cite{Labbe:2009}), the \abbdef{EBL}{Extragalactic Background Light}, radiation emitted from the stars and galaxies, starts playing a role. The \abb{EBL} covers mostly infrared and optical frequencies, exhibiting a peak at each range. There exist additional photon backgrounds with energies that go from the radio up to the x-rays and gamma rays (see Chapter~\ref{chap:gammas} for more information). 
However, the \abb{EBL} is the second most intense radiation background, only below \abb{CMB}. Additionally, for the processes and energies studied in this work, the \abb{CMB} and \abb{EBL} are the only relevant backgrounds.

\begin{figure}[tb]
    \centering
    \includegraphics[width=0.55\textwidth]{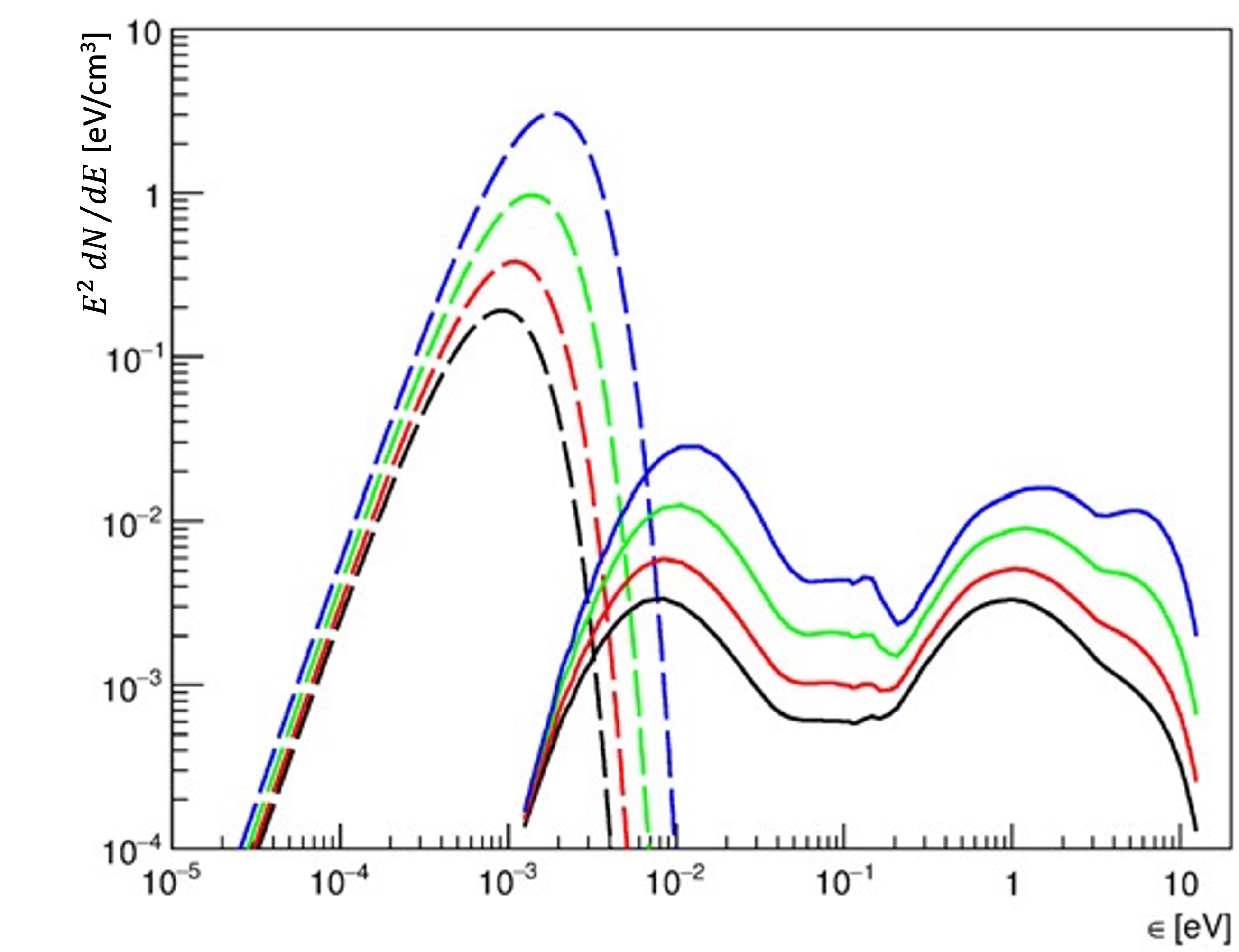}
    \caption{\abb{CMB} (dashed) and \abb{EBL} (solid) spectrum for different epochs (black $z=0$, red $z=0.186$, green $z=0.5$, and blue $z=1$). From~\cite{Dzhatdoev:2017xnk} using the \abb{EBL} model of~\cite{Gilmore:2011ks}.}
    \label{fig:CMB_EBL}
\end{figure}

Let us note that, while the \abb{CMB} spectrum can be computed analytically as a black body thermal emission at a certain temperature (which decreases with the time), the \abb{EBL} still have lots of uncertainties and several models have been proposed~\cite{Stecker:2005qs,Kneiske:2003tx,Dominguez:2010bv,Gilmore:2011ks}. In Fig.~\ref{fig:CMB_EBL} we can see the relative importance of the \abb{EBL} respect to the \abb{CMB} (using the \abb{EBL} model of~\cite{Gilmore:2011ks}), for different values of the redshift. For more information about the other photon backgrounds see Fig.~\ref{fig:debr}, and the corresponding discussion, in Chapter~\ref{chap:gammas}.

\subsection{Limitations of the Standard Model}
\label{sec:beyond}

We have seen how some ingredients required by the \abbs{$\Lambda$CDM}{LCDM} model to explain the cosmological observations are missing in the \abb{SM}. Additionally, there are also laboratory measurements, like those regarding neutrino oscillations (which lead to a non-null mass for the neutrino~\cite{Drexlin:2013lha}), which cannot be explained within the \abb{SM}. This shows us the need to go beyond it.

One can add additional ingredients to the \abb{SM} to try fix some of the previously mentioned problems (such as axions or sterile neutrinos). However, in the end, we expect both, the \abb{SM} and \abb{GR}, to be replaced by a new theory which will surpass these problems and contain all the ingredients of the \abb{SM} of particle physics and \abb{GR}. This new unified theory would be necessary to describe systems in which we cannot disregard neither quantum effects nor gravity (for instance, the interior of the black-holes or the first stages of the early universe). However, to describe such systems, one would need to replace \abb{GR} with a quantum description, i.e., a \abbdef{QG}{Quantum Gravity} theory.

The attempts to quantize gravity have led to non-renormalizable~\cite{Carlip:2001wq} or non-predictable theories (such as string theory~\cite{Mukhi:2011zz,Aharony:1999ks,Dienes:1996du} or loop quantum gravity~\cite{Sahlmann:2010zf,Dupuis:2012yw}). This leads us to look at the problem in another way: since it is not yet possible to develop and test a complete theoretical framework of \abb{QG}, we can start from modifying the fundamental ingredients in which both the \abb{SM} and \abb{GR} are based on, and then look for simple phenomenological consequences that could be tested; the resulting information may guide us in the correct direction to construct the complete theoretical framework. This strategy is called a \textit{bottom-up} approach, and it is historically similar to what happened in the development of \abbdef{QM}{Quantum Mechanics}.

Following this approach, one fundamental ingredient in both theories is the concept of spacetime, described by \abbdef{SR}{Special Relativity}. On the one hand, in \abb{GR} the \abb{SR} spacetime is generalized to a dynamic variable which interacts with matter; on the other hand, in \abb{QFT} the \abb{SR} spacetime is a fixed scenario in which the quantum evolution happens. These two visions are incompatible, and it is natural to think that the new theory of \abb{QG} will surpass both conceptions in favor of a new one. One way to change our current concept of spacetime is to modify the fundamental symmetry of \abb{SR} which defines it, i.e., the \abbdef{LI}{Lorentz Invariance}.

One can find two different ways to modify \abb{LI}. Historically, the first one to be studied was the \abbdef{LIV}{Lorentz Invariance Violation}~\cite{Colladay:1998fq,Kostelecky:2008ts}, in which one assumes that the symmetry under \abb{SR} Lorentz transformations is not exact, but only partial, and it is manifestly broken at high enough energies. The second one is to consider that \abb{LI} is still an exact symmetry of spacetime, but that the Lorentz transformations are different from the ones of \abb{SR}; the most studied realization is called \abbdef{DSR}{Doubly Special Relativity}~\cite{Amelino-Camelia:2000cpa,Amelino-Camelia:2000stu}. Both approaches have different phenomenological and conceptual implications, which will be discussed in more detail in Chapter~\ref{chap:LIV-DSR}.

\section{Objectives, scope, and structure of the dissertation}
\label{sec:scope}

When considering deviations from \abb{LI}, one has to keep in mind that they should still be consistent with the fact that we do not observe any obvious deviation in our current observations or experiments. This fact implies that the deviations have to be small enough at the current explored energies. Additionally, as said in the previous section, one expects the effects of a \abb{QG} theory to manifest where both the quantum nature and the gravitational interaction are necessary to describe the system; i.e., when the Compton wavelength is of the same order as the Schwarzschild radius of the system,
\be 
    \lambda_C \sim r_S \srarrow \frac{\hbar}{mc} \sim \frac{Gm}{c^2} \srarrow m_P \sim \sqrt{\frac{\hbar c}{G}} \,.
\ee

The objective of this dissertation is to explore possible windows of \abb{LIV} and \abb{DSR} below the Planck scale, i.e. without the need to reach energies of the order of $E_\text{P}\sim m_P\,c^2 \sim 10^{28} \unit{eV}$, which is orders of magnitude above any experiment performed on Earth ($\sim 10^{13}$ eV) or even the most energetic astrophysical observations ($\sim 10^{20}$ eV).

\begin{figure}[tb]
    \centering
    \includegraphics[width=0.65\textwidth]{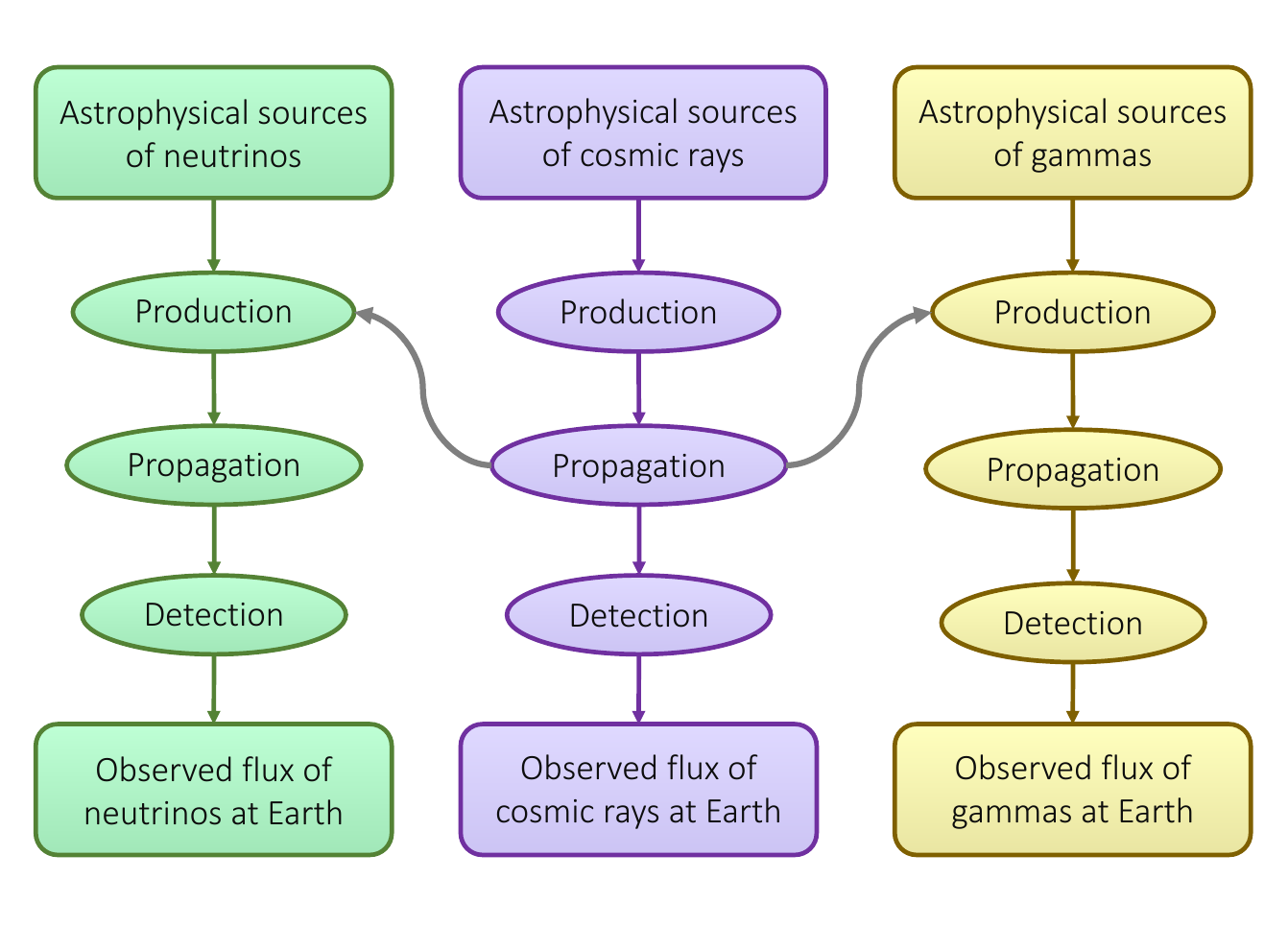}
    \caption{The three astroparticle messengers and its correlations.}
    \label{fig:astro_messengers}
\end{figure}

The effects of new physics are supposed to increase with the energy (until being completely manifest at the Planck scale). This way, the best chance to detect an effect of new physics is to use the most possible energetic observations, the very high-energy astroparticles. There exist three astroparticles messengers\footnotemark: the neutrino, the gamma rays (high-energy photons), and the cosmic rays (mostly high-energy nuclei). Neutrinos travel freely though the space; however, gamma rays and cosmic rays interact with the photon backgrounds mentioned in Sec.~\ref{sec:cosmology}. Additionally, cosmic rays can produce additional neutrinos and gamma rays during their propagation. This hierarchy is shown in Fig.~\ref{fig:astro_messengers}.
\footnotetext{Gravitational waves are an additional messenger, but in this work we will focus on the astroparticle messengers.}

In this way, when one considers neutrinos and gamma rays which are produced in the propagation of cosmic rays, the produced diffuse spectrum of secondary particles is highly correlated with the cosmic ray one. This implies that any uncertainly in the production of cosmic rays will propagate to the flux of neutrinos and gammas. In the same way, any new physics affecting the cosmic rays, will also imply changes in the secondary particle fluxes. This opens a powerful window towards multimessenger studies, but at the same time makes it difficult to study separately the phenomenological consequences of the new physics for each messenger. Therefore, in order to study each messenger separately, we will focus on the study of neutrinos and gamma rays produced at astrophysical sources. However, in Chapter~\ref{chap:cosmicrays} we will briefly study the effects of new physics in the neutrinos produced by cosmic rays, considering different astrophysical assumptions and standard physics for the cosmic rays.

On another note, we should also take into account that the effects of new physics could manifest in three completely different stages for each messenger: in their production, propagation and detection. The methods of detection are very subtle, and these techniques are out of the scope of this dissertation. However, it is usual to make the assumption that all the techniques used to detect astroparticles at Earth are low energy processes, at least in comparison with the production and propagation. This way, one does not expect to see a strong influence of the new physics in the process of detection. Instead, we do expect to see effects in both the production and propagation of astroparticles. Specially in the propagation, where the effects of new physics can be accumulative with the travelled distance. Therefore, even if the effects of new physics are small, for energetic and distant enough particles, we will be able to access a phenomenology that would be undetectable in other circumstances. Consequently, in this dissertation we will focus on the possible effects of new physics in the propagation of astroparticles. The study of effects of new physics at the level of the production is broad enough to cover another dissertation, and we hope both kind of studies could be combined in the future.

This dissertation is organized as follows. In Chapter~\ref{chap:LIV-DSR},
we will present the two ways to go beyond \abb{LI}: the Lorentz Invariance Violation and the Doubly Special Relativity. We will discuss the main characteristics of each approach, as well as their phenomenological and conceptual consequences. In the following chapter we focus on neutrinos; however, the expected modifications in the neutrino flux due to \abb{LIV} are stronger than the ones expected due to \abb{DSR}. Accordingly, in Chapter~\ref{chap:neutrinos} we will focus on the effects of \abb{LIV} in the propagation of astrophysical neutrinos. We extend the analysis to cosmogenic neutrinos in Chapter~\ref{chap:cosmicrays}, with a previous discussion about the propagation and the neutrino production of the cosmic rays. In Chapter~\ref{chap:gammas} we focus on gamma rays; however, the scenarios of gamma rays with \abb{LIV} have been extensively studied in the literature. Consequently, we will focus on the study of gamma rays in a \abb{DSR} scenario, which may produce effects both at the level of the transparency of the universe, and of the time of flight of photons.

\printbibliography[heading=subbibintoc,title={References}]
\end{refsection}

\begin{refsection}
\chapter{Lorentz Invariance Violation and Doubly Special Relativity}
\graphicspath{{Chapter2/Figs/}}
\label{chap:LIV-DSR}

One can split the study of particle physics into two parts: the study of free propagation and the study of interactions. Free propagation represents the behaviour of the particle if no other interaction is present, and it is characterized by the relativistic energy-momentum relation, also called dispersion relation, which can be derived from the free part of the Lagrangian. In contrast, the study of interactions represents how the particles appear, disappear or exchange energy and momentum with other particles.

In order to study a certain interaction one needs to take into account two main ingredients. The first one is to check whether the interaction can be performed or not, and for that, one should check if the conservation laws hold before and after the interaction (conservation of the quantum numbers, energy and momentum). In order to check the conservation of energy and momentum, the dispersion relation of the particles is again a crucial ingredient. As a result, one can obtain that the interaction is always allowed, always forbidden, or only allowed in a certain range of energies (i.e., the interaction has a threshold).

For the energies for which the interaction is allowed, the second ingredient, the rate of the interaction, starts playing a role. The rate of the interaction represents how frequently a certain interaction is expected, i.e., something proportional to the probability of the interaction. If the interaction is a decay, i.e., one particle spontaneously transforms into other particles, this rate is represented by the decay width of the interaction. Instead, if the interaction is a scattering, i.e., two particles interact to produce other particles, the rate is characterized by the cross section of the interaction. The decay widths and cross sections can be obtained from the transition amplitudes of the processes, which can be obtained using the dynamical framework provided by \abbdef{EFT}{Effective Field Theories}.

In models beyond \abb{LI}, one needs to see how the previously mentioned ingredients, the dispersion relation, cross sections, and decay widths, are modified. Let us start by studying the main characteristics of a modified energy-momentum relation or \abbdef{MDR}{Modified Dispersion Relation}. Let us remember that the usual energy-momentum relation of \abb{SR} is given by
\be
    E^2 = m^2 + p^2 \,.
    \label{eq:SR_dispersion}
\ee
If we now go beyond \abb{LI}, we expect changes in the dispersion relation that could cause that some reactions that violated the conservation of energy or momentum in some range of energies (or were forbidden for all the energies) in \abb{SR} are now allowed. One could also have the opposite effect, in which the dispersion relation disfavours the process in a certain range of energies. Or even the combination of both, a lower and an upper threshold which delimit a specific range of energies in which the process is allowed. The appearance or disappearance of these \textit{anomalous thresholds} is one of the phenomenological windows to physics beyond the \abb{SM}~\cite{Mattingly:2002ba,Addazi:2021xuf}.

Following the \textit{bottom-up} approach, one can propose new candidates for the dispersion relations of some particles and study their phenomenological consequences, even if we do not know whether they could emerge from a \abb{QG} theory or not. A generic modification of the \abb{SR} energy-momentum relation, Eq.~\eqref{eq:SR_dispersion}, for a certain particle can be written as
\be
    E^2 = m^2 + p^2 \Big[1\pm \epsilon(p,\delta)\Big] \,,
\ee
where $\epsilon(p,\delta)$ is a generic function of its momentum, $p=|\vec p|$, and $\delta$ represents some parameter which measures the magnitude of the new physics effects on that particle. If $\epsilon$ and $\delta$ are the same for every particle, the modification is called \textit{universal}. The choice of the sign that accompanies $\epsilon$ will lead to a superluminal or subluminal velocity of the particle.

As we explained in Sec.~\ref{sec:scope}, these effects of new physics should be small enough to be compatible with the current observations, that is to say, $\epsilon$ must be small, at least at the current energy scales. This allow us to expand the unknown function $\epsilon$ as a power series on the parameter $\delta$, which should also be small. In fact, it is convenient to write the parameter $\delta$ as the inverse of an energy scale $\Lambda$, which is usually associated with the Plank energy (energy at which the effects of the new physics should manifest in an evident way),
\be
    E^2 = m^2 + p^2 \qty[1\pm \sum_n a_n \qty(\frac{p}{\Lambda})^n] \,.
\ee

Usually, one restricts the analysis to the dominant term, i.e., the first non-null term. The case $n=0$ must be discarded if we want these effects to grow with the energy, until being clearly manifest at the Plank scale. Then, it is common to consider just the cases $n=1$ and $n=2$,
\be
    E^2 \approx m^2 + p^2 \qty[1\pm \qty(\frac{p}{\Lambda})^n] \,,
    \label{eq:mdr}
\ee
where in the last step we have absorbed the constant $a_n$ into $\Lambda$ by a redefinition. 

Let us note that a direct consequence of a \abb{MDR} of the kind of Eq.\eqref{eq:mdr}, is that massless particles with different energies will propagate at different velocities,
\be
    v\coloneqq\frac{dE}{dp}\approx 1+\frac{(n+1)}{2} \qty(\frac{p}{\Lambda})^n \,.
\ee
This phenomenon can produce anomalies in the times of flight of photons, also called ``time delays''~\cite{Jacob:2008bw,Pan:2020zbl,Martinez-Huerta:2020cut}, or ``lateshifts''~\cite{Amelino-Camelia:2013uya,Pfeifer:2018pty,Barcaroli:2016yrl,Barcaroli:2017gvg} in the \abb{DSR} frameworks. The study of times of flight anomalies open another phenomenological window to physics beyond the \abb{SM}.

In the following sections we will study how this \abb{MDR} could emerge in each model, \abb{LIV} or \abb{DSR}. In the same way, we will discuss in each case the availability of a dynamical framework, necessary to compute the decay widths or cross sections of the interactions.

\section{Lorentz Invariance Violation}
\label{sec:LIV}

The violation of \abb{LI} is the loss of the symmetry under Lorentz transformations (rotations and boosts). One important conceptual consequence it is the loss of the relativity principle, i.e., now the laws of physics are different for different observers. Then, if one writes a \abb{MDR} for a particle, like Eq.~\eqref{eq:mdr}, it is fair to ask which observer \textit{sees} that energy-momentum relation.
%This force us to chose some observer and all agree on describe the laws of physics in that observer.
This  privileged observer is usually taken as the one which sees the \abb{CMB} isotropic; however, the fact that we measure the \abb{CMB} to a large extent as isotropic shows us that we can use the Earth as a good approximation to that observer.

In the framework of the \abb{EFT}, one can build a model of \abb{LIV} adding terms which do not preserve \abb{LI} to the Lagrangian corresponding to a Lorentz-invariant model (like the \abb{SM}). The added terms, which violate \abb{LI}, can also (but not necessarily) break other properties of the \abb{SM} like the \abbdef{CPT}{Charge, Parity and Time reversal} symmetry. The model which considers all the possible terms that violate \abb{LI}, preserving gauge invariance and energy-momentum conservation, but still considering both \abb{CPT}-violating and \abb{CPT}-preserving terms, is called the \abbdef{SME}{Standard Model Extension}~\cite{Colladay:1998fq}. If one restricts the study to renormalizable terms, i.e., with operators with mass dimension $[d]\leq 4$\footnotemark, the model is called \abbdef{mSME}{minimal Standard Model Extension}. The study of \abb{LIV} within the \abb{SME} allows us to use the framework provided by the \abb{EFT} to calculate transition amplitudes, decay widths, and cross sections.
\footnotetext{$[d]$ is defined as the mass dimension of the \abb{LIV} operator, which must be multiplied by a coefficient of dimension $4-[d]$ in order to ensure that the action is adimensional.}

Additionally, \abb{LIV} is closely related with the violation of the \abb{CPT} symmetry. In fact, in~\cite{Greenberg:2002uu} it was shown that a \abb{CPT} violation also implies \abb{LIV} when one considers models with a local\footnotemark \abb{QFT} description. Let us notice that the reciprocal is not true, one can consider \abb{LIV} without violating \abb{CPT} invariance. In fact, one can check that the \abb{CPT}-violating and \abb{CPT}-preserving models of \abb{LIV} correspond to considering an odd or even order of correction $n$ (respectively) in the \abb{MDR} (Eq.~\eqref{eq:mdr}). As a consequence, particles and antiparticles can have different behaviours beyond \abb{LI}, and the \abb{CPT} symmetry breaking could be a key ingredient to explain the apparent matter-antimatter asymmetry of the universe~\cite{Bertolami:1996cq}.
\footnotetext{This does not hold in nonlocal models; for instance, those that consider noncommutative spacetime such as some \abb{DSR} realizations~\cite{Chaichian:2011fc}.}

The effects of \abb{LIV} on photons have been extensively studied in the literature (see, for example, \cite{Terzic:2021rlx,Martinez-Huerta:2020cut,Bolmont:2022yad,PerezdelosHeros:2022izj} and the references therein). On the one hand, one can find differences between the times of flight of high and low energy photons due to the dependency of the velocity on the energy (see~\cite{MAGIC:2020egb,Pan:2020zbl,Levy:2021oec} for some examples). On the other hand, one can also have phenomenological consequences in the photon-photon interaction, which produces a change in the transparency of the universe to gamma rays. For the superluminal case one finds a premature decrease of the flux with respect to the case of \abb{SR}, and for the subluminal case one finds instead a recovery of the flux (see~\cite{Fairbairn:2014kda,Lang:2018yog,Tavecchio:2015rfa,Abdalla:2018sxi} for more information).

On another note, the studies of the effects of \abb{LIV} on neutrinos have mainly focused in the modification of neutrino oscillations. It has been shown that a \abb{MDR} like Eq.\eqref{eq:mdr} produces an additional energy dependence in the oscillations which is proportional to the energy, $L\times E$, instead of the usual inverse dependence, $L/E$ (where $E$ is the energy of the neutrino and $L$ the travelled distance). However, the magnitude of this additional dependence is heavily constrained by the current data. In this work we will assume no flavour dependence in the \abb{LIV} terms, avoiding the stringent constraints from neutrino oscillations~\cite{Mattingly:2005re,Torri:2020dec,Arias:2006vgq}.

In contrast to the case of gamma rays, the effects of \abb{LIV} on the neutrino spectrum have been less studied. Accordingly, in this dissertation we will focus the study of \abb{LIV} on the phenomenological consequences in the neutrino flux. We will consider a non-universal \abb{LIV} model, so that we can consider effects of new physics only in the free part of the Lagrangian involving the neutrino fields. This assumption does not only allow us to simplify the study, but it is also crucial to escape the very restrictive bounds on the scale $\Lambda$ that already exist for charged leptons~\cite{Jacobson:2002ye,Maccione:2007yc,Maccione:2011fr,Stecker:2013jfa,Rubtsov:2016bea,Altschul:2021wrm}. Let us note, however, that the neutrino is part of an $\mathrm{SU}(2)_\mathrm{L}$ doublet along with the corresponding lepton, $f_{lL}=(\nu_{lL} \; l_L)^T$ (where the subindex $L$ refers to the left-handed chirality of the fields and $l$ refers to the three lepton flavours). Then, considering a scenario of \abb{LIV} that only affect the neutrinos, but not the leptons, would break the $\mathrm{SU}(2)_\mathrm{L}$ gauge invariance, which is outside of the scope of the \abb{SME}.

A way to avoid this is to consider that the added \abb{LIV} terms only involve the fermion doublet through its product with a complex scalar field doublet, related with the Higgs doublet $\Phi=(\Phi^+\; \Phi^0)^T$. This complex scalar field doublet can be defined as
\be 
    \tilde\Phi \coloneqq i \sigma_2 \Phi^* = \begin{pmatrix} \Phi^{0*} \\ - \Phi^- \end{pmatrix} \,,
\ee
with $\sigma_2$ the second Pauli matrix, and where we should notice that the neutral and charged components have exchanged places. This way, one can construct the following gauge invariant products,
\be
     \bar f_{lL} \tilde\Phi = \begin{pmatrix}\bar\nu_{lL} & \bar l_{L}\end{pmatrix} \cdot  \begin{pmatrix}\Phi^{0*} \\ -\Phi^{-}\end{pmatrix} \,, \eqtext{and} \tilde\Phi^\dag f_{lL} = \begin{pmatrix}\Phi^0 & \Phi^{+}\end{pmatrix} \cdot \begin{pmatrix}\nu_{lL} \\ l_{L}\end{pmatrix}
    \,.
    \label{eq:product_fields}
\ee
where $\bar\phi\coloneqq \phi^\dag \gamma^0$ is the Dirac adjoint. The use of these products ensures the $\mathrm{SU}(2)_\mathrm{L}\times \mathrm{U}(1)_\mathrm{Y}$ gauge invariance of the model, but after the Higgs field acquire its \abbdef{VEV}{Vacuum Expectation Value}, the charged component will go to zero and the neutral one will produce a quadratic factor in the neutrino field,
\be 
    \avg{\tilde\Phi}\approx \begin{pmatrix} v/\sqrt{2} \\ 0\end{pmatrix} \srarrow (\bar f_{lL} \tilde\Phi)\; (\tilde\Phi^\dag f_{lL}) \approx \;\frac{v^2}{2}\, \bar\nu_{lL} \,\nu_{lL} \,,
\ee
which will only affect the free part of the neutrino Lagrangian. As we will see in Sec.~\ref{sec:LIV_neu} of Chapter~\ref{chap:neutrinos}, this implies going beyond the \abb{mSME}, considering an additional term with mass dimension $[d]=6+n$, with $n$ the order of correction that appears in the \abb{MDR} (Eq.~\eqref{eq:mdr}). This term must be accompanied by a coefficient with negative mass dimension $-(2+n)$, so we can call it $\qty(1/\tilde\Lambda)^{2+n}$; however, we can absorb the coefficient $\tilde\Lambda$ and the square of the \abb{VEV} of the Higgs field in the same constant, such that
\be 
    \frac{v^2}{2} \qty(\frac{1}{\tilde\Lambda})^{2+n} \eqqcolon \qty(\frac{1}{\Lambda})^n \,,
\ee
where $\Lambda$ is the scale of new physics that appears in Eq.~\eqref{eq:mdr}.

\section{Doubly Special Relativity}
\label{sec:DSR}

In \abb{DSR}, in contrast to \abb{LIV}, one still has a relativity principle and \abb{LI}, but under different (\textit{deformed}) Lorentz transformations. These deformed Lorentz transformations are part of a group of deformed Poincaré transformations, which are deformed in such a way that they preserve a certain scale of energy $\Lambda$, or its inverse, a certain scale of length $\ell\sim(1/\Lambda)$ (historically associated with a \textit{minimum length}~\cite{Amelino-Camelia:2000cpa,Amelino-Camelia:2000stu}). In the context of \abb{DSR}, the parameter $\Lambda$ has been historically named as $\kappa$~\cite{Kowalski-Glikman:2004fsz}, and, as a consequence, the deformed Lorentz transformations are known as $\kappa$-Lorentz transformations and the deformed Poincaré transformations as $\kappa$-Poincaré transformations. In this work we will respect those historical names, but we will still name the scale of new physics as $\Lambda$.

The Poincaré transformations are composed by the Lorentz transformations (rotations and boosts) and the space and time translations. If we consider isotropic \abb{DSR} models (i.e., without any deformation at the level of rotations), the $\kappa$-Poincaré transformations are characterized by the infinitesimal generators of the boosts, $N$, and space and time translations, $P$ and $E$ respectively\footnotemark. The generators of the $\kappa$-Poincaré transformations fulfil a deformed Poincaré algebra, which is not a Lie algebra anymore but a Hopf algebra (see~\cite{Arzano:2022ewc,Agostini:2003vg} for more information), with non-trivial Poisson brackets $\{N,E\}$ and $\{N,P\}$. The different models of \abb{DSR} correspond to different choices of this algebra and are called the different \textit{basis} of $\kappa$-Poincaré. Each basis represents a different \abb{DSR} model with different physical and phenomenological implications.
\footnotetext{Note the change in the notation in this section. Now $P$ and $E$ represents generators of infinitesimal space and time translations.}

In \abb{SR}, one identifies the energy and momentum variables of the one-particle phase space with the generators of the time and space translations. In \abb{DSR}, one can follow a similar approach and relate the deformed generators $E$ and $P$ with the energy and momentum variables, or a function of them. This association has been a longstanding source of confusion in the literature, because a change in the choice of energy-momentum variables, which should not affect the physics, can be mistaken for a change of the deformed generators, i.e., a change of the Poincaré basis with physical implications. We will study this issue in more depth in Sec.~\ref{sec:time_delays}, while we discuss the existence of time delays in \abb{DSR}.

The freedom in the choice of energy-momentum variables, allows us to relate different, but equivalent, descriptions of \abb{DSR}. If we identify the generators $E$ and $P$ with the energy and momentum variables of the one-particle phase space, we can describe the different \abb{DSR} models using their corresponding non-trivial composition law of momenta.

In the transition from Galilean relativity to \abb{SR}, the uniform-motion transformations (relating observers with relative constant velocity) were substituted by the Lorentz boosts, which are just a generalization such that they preserve the invariance of the speed of light $c$. This invariance also requires to replace the usual sum of velocities by a non-trivial composition of velocities,
\be
    \vec v + \vec u  \srarrow \vec v \oplus \vec u\,
\ee
where the composition of velocities can be written as
\be 
    v \oplus \vec u \equiv \frac{1}{1+ \frac{\vec u \cdot\vec v}{c^2}} \qty(\vec v+ \frac{\vec u}{\gamma_v} + \frac{1}{c^2}\frac{\gamma_v}{1+\gamma_v} (\vec v \cdot \vec u)\cdot \vec v)\,, \eqtext{with} \gamma_v\coloneqq\sqrt{1-\frac{v^2}{c^2}}\,.
\ee
Similarly, the transition from \abb{SR} to \abb{DSR} can be interpreted as a substitution of the Lorentz boosts by deformed boost transformations, such that they preserve the invariance of a certain scale of energy $\Lambda$. The invariance of the scale of new physics $\Lambda$ also requires to replace the sum of momenta by a non-trivial composition,
\be
    p + q \srarrow p \oplus q\,,
    \label{eq:g_comp_law}
\ee
where the specific form of the composition law is what characterizes the \abb{DSR} model in study.

Let us note that Eq.~\eqref{eq:g_comp_law} cannot be commutative in $p$ and $q$, otherwise a change of energy-momentum variables can reduce the composition of momenta to a sum, which means that we are still in \abb{SR}. This produces two possible definitions of the ``total momentum'' for a system of two particles, $(p\oplus q)$ or $(q\oplus p)$. In fact, Eq.~\eqref{eq:g_comp_law} may not even be associative (like in the Snyder-deformed \abb{DSR}~\cite{Battisti:2010sr}), introducing even more uncertainty in the definition of a total momentum. The most studied \abb{DSR} models correspond to noncommutative, but still associative, composition laws. This allows one to completely fix the multiparticle sector once the two-particle sector is known. This topic is still in an early phase of study.

Imposing a relativity principle over the composition of momenta, one can obtain the corresponding deformed energy-momentum relation. Since the composition of momenta is a universal rule, the obtained \abb{MDR} is also universal, i.e., the same for all particles. However, a problem arises when one tries to apply the generic modified dispersion relation, Eq.~\eqref{eq:mdr}, to macroscopic objects. We find that the correction to the energy-momentum relation of \abb{SR} would be of the order of $(p_\text{macro}/\Lambda)$. In order to understand how large this contribution is let us take $\Lambda$ around the Planck scale. The momentum of the macroscopic object is around the order of its mass, let us say around the kilogram; however, let us remember that the Planck mass in these units is around $10^{-8}$ kg. This contradicts the necessary condition that the \abb{QG} effects should be very small at the current studied energies. One finds the same kind of paradox when considering the composition law applied to two macroscopic objects, or even when one tries to compute the momentum of a macroscopic object, considering that it is composed by a very large number (around the Avogadro number) of constituents, i.e., $p_\text{macro}=p_1\oplus \cdots \oplus p_N$. Possible solutions to this problem, known as the \textit{soccer ball problem}~\cite{Hossenfelder:2007fy,Amelino-Camelia:2011dwc,Kowalski-Glikman:2022xtr}, have been discussed in the literature~\cite{Amelino-Camelia:2011dwc,Amelino-Camelia:2014gga}; however, how this problem is solved in DSR is still an open question.

\begin{figure}[tb]
    \centering
    \includegraphics[width=0.25\textwidth]{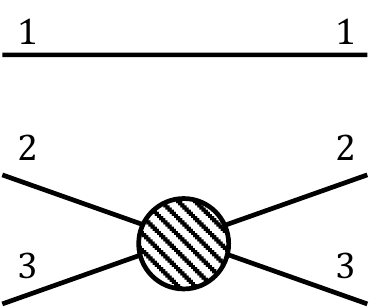}
    \caption{Interaction of two particles and one spectator particle.}
    \label{fig:interaction}
\end{figure}

Another conceptual problem arises from the ambiguity in the definition of the total momentum that we mentioned before. Let us consider two interacting particles and one additional particle just passing by, as ``spectator''. This set up is shown in Fig.~\ref{fig:interaction}, where the spectator particle is numbered as 1, and the interacting particles as 2 and 3.
Let us call the momentum of the particles 2 and 3 before and after the interaction as $p_2$ and $p_3$, and $p_2'$ and $p_3'$, respectively. One can define the ``total momentum'' of the system (of the three particles) before and after the interaction and impose the conservation of energy and momentum. There exists an ambiguity in the ordering of the composition of the three momenta, so one can find situations like, 
\be 
    p_1 \oplus p_2 \oplus p_3 = p_2' \oplus p_1 \oplus p_3' \,,
    \label{eq:spec}
\ee
where one cannot disentangle the influence of the particle 1 in the conservation law of the interaction. In other words, the properties of the interaction can depend on particles that do not participate in the process, i.e., on the ``spectator'' particles. This problem is known as the \textit{spectator problem}~\cite{Carmona:2011wc,Amelino-Camelia:2011gae,Gubitosi:2019ymi}, and if it is applied recursively leads to a lost of the cluster property, i.e., one would need to consider \textit{the whole universe} in order to study a single interaction.

The soccer ball and spectator problems are still open questions that need to be clarified in order to consider \abb{DSR} as a consistent theory. In Appendix~\ref{sec:new_perspective} we propose a new perspective of \abb{DSR} that can avoid these two issues, and we also discuss its phenomenological implications.

\begin{figure}[tb]
    \centering
    \includegraphics[width=0.6\textwidth]{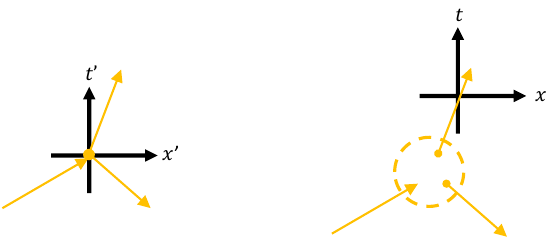}
    \caption{Simple diagram of the relative locality on \abb{DSR} scenarios.}
    \label{fig:relative_locality}
\end{figure}

Another different phenomenon arises when one consider the effects of \abb{DSR} at the level of interactions. In \abb{SR} one would say that a group of particles interact if their worldlines meet at some point, being that point the location where the interaction takes place. Let us now consider a \abb{DSR} scenario and an interaction taking place at the origin of the space-time coordinates. If one now performs a spatial translation generated by the total momentum of the system\footnotemark, one can check how the worldlines of the particles participating in the interaction do not meet anymore (see Appendix~\ref{sec:relative_locality} for a derivation from an action formulation). Only the observer for which the interaction takes place at its origin of coordinates sees it as local; any other observer does not see the worldlines crossing at the same point anymore (Fig.~\ref{fig:relative_locality}). This phenomenon is known as \textit{relative locality} of the interactions~\cite{Amelino-Camelia:2011lvm,AmelinoCamelia:2011pe,AmelinoCamelia:2011bm}.
\footnotetext{Let us note that, as we discussed before, there exists an ambiguity in the definition of the total momentum. This way, the discussion has to be done separately for each choice of the total momentum.}

In order to avoid this, some authors in the literature~\cite{Carmona:2017oit,Carmona:2019oph,Amelino-Camelia:2011ebd,Mignemi:2016ilu} have proposed that the ``physical spacetime'', where the trajectories of the particles should be described, is not defined by the canonical space-time coordinates, but by a new set of noncommutative space-time coordinates, which can be associated with the generators of the translations in momentum space. However, along this work we will consider the canonical spacetime as the physical spacetime, and we will treat relative locality as an inescapable feature of \abb{DSR}, which should be taken into account in the computations of time delays.

Let us note that, even if certain \abb{DSR} and \abb{LIV} models can consider a similar \abb{MDR}, the conceptual and phenomenological implications will be very different. Let us use the neutrino as an example. A \abb{MDR} like Eq.~\eqref{eq:mdr} will cause that some decays that were forbidden in \abb{SR}, like the electron-positron pair emission, now do not violate the conservation rules in a certain range of energies. In the framework of \abb{LIV} this will allow the decay if the energy of the neutrino is above some threshold. In contrast, in \abb{DSR}, even if the decay does not violate the conservation rules in a certain energy range, there cannot exist a threshold for the reaction due to the existence of a relativity principle. This way, the effects of \abb{DSR} in neutrinos are expected to be softer that in the case of \abb{LIV}. Accordingly, we will focus on \abb{LIV} for the case of the neutrino. In contrast, we will focus on the study of \abb{DSR} for the case of gamma rays since, even though a photon cannot either disintegrate in \abb{DSR}, the interactions of high-energy photons during their cosmic propagation can be affected.

\section{The collinear approximation}

In \abb{SR}, the \abb{LI} simplifies the technical computation of decay widths and cross sections. One can work in the center of mass reference frame and then perform a boost to get the result in the laboratory frame. Unfortunately, this is not valid anymore in a \abb{LIV} scenario due to the lost of \abb{LI}. In fact, even in \abb{DSR} scenarios, where one still has invariance under $\kappa$-Lorentz transformations, the deformed boosts are so complicated that the use of the center of mass does not simplify the computation anymore. We see then that when one goes beyond \abb{LI}, the computations must be done directly in the laboratory reference frame, which in general increases the complexity of the calculations. 

However, along this work we will study the effects of new physics in very high-energy astroparticles, for which we can use a different property to simplify the computations: the collinear approximation~\cite{Carmona:2022dtp}. The collinear approximation takes the advantage of the fact that when a very high-energy particle interacts, the produced particles follow almost the same direction as the primary particle. In the following we show an example of the use of the collinear approximation to compute the decay width of a neutrino three-body decay, in the case of a superluminal \abb{LIV} scenario. In Chapter~\ref{chap:neutrinos} we will see how the assumed decay amplitude can be particularized for the cases of our interest.

\subsection{LIV neutrino three-body decay}
\label{sec:collinear}

\subsubsection{Decay Amplitude}

We will assume that the (spin averaged) neutrino decay amplitude has the following form:
\be 
    \abs{A_\nu}^2= N (p_0\cdot p_1)(p_2\cdot p_3) \,,
    \label{eq:ansatz}
\ee
where $p_0$ and $p_i$ are the four-momentum of the primary neutrino and  the product particles, respectively.
Let us use it as an \textit{ansatz} for now; however, we will prove it for the different processes of interest in Chapter~\ref{chap:neutrinos}.

Expanding the product of four-momenta, we get that
\be 
    \abs{A_\nu}^2= N \abs{\vec p_0}\abs{\vec p_1}\abs{\vec p_2}\abs{\vec p_3} (1-\hat p_0\hat p_1)(1-\hat p_2\hat p_3) \,,
\ee
where $\hat p_i\hat p_j=\cos\theta_{ij}$. Under the collinear approximation, we expect all the angles between the particles to be very small. In fact, as we will see later, $(1-\hat p_i\hat p_j) \sim \theta_{ij}^2 \sim O\qty(1/\Lambda^n)$. Hence, the factor $(1-\hat p_0\hat p_1) (1-\hat p_2\hat p_2)$ is of order $O\qty((1/\Lambda^n)^2)$, and every other factor in the decay amplitude can be approximated at dominant order. Therefore, we can write $\abs{\vec p_i}\approx E_i$. So we get
\be 
    \abs{A_\nu}^2\approx N E_0 E_1 E_2 E_3 (1-\hat p_0\hat p_1)(1-\hat p_2\hat p_3) \,.
\ee

Now, let us note that, due to the conservation of momentum, the quantities $(1-\hat p_0\hat p_1)$ and $(1-\hat p_2\hat p_3)$ are related. Firstly, one can see that 
\be 
    \vec p_0=\vec p_1+\vec p_2+\vec p_3 \srarrow (\vec p_0-\vec p_1)^2=(\vec p_2+\vec p_3)^2 \,.
    \label{eq:col18}
\ee
But these two squared quantities can also be related with $(1-\hat p_0\hat p_1)$ and $(1-\hat p_2\hat p_3)$,
\be 
    (\vec p_0-\vec p_1)^2= (\abs{\vec p_0}-\abs{\vec p_1})^2 + 2\abs{\vec p_0}\abs{\vec p_1}(1-\hat p_0 \hat p_1) \,,\\
    (\vec p_2+\vec p_3)^2= (\abs{\vec p_2}+\abs{\vec p_3})^2 - 2\abs{\vec p_2}\abs{\vec p_3}(1-\hat p_2 \hat p_3) \,.
    \label{eq:col19}
\ee
Then, from the two previous equations we find that
\be 
    (1-\hat p_2 \hat p_3)=\frac{(\abs{\vec p_2}+\abs{\vec p_3})^2-(\abs{\vec p_0}-\abs{\vec p_1})^2}{2\abs{\vec p_2}\abs{\vec p_3}}-\frac{\abs{\vec p_0}\abs{\vec p_1}}{\abs{\vec p_2}\abs{\vec p_3}} (1-\hat p_0 \hat p_1) \,.
    \label{eq:col20}
\ee

We can simplify the numerator of Eq.~\eqref{eq:col20} using that $A^2-B^2=(A+B)(A-B)$:
\be
    \big(\abs{\vec p_2}+\abs{\vec p_3}\big)^2-\big(\abs{\vec p_0}-\abs{\vec p_1}\big)^2= \Big[(\abs{\vec p_2}+\abs{\vec p_3})+(\abs{\vec p_0}-\abs{\vec p_1})\Big]\;\Big[(\abs{\vec p_2}+\abs{\vec p_3})-(\abs{\vec p_0}-\abs{\vec p_1})\Big] \,.
\ee
Now, from Eq.~\eqref{eq:col18}, we get that $\abs{\vec p_0-\vec p_1}=\abs{\vec p_2+\vec p_3}$. As we expect all the momenta to be almost parallel under the collinear approximation, we can write
\be 
    \abs{\vec p_0-\vec p_1}=\abs{\vec p_2+\vec p_3} \srarrow 
    \abs{\vec p_0}-\abs{\vec p_1}\approx\abs{\vec p_2}+\abs{\vec p_3} \,,
\ee
so
\bea
    \Big[(\abs{\vec p_2}+\abs{\vec p_3})+(\abs{\vec p_0}-\abs{\vec p_1})\Big] &\approx 2 (\abs{\vec p_0}-\abs{\vec p_1}) \,, \\
    \Big[(\abs{\vec p_2}+\abs{\vec p_3})-(\abs{\vec p_0}-\abs{\vec p_1})\Big] &\sim O\qty(1/\Lambda^n) \ll 1 \,,
\eea
and then we get
\be 
    (1-\hat p_2 \hat p_3)=\frac{(\abs{\vec p_0}-\abs{\vec p_1})}{\abs{\vec p_2}\abs{\vec p_3}}\qty[(\abs{\vec p_2}+\abs{\vec p_3})-(\abs{\vec p_0}-\abs{\vec p_1})]-\frac{\abs{\vec p_0}\abs{\vec p_1}}{\abs{\vec p_2}\abs{\vec p_3}} (1-\hat p_0 \hat p_1) \,.
\ee

Recalling that both $(1-\hat p_0 \hat p_1)$ and $\qty[(\abs{\vec p_2}+\abs{\vec p_3})-(\abs{\vec p_0}-\abs{\vec p_1})]$ are of order $O(1/\Lambda^n)$, we can expand the other multiplicative factors until dominant order, i.e., we can write $\abs{\vec p_i}\approx E_i$, so
\be 
    (1-\hat p_2 \hat p_3)=\frac{(E_0-E_1)}{E_2 E_3}\qty[(\abs{\vec p_2}+\abs{\vec p_3})-(\abs{\vec p_0}-\abs{\vec p_1})]-\frac{E_0 E_1}{E_2 E_3} (1-\hat p_0 \hat p_1) \,.
\ee

In order to continue, we will use a generic modified dispersion relation of the kind
\be
    |\vec p_i| = E_i \qty[1+\alpha_i \qty(\frac{E_i}{\Lambda})^n]\,,
\ee
with $\alpha_i=-1$ or $(-1)^{n+1}$ if it is a neutrino or antineutrino, respectively, and zero for every other particle (see Sec.~\ref{sec:LIV_neu} for more information). Introducing the energy fractions $x_i=E_i/E_0$, one can write 
\be 
    \Big[(\abs{\vec p_2}+\abs{\vec p_3})-(\abs{\vec p_0}-\abs{\vec p_1})\Big]=E_0 \qty(\frac{E_0}{\Lambda})^n \qty[1+\sum_{i=1}^{3}\alpha_i x_i^{n+1}] \,.
\ee

Then, we get that $(1-\hat p_2 \hat p_3)$ and $(1-\hat p_0 \hat p_1)$ are related by
\be 
    (1-\hat p_2 \hat p_3)=\frac{(1-x_1)}{x_2x_3} \qty(\frac{E_0}{\Lambda})^n \qty[1+\sum_{i=1}^{3}\alpha_i x_i^{n+1}]-\frac{x_1}{x_2x_3} (1-\hat p_0 \hat p_1) \,.
    \label{eq:col28}
\ee

And then, the squared decay amplitude can be written as
\be 
    \abs{A_\nu}^2\approx N E_0^4\; x_1 x_2 x_3 \qty(\frac{(1-x_1)}{x_2x_3} \qty(\frac{E_0}{\Lambda})^n \qty[1+\sum_{i=1}^{3}\alpha_i x_i^{n+1}](1-\hat p_0 \hat p_1)-\frac{x_1}{x_2x_3} (1-\hat p_0 \hat p_1)^2) \,.
    \label{eq:col29}
\ee

\subsubsection{Phase space}

Generically, a differential part of the decay width is given by
\be
    d\Gamma=\frac{1}{2E_0} \abs{A_\nu}^2 d\Phi_\text{PS} \,,
\ee
with $E_0$ the energy of the primary particle, $\abs{A_\nu}^2$ the squared decay amplitude, and $d\Phi_\text{PS}$ a differential section of the phase space.

We have seen that $\abs{A_\nu}^2$ is of order $O((1/\Lambda^n)^2)$. Hence, it is enough to compute $d\Phi_\text{PS}$ up to dominant order. Then, the differential of the phase space is given by 
\bea 
    d\Phi_\text{PS} &=(2\pi)^4 \delta\qty(E_0-\sum_{i=1}^{3}E_i) \delta^3\qty(\vec p_0-\sum_{i=1}^{3}\vec p_i) \qty(\prod_{i=1}^{3} \frac{d^3\vec p_i}{(2\pi)^3 2E_i}) \\ 
    &= \frac{1}{8} \frac{1}{(2\pi)^5} \delta\qty(E_0-\sum_{i=1}^{3}E_i) \delta^3\qty(\vec p_0-\sum_{i=1}^{3}\vec p_i) \frac{d^3\vec p_1 d^3\vec p_2 d^3\vec p_3}{E_1 E_2 E_3}\,.
\eea
The triple Dirac delta can be used to remove 3 variables of integration, for example, the tree components of $d^3\vec p_3$, so from now on, $\vec p_3$ will be only a notation for $\vec p_3=\vec p_0-\vec p_1-\vec p_2$.

\begin{figure}[tb]
    \centering
    \includegraphics[width=0.6\textwidth]{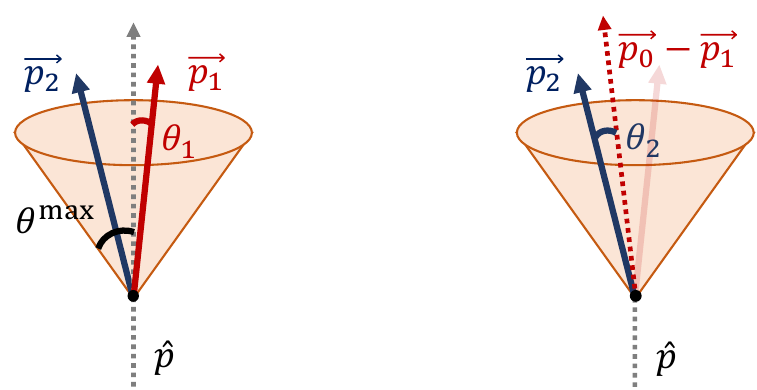}
    \caption{Cones of the possible directions of $\vec{p}_1$ (left) and $\vec{p}_2$ (right).}
    \label{fig:disin_cone}
\end{figure}

When performing the integration, we should integrate the other two vectors, $\vec p_1$ and $\vec p_2$, within the disintegration cone of aperture $\theta^\text{max}$ (see Fig.~\ref{fig:disin_cone}).
In order to do that, let us make a change of variables to spherical coordinates, but a different change of variables for each particle. For the first particle, let us choose the direction of the initial particle as the $z$ axis. Then, the change of variables for the first particle is given by
\be
    d^3\vec p_1 = d\abs{\vec p_1} d\theta_1 d\phi_1 \abs{\vec p_1}^2 \sin\theta_1 \approx dE_1 d\theta_1 d\phi_1 E_1^2 \sin\theta_1 \,,
\ee
where in the last step we have used that at dominant order $\abs{\vec p_i}\approx E_i$. Let us note that, due to the symmetry of the system with respect to the direction of the parent particle (chosen as the $z$ axis), $\phi_1\in [0,2\pi)$ and $\theta_1\in [0,\theta^\text{max}]$.

For the second particle we will use the direction of the vector $(\vec p_0-\vec p_1)$ as the $z$ axis, for each direction of the first particle. This implies that we should perform firstly the integral of the second particle, as a function of the direction of the first one, and later over the first one. The reason of this choice will become clear afterwards. The change of variables for the second particle is given by
\be
    d^3\vec p_2 = d\abs{\vec p_2} d\theta_2 d\phi_2 \abs{\vec p_2}^2 \sin\theta_2 = dE_2 d\theta_2 d\phi_2 E_2^2 \sin\theta_2 \,,
\ee
where $\theta_2$ and $\phi_2$ are the polar and azimuthal angles respect to the the $(\vec p_0-\vec p_1)$ axis. With this selection of $z$ axis, we do not have such a symmetry for the limits of integration of $\theta_2$ and $\phi_2$, so they cannot be integrated independently. Instead we can cover all the cone fixing that $\phi_2\in [0,2\pi)$ and letting $\theta_2\in [\theta_2^\text{bot}(\phi_2),\theta_2^\text{up}(\phi_2)]$, where $\theta_2^\text{bot}(\phi_2)$ and $\theta_2^\text{up}(\phi_2)$ are some complicated functions of $\phi_2$.

Making an additional change of variables from $E_i$ to the energy fractions $x_i$\footnotemark, we can write the differential of phase space as 
\footnotetext{In order to make the change of variable in the energy Dirac delta one have to use the identity \mbox{$\delta(f(x))=\delta(x-x_0)/\abs{f'(x_0)}$}, with $x_0$ the root of $f(x)$.}
\be
    d\Phi_\text{PS} = \frac{1}{8} \frac{E_0^2}{(2\pi)^5} \delta\qty(1-\sum x_i) \frac{x_1^2 x_2^2}{x_1 x_2 x_3} \sin\theta_1 \sin\theta_2 dx_1 dx_2 d\theta_1 d\theta_2 d\phi_1 d\phi_2 \,.
    \label{eq:col34}
\ee

\subsubsection{Decay width}

Gathering Eqs.~\eqref{eq:col29} and \eqref{eq:col34} together, we find that the differential of the decay width can be written as 
\bea
    d\Gamma =&\frac{1}{2E_0} \abs{A_\nu}^2 d\Phi_\text{PS} \\
    =& \frac{N}{16} \frac{E_0^5}{(2\pi)^5} \delta\qty(1-\sum x_i) \sin\theta_1 \sin\theta_2 x_1^2 x_2^2 dx_1 dx_2 d\theta_1 d\theta_2 d\phi_1 d\phi_2 \\ &\times \qty(\frac{(1-x_1)}{x_2x_3} \qty(\frac{E_0}{\Lambda})^n \qty[1+\sum_{i=1}^{3}\alpha_i x_i^{n+1}](1-\hat p_0 \hat p_1)-\frac{x_1}{x_2x_3} (1-\hat p_0 \hat p_1)^2) \,,
\eea
where we should remember to integrate the second particle, prior to the integration of the first one. Let us also define the variables $\omega_i=(1-\cos\theta_i)$, so $d\omega_i=\sin\theta_i d\theta_i$. Let us also notice that $\omega_1=(1-\cos\theta_1)=(1-\hat p_0 \hat p_1)$, and so
\bea
    \Gamma =& \frac{N}{16} \frac{E_0^5}{(2\pi)^5} \idotsint \delta\qty(1-\sum x_i) dx_1 dx_2 d\omega_1 d\omega_2 d\phi_1 d\phi_2 \\ & \times x_1^2 x_2^2 \qty(\frac{(1-x_1)}{x_2x_3} \qty(\frac{E_0}{\Lambda})^n \qty[1+\sum_{i=1}^{3}\alpha_i x_i^{n+1}]\omega_1-\frac{x_1}{x_2x_3} \omega_1^2) \,.
\eea

Let us perform the integration over the second particle. The integral over $\phi_2$ will give a factor $2\pi$. On the other hand, the integral over $\omega_2$ is complicated due to its limits of integration. Fortunately, it can be done using the energy Dirac delta in a non-obvious way. First, let us note that for a fixed $\vec p_1$, we can relate $\omega_2$ with the energy fraction of the third particle $x_3$. Using the momentum conservation,
\be 
    \abs{\vec p_3}^2 = \abs{(\vec p_0 -\vec p_1) - \vec p_2}^2 = \abs{\vec p_0 - \vec p_1}^2 + \abs{\vec p_2}^2 - 2 \abs{\vec p_0 - \vec p_1} \abs{\vec p_2} \cos\theta_2 \,,
\ee
where we should remember that $\theta_2$ is the angle between the vector $\vec p_2$ and the axis $(\vec p_0 - \vec p_1)$. Then, differentiating the previous expression one gets
\be 
    2\abs{\vec p_3} d\abs{\vec p_3}= 2 \abs{\vec p_0 - \vec p_1} \abs{\vec p_2} \sin\theta_2 d\theta_2 \,,
\ee
and then 
\be 
    d\omega_2=\frac{\abs{\vec p_3}}{\abs{\vec p_0 - \vec p_1} \abs{\vec p_2}} d\abs{\vec p_3} \approx \frac{x_3}{(1-x_1)x_2} dx_3 \,,
\ee
where in the last step we have used that $\abs{\vec p_i}\approx E_i$ at dominant order. In this way we can relate $\omega_2$ to $x_3$, and, using the energy Dirac delta, the integral over $x_3$ is simply equivalent to take into account that $x_3=(1-x_1-x_2)$. From now on $x_3$ is just a notation for $(1-x_1-x_2)$. Then,
\bea
    \Gamma =& \frac{N}{16} \frac{E_0^5}{(2\pi)^4} \idotsint \delta\qty(1-\sum x_i) dx_1 dx_2 dx_3 d\omega_1 d\phi_1 \\ &\times \frac{x_1^2}{(1-x_1)} \qty((1-x_1) \qty(\frac{E_0}{\Lambda})^n \qty[1+\sum_{i=1}^{3}\alpha_i x_i^{n+1}]\omega_1-x_1 \omega_1^2) \,.
\eea

Finally, we should perform the integration over the first particle. Again, the azimuthal angle will give a factor $2\pi$. On the other hand, the values of $\omega_1$ will go from zero to some maximum value $\omega_1^\text{max}$, that can be read from Eq.~\eqref{eq:col28}. Solving for $(1-\hat p_0 \hat p_1)\equiv \omega_1$ in Eq.~\eqref{eq:col28} we get
\be 
    \omega_1=\frac{(1-x_1)}{x_1} \qty[1+\sum_{i=1}^{3}\alpha_i x_i^{n+1}] \qty(\frac{E_0}{\Lambda})^n-\frac{x_2x_3}{x_1} (1-\hat p_2 \hat p_3) \,,
\ee
which is maximum for $(1-\hat p_2 \hat p_3)=0$, 
\be 
    \omega_1^\text{max}=\frac{(1-x_1)}{x_1} \qty[1+\sum_{i=1}^{3}\alpha_i x_i^{n+1}] \qty(\frac{E_0}{\Lambda})^n \,.
\ee
Then, integrating in $\omega_1$ from 0 to $\omega_1^\text{max}$ we get that the total decay width of a generic neutrino three-body decay in a superluminal \abb{LIV} scenario is given by
\be
    \Gamma = \frac{N}{96} \frac{E_0^5}{(2\pi)^3} \qty(\frac{E_0}{\Lambda})^{3n} \iiint \delta\qty(1-\sum x_i) dx_1 dx_2 dx_3 (1-x_1)^2 \qty[1+\sum_{i=1}^{3}\alpha_i x_i^{n+1}]^3 \,.
    \label{eq:decay_width}
\ee

\begin{subappendices}
    \section{DSR as a departure from locality on elementary interactions}
\label{sec:new_perspective}

\begin{figure}[ht]
    \centering
    \includegraphics[width=0.6\textwidth]{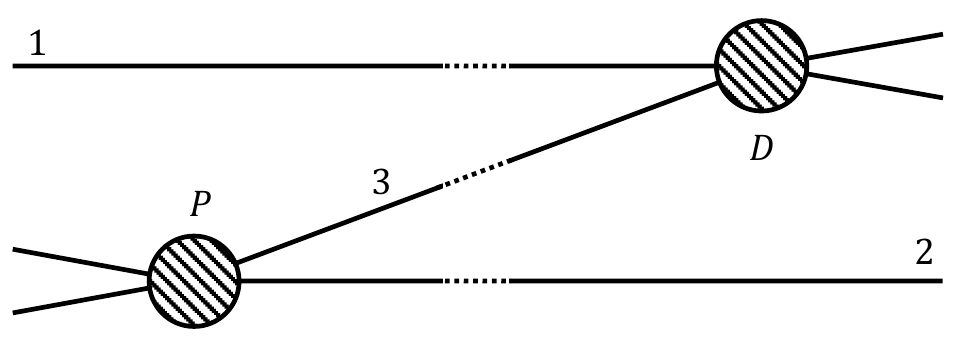}
    \caption{A particle, labelled 3, is produced in an interaction $P$, propagates over a large distance, and is detected in a second interaction $D$. Particles labelled by 1 and 2 only participate in the interactions $D$ and $P$, respectively.}
    \label{fig:prod-prop-dete}
\end{figure}

We present a new interpretation of \abb{DSR}~\cite{Carmona:2023luz}, in which it is considered as a way to implement a departure from the locality of the interactions of elementary particles (leptons, quarks, and mediators of the interactions of the \abb{SM}). To better understand the characteristics of the new proposed perspective of \abb{DSR}, let us consider a process with two interactions in regions of spacetime separated by a large distance. We consider an interaction producing a particle which propagates and is detected by a second interaction, as shown in Fig.~\ref{fig:prod-prop-dete}.

In this new perspective, when one considers an interaction of particles which are seen as elementary when explored at distances larger than the scale $\ell\sim (1/\Lambda)$ (with $\Lambda$ the energy scale of new physics), one is able to see the effects of the deviation from the locality of interactions, i.e., the change of momenta due to the interaction is determined by the kinematics of \abb{DSR}. On the other hand, when one considers an interaction with composite particles whose size is much larger than the scale $\ell$, one can neglect the deviations from locality, and the change of momenta produced by the interaction is determined by the kinematics of \abb{SR}.

We can apply this feature, which defines our model, to the process shown in Fig.~\ref{fig:prod-prop-dete}. Let us assume that particles $2$ and $3$ are elementary but particle $1$ is a composite particle, so that the interaction producing particle $3$ is between elementary particles, and the interaction in its detection involves a composite particle. A possible relation for the change of momenta in the interaction $D$ is
\be
    p_1 - p_1' \,=\, p_3'\oplus \hat{p}_3\,,
    \label{mcl-D}
\ee
where $\hat{p}_3$ is the antipode of the momentum of particle $3$ before its detection (defined such that $p_3\oplus \hat{p}_3$ = 0). Let us note that we have used the fact that particle 1 is not elementary to replace the composition with the antipode of $p_1'$ by the sum with $-p_1'$. The left-hand side is a difference of momentum variables of the composite particle $1$, whose kinematics is not affected by the deformation of \abb{SR}; then, they transform linearly under Lorentz transformations. This implies that the composition of momentum variables on the right-hand side also has to transform linearly. More generally, in this new perspective of \abb{DSR}, the composition of two momentum variables (or antipodes or a mixture of both), as well as a single momentum variable, has to transform linearly under Lorentz transformations. Therefore, there is no possible modification of the dispersion relation with respect to \abb{SR}. Moreover, this also implies that there is no signal of the deformation in the propagation of a particle. This agrees with the interpretation of \abb{DSR} as a modification of the interaction between elementary particles instead of an effect in the propagation of particles in a quantum spacetime.

Let us now analyze the state of the soccer ball problem and spectator problem in the new perspective of \abb{DSR}. Let us start with the spectator problem. In the usual interpretation of \abb{DSR}, if one considers the total momentum of the system before and after each interaction, from the conservation of the total momentum during the interaction, one could end up with situations in which one cannot disentangle the influence of the spectator particles on the conservation law. For example, if interaction $P$ is a scattering of the particles 2 and 3, then
\be 
    p_1 \oplus p_2 \oplus p_3 = p_2' \oplus p_1 \oplus p_3' \,.
    \tag{See Eq.~\eqref{eq:spec}}
\ee
Following the same argument for the second interaction, now the conservation law of the interaction between particle 1 and 3 will depend on the spectator particle 2. We see then how the spectator problem arises. Moreover, as both interactions share particle 2, we see that we cannot treat the two interactions independently. Therefore, \abb{DSR} leads to a violation of the cluster property of interactions, which is at the basis of relativistic \abb{QFT}.

However, in the new perspective of \abb{DSR}, the deviations from \abb{SR} are only present in the interactions of elementary particles, as a modification of the conservation law of the momenta of the particles involved in the interaction. That is to say, the spectator particles do not play any role in the interaction and therefore the spectator problem does not exist in this perspective.

The other conceptual problem of the usual perspective of \abb{DSR} is the soccer ball problem, which arises in two different ways. The first one is when one tries to apply the modified dispersion relation to macroscopic objects. Given that the characteristic energies and momenta of macroscopic objects are much greater that scale $\Lambda$, the modification adds a very large contribution, which, of course, is not compatible with our daily life experience. In the new perspective of \abb{DSR} one is able to avoid this problem because, as we have shown before, there is no deviation of the \abb{SR} dispersion relation. The second way the soccer ball problem can arise is when one tries to use the composition law of momentum to add the momentum of two macroscopic objects. However, in the new perspective of \abb{DSR}, the macroscopic objects are composed objects, with a characteristic scale much larger than $\ell\sim (1/\Lambda)$, so the kinematics we should use is the one of \abb{SR}, i.e. the usual sum of momenta.

Finally, let us briefly discuss the phenomenological implications of the new perspective of \abb{DSR}. Firstly, since the dispersion relation is the one of \abb{SR}, time delays are absent~\cite{Carmona:2017oit,Carmona:2022pro}. This means that the strong constraints on the high-energy scale based on the possibility of time delays~\cite{Martinez:2008ki,HESS:2011aa,Vasileiou:2013vra,MAGIC:2017vah,HESS:2019rhe,MAGIC:2020egb,Du:2020uev} do not apply in this scheme. As a consequence, one can consider values of the scale $\Lambda$ much lower than the Planck scale, increasing the predictive power of the high-energy astrophysical observations. In the same line, lowering the scale of new physics also opens the possibility to search for footprints of \abb{DSR} in interactions at the highest energies in laboratory experiments. We hope this novel perspective on \abb{DSR}, recently published in~\cite{Carmona:2023luz}, can encourage the investigation of a \abb{DSR} with a lower characteristic energy scale.

\clearpage
\section{Action formulation of relative locality}
\label{sec:relative_locality}

A realization of the relative locality of interactions in \abb{DSR} can be derived from the action formulation of an interaction~\cite{AmelinoCamelia:2011bm}. A simple action in 1+1 dimension that represents an interaction that happens at $\tau=0$ can be written as the sum of the free Lagrangian of incoming and outgoing particles, and the interaction Lagrangian, where one imposes the conservation of energy and momentum. 

On the one hand, the free Lagrangian can be written as
\be
    S_\text{free}\,=\, \sum_i \qty(\int_{-\infty}^{0} d\tau \,(x_i \dot{\Pi}_i - t_i \dot{\Omega}_i + \mathcal{N}_i C_i)) +
    \sum_j \qty(\int_{0}^{\infty} d\tau \,(x_j \dot{\Pi}_j - t_j \dot{\Omega}_j + \mathcal{N}_j C_i)) \,, \\
\ee
where $i$ labels the incoming particles, $j$ the outgoing ones, $x$ and $t$ are their space and time coordinates, $\Omega$ and $\Pi$ the energy and momentum of the particles, and finally the $\mathcal{N}_i$ are just Lagrange multipliers that impose the modified dispersion relations given by $C_i=0$.

On the other hand, the interaction Lagrangian is given by
\begin{equation}
    S_\text{int}\,=\,\,z_1[\mathcal{P}^\text{\emph{in}}(0)-\mathcal{P}^\text{\emph{out}}(0)] - z_0[\mathcal{E}^\text{\emph{in}}(0)-\mathcal{E}^\text{\emph{out}}(0)] \,,
    \label{eq:action}
\end{equation}
where $(z_1, z_0)$ are also Lagrange multipliers that impose, at $\tau=0$, the conservation of the total energy and momentum, $\mathcal{E}$ and $\mathcal{P}$, obtained from the modified composition law of the incoming and outgoing momenta.

Applying the variational principle, one obtains that the worldlines of the incoming and outgoing particles are characterized by constant momenta (independent of $\tau$), and the space-time coordinates of the end or starting points of their worldlines, respectively, are
\be
    t_i(0)\,=\,-z_1\frac{\partial \mathcal{P}}{\partial \Omega_i} + z_0\frac{\partial \mathcal{E}}{\partial \Omega_i}\,, \qquad
    x_i(0)\,=\, z_1\frac{\partial \mathcal{P}}{\partial \Pi_i} - z_0\frac{\partial \mathcal{E}}{\partial \Pi_i}\,.  
    \label{eq:RLpoint}
\ee
We see then that the set of momenta and the values of $(z_1,z_0)$ determine the end or starting points of the incoming and outgoing worldlines. As a consequence, the interaction is seen as local (all the worldlines meet at the same point) only for the observer which establishes the origin of space-time coordinates at the interaction vertex (corresponding to $z_1=z_0=0$). Let us also notice that Eq.~\eqref{eq:RLpoint} can be considered as a space-time translation generated by the total energy and momentum, $\mathcal{E}$ and $\mathcal{P}$, with parameters $(z_1,z_0)$, i.e.
\be
    t_i(0)\,=\,
-z_1 \{\mathcal{P},t_i\} + z_0 \{\mathcal{E},t_i\}\,, \quad
    x_i(0)\,=\,
z_1 \{\mathcal{P},x_i\} - z_0 \{\mathcal{E},x_i\}\,.  
    \label{eq:RLpoint_tras}
\ee
An observer who assigns a different value of the pair $(z_1,z_0)$ to the interaction is related with the local observer by a translation generated by the total energy and momentum with parameters $(z_1,z_0)$. Translations correspond then to constant displacements in the Lagrange multipliers, and translated observers with respect to the local one do not see the worldlines meet at a single space-time point. This feature of \abb{DSR} is what we call Relative Locality of interactions.
\end{subappendices}
\printbibliography[heading=subbibintoc,title={References}]
\end{refsection}

\begin{refsection}
\chapter{Neutrinos}
\graphicspath{{Chapter3/Figs/}}
\label{chap:neutrinos}

Neutrinos are very special particles that only interact through the weak (and gravitational) interactions. In this way, they are able to travel cosmological distances in straight lines, because they are neither affected by the cosmological backgrounds mentioned in Sec.~\ref{sec:cosmology}, nor by the galactic magnetic fields. The universe is almost transparent to them. This gives us a privileged messenger that can reach the Earth almost unaffected.

In the \abb{SM} they are considered massless particles; however, the discovery of the neutrino oscillations led to the conclusion that they have non-zero mass. Neutrino oscillations occur due to the fact that the interaction eigenstates, in which the neutrinos are created and detected (electron, muon, and tau neutrinos), are a superposition of mass eigenstates, whose quantum phases, due to the difference in their masses, oscillate at different speeds during the propagation~\cite{Kajita:2010zz,Gonzalez-Garcia:2007dlo}. The existence of neutrino oscillations provided one of the first clear indications of the need to go beyond the \abb{SM}. 

\begin{figure}[tb]
    \centering
    \includegraphics[width=0.8\textwidth]{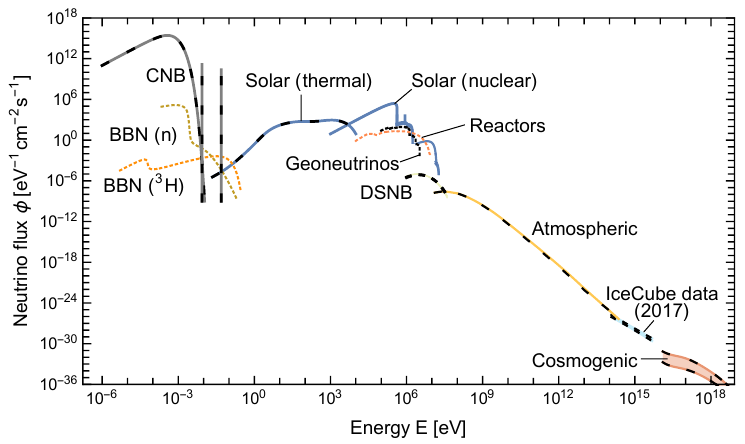}
    \caption{Spectrum of detected neutrinos at Earth. From~\cite{Vitagliano:2019yzm}.}
    \label{fig:nu_spec}
\end{figure}

The detected flux of neutrinos at Earth has a very different origin, depending on the energy range (see Fig.~\ref{fig:nu_spec}). For example, the lowest energy spectrum is mainly composed by the \abbs{C$\nu$B}{CnB} we introduced in Sec.~\ref{sec:cosmology}. Instead, the spectrum of medium energies is mostly composed by the neutrinos produced at the sun. For the case of the high-energy spectrum, the main contribution comes from the interaction of the cosmic rays with the atmosphere, which produce neutrinos as secondary particles. However, as discussed in Sec.~\ref{sec:scope}, we are interested in even more energetic neutrinos, i.e., we are interested in the very high-energy (from $10^{14}\unit{eV}$ to $10^{16}\unit{eV}$) and ultra high-energy (above $10^{16}\unit{eV}$) part of the spectrum.

To this day, the only known mechanism to produce neutrinos with energies above $10^{14}\unit{eV}$ is as secondary particles from the interaction of the ultra high-energy cosmic rays (see Chapter~\ref{chap:cosmicrays}). A part of this neutrino flux can be produced within or in the vicinity of the sources, where the cosmic rays undergo complex interactions with matter. Then, these neutrinos can escape from the sources and reach the Earth. This kind of neutrinos are known as \textit{astrophysical neutrinos}, and they are supposed to populate the very high-energy range of the neutrino spectrum. The sources of astrophysical neutrinos are still unclear, but the main candidates are extreme energetic objects, like \abbdef{GRB}{Gamma Ray Bursts} or \abbdef{AGN}{Active Galactic Nuclei}.

However, if the neutrinos are not produced within the source, but from the interaction of the cosmic rays with the diffuse photon background (mainly \abb{CMB} and \abb{EBL}) during their propagation, they are called \textit{cosmogenic neutrinos}. These neutrinos can reach even higher energies, forming the ultra high-energy range. They complete the highest part of the neutrino spectrum and will be discussed in Chapter~\ref{chap:cosmicrays}.

Neutrinos from the high-energy range and above are very difficult to detect, so very large detectors like the \abbdef{IC}{IceCube Neutrino Observatory} or the \abbdef{PA}{Pierre Auger Observatory} are required. Neutrinos cannot be directly detected, but indirect methods are necessary for their identification. \abb{IC} uses the Cherenkov light produced by the charged lepton generated in the interaction of the neutrino with a nucleus of the ice~\cite{IceCube:2013dkx}. Instead, \abb{PA} uses the low showers produced by neutrinos when interacting with the nuclei of the atmosphere~\cite{PierreAuger:2013wqu}.

The objective of this chapter is to develop the \abb{LIV} model discussed in Sec.~\ref{sec:LIV} for the neutrinos and study its phenomenological consequences in the astrophysical neutrino flux. A discussion about the effects in the cosmogenic neutrino flux is held in Chapter~\ref{chap:cosmicrays}.

\section{LIV effects in neutrinos: theoretical description of the decays}
\label{sec:LIV_neu}

We are going to consider the \abb{LIV} model discussed in Sec.~\ref{sec:LIV}, i.e., a non-universal \abb{LIV} which only affects the neutrino. In order to do that, we will add a new term to the \abb{SM} Lagrangian which involves the neutrino matter fields (a way to do this without breaking gauge invariance is shown in Sec.~\ref{sec:LIV}) and an operator which violates \abb{LI}.

We can produce an operator that violates the \abb{LI} if we include an asymmetry between temporal and spatial derivatives. As we will see in the following, in order to get a \abb{MDR} with a correction of order $n$, we need a term with $n+1$ derivatives. As a consequence, we need consider all the \abb{LIV} terms that come from considering the different asymmetric combinations of $n+1$ spatial and temporal derivatives,
\be
    \mathcal{L}_{\text{free}}= \bar{\nu}_{lL}(i\gamma^\mu\partial_\mu)\nu_{lL} - \sum_{\substack{k=0 \\ k\neq n+1-k}}^{n+1} \qty(\frac{1}{\Lambda})^n \bar\nu_{lL} \qty(i\gamma^0 \partial_0)^k \qty(i\gamma^i \partial_i)^{n+1-k} \nu_{lL} \,,
    \label{eq:LIV_Lagr}
\ee
where the subindex $L$ refers to the left-handed chirality of the neutrino fields and $l$ refers to the three types of neutrinos ($e$, $\mu$, $\tau$). However, let us note that the derivatives are suppressed by a factor $(1/\Lambda)^n$. Therefore, we can relate the spatial and temporal derivatives using the solution of the equations of motion of the Lagrangian~\eqref{eq:LIV_Lagr} at the dominant zeroth order; that is to say, using the solutions of the \abb{SM} neutrino Dirac equation,
\be
    \qty(i\gamma^\mu \partial_\mu)\nu_{lL}=0 \srarrow \qty(i\gamma^i \partial_i) \nu_{lL} =  \qty(i\gamma^0 \partial_0) \nu_{lL} \,.
\ee 
As a consequence, we see that at order $(1/\Lambda)^n$ all the new \abb{LIV} terms are equivalent, and we only need to consider one additional term in the free Lagrangian,
\be
    \mathcal{L}_{\text{free}}= 
    \bar{\nu}_{L}(i\gamma^\mu\partial_\mu)\nu_{L}-\frac{1}{\Lambda^n}\bar{\nu}_{L} \gamma^0(i\partial_0)^{n+1}\nu_{L} \,.
    \label{eq:LIV-nu}
\ee
The choice of the sign of the \abb{LIV} term is arbitrary and will be discussed later. Let us note that in Eq.~\eqref{eq:LIV-nu}, and from now on, we assume massless neutrinos and no flavour dependence.

Solving the equations of motion, i.e., the Euler-Lagrange equations, we get a modified Dirac equation,
\be
    \qty(i\gamma^\mu \partial_\mu-\frac{1}{\Lambda^n}\gamma^0(i\partial_0)^{n+1})\nu_{lL}=0 \,.
\ee
Introducing a plane-wave expansion with positive and negative energy solutions,
\be
\nu_{L}(t, \vec{x}) \,=\, \int d^3\vec{p} \: \left[\tilde{b}_{\vec{p}} \, \tilde{u}(\vec{p}) \, e^{-i E_- t + i \vec{p}\cdot\vec{x}} + \tilde{d}^\dagger_{\vec{p}} \,\tilde{v}(\vec{p})\, e^{i E_+ t - i \vec{p}\cdot\vec{x}}\right]\,,
\ee
we find that the spinors $\tilde{u}$, $\tilde{v}$ have to satisfy the equations 
\bea
\left[\gamma^0 E_- \,-\,\vec{\gamma}\cdot\vec{p} \,-\, \gamma^0 \frac{E_-^{n+1}}{\Lambda^n}\right] \,&\tilde{u}(\vec{p}) \,=\, 0\,, \\
\left[\gamma^0 E_+ \,-\,\vec{\gamma}\cdot\vec{p} \,+\, (-1)^{n+1} \gamma^0 \frac{E_+^{n+1}}{\Lambda^n}\right] \,&\tilde{v}(\vec{p}) \,=\, 0\,.
\eea
In the chiral representation for the Dirac matrices, the spinors $\tilde{u}$, $\tilde{v}$ can be written as $\tilde{u}=(\chi,0)^T$, $\tilde{v}=(\eta,0)^T$, and the bi-spinors $\chi$, $\eta$ satisfy the equations
\bea
(\vec{\sigma}\cdot\vec{p}) \, \chi(\vec{p}) \,&=\, - E_- \qty[1 - \qty(\frac{E_-}{\Lambda})^n ] \, \chi(\vec{p}) \,, \\
(\vec{\sigma}\cdot\vec{p}) \,\eta(\vec{p}) \,&=\, - E_+ \qty[1 + (-1)^{n+1} \qty(\frac{E_+}{\Lambda})^n ] \, \eta(\vec{p}) \,.
\eea
The matrix $(\vec{\sigma}\cdot\vec{p})$ has two eigenvalues $\pm |\vec{p}|$. Then we conclude that the relation between the momentum ($\vec{p}$) and the energy ($E_-$) for the neutrino and the antineutrino ($E_+$) are, respectively, 
\bea
|\vec{p}| \,&=\, E_- \qty[1- \qty(\frac{E_-}{\Lambda})^n] \,, \\
|\vec{p}| \,&=\, E_+ \qty[1+ (-1)^{n+1} \qty(\frac{E_+}{\Lambda})^n] \,.
\label{eq:mdr_neu}
\eea
Additionally, even if the norm of a four-momentum is not a \abb{LI} quantity anymore, it is still convenient to define a four-momentum for the neutrino as $p \equiv (|\vec p|,\vec p)$.

We see that, for the minus sign choice in Eq.~\eqref{eq:LIV-nu}, when $n$ is even, both neutrinos and antineutrinos are superluminal, while in the case of odd $n$ the neutrino is superluminal and the antineutrino is subluminal. However, if instead one considers a positive coefficient in Eq.~\eqref{eq:LIV-nu}, any superluminal state would become subluminal and vice versa. This information is summarized in Tab.~\ref{tab:s}.
\begin{table}[htb]
    \centering
    \caption{Choice of the sign of \abb{LIV} term.}
    \begin{tabular}{ccccc} \toprule
        & \multicolumn{2}{c}{($-$)} & \multicolumn{2}{c}{($+$)} \\ \midrule
        & neutrinos & antineutrinos & neutrinos & antineutrinos \\ \midrule
       $n=1$ & superluminal & subluminal & subluminal & superluminal \\ 
       $n=2$ & superluminal & superluminal & subluminal & subluminal \\ \bottomrule
    \end{tabular}
    \label{tab:s}
\end{table}

If (anti)neutrinos are superluminal, they become unstable particles able to decay in vacuum through two new mechanisms: \abbdef{VPE}{electron-positron Vacuum Pair Emission} and neutrino-antineutrino Pair Emission (also called \abbdef{NSpl}{Neutrino Splitting}). As a consequence, these new two mechanisms of decay will produce a strong suppression of the high-energy spectrum. In contrast, if (anti)neutrinos are subluminal, no new decay is allowed; instead, the departures from \abb{SR} come from the modification of the rates of the neutrino interactions due to the \abb{MDR}. Then one can conclude that the scenario with the strongest phenomenological consequences is the one with more superluminal particles. Checking Tab.~\ref{tab:s}, we see that this scenario corresponds to the negative sign in the \abb{LIV} term of Eq.~\eqref{eq:LIV-nu}. Therefore, this will be our choice of study along this work.

The possibility of a \abb{LI} violating model with superluminal neutrinos was a topic of interest, briefly studied in the literature, after the OPERA experiment claimed a possible detection of a superluminal neutrino velocity~\cite{OPERA:2011ijq}. Some approximations, like considering only the $Z$-mediated channel, were followed~\cite{Cohen:2011hx,Bezrukov:2011qn} in order to get the decay width of the \abb{VPE} neutrinos. Those studies concluded that in a \abb{LIV} scenario with superluminal neutrinos, the neutrinos detected by OPERA would never reach the detector with the claimed energies. Later on, more explicit calculations of the neutral current \abb{VPE} decay width were done~\cite{Carmona:2012tp}, assuming different hypothesis for the decay amplitude. However, explicit calculations of the \abb{VPE} decay width for the $W$-mediated channel (only relevant for the electron neutrino) and \abb{NSpl} were still missing in the literature until the results shown in this chapter were published~\cite{Carmona:2022dtp}.

In order to compute explicitly the desired decay widths, we need to consider, in addition to the modified free Lagrangian, also the neutrino interaction part. Since we assume the additional \abb{LIV} term only affects the free part of the Lagrangian, the interaction part will be the one of the \abb{SM}, where some vertices, which were forbidden by energy-momentum conservation in \abb{SR}, will be now allowed leading to the new two decay mechanisms (\abb{VPE} and \abb{NSpl}),
\bea
{\cal L}_{\text{int}} \,=\, &- \frac{g}{\sqrt{2}} \left[W_\mu^- \, \overline{e}_L \gamma^\mu \nu_{eL} + W_\mu^+ \, \overline{\nu}_{eL} \gamma^\mu e_L\right] - \frac{g}{2 c_W} Z_\mu \, \overline{\nu}_{lL} \gamma^\mu \nu_{lL} \\
&- \frac{g}{c_W} (s_W^2 - 1/2) Z_\mu \, \overline{e}_L \gamma^\mu e_L - \frac{g}{c_W} s_W^2 Z_\mu \, \overline{e}_R \gamma^\mu e_R\,.
\label{eq:lagrangian_int}
\eea

In the following sections we study each decay in more detail and we will perform the computation of decay widths, using the results of the collinear approximation we found in Sec.~\ref{sec:collinear}.

\subsection{Electron-positron pair emission}
\label{sec:neu_VPE}

\begin{figure}[htb]
    \centering
    \begin{feynman}{180,100}
        \fmfleft{i1}
        \fmfright{o1,o2,o3}
        \fmf{fermion,label=$\nu_\alpha$}{i1,v1}
        \fmf{fermion,label=$\nu_\alpha$}{v1,o1}
        \fmf{boson,label=$Z^0$}{v1,v2}
        \fmf{fermion,label=$e^-$}{v2,o2}
        \fmf{fermion,label=$e^+$}{o3,v2}
        \fmfforce{(0.00w,0.33h)}{i1}
        \fmfforce{(0.33w,0.33h)}{v1}
        \fmfforce{(0.66w,0.00h)}{o1}
    \end{feynman}
    \begin{feynman}{180,100}
        \fmfleft{i1}
        \fmfright{o1,o2,o3}
        \fmf{fermion,label=$\nu_e$}{i1,v1}
        \fmf{fermion,label=$e^-$}{v1,o1}
        \fmf{boson,label=$W^+$}{v1,v2}
        \fmf{fermion,label=$\nu_e$}{v2,o2}
        \fmf{fermion,label=$e^+$}{o3,v2}
        \fmfforce{(0.00w,0.33h)}{i1}
        \fmfforce{(0.33w,0.33h)}{v1}
        \fmfforce{(0.66w,0.00h)}{o1}
    \end{feynman}
    \caption{Neutral (left) and charged (right) channels for the electron-positron pair emission.}
    \label{fig:feynman_vpe}
\end{figure}
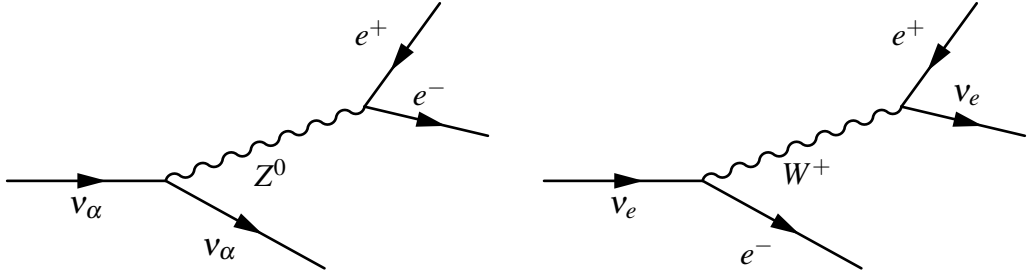

Let us start by considering the neutrino electron-positron pair emission ($\nu\rarrow\nu e^- e^+$), or \abb{VPE}. The neutrinos can follow a neutral (the three flavours) or charged (electronic flavour only) channel. The Feynman diagrams of each channel are shown in Fig.~\ref{fig:feynman_vpe}.

Additionally, as the particles of the final state are massive, the process requires a minimum energy of the initial state in order to produce the two electrons of the final state. This threshold energy for the initial neutrino can be found to be
\be
    E_\text{th}^{(e)} \coloneqq \qty(2 m_e^2 \Lambda^n)^{1/(2+n)} \,.
\ee

We can start with the simplest case, muon and tau neutrinos, which can only decay trough the neutral channel. Using the left Feynman diagram of Fig.~\ref{fig:feynman_vpe} and the interaction Lagrangian of Eq.~\eqref{eq:lagrangian_int}, we can compute the decay amplitude to left-handed and right-handed electrons,
\begin{small}
\bea
    A_{\nu_{\mu,\tau}\rarrow \nu_{\mu,\tau} e_L^- e_L^+} =& \\ &\qty[\bar{\tilde u}_L(\vec p')\qty(-i\frac{g}{2c_W}\gamma^\mu)\tilde u_L(\vec p)]
    \qty{-i\frac{\eta_{\mu\nu}}{M_Z^2}}
    \qty[\bar u_L(\vec p^-)\qty(-i\frac{g}{c_W}(s_W^2-1/2)\gamma^\nu) v_L(\vec p^+)] \,, \\
    A_{\nu_{\mu,\tau}\rarrow \nu_{\mu,\tau} e_R^- e_R^+} =& \\ &\qty[\bar{\tilde u}_L(\vec p')\qty(-i\frac{g}{2c_W}\gamma^\mu)\tilde u_L(\vec p)]
    \qty{-i\frac{\eta_{\mu\nu}}{M_Z^2}}
    \qty[\bar u_R(\vec p^-)\qty(-i\frac{g}{c_W} s_W^2 \gamma^\nu) v_R(\vec p^+)] \,.
    \label{eq:A_muontau}
\eea
\end{small}
After this, one should continue with the process of squaring the amplitude, summing over spins, writing the result in terms of traces, and contracting the indices. In order to not interrupt the reading flow with extensive calculations, we have moved this computation to Appendix~\ref{sec:amplitude_VPE_muon}, and the final result obtained is
\bea
    \abs{A_{\nu_{\mu,\tau}\rarrow \nu_{\mu,\tau} e_L^- e_L^+}}^2 & = 4 \frac{g^4}{M_W^4} (s_W^2-1/2)^2 (p\cdot p^+) (p'\cdot p^-) \,, \\ 
    \abs{A_{\nu_{\mu,\tau}\rarrow \nu_{\mu,\tau} e_R^- e_R^+}}^2 &= 4 \frac{g^4}{M_W^4} (s_W^2)^2 (p\cdot p^-) (p'\cdot p^+) \,,
    \label{eq:A2_muontau}
\eea
where $p$ stands for the four-momentum of the primary neutrino, $p'$ for the four-momentum of the produced neutrino, and $p^-$ and $p^+$ for the electron and positron ones, respectively.

We can check that the obtained decay amplitudes fit the \textit{ansatz} assumed in the example of the Sec.~\ref{sec:collinear}, where we computed the decay width of a neutrino three-body decay, using the collinear approximation, and assuming a generic decay amplitude of the form
\be 
    \abs{A_\nu}^2= N (p_0\cdot p_1)(p_2\cdot p_3) \,.
    \tag{See Eq.~\eqref{eq:ansatz}}
\ee
The result obtained was a decay amplitude of the form
\be
    \Gamma = \frac{N}{96} \frac{E_0^5}{(2\pi)^3} \qty(\frac{E_0}{\Lambda})^{3n} \iiint \delta\qty(1-\sum x_i) dx_1 dx_2 dx_3 (1-x_1)^2 \qty[1+\sum_{i=1}^{3}\alpha_i x_i^{n+1}]^3 \,.
    \tag{See Eq.~\eqref{eq:decay_width}}
\ee
Then, we can apply Eq.~\eqref{eq:decay_width} to Eq.~\eqref{eq:A2_muontau}, in order to get the muon and tau \abb{VPE} decay widths. The results obtained for left and right handed electrons, respectively, are
\bea
    \Gamma_{\nu_{\mu,\tau}\rarrow \nu_{\mu,\tau} e_L^- e_L^+} = \frac{E_0^5}{192\pi^3} \frac{g^4}{M_W^4} (s_W^2-1/2)^2 \qty(\frac{E_0}{\Lambda})^{3n} \iiint & \delta\qty(1-x_+-x'-x_-) dx_+ dx' dx_- \\ & \times (1-x_+)^2 \qty(1-x'\,^{n+1})^3 \,, \\
    \Gamma_{\nu_{\mu,\tau}\rarrow \nu_{\mu,\tau} e_R^- e_R^+} = \frac{E_0^5}{192\pi^3} \frac{g^4}{M_W^4} (s_W^2)^2 \qty(\frac{E_0}{\Lambda})^{3n} \iiint & \delta\qty(1-x_--x'-x_+) dx_- dx' dx_+ \\ & \times (1-x_-)^2 \qty(1-x'\,^{n+1})^3 \,,
\eea
with $x'$, $x_-$ and $x_+$ the energy fractions of the produced neutrino, electron, and positron, respectively.

In order to get the total decay width, we should now sum the left-handed and right-handed contributions. Then, the total decay width of a muon or tau neutrino to an electron-positron pair can be written as
\bea
    \Gamma_{\nu_{\mu,\tau}\rarrow \nu_{\mu,\tau} e^- e^+} = \frac{E_0^5}{192\pi^3} \frac{g^4}{M_W^4} & \qty(\frac{E_0}{\Lambda})^{3n} \iiint \delta\qty(1-x'-x_+-x_-) dx' dx_+ dx_- \\
    & \times (1-x'\,^{n+1})^3 \qty[(s_W^2-1/2)^2 (1-x_+)^2 + (s_W^2)^2 (1-x_-)^2] \,,
    \label{eq:GammaVPE_muon}
\eea
from where one can also read the energy distribution of the produced particles,
\bea
    {\cal P}_{\nu_{\mu,\tau}\rarrow \nu_{\mu,\tau} e^- e^+} (x',x_-,x_+) \,=&\, \frac{1}{\,c_n^{(e)}} \,\delta(1-x'-x_+-x_-)\,(1-{x'\,}^{n+1})^3 \\ &\times \qty[\cos^2(\theta_{\mu,\tau}) (1-x_+)^2 + \sin^2(\theta_{\mu,\tau}) (1-x_-)^2] \,,
    \label{eq:energy_dist_VPE_muon}
\eea
where $c_n^{(e)}$ is a normalization constant, whose value will be shown in short, and $\cos^2(\theta_{\mu,\tau})$ and $\sin^2(\theta_{\mu,\tau})$ are just shortcuts for
\bea
    \cos^2(\theta_{\mu,\tau}) &\coloneqq \frac{(s_W^2 - 1/2)^2}{(s_W^2 - 1/2)^2+(s_W^2)^2} \approx 0.61 \,, \quad\text{and} \\
    \sin^2(\theta_{\mu,\tau}) &\coloneqq \frac{(s_W^2)^2}{(s_W^2 - 1/2)^2+(s_W^2)^2} \approx 0.39\,.
    \label{eq:energy_dist_VPE_muon_cos}
\eea
The value of the squared sine of the Weinberg angle ($s_W^2\approx 0.22$), as well as other numerical constants used along this work, has been obtained from~\cite{ParticleDataGroup:2020}.

The integral over the final energy fractions of Eq.~\eqref{eq:GammaVPE_muon} can be done as follows. First, one can use the Dirac delta to remove the energy fraction of the positron and then one can integrate the electron energy fraction from zero to $(1-x')$, where $x'$ is the energy fraction of the produced neutrino. Then we get
\be
    \Gamma_{\nu_{\mu,\tau}\rarrow \nu_{\mu,\tau} e^- e^+} = \frac{E_0^5}{192\pi^3} \frac{g^4}{M_W^4} \qty[(s_W^2-1/2)^2 + (s_W^2)^2] \qty(\frac{E_0}{\Lambda})^{3n} \;\frac{1}{3}\int dx' (1-x'\,^{n+1})^3 (1-x'\,^{3}) \,.
    \label{eq:Gamma_VPE_nu}
\ee
In this middle step one can read from the integrand of Eq.~\eqref{eq:Gamma_VPE_nu} the energy distribution of the final neutrino (integrated to the positron and electron).
Finally, integrating the energy fraction of the produced neutrino from zero to one we get that the total neutrino decay width of muon and tau neutrinos to an electron-positron pair is given by
\be
    \Gamma_{\nu_{\mu,\tau}\rarrow \nu_{\mu,\tau} e^- e^+} = \frac{E_0^5}{192\pi^3} \frac{g^4}{M_W^4} \qty[(s_W^2-1/2)^2 + (s_W^2)^2]  \qty(\frac{E_0}{\Lambda})^{3n} c_n^{(e)} \,,
    \label{eq:total_decay_muon_tau}
\ee
where the normalization constant $c_n^{(e)}$, which only depends on $n$, is given by
\bea
    c_n^{(e)} \coloneqq& \frac{1}{3} \int_0^1 d{x'} (1-{x'\,}^{n+1})^3 (1-{x'\,}^{3}) \\ =&\frac{1}{4} - \frac{3}{(n+2)(n+5)}+ \frac{3}{(2n+3)(2n+6)}-\frac{1}{(3n+4)(3n+7)} \,.
    \label{eq:cen}
\eea

The result obtained is compatible with the $Z$-mediated decay widths (up to a model dependent constant), used by the different authors in the literature~\cite{Cohen:2011hx,Bezrukov:2011qn} after the OPERA claim of a possible detection of superluminal neutrino velocity~\cite{OPERA:2011ijq}. In fact, the result coincides with one of the posterior explicit calculations of the muon and tau decay width; more specifically, with the second example of~\cite{Carmona:2012tp}. However, the probability distribution of the final energy fractions, together with all the calculations below, are a novel result of this work.

We should now repeat all the process, but for the electron flavoured neutrinos. The matrix element corresponding to right-handed electrons is the same as in the case of muon and tau neutrinos. In contrast, when one considers the decay to left-handed neutrinos, now the charged channel has to be taken into account,
\bea
    A_{\nu_e\rarrow \nu_e e_L^- e_L^+} =&\qty[\bar{\tilde u}_L(\vec p')\qty(-i\frac{g}{2c_W}\gamma^\mu)\tilde u_L(\vec p)]
    \qty{-i\frac{\eta_{\mu\nu}}{M_Z^2}}
    \qty[\bar u_L(\vec p^-)\qty(-i\frac{g}{c_W}(s_W^2-1/2)\gamma^\nu) v_L(\vec p^+)] \\ +&
    \qty[\bar u_L(\vec p^-)\qty(-i\frac{g}{\sqrt{2}}\gamma^\mu) \tilde u_L(\vec p)]
    \qty{-i\frac{\eta_{\mu\nu}}{M_W^2}}
    \qty[\bar{\tilde u}_L(\vec p')\qty(-i\frac{g}{\sqrt{2}}\gamma^\nu) v_L(\vec p^+)] \,, \\
    A_{\nu_{e}\rarrow \nu_{e} e_R^- e_R^+} =& \\ &\qty[\bar{\tilde u}_L(\vec p')\qty(-i\frac{g}{2c_W}\gamma^\mu)\tilde u_L(\vec p)]
    \qty{-i\frac{\eta_{\mu\nu}}{M_Z^2}}
    \qty[\bar u_R(\vec p^-)\qty(-i\frac{g}{c_W} s_W^2 \gamma^\nu) v_R(\vec p^+)] \,.
    \label{eq:A_elec}
\eea
Again, we move the rest of the computation to the Appendix~\ref{sec:amplitude_VPE_electron}. Skipping the development, the final results obtained for the squared amplitudes are given by
\bea
    \abs{A_{\nu_e\rarrow \nu_e e_L^- e_L^+}}^2=& 4 \frac{g^4}{M_W^4} (s_W^2-3/2)^2 (p\cdot p^+) (p'\cdot p^-) \,,  \\
    \abs{A_{\nu_e\rarrow \nu_e e_R^- e_R^+}}^2 =& 4 \frac{g^4}{M_W^4} (s_W^2)^2 (p\cdot p^-) (p'\cdot p^+) \,.
    \label{eq:A2_elec}
\eea
Let us note that the right-handed contribution is the same as in the case of muon and tau neutrinos. For the left-handed contribution, the result is also similar, but substituting \mbox{$(s_W^2-1/2)^2$} by \mbox{$(s_W^2-3/2)^2$}. Just from this, we can anticipate that the total decay width for electron neutrinos will be the same as in the case of muon and tau neutrinos, but making the aforementioned substitution in the left-handed contribution. That is to say, the total decay width of an electron neutrino to an electron-positron pair can be written as
\bea
    \Gamma_{\nu_{e}\rarrow \nu_{e} e^- e^+} = \frac{E_0^5}{192\pi^3} \frac{g^4}{M_W^4} & \qty(\frac{E_0}{\Lambda})^{3n} \iiint \delta\qty(1-x'-x_+-x_-) dx' dx_+ dx_- \\
    & \times (1-x'\,^{n+1})^3 \qty[(s_W^2-3/2)^2 (1-x_+)^2 + (s_W^2)^2 (1-x_-)^2] \,,
    \label{eq:GammaVPE_electron}
\eea
from where one can also read the energy distribution of the final energies
\bea
    {\cal P}_{\nu_{e}\rarrow \nu_{e} e^- e^+} (x',x_-,x_+) \,=&\, \frac{1}{\,c_n^{(e)}} \,\delta(1-x'-x_+-x_-)\,(1-{x'\,}^{n+1})^3 \\ &\times \qty[\cos^2(\theta_{e}) (1-x_+)^2 + \sin^2(\theta_{e}) (1-x_-)^2] \,,
    \label{eq:energy_dist_VPE_electron}
\eea
where $c_n^{(e)}$ is the same normalization constant as before, Eq.~\eqref{eq:cen}, and $\cos^2(\theta_{e})$ and $\sin^2(\theta_{e})$ are just shortcuts for
\bea 
    \cos^2(\theta_{e}) &\coloneqq \frac{(s_W^2 - 3/2)^2}{(s_W^2 - 3/2)^2+(s_W^2)^2} \approx 0.97 \,, \quad\text{and} \\
    \sin^2(\theta_{e}) &\coloneqq \frac{(s_W^2)^2}{(s_W^2 - 3/2)^2+(s_W^2)^2} \approx 0.03\,.
    \label{eq:energy_dist_VPE_electron_cos}
\eea

Integrating over the positron and electron energy fraction we obtain,
\be
    \Gamma_{\nu_{e}\rarrow \nu_{e} e^- e^+} = \frac{E_0^5}{192\pi^3} \frac{g^4}{M_W^4} \qty[(s_W^2-3/2)^2 + (s_W^2)^2] \qty(\frac{E_0}{\Lambda})^{3n} \;\frac{1}{3}\int dx' (1-x'\,^{n+1})^3 (1-x'\,^{3}) \,,
\ee
from where we can check, comparing with Eq.~\eqref{eq:Gamma_VPE_nu}, that the energy distribution of the final neutrino is flavour independent,
\be
    {\cal P}_{\nu_e} (x') \,=\,{\cal P}_{\nu_\mu} (x') \,=\, {\cal P}_{\nu_\tau} (x') \,=\,\frac{1}{3\,c_n^{(e)}} \,(1-{x'\,}^{n+1})^3 \,(1-{x'\,}^3)\,.
\ee
Finally, integrating the energy fraction of the produced neutrino from zero to one we get that the total neutrino decay width of electron neutrinos to an electron-positron pair is given by
\be
    \Gamma_{\nu_{e}\rarrow \nu_{e} e^- e^+} = \frac{E_0^5}{192\pi^3} \frac{g^4}{M_W^4} \qty[(s_W^2-3/2)^2 + (s_W^2)^2]  \qty(\frac{E_0}{\Lambda})^{3n} c_n^{(e)} \,,
    \label{eq:total_decay_electron}
\ee
with $c_n^{(e)}$ given by Eq.~\eqref{eq:cen}.
 
\subsection{Neutrino-antineutrino pair emission}
\label{sec:neu_NSpl}

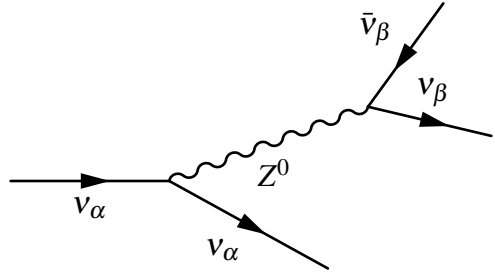
\begin{figure}[htb]
    \centering
    \begin{feynman}{180,100}
        \fmfleft{i1}
        \fmfright{o1,o2,o3}
        \fmf{fermion,label=$\nu_\alpha$}{i1,v1}
        \fmf{fermion,label=$\nu_\alpha$}{v1,o1}
        \fmf{boson,label=$Z^0$}{v1,v2}
        \fmf{fermion,label=$\nu_\beta$}{v2,o2}
        \fmf{fermion,label=$\bar\nu_\beta$}{o3,v2}
        \fmfforce{(0.00w,0.33h)}{i1}
        \fmfforce{(0.33w,0.33h)}{v1}
        \fmfforce{(0.66w,0.00h)}{o1}
    \end{feynman}
    \caption{Neutrino-antineutrino pair emission.}
    \label{fig:feynman_NSpl}
\end{figure}

Let us now consider the neutrino-antineutrino pair emission ($\nu\rarrow\nu\nu\bar\nu$) or \abb{NSpl}. More specifically, let us consider the process in which a neutrino of flavour $\alpha$ emits a  neutrino-antineutrino pair of flavour $\beta$. This process is mediated by a $Z^0$ boson and the corresponding Feynman diagram is shown in Fig.~\ref{fig:feynman_NSpl}.

Given the smallness of the neutrino mass, the threshold of this process is negligible. Using the Feynman diagram in Fig.~\ref{fig:feynman_NSpl} and the interaction Lagrangian of Eq.~\eqref{eq:lagrangian_int} we can write the decay amplitude of the process,
\be
    A_{\nu_\alpha\rarrow \nu_\alpha \nu_\beta \bar\nu_\beta} =\qty[\bar{\tilde u}_L(\vec p')\qty(-i\frac{g}{2c_W}\gamma^\mu)\tilde u_L(\vec p)]
    \qty{-i\frac{\eta_{\mu\nu}}{M_Z^2}}
    \qty[\bar{\tilde u}_L(\vec p^-)\qty(-i\frac{g}{2c_W}\gamma^\nu) \tilde v_L(\vec p^+)] \,.
    \label{eq:A_nu}
\ee
Let us note that for the case $\beta=\alpha$ one should consider two contributions accounting for the exchange of the momenta of the final-state neutrinos and add a factor $(1/\sqrt{2})$ from the antisymmetrization of the states.

We follow the same procedure as in the case of the electron-positron pair emission, and it is shown in Appendix~\ref{sec:amplitude_NSpl}. There we get that the decay amplitude of the neutrino-antineutrino vacuum emission is given by
\be
    \abs{A_{\nu_\alpha\rarrow \nu_\alpha \nu_\beta \bar\nu_\beta}}^2=\frac{g^4}{M_W^4} (p\cdot p^+)(p'\cdot p^-) \,,
    \label{eq:A2_nu}
\ee
which is independent of both, the initial and final neutrino flavours.

Again, this decay amplitude fits the \textit{ansatz} assumed in the example developed in Sec.~\ref{sec:collinear}, then the decay width can be written as
\bea
    \Gamma_{\nu_\alpha\rarrow \nu_\alpha \nu_\beta \bar\nu_\beta} = \frac{E_0^5}{192\pi^3} \frac{g^4}{M_W^4} \qty(\frac{E_0}{\Lambda})^{3n} \, \frac{1}{4}\iiint &\delta\qty(1-x'-x_+-x_-) dx' dx_+ dx_- \\
    & \times (1-x'\,^{n+1}-(-1)^{n}x_+\,^{n+1}-x_-\,^{n+1})^3 (1-x_+)^2 \,,
\eea
with $x'$ and $x_-$ the energy fractions of the produced neutrinos, and $x_+$ the one of the antineutrino.

This expression is a bit more complicated than for the case of \abb{VPE}, but particularizing for the cases $n=1,2$ one can find the following simplification\footnote{This simplification has to be derived and checked for $n=1$ and $n=2$ separately. Eq.~\eqref{eq:simplification} is just a compact way of writing both cases.}
\bea
    &\delta\qty(1-x'-x_+-x_-) (1-{x'\,}^{n+1}-(-1)^{n}{x_+\,}^{n+1}-{x_-\,}^{n+1}) \\
    = &\delta\qty(1-x'-x_+-x_-) (n+1) (1-{x'})(1-{x_-})(1-{x_+})^{n-1} \,.
    \label{eq:simplification}
\eea
Then we can rewrite the decay width (only for the cases $n=1$ and $n=2$) as
\bea
    \Gamma_{\nu_\alpha\rarrow \nu_\alpha \nu_\beta \bar\nu_\beta} = \frac{E_0^5}{192\pi^3} \frac{g^4}{M_W^4} \qty(\frac{E_0}{\Lambda})^{3n}\, \frac{(n+1)^3}{4} & \iiint \delta\qty(1-x'-x_+-x_-) dx' dx_+ dx_- \\
    & \times (1-x')^3 (1-x_-)^3 (1-x_+)^{3n-1} \,,
    \label{eq:Gamma_NSpl}
\eea
from which one can read the energy distribution of the produced particles
\be
    {\cal P}_{\nu_\alpha\rarrow \nu_\alpha \nu_\beta \bar\nu_\beta}(x', x_-, x_+) \,=\, \frac{(n+1)^3}{4\,c_n^{(\nu)}} \,\delta(1-x'-x_--x_+)\,
    (1-x')^3 \,(1-x_-)^3 \, (1-x_+)^{3n-1}\,,
    \label{eq:energy_dist_NSpl}
\ee
where $c_n^{(\nu)}$ is a normalization constant, whose value will be shown in short

Performing the successive integrals over the energy fractions of the product particles, we finally get that the decay width of the emission of a neutrino-antineutrino pair of flavour $\beta$ by a neutrino of flavour $\alpha$ is given by
\be
    \Gamma_{\nu_\alpha\rarrow \nu_\alpha \nu_\beta \bar\nu_\beta} = \frac{E_0^5}{192\pi^3} \frac{g^4}{M_W^4} \qty(\frac{E_0}{\Lambda})^{3n} c_n^{(\nu)} \,,
    \label{eq:Gamma_NSpl_beta}
\ee
where the normalization constant $c_n^{(\nu)}$, which only depends on $n$, is given by
\begin{small}
\bea
    c_n^{(\nu)} &\coloneqq  \frac{(n+1)^3}{4} \iiint \delta \qty(1-{x'}-{x_+}-{x_-}) \; d{x'} d{x_+} d{x_-}
    (1-{x'})^3 (1-{x_-})^3 (1-{x_+})^{3n-1} \\ &=  \frac{(n+1)^3}{4} \,\bigg[ \frac{1}{(3n+1)}-\frac{3}{(3n+2)}+ \frac{7}{2(3n+3)}-\frac{2}{(3n+4)} + \frac{3}{5(3n+5)}-\frac{1}{10(3n+6)} +\frac{1}{140(3n+7)} \bigg] \,.
    \label{eq:cnn}
\eea
\end{small}
Given that the result is flavour independent, the total \abb{NSpl} decay width, result of summing over all the possible flavors of the final state, is just three times the one corresponding to the emission of a neutrino-antineutrino pair of flavour $\beta$ (Eq.~\eqref{eq:Gamma_NSpl_beta}).

\subsection{Numerical comparison of the different decays}
\label{sec:comparison}

The three different decay widths we have computed for the superluminal neutrino can be expressed as a product of four factors
\be
\Gamma_{\nu_f \rightarrow A}(E) \,=\, c_0\times c_{\nu_f}^{(A)}\times \qty[E^5 \left(\frac{E}{\Lambda}\right)^{3n}]\times c_n^{(A)}\,.
\ee
The first factor, $c_0$, is a common constant, given by
\be
c_0 \,\coloneqq\, \frac{1}{192\pi^3} \frac{g^4}{M_W^4} = \frac{32}{192\pi^3} G_F^2 \,\approx\, 7.31 \times 10^{-13} \,\text{GeV}^{-4}\,,
\ee
with $G_F\approx 1.17\cdot 10^{-5} \unit{GeV^{-2}}$ the Fermi coupling constant.

The second factor is an adimensional constant, $c_{\nu_f}^{(A)}$, which depends on the flavour of the initial neutrino ($\nu_f$) and the particles of the final state ($A$),
\be
c_{\nu_{\mu,\tau}}^{(e)} \,\coloneqq\, \left[\left(s_W^2 - \frac{1}{2}\right)^2 + (s_W^2)^2\right] \,\approx\,0.13\,, \quad
c_{\nu_e}^{(e)} \,\coloneqq\, \left[\left(s_W^2 - \frac{3}{2}\right)^2 + (s_W^2)^2\right]\,\approx \,1.68 \,, \\
\text{and}\quad c_{\nu_\alpha}^{(\nu)} \,\coloneqq\, 1 \quad\text{for every flavour} \,.
\ee

Finally, we have the energy-dependent factor $\qty[E^5 (E/\Lambda)^{3n}]$, followed by the factor $c_n^{(A)}$, given by Eqs.~\eqref{eq:cen} and \eqref{eq:cnn}, which depends on the produced lepton pair ($A$) and on the order $n$. The numerical values for $n=1,2$ are the following:
\bea
&c_{1}^{(e)} \coloneqq \frac{121}{840} \,\approx\, 0.144 \,, \quad
c_{2}^{(e)} \coloneqq \frac{81}{455} \,\approx\, 0.178 \,, \quad\text{and} \\
&c_{1}^{(\nu)} \coloneqq \frac{11}{450} \approx \, 0.024 \,, \quad
c_{2}^{(\nu)} \coloneqq \frac{237}{10010} \,\approx\, 0.024 \,.
\eea

The numerical values we have just computed will be very useful to do some easy calculations in the following. In fact, we can check that the order of magnitude of the emission of a neutrino-antineutrino pair (of certain flavour $\beta$) is very similar to the muon and tau \abb{VPE} decay width. In fact, one has the quotient
\be 
    \frac{\Gamma_{\nu_\alpha\rarrow \nu_\alpha \nu_\beta \bar\nu_\beta}}{\Gamma_{\nu_{\mu,\tau}\rarrow \nu_{\mu,\tau} e^- e^+}} = \frac{c_{\nu_\alpha}^{(\nu)} \times c_{n}^{(\nu)}}{c_{\nu_{\mu,\tau}}^{(e)} \times c_{n}^{(e)}} \approx 
    \begin{cases} 
        1.34 & \text{for } n=1\\
        1.05 & \text{for } n=2
    \end{cases} \,,
    \label{eq:quotient_nu_muon}
\ee
which does not depend on the energy or the value of the scale of new physics $\Lambda$, but only on the order of correction $n$. This can serve as a proof of the assumption usually made in the literature~\cite{Stecker:2014oxa,Stecker:2017gdy,Stecker:2022tzd} that the total \abb{NSpl} decay width, after summing over the three possible flavors of the produced neutrinos, is approximately three times the muon or tau \abb{VPE} total decay width,
\be 
    \sum_{\beta=\mu,\tau,e}  \Gamma_{\nu_\alpha\rarrow \nu_\alpha \nu_\beta \bar\nu_\beta} \approx 3 \times \Gamma_{\nu_{\mu,\tau}\rarrow \nu_{\mu,\tau} e^- e^+} \,.
\ee
This approximation was usually taken due to the lack of an explicit computation of the \abb{NSpl} decay width in the literature; and we find that it is a reasonable approximation, specially in the case $n=2$.

\begin{figure}[tbp]
    \centering
    \includegraphics[width=0.85\textwidth]{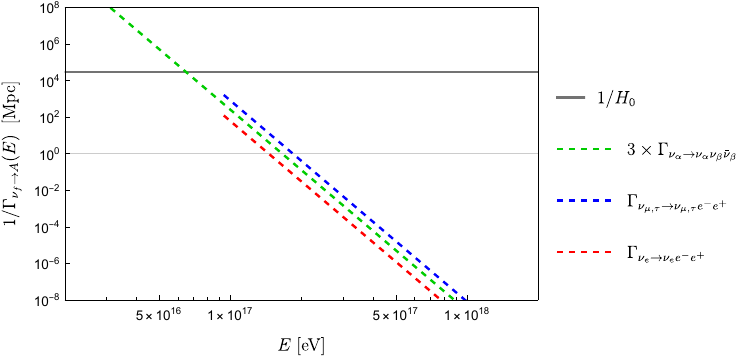}
    \caption{Decay lengths in Mpc of the different processes for $\Lambda=M_P$ and $n=2$. The \abb{VPE} is only defined above the energy threshold $E_\text{th}^{(e)}$. The characteristic length of the expansion of the universe, $(1/H_0)$, is also shown for comparison (gray line).}
    \label{fig:decay_length}
\end{figure}

One can also check that for energies above the \abb{VPE} threshold, the dominant effect is given by the electron-positron pair emission of electron neutrinos. In fact,
\be 
    \frac{\Gamma_{\nu_{e}\rarrow \nu_{e} e^- e^+}}{\Gamma_{\nu_{\mu,\tau}\rarrow \nu_{\mu,\tau} e^- e^+}} = \frac{c_{\nu_{e}}^{(e)}}{c_{\nu_{\mu,\tau}}^{(e)}} \approx 13.29 \,,
\ee
so in a equally flavoured flux, the electron neutrino \abb{VPE} would be around 13 times more frequent that the muon or tau \abb{VPE}. In order to get a better idea of the order of magnitudes we are working with, we show in Fig.~\ref{fig:decay_length}, for $\Lambda=M_P$ and $n=2$, the corresponding decay lengths to each process, i.e., the inverse of each total decay width.
Let use note that for the \abb{NSpl} we have already summed over the 3 final flavours.

Regarding the probability distributions obtained in Eqs.~\eqref{eq:energy_dist_VPE_muon}, \eqref{eq:energy_dist_VPE_electron} and \eqref{eq:energy_dist_NSpl}; they can be used to compute the mean values of the energy fractions for the final particles in the different decays. We show the obtained values in Tabs.~\ref{tab:mean_values_VPE_muon}, \ref{tab:mean_values_VPE_electron} and \ref{tab:mean_values_NSpl}.
\begin{table}[htbp]
\centering
\begin{minipage}{0.48\textwidth}
    \centering
    \caption{Mean value for the final energies after a muon or tau neutrino electron-positron pair emission.}
    \begin{tabular}{cccc} \toprule
       & $\avg{x'}$ & $\avg{x_-}$ & $\avg{x_+}$ \\ \midrule
      $n=1$ & 0.26 & 0.40 & 0.34 \\
      $n=2$ & 0.30 & 0.38 & 0.32 \\ \bottomrule
    \end{tabular}
    \label{tab:mean_values_VPE_muon}
\end{minipage}
\hfill
\begin{minipage}{0.48\textwidth}
    \centering
    \caption{Mean value for the final energies after an electron neutrino electron-positron pair emission.}
    \begin{tabular}{cccc} \toprule
       & $\avg{x'}$ & $\avg{x_-}$ & $\avg{x_+}$ \\ \midrule
      $n=1$ & 0.26 & 0.50 & 0.24 \\
      $n=2$ & 0.30 & 0.47 & 0.23 \\ \bottomrule
    \end{tabular}
    \label{tab:mean_values_VPE_electron}
\end{minipage}\\
\vspace{1.5em}
\begin{minipage}{0.48\textwidth}
    \centering
    \caption{Mean value for the final energies after a neutrino-antineutrino pair emission.}
    \begin{tabular}{cccc} \toprule
       & $\avg{x'}$ & $\avg{x_-}$ & $\avg{x_+}$ \\ \midrule
      $n=1$ & 0.30 & 0.30 & 0.40 \\
      $n=2$ & 0.38 & 0.38 & 0.24 \\ \bottomrule
    \end{tabular}
    \label{tab:mean_values_NSpl}
\end{minipage}
\end{table}

We see then how the approximation of equipartition of the energy, also assumed in~\cite{Stecker:2014oxa,Stecker:2017gdy,Stecker:2022tzd}, would not be enough depending on the accuracy of the conclusions to be drawn. This approximation was usually taken due to the lack of an explicit computation of the probability distributions and mean values in the literature, prior to the publication of these results~\cite{Carmona:2022dtp}.

\section{LIV effects in neutrinos: study of the propagation and the flux}

\subsection{Model of continuous energy loss}
\label{sec:neu_prop_cont}

We want to study how these two new decays affect the propagation of neutrinos. For that, let us consider first an individual neutrino traveling freely from an extragalactic source to our detector at Earth. In order to deal with cosmological distances, we are going to use the redshift $z$ as a way to follow the trajectory of the neutrino. Then, the neutrino is emitted at $z_e$ and detected at $z_d=0$. During this trip, the neutrino will suffer a continuous adiabatic loss of energy due to the expansion of the Universe. This energy loss can be characterized by a differential equation
\be
    \frac{1}{E}\frac{dE}{dz}=\frac{1}{1+z} \,.
    \label{eq:E_evol_classic}
\ee
This process is deterministic, so the energy of the neutrino can be followed all along the trajectory. If this is the only mechanism of energy loss, like in \abb{SR}, its propagation is trivial and the relation between the emission and detection energies is given by
\be 
    E_e= (1+z_e) E_d\,.
    \label{eq:Ee_Ed_classic}
\ee

If we consider now the \abb{LIV} framework previously discussed, the neutrino can suffer additionally two mechanisms of decay. For this first analysis, let us consider only muon and tau neutrinos and only the \abb{VPE} decay. In the \abb{VPE} process, there is only one neutrino in the initial and final states, i.e., the number of neutrinos is conserved. This way, we can picture the pair emission as a process of energy loss. In contrast to the expansion of the universe, the decays are intrinsically stochastic processes. Therefore, during the propagation, the neutrino has some probability of losing energy, which is proportional to its total decay width,
\be
    \Gamma_{\nu_{\mu,\tau}}^{(e)} (E)= \frac{E^5}{192\pi^3} \frac{g^4}{M_W^4} \qty[(s_W^2-1/2)^2 + (s_W^2)^2]  \qty(\frac{E}{\Lambda})^{3n} c_n^{(e)} \,,
    \tag{See Eq.~\eqref{eq:total_decay_muon_tau}}
\ee
where we have shortened the notation using $\Gamma_{\nu_{\mu,\tau}}^{(e)} \equiv \Gamma_{\nu_{\mu,\tau}\rarrow \nu_{\mu,\tau} e^- e^+}$. Then, the energy of each individual neutrino becomes undetermined during its propagation making impossible to follow its evolution. However, if instead of considering the propagation of one neutrino we consider a large enough number of them, these probabilities should allow us to study the typical (mean) evolution of the energy along the trajectory. The differential variation of the energy $dE$ in a differential time $dt$ is given by the product of the decay width, $\,\Gamma_{\nu_{\mu,\tau}}^{(e)}(E)$, times the energy of the neutrino, $E$, and the mean fraction of energy loss, $\avg{1-x}$ (with $x$ the fraction of energy of the produced neutrino). Then,
\be
    \frac{dE}{dt} = - E\, \Gamma_{\nu_{\mu,\tau}}^{(e)}(E) \avg{1-x} \,.
    \label{eq:vpe_rate}
\ee
One can write the previous equation in terms of the redshift using the relation between $dt$ and $dz$, given by Eqs.~\eqref{eq:dtdz} and \eqref{eq:H(z)}. Hence, the evolution of the energy due to electron-positron pair emission per differential redshift step is given by
\be
    \frac{1}{E}\frac{dE}{dz} = \frac{\Gamma_{\nu_{\mu,\tau}}^{(e)}(E)}{(1+z)H(z)} \avg{1-x} \equiv \qty(\frac{E}{E_{\nu_{\mu,\tau}}^{(e)}})^{5+3n} \frac{\avg{1-x}}{(1+z)\sqrt{(1+z)^3 \Omega_m+ \Omega_\Lambda}} \,,
    \label{eq:E_evol_VPE}
\ee
where we have defined
\be 
    E_{\nu_{\mu,\tau}}^{(e)} \coloneqq \qty(\frac{32}{192\pi^3} \frac{G_F^2}{H_0 \Lambda^{3n}} \qty[(s_W^2-1/2)^2 + (s_W^2)^2] c_n^{(e)})^{-1/(5+3n)} \,,
    \label{eq:E_r_VPE}
\ee
whose value can also be easily computed in terms of the constants presented in Sec.~\ref{sec:comparison}:  \mbox{$E_{\nu_{\mu,\tau}}^{(e)}=[(c_0\times c_{\nu_{\mu,\tau}}^{(e)}\times c_n^{(e)})/(H_0 \Lambda^{3n})]^{-1/(5+3n)}$}, and where we have used $G_F=(\sqrt{2}\,g^2/8M_W^2)$.

Let us note that the energy $ E_{\nu_{\mu,\tau}}^{(e)}$ acts as an effective threshold for the electron-positron pair emission, because if the energy is above that energy scale, then
\be 
    \qty(E/E_{\nu_{\mu,\tau}}^{(e)})^{5+3n} = \Gamma_{\nu_{\mu,\tau}}^{(e)}(E)/H_0 \gg 1 \,,
    \label{eq:vpe_strong}
\ee
because of the large power $(5+3n)$ of the quotient of energies, and so the energy loss in Eq.~\eqref{eq:E_evol_VPE} will be very strong. We can call this the strong regime. On the contrary, if the energy is below the effective threshold, the energy loss by pair emission is negligible with respect to the energy loss due to the expansion of the universe, recovering Eq.~\eqref{eq:E_evol_classic}. We can call this the weak regime. In this way, the \abb{VPE} process has two different thresholds: a kinematical one, $E_\text{th}^{(e)}$, and an effective or dynamical one, $E_{\nu_{\mu,\tau}}^{(e)}$. The greater of the two will control at which energy the pair emission stops or is negligible. However, the relative magnitude between both is a function of the scale of new physics and the order of the correction,
\be 
    \qty(E_{\nu_{\mu,\tau}}^{(e)}/E_\text{th}^{(e)})^{5+3n} \propto \Lambda^{n/(n+2)} \,.
\ee

\begin{figure}[tb]
    \centering
    \includegraphics[width=0.75\textwidth]{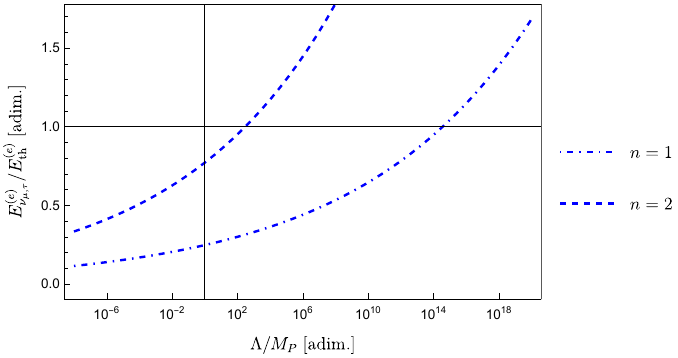}
    \caption{Quotient between the dynamical and kinematical thresholds of the emission of an electron-positron pair by a muon or tau neutrino for $n=1$ (dot-dashed) and $n=2$ (dashed). The grid lines cross at (1,1).}
    \label{fig:threshold_comp}
\end{figure}
The evolution of the quotient $(E_{\nu_{\mu,\tau}}^{(e)}/E_\text{th}^{(e)})$ as a function of the scale of new physics $\Lambda$ is also shown in Fig.~\ref{fig:threshold_comp}. We see there that for values of $\Lambda$ below or equal the Planck scale, $E_\text{th}^{(e)}$ is always greater than $E_{\nu_{\mu,\tau}}^{(e)}$. This implies that for $\Lambda\leq M_P$, the kinematical threshold is the relevant scale to study the stop of the electron-positron pair emission; additionally, it implies that the energy loss is always in the strong regime because, for every neutrino producing an electron-positron pair, it holds that $E \geq E_\text{th}^{(e)} > E_{\nu_{\mu,\tau}}^{(e)}$ (see Eq.~\eqref{eq:vpe_strong}).

Let us consider neutrinos emitted with energy above the threshold, so they are affected by \abb{VPE} in some part of their trajectory. Combining the decay energy loss, Eq.~\eqref{eq:E_evol_VPE}, with the standard adiabatic energy loss, Eq.~\eqref{eq:E_evol_classic}, we can get a differential equation for the evolution of the energy of a neutrino, while its energy is above $E_\text{th}^{(e)}$,
\be 
    \frac{1}{E} \frac{dE}{dz} = \frac{1}{1+z} + \qty(\frac{E}{E_{\nu_{\mu,\tau}}^{(e)}})^{5+3n} \frac{\avg{1-x}}{(1+z)\sqrt{(1+z)^3 \Omega_m + \Omega_\Lambda}} \,.
    \label{eq:E_evol}
\ee
This equation will rule the evolution of the energy until the neutrinos reach the kinematical threshold energy or are detected. In the last case, one can integrate\footnote{To integrate Eq.~\eqref{eq:E_evol} one can make a change of variables from $E$ to the redshift-invariant variable \mbox{$\tilde E=E/(1+z)$}.} Eq.~\eqref{eq:E_evol} from the emission until the detection, hence obtaining that the relation between the emission and detection energies is
\be 
    E_e = (1+z_e) E_d \qty[1-\qty(\frac{E_d}{E_{\nu_{\mu,\tau}}^{(e)}})^{5+3n} \avg{1-x}\,J_n\qty(z_e,0)]^{-1/(5+3n)}\,, \quad (E_d\ge E_\text{th}^{(e)})\,,
    \label{eq:Ee_Ed_th}
\ee
where $J_n(z_i,z_f)$ is a shortcut for
\be 
    J_n(z_i,z_f) \coloneqq \int_{z_f}^{z_i} dz \frac{(1+z)^{4+3n}}{\sqrt{(1+z)^3 \Omega_m + \Omega_\Lambda}} \,.
\ee

Let us note that, contrary to the trivial case shown in Eq.~\eqref{eq:Ee_Ed_classic}, now the equation relating emission and detection does not always have solution for every combination of the energy of emission, detection, and location of the source. In fact, we are restricted to the parameters that make the term inside of the square brackets larger than zero, i.e. 
\be
     J_n(z_e,0) < \frac{1}{\avg{1-x}} \, \qty(\frac{E_{\nu_{\mu,\tau}}^{(e)}}{E_d})^{5+3n} \ll 1\,, \quad (E_d\ge E_\text{th}^{(e)}) \,.
     \label{eq:condition_below}
\ee
Recalling that we are in the case $E_d\ge E_\text{th}^{(e)}>E_{\nu_{\mu,\tau}}^{(e)}$, the value of the right part of the inequality is very small, and so we expect the limits of the integral $J_n(z_i,z_f)$ to be very close: $z_e\approx 0$. In other words, the number of sources capable of emit neutrinos that will be detected with energies above the \abb{VPE} threshold is close to zero, and so, we do not expect to detect neutrinos with energies above $E_\text{th}^{(e)}$.

If we now consider neutrinos detected with energies below $E_\text{th}^{(e)}$, then Eq.~\eqref{eq:E_evol} will rule the evolution of their energy until they reach the energy threshold at some point $z_\text{th}^{(e)}$ of the trajectory. From that point to the detector the evolution is trivial, so we can solve $z_\text{th}^{(e)}$ as a function of the detection energy, 
\be 
    E_d = E_\text{th}^{(e)}/(1+z_\text{th}^{(e)} ) \srarrow z_\text{th}^{(e)} (E_d) \coloneqq E_\text{th}^{(e)}/E_d-1 \,, \quad (E_d\le E_\text{th}^{(e)})\,.
    \label{eq:zth}
\ee
Integrating Eq.~\eqref{eq:E_evol} from the emission to $z_\text{th}^{(e)}(E_d)$, we get that the relation between the emission and detection energies is now
\be 
    E_e = (1+z_e) E_d \qty[1-\qty(\frac{E_d}{E_{\nu_{\mu,\tau}}^{(e)}})^{5+3n} \avg{1-x}\,J_n\qty(z_e,z_\text{th}^{(e)}(E_d))]^{-1/(5+3n)} \,, \quad (E_d\le E_\text{th}^{(e)})\,.
    \label{eq:Ee_Ed_th_2}
\ee
One more time, the existence of solution of the previous equation imposes
\be
     J_n\qty(z_e,z_\text{th}^{(e)}(E_d)) < \frac{1}{\avg{1-x}} \, \qty(\frac{E_{\nu_{\mu,\tau}}^{(e)}}{E_d})^{5+3n}\,, \quad (E_d\le E_\text{th}^{(e)})\,,
     \label{eq:condition_above}
\ee
but in contrast to the previous case, we now have to possibilities: \mbox{$E_{\nu_{\mu,\tau}}^{(e)}<E_d<E_\text{th}^{(e)}$} or  \mbox{$E_d<E_{\nu_{\mu,\tau}}^{(e)}<E_\text{th}^{(e)}$}. In the last case, the condition \eqref{eq:condition_above} is easily satisfied without imposing any condition over $z_e$. However, the most interesting case is to consider the most energetic neutrinos below the threshold, \mbox{$E_\text{th}^{(e)}>E_d>E_{\nu_{\mu,\tau}}^{(e)}$}, because in that case Eq.~\eqref{eq:Ee_Ed_th_2} only has solution if $z_e$ is below some critical value $z_c(E_d)$, whose value is defined as the solution of imposing an equality in Eq.~\eqref{eq:condition_above},
\be
     J_n\qty(z_c(E_d),z_\text{th}^{(e)}(E_d)) \equiv \frac{1}{\avg{1-x}} \, \qty(\frac{E_{\nu_{\mu,\tau}}^{(e)}}{E_d})^{5+3n}\,, \quad (E_d\le E_\text{th}^{(e)})\,.
     \label{eq:zc}
\ee

Let us note that $z_c(E_d)$ has to be a monotonic decreasing function on $E_d$. Then, there exists some energy $E_\text{cut}^{(e)}$ for which $z_c(E_\text{cut}^{(e)})$ will coincide with the closest source $z_\text{min}$ of the neutrino source distribution. This means that we are not going to find any neutrino with energy above $E_\text{cut}^{(e)}$, since it would require a source closer than $z_\text{min}$.

Using the same arguments as in Eq.~\eqref{eq:condition_below}, we can conclude that $z_c(E_d)$ is close to $z_\text{th}^{(e)}(E_d)$; but let us note how the condition is not as strong as in Eq.~\eqref{eq:condition_below}. The difference between $z_c(E_d)$ and $z_\text{th}^{(e)}(E_d)$ is very important to determine the shape of the detected neutrino flux (see Appendix~\ref{sec:flux_continuous} for more details); however, if we are only interested in obtaining a rough approximation of the energy $E_\text{cut}^{(e)}$ one can approximate $z_c(E_d)\approx z_\text{th}^{(e)}(E_d)$. Then, in order to obtain the energy of the cutoff, instead of solving the difficult equation $z_c(E_\text{cut}^{(e)})\equiv z_\text{min}$, we can solve instead $z_\text{th}^{(e)}(E_\text{cut}^{(e)})\equiv z_\text{min}$, which allows us to approximate the energy of the cutoff produced by the emission of electron-positron pairs as
\be 
    E_\text{cut}^{(e)} \approx E_\text{th}^{(e)}/(1+z_\text{min}) \,.
\ee
Using that the threshold energy is a function of $\Lambda$ and $n$, we can write the cutoff energy as
\be 
    E_\text{cut}^{(e)} \approx \frac{\Lambda^{n/(n+2)}}{1+z_\text{min}} \; (2m_e^2)^{1/(n+2)} \,,
    \label{eq:cutoff_VPE}
\ee
which gives us a formula to predict an approximate location of the cutoff as a function of the parameters of the new physics. Let us note how the formula does not depend on the shape of the energy distribution, but only on the closest source. In the same line, the equation is independent on the model of emission, because all the neutrinos emitted above the threshold energy $E_\text{th}^{(e)}$  will lose energy almost instantaneously (without modifying their redshift), and will act as neutrinos emitted with energy $E_\text{th}^{(e)}$.

The relation given in Eq.~\eqref{eq:cutoff_VPE} can also be inverted, so that we can use it to put constraints on the possible values of $\Lambda$ given the observation of neutrinos of energy $E_d$, 
\be 
    \Lambda > \qty(E_d (1+z_\text{min}))^{(n+2)/n} (2m_e^2)^{-1/n} \,.
    \label{eq:Lambda_VPE}
\ee
Very recently, \abb{IC} reported~\cite{IceCube:2021rpz} an event compatible with the measurement of a neutrino of the Glashow resonance~\cite{Glashow:1960zz} (6.3 PeV). Eq.~\eqref{eq:Lambda_VPE} provides us an easy way to update the current constraints on the scale of new physics from the observation of high-energy neutrinos. The new constraints are shown in Tab.~\ref{tab:constraints_VPE}.
\begin{table}[htb]
    \centering
    \caption{Updated constraints of the scale of new physics from the detection of a neutrino at the Glashow resonance, considering a superluminal \abb{LIV} scenario with only \abb{VPE}.}
    \begin{tabular}{ccc} \toprule
       & $n=1$ & $n=2$ \\ \midrule
      $\Lambda/M_p>$ & $3.92\E{7}$ & $4.50\E{-3}$ \\ \bottomrule
    \end{tabular}
    \label{tab:constraints_VPE}
\end{table}

If one wants to make additional predictions, we need a model for the neutrino flux. One attempt to obtain a prediction for the detected neutrino flux has been developed in Appendix~\ref{sec:flux_continuous}. However, we have found that the shape of the flux depends strongly on the value of $z_c(E_d)$. Given that we cannot determine $z_c(E_d)$ analytically, and that the numerical solutions rely on approximations that can affect the flux to a large extent, we found that the computation of the flux through this method is not reliable. Additionally, one cannot include the effect of the neutrino-antineutrino pair emission in this model (as the number of neutrinos is not conserved). All these reasons show us the need to move towards a different model.

\subsection{Instantaneous cascade model}
\label{sec:neu_prop_ins}

In the previous section we have seen that, due to the peculiar energy dependence of the \abb{VPE} decay width, particles emitted with energies above $E_\text{th}^{(e)}$ reach the threshold energy almost instantaneously and are detected with energy $E_d\approx E_\text{th}^{(e)}/(1+z_e)$. As the neutrino-antineutrino emission decay width has the same energy dependence, this encourages us to approximate both decays as instantaneous effects. For that, we are again interested in a regime in which the effects are strong compared to the expansion of the universe. If we focus anew on the muon and tau neutrinos, this means both $\Gamma_{\nu_{\mu,\tau}}^{(e)}(E)$  and $\Gamma_{\nu_{\alpha}}^{(\nu)}(E)$ should be much greater than $H_0$, where 
\be 
    \Gamma_{\nu_{\alpha}}^{(\nu)}(E) \coloneqq \sum_{\beta=\mu,\tau,e} \Gamma_{\nu_\alpha\rarrow \nu_\alpha \nu_\beta \bar\nu_\beta} = 3\times \frac{E^5}{192\pi^3} \frac{g^4}{M_W^4}  \qty(\frac{E}{\Lambda})^{3n} c_n^{(\nu)}\,.
    \label{eq:Gamma_nu}
\ee

The condition for the strong regime for the \abb{VPE} has already been studied in the previous section for the case of muon and tau neutrinos, Eq.~\eqref{eq:vpe_strong}. Imposing the same condition for the \abb{NSpl} we get that
\be 
    \Gamma_{\nu_\alpha}^{(\nu)}(E)/H_0 \equiv \qty(E/E_{\nu_\alpha}^{(\nu)})^{5+3n} \gg 1 \,,
    \label{eq:nspl_strong}
\ee
which allows us to define an effective threshold for neutrino-antineutrino pair emission
\be 
    E_{\nu_\alpha}^{(\nu)} \coloneqq \qty(\frac{96}{192\pi^3} \frac{G_F^2}{H_0 \Lambda^{3n}}\, c_n^{(\nu)})^{-1/(5+3n)} \,,
    \label{eq:E_r_NSpl}
\ee
whose value can also be easily computed in terms of the constants presented in Sec.~\ref{sec:comparison}: \mbox{$E_{\nu_\alpha}^{(\nu)}=[(3\times c_0\times c_{\nu_{\alpha}}^{(\nu)}\times c_n^{(\nu)})/(H_0 \Lambda^{3n})]^{-1/(5+3n)}$}.

We see then that an effective dynamical threshold emerges naturally for the neutrino-antineutrino pair emission as well. Additionally, in contrast to the \abb{VPE}, for the neutrino-antineutrino decay there is no kinematical threshold\footnotemark, so the effective threshold $E_{\nu_\alpha}^{(\nu)}$ will also control when the \abb{NSpl} can be disregarded. Then, every neutrino emitted with energy above $E_{\nu_\alpha}^{(\nu)}$ will decay very fast (instantaneously) until reaching the effective threshold. \footnotetext{As previously mentioned, actually there exists a kinematical threshold for the neutrino-antineutrino pair emission; however, it can be disregarded due to the small value of the neutrino mass.}

\begin{figure}[tb]
    \centering
    \includegraphics[width=0.85\textwidth]{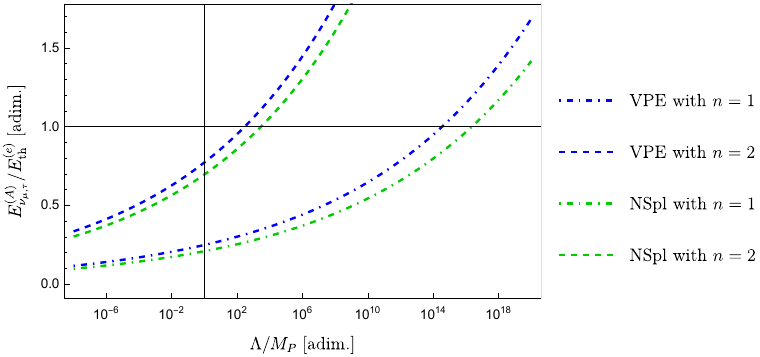}
    \caption{Quotient between the dynamical thresholds (of \abb{VPE} in blue and \abb{NSpl} in green) and the kinematical threshold of \abb{VPE}, by a muon or tau neutrino, and for $n=1$ (dot-dashed) and $n=2$ (dashed). The grid lines cross at (1,1).}
    \label{fig:threshold_comp_2}
\end{figure}

If we compare the two effective thresholds, of electron-positron and neutrino-antineutrino pair emission, we will see that $E_{\nu_\alpha}^{(\nu)}$ is always smaller than $E_{\nu_{\mu,\tau}}^{(e)}$. In fact, the relation does not depend on the scale of new physics $\Lambda$, but only on $n$,
\be 
    \frac{E_{\nu_\alpha}^{(\nu)}}{E_{\nu_{\mu,\tau}}^{(e)}} = \qty[\frac{3\times c_{\nu_{\alpha}}^{(\nu)}\times c_n^{(\nu)}}{c_{\nu_{\mu,\tau}}^{(e)}\times c_n^{(e)}}]^{-1/(5+3n)} \approx
    \begin{cases}
        0.84 \quad\text{for }n=1 \\
        0.90 \quad\text{for }n=2
    \end{cases} \,.
    \label{eq:VPE_thresholds_comp}
\ee
This assures us that the effective threshold of the neutrino-antineutrino pair emission, $E_{\nu_\alpha}^{(\nu)}$, is smaller than the kinematical threshold of the electron-positron pair emission, $E_\text{th}^{(e)}$, for all the values of $\Lambda$ below or equal to the Planck scale (see Fig.~\ref{fig:threshold_comp_2}).

As $E_{\nu_\alpha}^{(\nu)}<E_\text{th}^{(e)}$, when the neutrinos fall below the kinematical threshold of the \abb{VPE} process, they still will produce neutrino-antineutrino pairs. This way, the last threshold before the classical propagation is $E_{\nu_\alpha}^{(\nu)}$. Every neutrino emitted above $E_{\nu_\alpha}^{(\nu)}$  (independently if above or below the kinematical threshold of the \abb{VPE}) will lose energy until reaching the \abb{NSpl} effective threshold. Then, the neutrino-antineutrino pair emission is the most restrictive effect and will control at which energy we do expect a cutoff in the neutrino flux,
\be 
    E_\text{cut}^{(\nu)} \approx \frac{E_{\nu_\alpha}^{(\nu)}}{1+z_\text{min}} =  \frac{\Lambda^{3n/(5+3n)}}{1+z_\text{min}} \qty(\frac{96}{192\pi^3} \frac{G_F^2}{H_0}\,c_n^{(\nu)})^{-1/(5+3n)} \,,
    \label{eq:cutoff_NSpl}
\ee
or written the other way around, we can get a bound on $\Lambda$,
\be 
    \Lambda > \qty(E_d (1+z_\text{min}))^{(5+3n)/(3n)} \qty(\frac{96}{192\pi^3} \frac{G_F^2}{H_0}\,c_n^{(\nu)})^{1/3n} \,,
    \label{eq:Lambda_NSpl}
\ee
from the observation of neutrinos with an energy $E_d$. 

We can again put constraints on the scale of new physics using the detection by \abb{IC} of a neutrino of the Glashow resonance. The results are shown in Tab.~\ref{tab:constraints_NSpl}.
\begin{table}[htb]
    \centering
    \caption{Updated constraints of the scale of new physics from the detection of a neutrino of the Glashow resonance, considering a superluminal \abb{LIV} scenario.}
    \begin{tabular}{ccc} \toprule
       & $n=1$ & $n=2$ \\ \midrule
      $\Lambda/M_p>$ & $3.71\E{8}$ & $1.38\E{-2}$ \\ \bottomrule
    \end{tabular}
    \label{tab:constraints_NSpl}
\end{table}
Let us note that these new constraints are more restrictive than the ones we found using the \abb{VPE} scenario.

Let us recall that one can use Eqs.~\eqref{eq:cutoff_NSpl} and \eqref{eq:Lambda_NSpl}, only when the \abb{NSpl} is the most restrictive effect, i.e., when $E_{\nu_\alpha}^{(\nu)}<E_\text{th}^{(e)}$. Once this relation is inverted, the kinematical threshold of the \abb{VPE} will control the cutoff, and we should use Eqs.~\eqref{eq:cutoff_VPE} and \eqref{eq:Lambda_VPE} instead. One can identify for which value of $\Lambda$ this inversion occurs,
\be 
    \qty(E_{\nu_{\alpha}}^{(\nu)}/E_\text{th}^{(e)})^{5+3n} \equiv \qty(\Lambda/\Lambda_{\nu_{\alpha}}^{(\nu)})^{n/(n+2)} \,,
\ee
with $\Lambda_{\nu_{\alpha}}^{(\nu)}$ a constant given by
\be 
    \Lambda_{\nu_{\alpha}}^{(\nu)} \coloneqq (2m_e^2)^{(5+3n)/n} \times \qty(\frac{96}{192\pi^3} \frac{G_F^2}{H_0} c_n^{(\nu)})^{(n+2)/n} \,,
\ee
whose value can also be easily computed in terms of the constants presented in Sec.~\ref{sec:comparison}: \mbox{$\Lambda_{\nu_{\alpha}}^{(\nu)}=(2m_e^2)^{(5+3n)/n} \times [(3\times c_0\times c_{\nu_{\alpha}}^{(\nu)}\times c_n^{(\nu)})/H_0)]^{(n+2)/n}$}. Numerical values of this constant, for $n=1,2$, are given below
\be
    \Lambda_{\nu_{\alpha}}^{(\nu)} = 
    \begin{cases} 
        2.37 \times 10^{16} \;M_p & \text{for } n=1\\
        3.02 \times 10^{3} \;M_p & \text{for } n=2
    \end{cases} \,.
\ee

\begin{figure}[tb]
    \centering
    \includegraphics[width=0.8\textwidth]{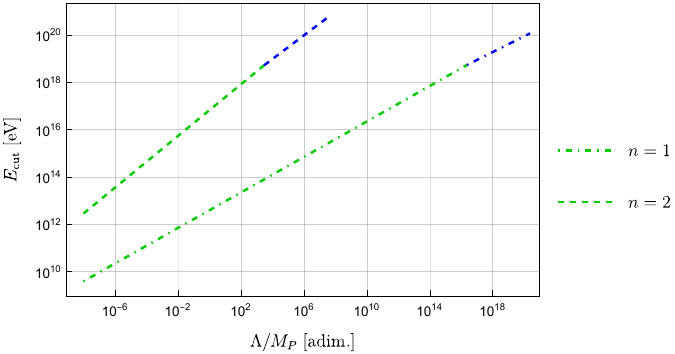}
    \caption{Approximate energy of the cutoff (for $z_\text{min}=0$) in the detected spectrum of muon and tau neutrinos, for $n=1$ (dashed) and $n=2$ (dot-dashed). The green and blue values are controlled by the \abb{NSpl} and \abb{VPE}, respectively.}
    \label{fig:Ecut}
\end{figure}

In Fig.~\ref{fig:Ecut} we show the predicted energies for the cutoff, taking into account the change in the dominant threshold we have just discussed above, for different values of the scale $\Lambda$.

In order to make additional predictions we need a model for the detected neutrino flux. We did another attempt to develop an analytical model for the neutrino flux applying equipartition of the energy to the instantaneous cascade model; however, the approximations used to build the model lead to bad predictions (discontinuities) close to the cutoff (for more information see Appendix~\ref{sec:flux_instantaneous}). 

This way, we conclude that the best way to proceed is to implement the instantaneous approximation in a Monte Carlo software which simulates the propagation of neutrinos from the sources to the Earth. In the next section we carry out this approach.

\subsection{Neutrino flux using SimProp}
\label{sec:monte_carlo}

The use of Monte Carlo software is well established in the study of the propagation of ultra high-energy astroparticles. There are very powerful alternatives like CRPropa~\cite{AlvesBatista:2022vem} or SimProp~\cite{Aloisio:2017iyh}; however, they usually focus on simulate particle propagation for different astrophysical scenarios in the framework of conventional physics. In order to study the propagation of neutrinos in a superluminal \abb{LIV} scenario, we can make use of the machinery these programs have already included to simulate the propagation of particles, but adding the ingredients of the new physics. We have developed our implementation of the new physics based on SimProp. In order to do that, we initiated a fruitful collaboration with Denise Boncioli, one of the coauthors of the software. In the future, we would also like to crosscheck our results using another program like CRPropa.

SimProp is a Monte Carlo software mostly focused in the propagation of cosmic rays. During their propagation, they interact with the low energy photons of the backgrounds (\abb{CMB} and \abb{EBL}) producing secondary particles, which are also propagated. Neutrinos are particles already included in SimProp, as a secondary product of the propagation of the cosmic rays and with a trivial propagation to Earth; however, in this chapter we are interested in neutrinos as primary particles (astrophysical neutrinos). Then, in order to make SimProp capable of simulating the propagation of astrophysical neutrinos on a superluminal \abb{LIV} scenario we have done two main modifications.

The first one is to add the possibility to use neutrinos as a primary particles. This implementation is merely a technical issue, so we will omit the details here. However, once this feature is implemented, specifying a certain source distribution and a certain emission spectrum, we can simulate the emission of neutrinos from astrophysical sources. The second implementation regards the modified propagation of neutrinos. In order to include the new two decays, we must replace the current subroutine, in charge of the trivial propagation, with a new one. The technical implementation of the main ingredients are shown in Appendix~\ref{sec:simprop}.

In SimProp, the propagation of cosmic rays is done in redshift steps from their emission to the Earth; however, the decays, as beta decay or pion decay, are treated as instantaneous. We have seen in the previous sections that the neutrino decays also have an instantaneous description (the instantaneous cascade model), so this is the kind of implementation we are going to consider.

Each time a neutrino is produced, the new subroutine is called. This script has to check whether the neutrino energy is above the different thresholds and randomly choose, according to the probabilities assigned by the total decay widths, a process to undergo. Additionally, once the process is chosen, the script also has to characterize the produced particles.

\begin{figure}[tbp]
    \centering
    \begin{minipage}{0.49\textwidth}
        \centering
        \includegraphics[width=\textwidth]{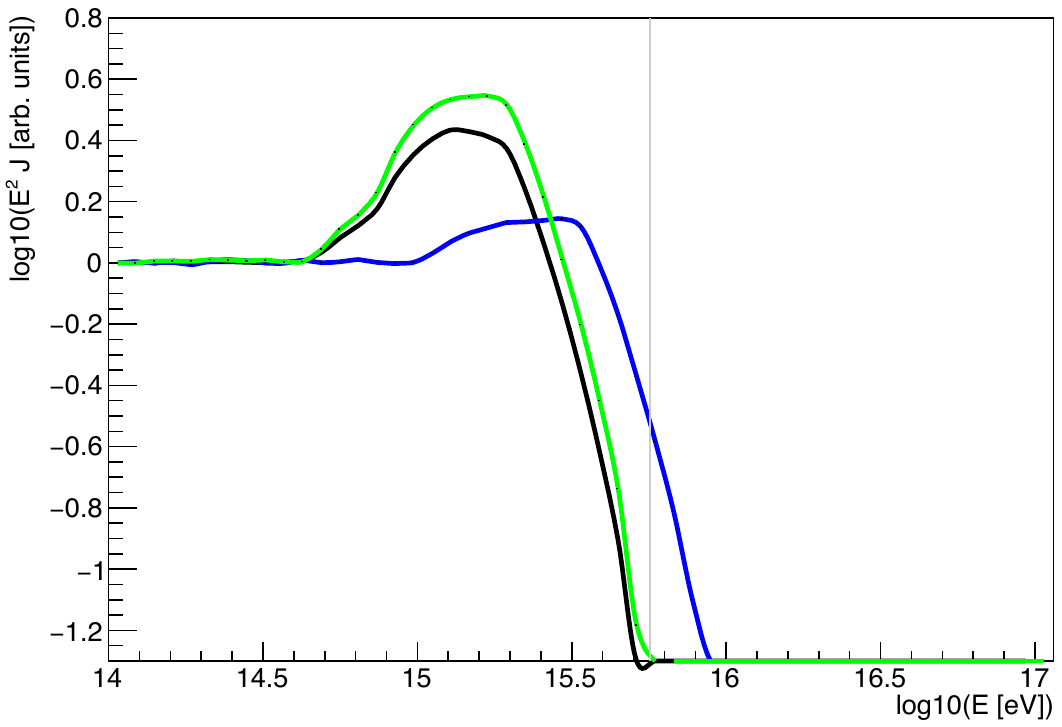}
        \caption{Neutrino flux at Earth (black) using mean values to compute the energies of the product particles. The only \abb{VPE} (blue) and only \abb{NSpl} (green) scenarios are also presented. The parameters of the simulation are discussed in the following paragraphs.}
        \label{fig:flux_mean}
    \end{minipage}%
    \hfill
    \begin{minipage}{0.49\textwidth}
        \centering
        \includegraphics[width=\textwidth]{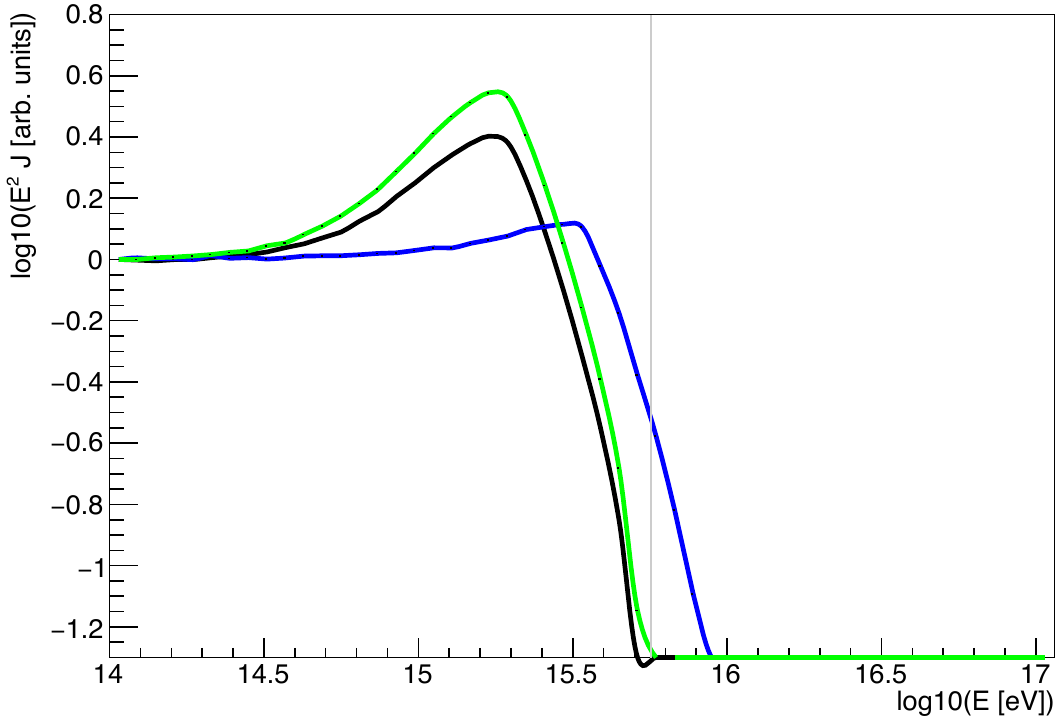}
        \caption{Neutrino flux at Earth (black) using probability distributions to compute the energies of the product particles. The only \abb{VPE} (blue) and only \abb{NSpl} (green) scenarios are also presented. The parameters of the simulation are discussed in the following paragraphs.}
        \label{fig:flux_no_mean}
    \end{minipage}
\end{figure}

Let us note that there are two different approaches regarding the produced particles. On the one hand, we have computed explicitly the probability distributions of the final energies (see Eqs.~\eqref{eq:energy_dist_VPE_muon}, \eqref{eq:energy_dist_VPE_electron}, and \eqref{eq:energy_dist_NSpl}), so we can randomly sample the final energies from the distributions. This approach is more precise at the expense of more computational weight and time. On the other hand, in Sec.~\ref{sec:comparison} we also computed the mean energy fractions of the product particles. Following this approach one can get approximated final energies using the mean values (see Tabs.~\ref{tab:mean_values_VPE_muon}, \ref{tab:mean_values_VPE_electron}, and \ref{tab:mean_values_NSpl}). The larger the number of neutrinos in the flux, the better approximation this last approach will be. An example of comparison between these two approaches is shown in Figs.~\ref{fig:flux_mean} and \ref{fig:flux_no_mean}.
There, one can see how the use of the mean values produce a softening of the bump, even producing a slight shift of the maximum of the bump in some cases. The cutoff is mostly unaffected.

\begin{figure}[tbp]
    \centering
    \includegraphics[width=0.65\textwidth]{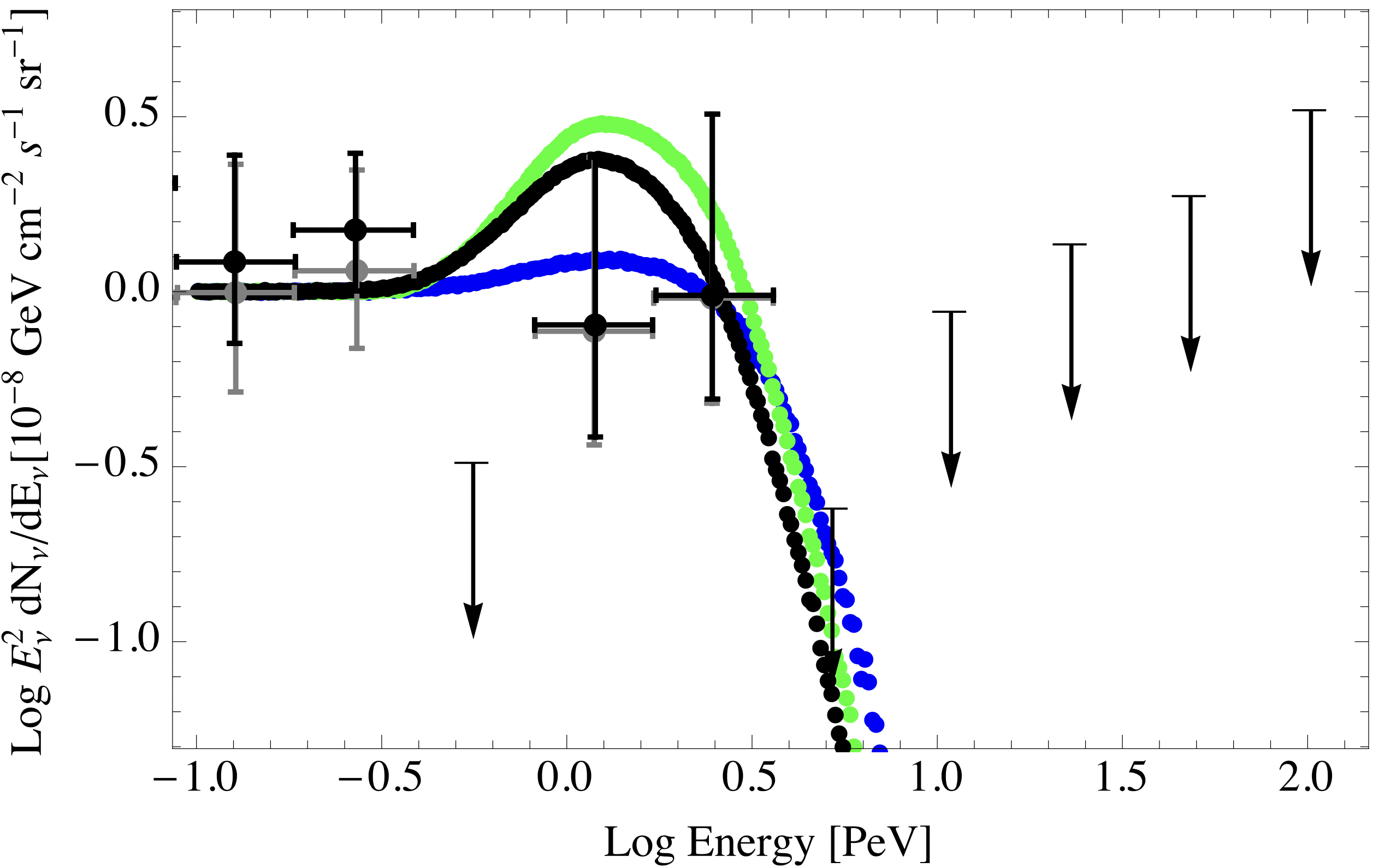}
    \caption{Neutrino flux at Earth (black) computed in~\cite{Stecker:2014oxa} using a Monte Carlo simulation, and \abb{IC} three-years data~\cite{IceCube:2014stg}. The only \abb{VPE} (blue) and only \abb{NSpl} (green) scenarios are also presented.}
    \label{fig:flux_stecker}
\end{figure}

Before the end of 2021 (when \abb{IC} reported an event compatible with the measurements of a neutrino of the Glashow resonance at 6.3 PeV) the detected neutrino spectra showed an absence of events above 2 PeV. In~\cite{Stecker:2014oxa,Stecker:2014xja}, Stecker et al. proposed a cutoff in the detected spectrum to explain this lack of events, with possible origin in \abb{LIV}. They found that, for certain astrophysical assumptions (that will be named in the next paragraph), a value of $\Lambda$ such that $E_\text{th}^{(e)}=10$ PeV and $n=2$ could satisfactorily explain the mentioned absence of events and also be compatible with measured data by \abb{IC} at that time~\cite{IceCube:2014stg} (see Fig.~\ref{fig:flux_stecker}).

We have performed a simulation using our modified version of SimProp and the same parameters of new physics used in~\cite{Stecker:2014oxa,Stecker:2014xja}, i.e., $n=2$ and $\Lambda\approx 1.13\cdot 10^{-2} M_P$, such that $E_\text{th}^{(e)}=10\unit{PeV}$. We are using also the same astrophysical assumptions, i.e., we are taking the source distribution as the \abbdef{SFR}{Stellar Formation Rate}, sampled from~\cite{Behroozi:2012iw}, within $z=0.5$ and $z=2$.
Additionally, we assume sources with the same emission spectrum: 50\% of emitting a neutrino or antineutrino, and an energy dependence given by a single inverse power law with index $\gamma=2$ from 100 TeV to 100 PeV, that is to say $\Phi_e(E_e)=(E_0/E_e)^2$, with $E_0$ a constant related with the luminosity of the sources. Finally, the value of $E_0$ is absorbed in a global normalization constant which is set to $\approx 10^{-8} \unit{GeV\;cm^{-2}\;s^{-1}\; sr^{-1}}$. The results of these simulations have been already shown in Figs.~\ref{fig:flux_mean} and \ref{fig:flux_no_mean}, where the gray vertical line shows the location of the cutoff predicted by Eq.~\eqref{eq:cutoff_NSpl}.

Comparing Figs.~\ref{fig:flux_no_mean} and \ref{fig:flux_stecker}, we can see how our performed simulation reproduces the global behaviour of the flux, i.e., a bump prior to a cutoff in the spectrum. However, one can also see some differences between both simulations. In the first place, given that in~\cite{Stecker:2014oxa} the performed simulation is using mean values, we see how the shape of the bump computed by Stecker et al. is closer to Fig.~\ref{fig:flux_mean} than to Fig.~\ref{fig:flux_no_mean}. This can also contribute to the slight shift of the energy of the maximum of the bump in the green and black curves (which correspond to the  \abb{NSpl} and \abb{NSpl} + \abb{VPE} scenarios, respectively). However, no one of the previously mentioned approximations can explain the much larger shift of the blue curve (corresponding to the only \abb{VPE} case), which is a direct consequence of the instantaneous cascade model. Under this approximation, the energy of the cutoff for each scenario can be predicted by dividing the relevant threshold by $(1+z_\text{min})$, and since the \abb{VPE} and \abb{NSpl} have different relevant thresholds, $E_\text{th}^{(e)}$ and $E_{\nu_{\alpha}}^{(\nu)}$ respectively, the cutoff will appear at different energies. This behaviour is reproduced by Figs.~\ref{fig:flux_mean} and \ref{fig:flux_no_mean}, but is not present in Fig.~\ref{fig:flux_stecker}. Additionally, the use of the instantaneous approximation may lead to a slight overestimation of the flux in the bump, given that it is assumed that all the neutrinos above the threshold will decay. This behaviour is also apparent when comparing Figs.~\ref{fig:flux_no_mean} and \ref{fig:flux_stecker}. In order to address these differences, and understand better the limitations of the instantaneous cascade model, we have initiated a collaboration with Floyd Stecker and Sean Scully, authors of~\cite{Stecker:2014oxa,Stecker:2014xja}. This work is currently in progress, and we hope both codes can be improved as a result of this collaboration.

\begin{figure}[tbp]
    \centering
    \begin{minipage}{0.49\textwidth}
        \centering
        \includegraphics[width=\textwidth]{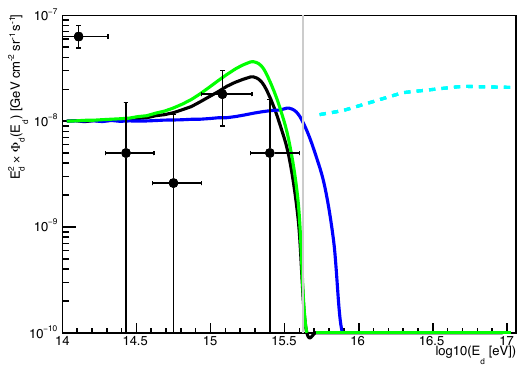}
        \caption{Neutrino flux at Earth (black) for $n=2$ and $\Lambda=1.38\E{-2} M_P$. The only \abb{VPE} (blue) and only \abb{NSpl} (green) scenarios are also presented. The \abb{IC} \abb{HESE} 7.5-year data (datapoints) and 90\% CL upper limits (dashed cyan) are taken from~\cite{IceCube:2020wum} and~\cite{IceCube:2018fhm}, respectively.}
        \label{fig:flux_glashow_n2}
    \end{minipage}%
    \hfill
    \begin{minipage}{0.49\textwidth}
        \centering
        \includegraphics[width=\textwidth]{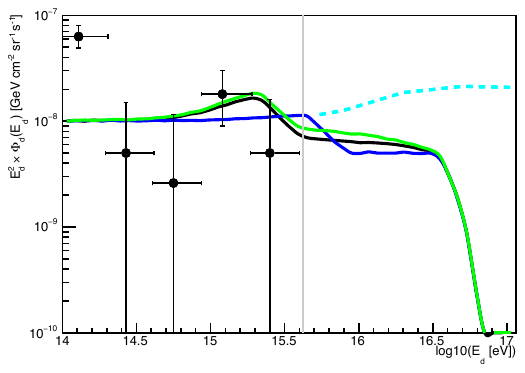}
        \caption{Neutrino flux at Earth (black) for $n=1$ and $\Lambda=3.71\E{8} M_P$. The only \abb{VPE} (blue) and only \abb{NSpl} (green) scenarios are also presented. The \abb{IC} \abb{HESE} 7.5-year data (datapoints) and 90\% CL upper limits (dashed cyan) are taken from~\cite{IceCube:2020wum} and~\cite{IceCube:2018fhm}, respectively.}
        \label{fig:flux_glashow_n1}
    \end{minipage}
\end{figure}

The analysis made in~\cite{Stecker:2014oxa} used the \abb{IC} data available at that time (the three-year data, which can be found in~\cite{IceCube:2014stg}). However, a few years ago the \abb{IC} collaboration released a new set of 7.5-year data and an analysis of \abbdef{HESE}{High Energy Starting Events}~\cite{IceCube:2020wum}. The \abb{IC} \abb{HESE} datasample only considers events with their interaction vertex contained inside the fiducial volume; this reduces the statistics respect to general studies, but it should help to disentangle the astrophysical flux from the background produced by atmospheric neutrinos.
Therefore, we can update the comparison made in~\cite{Stecker:2014oxa} using the \abb{HESE} data of~\cite{IceCube:2020wum}. However, this sample of events does not contain the recent Glashow resonance single event at 6.3 PeV (an update of the data is still pending release). This event constraints the scale of new physics to the values shown in Tab.~\ref{tab:constraints_NSpl}. As a consequence, we will consider the minimum value of $\Lambda$ compatible with the detection of a neutrino at 6.3 PeV for $n=2$ and $n=1$ (see Tab.~\ref{tab:constraints_NSpl}). The results  are shown in Figs.~\ref{fig:flux_glashow_n2} and \ref{fig:flux_glashow_n1}, respectively.

Firstly, let us discuss the spectrum above 250 TeV (last four datapoints). For the case $n=2$, Fig.~\ref{fig:flux_glashow_n2}, the relative increase of the detected flux around the PeV (second to last datapoint) looks to match the position of the bump of \abb{LIV}. For energies above $10$ PeV, there is an absence of events, represented by the 90\% CL upper limits for the detected flux (dashed cyan curve), obtained considering no detected events in that energy range~\cite{IceCube:2018fhm}. This lack of events is compatible with the presented cutoff at energy $E_\text{cut}$ (gray vertical line, obtained using Eq.~\ref{eq:cutoff_NSpl}). Instead, for the case $n=1$, Fig.~\ref{fig:flux_glashow_n1}, we do not have a cutoff at the energy $E_\text{cut}$, but only a decrease in the flux. This is due to the fact that for $n=1$ only the neutrinos are superluminal. Then, the flux of antineutrinos does not has a cutoff due to the propagation\footnotemark.
\footnotetext{The cutoff that appears at higher energies is just a consequence of the intrinsic cutoff in the emission caused by considering an emission spectrum only up to $100$ PeV.}

There exists an additional datapoint in Figs.~\ref{fig:flux_glashow_n2} and \ref{fig:flux_glashow_n1}, below 250 TeV, which is not compatible with the computed spectrum. The incompatibility stems from the fact that this datapoint does not follow the approximate general tendency $E^{-2}$ of the other \abb{HESE} data. This incompatibility have led \abb{IC} to consider new models to fit the data. In~\cite{IceCube:2020wum} some of them are discussed. One of them is still a single inverse power law, but with an spectral index closer to 3, instead of the previous fit $E^{-2}$. This would imply to revise the astrophysical assumptions, for example considering sources with an emission spectrum with $\gamma\approx 3$. This assumption has strong implications over the possible sources of astrophysical neutrinos. The visual comparison presented in this work does not intent to replace a parametric study using different combinations of parameters of new physics and different astrophysical assumptions; it, however, allows us to fulfill the purpose of this chapter, which is to understand the effects of a superluminal \abb{LIV} on the neutrino spectrum at the Earth.

\begin{subappendices}
    \section{Matrix elements of neutrino LIV processes}

Let us call $u$ to the spinor of the electron, $v$ the spinor of the positron, $\tilde u$ the spinor of the neutrino, and $\tilde v$ the spinor of the antineutrino.

\subsection*{Electron-positron pair emission (muon and tau neutrino)}
\label{sec:amplitude_VPE_muon}

Let us consider the decay of a muon or tau neutrino into an electron-positron pair. Then, it is only necessary to take into account the neutral channel. The matrix elements of the disintegration to left-handed and right-handed electrons are given by
\begin{small}
\begin{align}
    A_{\nu_{\mu,\tau}\rarrow \nu_{\mu,\tau} e_L^- e_L^+} =& \notag \\ &\qty[\bar{\tilde u}_L(\vec p')\qty(-i\frac{g}{2c_W}\gamma^\mu)\tilde u_L(\vec p)]
    \qty{-i\frac{\eta_{\mu\nu}}{M_Z^2}}
    \qty[\bar u_L(\vec p^-)\qty(-i\frac{g}{c_W}(s_W^2-1/2)\gamma^\nu) v_L(\vec p^+)] \,, \notag \\
    A_{\nu_{\mu,\tau}\rarrow \nu_{\mu,\tau} e_R^- e_R^+} =& \notag \\ &\qty[\bar{\tilde u}_L(\vec p')\qty(-i\frac{g}{2c_W}\gamma^\mu)\tilde u_L(\vec p)]
    \qty{-i\frac{\eta_{\mu\nu}}{M_Z^2}}
    \qty[\bar u_R(\vec p^-)\qty(-i\frac{g}{c_W} s_W^2 \gamma^\nu) v_R(\vec p^+)] \,.
    \tag{See Eq.~\eqref{eq:A_muontau}}
\end{align}
\end{small}
In order to get the square we should multiply by the complex conjugate, obtaining
\bea
    \abs{A_{\nu_{\mu,\tau}\rarrow \nu_{\mu,\tau} e_L^- e_L^+}}^2=\frac{1}{4} \frac{g^4}{M_W^4} (s_W^2-1/2)^2 \eta_{\mu\nu} \eta_{\rho\sigma} &\qty[\bar{\tilde u}_L(\vec p') \gamma^{\mu} \tilde u_L(\vec p)] \qty[\bar u_L(\vec p^-) \gamma^{\nu} v_L(\vec p^+)] \\ \times &\qty[\bar{\tilde u}_L(\vec p') \gamma^{\rho} \tilde u_L(\vec p)]^\dag \qty[\bar u_L(\vec p^-) \gamma^{\sigma} v_L(\vec p^+)]^\dag \,, \\
    \abs{A_{\nu_{\mu,\tau}\rarrow \nu_{\mu,\tau} e_R^- e_R^+}}^2=\frac{1}{4} \frac{g^4}{M_W^4} (s_W^2)^2 \eta_{\mu\nu} \eta_{\rho\sigma} &\qty[\bar{\tilde u}_L(\vec p') \gamma^{\mu} \tilde u_L(\vec p)] \qty[\bar u_R(\vec p^-) \gamma^{\nu} v_R(\vec p^+)] \\ \times &\qty[\bar{\tilde u}_L(\vec p') \gamma^{\rho} \tilde u_L(\vec p)]^\dag \qty[\bar u_R(\vec p^-) \gamma^{\sigma} v_R(\vec p^+)]^\dag \,.
\eea
Using that 
\be 
    \qty[\bar A \gamma^{\mu} B]^\dag = \qty[(A^\dag \gamma^0) \gamma^{\mu} B]^\dag = B^\dag (\gamma^\mu)^\dag \gamma^0 A = B^\dag (\gamma^0 \gamma^\mu \gamma^0) \gamma^0 A = (B^\dag \gamma^0) \gamma^\mu \mathbb{I}\, A = \qty[\bar B \gamma^\mu A] \,,
    \label{eq:conjugated}
\ee
we can rearrange the terms, getting
\bea
    \abs{A_{\nu_{\mu,\tau}\rarrow \nu_{\mu,\tau} e_L^- e_L^+}}^2=\frac{1}{4} \frac{g^4}{M_W^4} (s_W^2-1/2)^2 \eta_{\mu\nu} \eta_{\rho\sigma} &\qty[\bar{\tilde u}_L(\vec p') \gamma^{\mu} \tilde u_L(\vec p)] \qty[\bar{\tilde u}_L(\vec p) \gamma^{\rho} \tilde u_L(\vec p')] \\ \times & \qty[\bar u_L(\vec p^-) \gamma^{\nu} v_L(\vec p^+)]  \qty[\bar{v}_L(\vec p^+) \gamma^{\sigma} u_L(\vec p^-)] \,, \\
    \abs{A_{\nu_{\mu,\tau}\rarrow \nu_{\mu,\tau} e_R^- e_R^+}}^2=\frac{1}{4} \frac{g^4}{M_W^4} (s_W^2)^2 \eta_{\mu\nu} \eta_{\rho\sigma} &\qty[\bar{\tilde u}_L(\vec p') \gamma^{\mu} \tilde u_L(\vec p)] \qty[\bar{\tilde u}_L(\vec p) \gamma^{\rho} \tilde u_L(\vec p')] \\ \times & \qty[\bar u_R(\vec p^-) \gamma^{\nu} v_R(\vec p^+)]  \qty[\bar v_R(\vec p^+) \gamma^{\sigma} u_R(\vec p^-)] \,.
\eea
Now, we should sum over the final spin states and average over the initial ones. Given that the neutrinos are almost massless, we have a one-to-one correspondence between the spin states (so the helicity states) and the chirality of the particles. In addition, all neutrinos come from high-energy electroweak interactions, so they are all left-handed. Then, we are sure we only have one initial spin state, so we should not average. Summing over the final states and applying the completion rules we get 
\bea
    \abs{A_{\nu_{\mu,\tau}\rarrow \nu_{\mu,\tau} e_L^- e_L^+}}^2=\frac{1}{4} \frac{g^4}{M_W^4} (s_W^2-1/2)^2 \eta_{\mu\nu} \eta_{\rho\sigma}  
    &\Tr[\qty(\frac{1-\gamma^5}{2})\fsl p' \gamma^\mu \qty(\frac{1-\gamma^5}{2})\fsl p \gamma^\rho] \\ \times &\Tr[\qty(\frac{1-\gamma^5}{2})\fsl p^- \gamma^\nu \qty(\frac{1-\gamma^5}{2})\fsl p^+ \gamma^\sigma] \,, \\
    \abs{A_{\nu_{\mu,\tau}\rarrow \nu_{\mu,\tau} e_R^- e_R^+}}^2 =\frac{1}{4} \frac{g^4}{M_W^4} (s_W^2)^2 \eta_{\mu\nu} \eta_{\rho\sigma}  
    &\Tr[\qty(\frac{1-\gamma^5}{2})\fsl p' \gamma^\mu \qty(\frac{1-\gamma^5}{2})\fsl p \gamma^\rho] \\ \times &\Tr[\qty(\frac{1+\gamma^5}{2})\fsl p^- \gamma^\nu \qty(\frac{1+\gamma^5}{2})\fsl p^+ \gamma^\sigma] \,.
\eea
Recalling that $\gamma^5$ anticommute with the others gamma matrices and that the square of the projectors obey $((1\pm\gamma^5)/2)^2\equiv (1\pm\gamma^5)/2$, then
\begin{small}
\bea
   \abs{A_{\nu_{\mu,\tau}\rarrow \nu_{\mu,\tau} e_L^- e_L^+}}^2= \frac{1}{4} \frac{g^4}{M_W^4} (s_W^2-1/2)^2 \eta_{\mu\nu} \eta_{\rho\sigma}  
    \Tr[\qty(\frac{1-\gamma^5}{2})\fsl p' \gamma^\mu \fsl p \gamma^\rho] \Tr[\qty(\frac{1-\gamma^5}{2})\fsl p^- \gamma^\nu \fsl p^+ \gamma^\sigma] \,, \\
    \abs{A_{\nu_{\mu,\tau}\rarrow \nu_{\mu,\tau} e_R^- e_R^+}}^2 =\frac{1}{4} \frac{g^4}{M_W^4} (s_W^2)^2 \eta_{\mu\nu} \eta_{\rho\sigma}  
    \Tr[\qty(\frac{1-\gamma^5}{2})\fsl p' \gamma^\mu \fsl p \gamma^\rho] \Tr[\qty(\frac{1+\gamma^5}{2})\fsl p^- \gamma^\nu \fsl p^+ \gamma^\sigma] \,.
\eea
\end{small}
Using that
\bea
    \Tr[\fsl p' \gamma^\mu \fsl p \gamma^\rho] &= p_a p_b' \Tr[\gamma^b \gamma^\mu \gamma^a \gamma^\rho] = p_a p_b'\; 4 (\eta^{b\mu}\eta^{a\rho}-\eta^{ba}\eta^{\mu\rho}+\eta^{b\rho}\eta^{\mu a}) \,, \\
    \Tr[\gamma^5 \fsl p' \gamma^\mu \fsl p \gamma^\rho] &= p_a p_b' \Tr[\gamma^5 \gamma^b \gamma^\mu \gamma^a \gamma^\rho] = p_a p_b'\; (-4i) \epsilon^{b\mu a \rho} \,,
    \label{eq:traces}
\eea
we can write
\begin{small}
\bea
    \abs{A_{\nu_{\mu,\tau}\rarrow \nu_{\mu,\tau} e_L^- e_L^+}}^2= \frac{g^4}{M_W^4} (s_W^2-1/2)^2 p_a p_b' p_c^+ p_d^- \; \eta_{\mu\nu} \eta_{\rho\sigma} &\qty[\eta^{b\mu}\eta^{a\rho}-\eta^{ba}\eta^{\mu\rho}+\eta^{b\rho}\eta^{\mu a} + i\epsilon^{b\mu a \rho}] \\ \times &\qty[\eta^{d\nu}\eta^{c\sigma}-\eta^{dc}\eta^{\nu\sigma}+\eta^{d\sigma}\eta^{\nu c} + i\epsilon^{d\nu c \sigma}] \,, \\
    \abs{A_{\nu_{\mu,\tau}\rarrow \nu_{\mu,\tau} e_R^- e_R^+}}^2= \frac{g^4}{M_W^4} (s_W^2)^2 p_a p_b' p_c^+ p_d^- \; \eta_{\mu\nu} \eta_{\rho\sigma} &\qty[\eta^{b\mu}\eta^{a\rho}-\eta^{ba}\eta^{\mu\rho}+\eta^{b\rho}\eta^{\mu a} + i\epsilon^{b\mu a \rho}] \\ \times &\qty[\eta^{d\nu}\eta^{c\sigma}-\eta^{dc}\eta^{\nu\sigma}+\eta^{d\sigma}\eta^{\nu c} - i\epsilon^{d\nu c \sigma}] \, .\\
\eea
\end{small}
Developing the expression and contracting indices, we finally get 
\begin{small}
\begin{align}
    \abs{A_{\nu_{\mu,\tau}\rarrow \nu_{\mu,\tau} e_L^- e_L^+}}^2 &= \frac{g^4}{M_W^4} (s_W^2-1/2)^2 p_a p_b' p_c^+ p_d^- \; \qty[2\eta^{bd}\eta^{ca}+\cancel{2\eta^{ad}\eta^{bc}}+ 2\eta^{ac}\eta^{bd}-\cancel{2\eta^{bc}\eta^{ad}}] \notag \\ 
    &= 4 \frac{g^4}{M_W^4} (s_W^2-1/2)^2 p_a p_b' p_c^+ p_d^- \eta^{bd}\eta^{ac} = 4 \frac{g^4}{M_W^4} (s_W^2-1/2)^2 (p\cdot p^+) (p'\cdot p^-) \,, \notag \\ 
     \abs{A_{\nu_{\mu,\tau}\rarrow \nu_{\mu,\tau} e_R^- e_R^+}}^2 &= \frac{g^4}{M_W^4} (s_W^2)^2 p_a p_b' p_c^+ p_d^- \; \qty[\cancel{2\eta^{bd}\eta^{ca}}+2\eta^{ad}\eta^{bc}- \cancel{2\eta^{ac}\eta^{bd}}+2\eta^{bc}\eta^{ad}] \notag \\ 
    &= 4 \frac{g^4}{M_W^4} (s_W^2)^2 p_a p_b' p_c^+ p_d^- \eta^{ad}\eta^{bc} = 4 \frac{g^4}{M_W^4} (s_W^2)^2 (p\cdot p^-) (p'\cdot p^+) \,.
    \tag{See Eq.~\eqref{eq:A2_muontau}}
\end{align}
\end{small}

\subsection*{Electron-positron pair emission (electron neutrino)}
\label{sec:amplitude_VPE_electron}

Now let us focus in the decay of an electron neutrino into an electron-positron pair. The matrix element corresponding to the decay to right-handed electrons only contains the neutral current, as in the case of muon and tau neutrinos. If one consider instead the decay to left-handed neutrinos, both neutral and charged channel have to be taken into account,
\begin{align}
    A_{\nu_e\rarrow \nu_e e_L^- e_L^+} =&\qty[\bar{\tilde u}_L(\vec p')\qty(-i\frac{g}{2c_W}\gamma^\mu)\tilde u_L(\vec p)]
    \qty{-i\frac{\eta_{\mu\nu}}{M_Z^2}}
    \qty[\bar u_L(\vec p^-)\qty(-i\frac{g}{c_W}(s_W^2-1/2)\gamma^\nu) v_L(\vec p^+)] \notag \\ +&
    \qty[\bar u_L(\vec p^-)\qty(-i\frac{g}{\sqrt{2}}\gamma^\mu) \tilde u_L(\vec p)]
    \qty{-i\frac{\eta_{\mu\nu}}{M_W^2}}
    \qty[\bar{\tilde u}_L(\vec p')\qty(-i\frac{g}{\sqrt{2}}\gamma^\nu) v_L(\vec p^+)] \,, \notag \\
    A_{\nu_{e}\rarrow \nu_{e} e_R^- e_R^+} =& \qty[\bar{\tilde u}_L(\vec p')\qty(-i\frac{g}{2c_W}\gamma^\mu)\tilde u_L(\vec p)]
    \qty{-i\frac{\eta_{\mu\nu}}{M_Z^2}}
    \qty[\bar u_R(\vec p^-)\qty(-i\frac{g}{c_W} s_W^2 \gamma^\nu) v_R(\vec p^+)] \,.
    \tag{See Eq.~\eqref{eq:A_elec}}
\end{align}

Given that for the decay to right-handed electrons the matrix element is the same as in the case of muon and tau neutrinos, without doing any calculation we can anticipate that
\be
     \abs{A_{\nu_e\rarrow \nu_e e_R^- e_R^+}}^2 = 4 \frac{g^4}{M_W^4} (s_W^2)^2 (p\cdot p^-) (p'\cdot p^+) \,.
     \tag{See Eq.~\eqref{eq:A2_elec}}
\ee

For the case of decay to left-handed electrons, we have the additional contribution from the charged channel. Then, multiplying by the complex conjugate we get that the square of the matrix element is now given by 
\bea
    \abs{A_{\nu_e\rarrow \nu_e e_L^- e_L^+}}^2=\frac{1}{4}\frac{g^4}{M_W^4} \eta_{\mu\nu} \eta_{\rho\sigma} \bigg\{ (s_W^2-1/2) &\qty[\bar{\tilde u}_L(\vec p') \gamma^{\mu} \tilde u_L(\vec p)] \qty[\bar{u}_L(\vec p^-) \gamma^{\nu} v_L(\vec p^+)] \\ +
    &\qty[\bar{u}_L(\vec p^-) \gamma^{\mu} \tilde u_L(\vec p)] \qty[\bar{\tilde u}_L(\vec p') \gamma^{\nu} v_L(\vec p^+)] \bigg\} \\
    \times \bigg\{ (s_W^2-1/2) &\qty[\bar{\tilde u}_L(\vec p') \gamma^{\rho} \tilde u_L(\vec p)]^\dag \qty[\bar{u}_L(\vec p^-) \gamma^{\sigma} v_L(\vec p^+)]^\dag \\ +
    &\qty[\bar{u}_L(\vec p^-) \gamma^{\rho} \tilde u_L(\vec p)]^\dag \qty[\bar{\tilde u}_L(\vec p') \gamma^{\sigma} v_L(\vec p^+)]^\dag \bigg\} \,.
\eea
Using Eq.~\eqref{eq:conjugated}, expanding the product and rearranging terms we get
\bea
    \abs{A_{\nu_e\rarrow \nu_e e_L^- e_L^+}}^2=
    \frac{1}{4}\frac{g^4}{M_W^4} \eta_{\mu\nu} \eta_{\rho\sigma} \bigg\{
    (s_W^2-1/2)^2 &\qty[\bar{\tilde u}_L(\vec p') \gamma^{\mu} \tilde u_L(\vec p)] \qty[\bar{\tilde u}_L(\vec p) \gamma^{\rho} \tilde u_L(\vec p')] \\
    \times &\qty[\bar{u}_L(\vec p^-) \gamma^{\nu} v_L(\vec p^+)] \qty[\bar{v}_L(\vec p^+) \gamma^{\sigma} u_L(\vec p^-)] \\  +
    (s_W^2-1/2) &\qty[\bar{\tilde u}_L(\vec p') \gamma^{\mu} \tilde u_L(\vec p)] \qty[\bar{\tilde u}_L(\vec p) \gamma^{\rho} u_L(\vec p^-)] \\ 
    \times &\qty[\bar{u}_L(\vec p^-) \gamma^{\nu} v_L(\vec p^+)] \qty[\bar{v}_L(\vec p^+) \gamma^{\sigma} \tilde u_L(\vec p')] \\ +
    (s_W^2-1/2) &\qty[\bar{u}_L(\vec p^-) \gamma^{\mu} \tilde u_L(\vec p)] \qty[\bar{\tilde u}_L(\vec p) \gamma^{\rho} \tilde u_L(\vec p')] \\ 
    \times &\qty[\bar{\tilde u}_L(\vec p') \gamma^{\nu} v_L(\vec p^+)] \qty[\bar{v}_L(\vec p^+) \gamma^{\sigma} u_L(\vec p^-)] \\ +
    &\qty[\bar{u}_L(\vec p^-) \gamma^{\mu} \tilde u_L(\vec p)] \qty[\bar{\tilde u}_L(\vec p) \gamma^{\rho} u_L(\vec p^-)] \\ 
    \times &\qty[\bar{\tilde u}_L(\vec p') \gamma^{\nu} v_L(\vec p^+)] \qty[\bar{v}_L(\vec p^+) \gamma^{\sigma} \tilde u_L(\vec p')] \bigg\} \,.
\eea
Summing over all the final spin states, applying the completion rules and using the properties of the projectors, we get
\begin{small}
\bea
    \abs{A_{\nu_e\rarrow \nu_e e_L^- e_L^+}}^2=\frac{1}{4}\frac{g^4}{M_W^4} \eta_{\mu\nu} \eta_{\rho\sigma} \bigg\{
      (s_W^2-1/2)^2 & \Tr[\qty(\frac{1-\gamma^5}{2})\fsl p' \gamma^\mu \fsl p \gamma^\rho] \Tr[\qty(\frac{1-\gamma^5}{2})\fsl p^- \gamma^\nu \fsl p^+ \gamma^\sigma] \\
    + (s_W^2-1/2) & \Tr[\qty(\frac{1-\gamma^5}{2})\fsl p' \gamma^\mu \fsl p \gamma^\rho \fsl p^- \gamma^{\nu} \fsl p^+ \gamma^{\sigma}] \\
    + (s_W^2-1/2) & \Tr[\qty(\frac{1-\gamma^5}{2})\fsl p^- \gamma^\mu \fsl p \gamma^\rho \fsl p' \gamma^{\nu} \fsl p^+ \gamma^{\sigma}] \\
    +  & \Tr[\qty(\frac{1-\gamma^5}{2})\fsl p^- \gamma^\mu \fsl p \gamma^\rho] \Tr[\qty(\frac{1-\gamma^5}{2})\fsl p' \gamma^\mu \fsl p^+ \gamma^\rho] 
    \bigg\} \,.
\eea
\end{small}
Using the properties of the Clifford algebra $\qty(\{\gamma^\mu,\gamma^\nu\}=2\eta^{\mu\nu})$, the properties of the trace (linearity and cyclical indices), and contracting indices, we finally get
\be 
    \abs{A_{\nu_e\rarrow \nu_e e_L^- e_L^+}}^2= 4 \frac{g^4}{M_W^4} (s_W^2-3/2)^2 (p\cdot p^+) (p'\cdot p^-) \,,
    \tag{See Eq.~\eqref{eq:A2_elec}}
\ee
which is the same result as in the case of muon and tau neutrinos, Eq.\eqref{eq:A_muontau}, but substituting \mbox{$(s_W^2-1/2)$} by \mbox{$(s_W^2-3/2)$}.

\subsection*{Neutrino-antineutrino pair emission}
\label{sec:amplitude_NSpl}

Let us now consider the decay of a neutrino of flavour $\alpha$ into a neutrino-antineutrino pair of flavour $\beta$. The matrix element is given by 
\be
    A_{\nu_\alpha\rarrow \nu_\alpha \nu_\beta \bar\nu_\beta} =\qty[\bar{\tilde u}_L(\vec p')\qty(-i\frac{g}{2c_W}\gamma^\mu)\tilde u_L(\vec p)]
    \qty{-i\frac{\eta_{\mu\nu}}{M_Z^2}}
    \qty[\bar{\tilde u}_L(\vec p^-)\qty(-i\frac{g}{2c_W}\gamma^\nu) \tilde v_L(\vec p^+)] \,,
    \tag{See Eq.~\eqref{eq:A_nu}}
\ee
for every $\alpha$ and $\beta$. Similarly to the previous cases, we obtain the square of the decay amplitude multiplying by the complex conjugate,
\bea
    \abs{A_{\nu_\alpha\rarrow \nu_\alpha \nu_\beta \bar\nu_\beta}}^2=\frac{1}{16} \frac{g^4}{M_W^4} \eta_{\mu\nu} \eta_{\rho\sigma} &\qty[\bar{\tilde u}_L(\vec p')\gamma^\mu \tilde u_L(\vec p)] \qty[\bar{\tilde u}_L(\vec p^-)\gamma^\nu \tilde v_L(\vec p^+)] \\ \times &\qty[\bar{\tilde u}_L(\vec p')\gamma^\rho \tilde u_L(\vec p)]^\dag \qty[\bar{\tilde u}_L(\vec p^-)\gamma^\sigma \tilde v_L(\vec p^+)]^\dag \,.
\eea
Using Eq.~\eqref{eq:conjugated} and rearranging terms we get
\bea
    \abs{A_{\nu_\alpha\rarrow \nu_\alpha \nu_\beta \bar\nu_\beta}}^2 =\frac{1}{16} \frac{g^4}{M_W^4} \eta_{\mu\nu} \eta_{\rho\sigma} &\qty[\bar{\tilde u}_L(\vec p')\gamma^\mu \tilde u_L(\vec p)] \qty[\bar{\tilde u}_L(\vec p) \gamma^\rho \tilde u_L(\vec p')] \\ &\qty[\bar{\tilde u}_L(\vec p^-)\gamma^\nu \tilde v_L(\vec p^+)] \qty[\bar{\tilde v}_L(\vec p^+) \gamma^\sigma \tilde u_L(\vec p^-)] \,.
\eea
Summing over the spin states, applying the completion rules and using the properties of the projectors we get 
\be
    \abs{A_{\nu_\alpha\rarrow \nu_\alpha \nu_\beta \bar\nu_\beta}}^2=\frac{1}{16} \frac{g^4}{M_W^4} \eta_{\mu\nu} \eta_{\rho\sigma} \Tr[\qty(\frac{1-\gamma^5}{2})\fsl p' \gamma^\mu \fsl p \gamma^\rho] \Tr[\qty(\frac{1-\gamma^5}{2})\fsl p^- \gamma^\nu \fsl p^+ \gamma^\sigma] \,.
\ee
Using Eqs.~\eqref{eq:traces}, we can substitute the value of the traces
\bea
    \abs{A_{\nu_\alpha\rarrow \nu_\alpha \nu_\beta \bar\nu_\beta}}^2=\frac{1}{16} \frac{g^4}{M_W^4} p_a p'_b p^+_c p^-_d \eta_{\mu\nu} \eta_{\rho\sigma} &\qty[\eta^{b\mu}\eta^{a\rho}-\eta^{ba}\eta^{\mu\rho}+\eta^{b\rho}\eta^{\mu a} + i\epsilon^{b\mu a\rho}] \\ &\qty[\eta^{d\nu}\eta^{c\sigma}-\eta^{dc}\eta^{\nu\sigma}+\eta^{d \sigma}\eta^{\nu c} + i\epsilon^{d\nu c\sigma}] \,.
\eea
Contracting indices and combining the different terms, we finally get 
\be
    \abs{A_{\nu_\alpha\rarrow \nu_\alpha \nu_\beta \bar\nu_\beta}}^2=\frac{g^4}{M_W^4} (p\cdot p^+)(p'\cdot p^-) \,.
    \tag{See Eq.~\eqref{eq:A2_nu}}
\ee

\clearpage
\section{Neutrino flux using the continuous energy loss model}
\label{sec:flux_continuous}

Let us call $dN_e$ the differential number of emitted neutrinos by one source, with energies in-between $E_e$ and $E_e +dE_e$ in a differential time $dt_e$. Similarly, one can call $dN_d$ the differential number of detected neutrinos, with energies in-between $E_d$ and $E_d +dE_d$ in a differential time $dt_d$. Then, the differential flux of emitted and detected neutrinos can be defined as:
\be
    \Phi_e \coloneqq \frac{dN_e}{dt_e dE_e} \quad\text{and}\quad \tilde\Phi_d \coloneqq \frac{dN_d}{dt_d dE_d} \,.
    \label{eq:dNe_dNd}
\ee
Taking into account that the electron-positron pair emission does not change the number of neutrinos, we can certainly say that a geometrical fraction (given by the solid angle subtended by the detector from the source, $A_d/r_e^2$) of the number of emitted neutrinos, must be detected. Then
\be
    \Phi_e dE_e dt_e \cdot \frac{A_d}{r_e^2} = \tilde\Phi_d dE_d dt_d \srarrow
    \tilde\Phi_d = \frac{dt_e}{dt_d} \frac{A_d}{r_e^2} \frac{dE_e}{dE_d} \Phi_e \,.
\ee

Let us now consider a volumetric distribution of sources $\rho(\theta,\phi,r_e)$, all with same model of emission. Integrating to the relevant volume of the universe,
\be
    \Phi_d = \iiint d\theta d\phi dr_e r_e^2 \sin\theta \rho(\theta,\phi,r_e) \tilde\Phi_d
    \label{eq:sources} \,.
\ee
If one considers an homogeneous and isotropic source distribution, we can define a radial source distribution $\rho(r_e)$ as
\be 
    \iint d\theta d\phi \sin\theta \rho(\theta,\phi,r_e) \eqqcolon 4\pi \rho(r_e) \,.
    \label{eq:radial_rho}
\ee
Then, the detected flux from a distribution of sources is given by
\be
    \Phi_d = 4\pi \int dr_e r_e^2 \rho(r_e) \tilde\Phi_d = 4\pi A_d \int dz_e \frac{\rho(z_e)}{(1+z_e)H(z_e)} \pdv{E_e(E_d,z_e)}{E_d} \Phi_e(E_e(E_d,z_e)) \,,
\ee
where in the last step we have made a change of variables from the comoving radial distance $r_e$ to the redshift variable $z_e$. For that we have used that $dr_e=dz_e/H(z_e)$ and $dt_d=(1+z_e)dt_e$.

One should take into account that the relation between emission and detection energies changes if the detection energy is above or below the threshold energy $E_\text{th}^{(e)}$. As we already know that we do not expect much neutrinos above the threshold, i.e., $\Phi_d(E_d>E_\text{th}^{(e)})\approx 0$, let us focus in the case with $E_d<E_\text{th}^{(e)}$. In that case, the relation between the emitted and detected energies is given by
\be 
    E_e = (1+z_e) E_d \qty[1-\qty(\frac{E_d}{E_{\nu_{\mu,\tau}}^{(e)}})^{5+3n} \avg{1-x}\,J_n\qty(z_e,E_\text{th}^{(e)}/E_d-1)]^{-1/(5+3n)} \,, \quad (E_d\le E_\text{th}^{(e)})\,,
    \tag{See Eq.~\eqref{eq:Ee_Ed_th_2}}
\ee
with the condition
\be
     J_n\qty(z_e,E_\text{th}^{(e)}/E_d-1) < \frac{1}{\avg{1-x}} \, \qty(\frac{E_{\nu_{\mu,\tau}}^{(e)}}{E_d})^{5+3n}\,, \quad (E_d\le E_\text{th}^{(e)})\,.
     \tag{See Eq.~\eqref{eq:condition_below}}
\ee
This defines, for each fixed value of $E_d$, a maximum source redshift $z_c(E_d)$, given by imposing an equality in the previous equation,
\be
     J_n\qty(z_c(E_d),z_\text{th}^{(e)}(E_d)) \equiv \frac{1}{\avg{1-x}} \, \qty(\frac{E_{\nu_{\mu,\tau}}^{(e)}}{E_d})^{5+3n}\,, \quad (E_d\le E_\text{th}^{(e)})\,.
     \tag{See \eqref{eq:zc}}
\ee
Then, the integration over the sources is limited by the value of $z_c(E_d)$,
\be
    \Phi_d(E_d) = 4\pi A_d \int_{z_\text{min}}^{z_c(E_d)} dz_e \frac{\rho(z_e)}{(1+z_e)H(z_e)} \frac{dE_e(E_d,z_e)}{dE_d} \Phi_e(E_e(E_d,z_e)) \,.
\ee

In order to integrate this function we need to specify every function in the integrand. For this example we can use $\rho(z_e)$ proportional to the \abb{SFR} distribution, as done in~\cite{Stecker:2014xja,Stecker:2014oxa,Stecker:2017gdy,Stecker:2022tzd}. The emission spectrum $\Phi_e(E_e)$ can be taken as an inverse power law from 100 TeV to 100 PeV, that is to say $\Phi_e(E_e)=(E_0/E_e)^2$, with $E_0$ a constant related with the luminosity of the source. The function $E_e(E_d,z_e)$ is given by Eq.~\eqref{eq:Ee_Ed_th_2}, and its derivative respect to $E_d$ can be calculated analytically if the derivative of the function $J_n(z_i,z_f)$ is known.

Let us note how the factor in square brackets in Eq.~\eqref{eq:Ee_Ed_th_2} goes to infinity as $z_e$ gets closer to $z_c(E_d)$ (by definition of $z_c(E_d)$, Eq.~\eqref{eq:zc}),
\be 
    \lim_{z_e\rightarrow z_c(E_d)} \qty[1-\qty(\frac{E_d}{E_{\nu_{\mu,\tau}}^{(e)}})^{5+3n} \avg{1-x}\,J_n\qty(z_e,E_\text{th}^{(e)}/E_d-1)]^{-1/(5+3n)} = \infty \,.
\ee
This makes both, $E_e(E_d,z_e)$ and its derivative respect to $E_d$, to also go to infinity. The singularity in the energy shows the impossibility to consider any source farther than $z_c(E_d)$. The singularity in the derivative shows the fact that the neutrinos lose energy so fast that an arbitrary large range of energies of emission correspond to a arbitrary small range of energies of detection.

An analytic expression for $J_n$ can be found using Mathematica,
\begin{mmaCell}[addtoindex=1,morefunctionlocal={z}]{Input}
  Integrate[(1+z)^(4+3n)/Sqrt[\mmaUnd{\({\Omega}\)m}*(1+z)^3+\mmaUnd{\({\Omega\Lambda}\)}],z]
\end{mmaCell}
\begin{mmaCell}{Output}
  \mmaFrac{\mmaSup{(1+z)}{5+3 n} \mmaSqrt{1+\mmaFrac{\mmaSup{(1+z)}{3} \(\Omega\)m}{\(\Omega\Lambda\)}} Hypergeometric2F1[\mmaFrac{1}{2},\mmaFrac{5}{3}+n,\mmaFrac{8}{3}+n,-\mmaFrac{\mmaSup{(1+z)}{3}
\(\Omega\)m}{\(\Omega\Lambda\)}]}{(5+3 n) \mmaSqrt{\mmaSup{(1+z)}{3} \(\Omega\)m+\(\Omega\Lambda\)}}
\end{mmaCell}
this way, both $E_e(E_d,z_e)$ and its derivative can be computed explicitly, even though the form of the latter is complicated.

\begin{figure}[tb]
    \centering
    \includegraphics[width=0.7\textwidth]{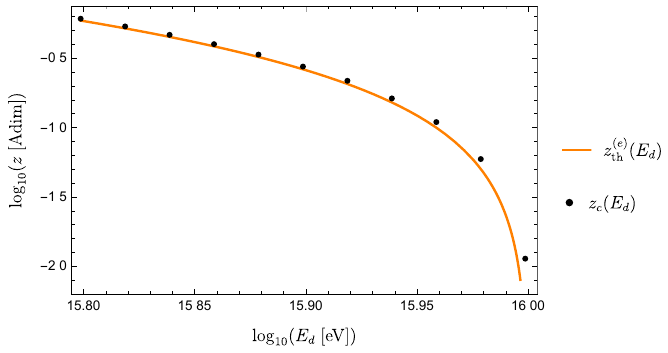}
    \caption{Location of the furthest source capable to contribute to the detected neutrino flux of energy $E_d$ (black dots). The function $z_\text{th}^{(e)}(E_d)$ is also shown for comparison (orange line).}
    \label{fig:zc_Ed}
\end{figure}

\begin{figure}[tb]
    \centering
    \includegraphics[width=0.8\textwidth]{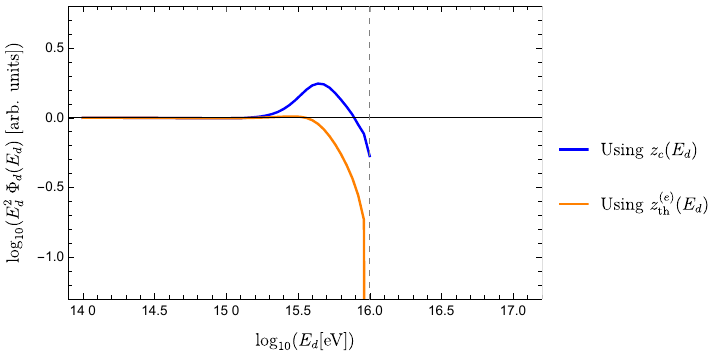}
    \caption{Neutrino flux at Earth (for energies below $E_\text{th}^{(e)}$) using the energy evolution model for $n=2$ and $\Lambda\approx 1.13\E{-2} M_P$ (such that $E_\text{th}^{(e)}=10^{16}$ eV). The result is highly sensitive to the computed values of $z_c(E_d)$.}
    \label{fig:flux_continuous}
\end{figure}

A different situation arises when we are trying to compute the function $z_c(E_d)$, whose definition is given implicitly by Eq.~\eqref{eq:zc}. There is no analytical expression of $z_c$ as a function of $E_d$, so approximated numerical methods are needed,
\begin{mmaCell}[addtoindex=1,moredefined={zc, x, Jn, zth},morepattern={Ed_, n_, Ed, n},morefunctionlocal={ze}]{Input}
  zc[Ed_,\mmaPat{\({\Lambda}\)_},n_]:=\mmaUnd{ze}/. \\ FindRoot[1-(Ed/\mmaDef{E\({\mu\tau}\)e}[\mmaPat{\({\Lambda}\)},n])^(5+3n)*(1-x[n])*\\ (Jn[ze,zth[Ed,\mmaPat{\({\Lambda}\)},n],n]//Normal),\{ze,10*zth[Ed,\mmaPat{\({\Lambda}\)},n]\}]
\end{mmaCell}
Using these numerical methods one can obtain an idea of the shape of this function (see Fig.~\ref{fig:zc_Ed}).
Let us note that the values of $z_c(E_d)$ are very similar to those of $z_\text{th}^{(e)}(E_d)$, justifying the approximation $z_c(E_d)\approx z_\text{th}^{(e)}(E_d)$ when one wants to compute the energy of the cutoff. However, the small differences between these two functions are of paramount importance when computing the shape of the flux (see Fig.~\ref{fig:flux_continuous}).

As we are using functions that have a singularity exactly at $z_c(E_d)$, any miscalculation, or just lack of precision, in the computation of $z_c(E_d)$ will change the flux to a large degree. In~Fig.~\ref{fig:flux_continuous} we compare, as an example, the differences in the expected flux if we approximate $z_c(E_d)$ by the function $z_\text{th}^{(e)}(E_d)$ (Eq.~\eqref{eq:zth}).
Given the strong sensitivity of the result to the determination of $z_c(E_d)$, we conclude that our computation using this method is not reliable, and we need another strategy to compute the flux. However, we can start anticipating some properties of the flux, as a confirmation of the strong suppression after $E_\text{cut}^{(e)}$ and an accumulation of neutrinos prior to it. This \textit{bump} could increase the flux of neutrinos in ranges of energies of interest for current and future experiments.

\clearpage
\section{Neutrino flux using a simplified instantaneous cascade model}
\label{sec:flux_instantaneous}

We will use the same definition of the emitted and detected flux introduced in Appendix~\ref{sec:flux_continuous},
\be
    \Phi_e \coloneqq \frac{dN_e}{dt_e dE_e} \quad\text{and}\quad \tilde\Phi_d \coloneqq \frac{dN_d}{dt_d dE_d} \,.
    \tag{See Eq.~\eqref{eq:dNe_dNd}}
\ee

The emitted neutrinos will propagate differently depending on their energy. For energies above both thresholds, both decays are possible until falling below the kinematical threshold of the \abb{VPE}. For energies below $E_\text{th}^{(e)}$, only neutrino-antineutrino pair emission is possible, and only until falling below the effective threshold $E_{\nu_\alpha}^{(\nu)}$. For energies below both thresholds the propagation is classical.

Let us start considering neutrinos emitted above both thresholds, i.e., $E_e>E_\text{th}^{(e)}>E_{\nu_\alpha}^{(\nu)}$. Let us assume that in average they produce $p'$ electron-positron and $q'$ neutrino-antineutrino pairs before falling below the kinematical threshold $E_\text{th}^{(e)}$, 
\be 
    E_e\; \qty(\avg{x}^{(e)})^{p'} \qty(\avg{x}^{(\nu)})^{q'} < E_\text{th}^{(e)}\,,
\ee
where $\avg{x}^{(e)}$ and $\avg{x}^{(\nu)}$ are the average fraction of energy of a secondary neutrino after an electron-positron or neutrino-antineutrino pair emission, respectively. The exact values of the energy fractions of the product particles for each decay can be read from Tables~\ref{tab:mean_values_VPE_muon}, \ref{tab:mean_values_VPE_electron} and \ref{tab:mean_values_NSpl}; however, for simplicity, we will take $\avg{x}^{(e)}$ and $\avg{x}^{(\nu)}\approx 1/3$ for this simple model.

In order to get the value of $p'$ and $q'$, we can approximate $p'$ and $q'$ by the number of decays necessary to get exactly $E_\text{th}^{(e)}$. That is to say,
\be 
    E_e\;(1/3)^{p'}\,(1/3)^{q'} \approx E_\text{th}^{(e)} \,.
    \label{eq:insta_88}
\ee
Let us recall that, although the number of decays a neutrino undergoes must be an integer, $p'$ and $q'$ are an average number of decays of a large enough number of neutrinos, so they can be any positive real number. Also, as we will see in the next paragraph, for a fixed emission energy, only a unique combination of $p'$ and $q'$ is possible.

The probability of decay through electron-positron and neutrino-antineutrino pair emission is proportional to their decay width, then we expect\footnotemark
\be 
    \frac{p'}{q'}\approx \frac{\Gamma_{\nu_{\mu,\tau}}^{(e)}(E)}{\Gamma_{\nu_{\alpha}}^{(\nu)}(E)} = \qty(\frac{E_{\nu_\alpha}^{(\nu)}}{E_{\nu_{\mu,\tau}}^{(e)}})^{5+3n} =
    \begin{cases}
        0.25 & \text{for }n=1 \\
        0.32 & \text{for }n=2
    \end{cases} \,.
    \label{eq:quotient_muon_nu}
\ee
\footnotetext{Let us note that  Eq.\eqref{eq:quotient_muon_nu} is not exactly the inverse of Eq.~\eqref{eq:quotient_nu_muon}, because $\Gamma_{\nu_{\alpha}}^{(\nu)}=3\times \Gamma_{\nu_\alpha\rarrow \nu_\alpha \nu_\beta \bar\nu_\beta}$ (see Eq.~\eqref{eq:Gamma_nu}), so there is difference in factor of 3.}
Let us call this ratio $r\coloneqq p'/q'$, and let us note that it is independent of the energy of the neutrino and of the scale of new physics. Inserting this relation in Eq.~\eqref{eq:insta_88}, we can solve for the value of $q'$,
\be 
    E_e\;(1/3)^{(1+r)q'} \approx E_\text{th}^{(e)} \srarrow 3^{q'}\approx\qty(\frac{E_e}{E_\text{th}^{(e)}})^{1/(1+r)} \,.
    \label{eq:insta_90}
\ee

After this, the neutrinos, which have now an energy exactly equal to the kinematical threshold of electron-positron pair emission, will produce only \abb{NSpl} until falling below the effective threshold $E_{\nu_\alpha}^{(\nu)}$. Let us assume that in average they produce $q''$ neutrino-antineutrino pairs before that, 
\be 
    E_\text{th}^{(e)}\qty(\avg{x}^{(\nu)})^{q''} < E_{\nu_\alpha}^{(\nu)} \,.
\ee
Again, let us approximate $q''$ by the number of decays necessary to go from the initial energy, in this case $E_\text{th}^{(e)}$, to the effective threshold $E_{\nu_\alpha}^{(\nu)}$. That is to say,
\be 
    E_\text{th}^{(e)}\;(1/3)^{q''} \approx E_{\nu_\alpha}^{(\nu)} \srarrow 3^{q''}\approx \qty(\frac{E_\text{th}^{(e)}}{E_{\nu_\alpha}^{(\nu)}}) \,.
    \label{eq:insta_92}
\ee

This way, following the instantaneous decay model, we can say that if $\Phi_e(E_e) dE_e dt_e$ neutrinos are emitted with energies above both thresholds, they will instantaneously decay in $3^{q'} 3^{q''} \Phi_e(E_e) dE_e dt_e$ neutrinos of energy $E_{\nu_\alpha}^{(\nu)}$. From that number of neutrinos, a fraction, given by the solid angle subtended by the detector from the source ($A_d/r_e^2$), will be detected, i.e.
\be 
    3^{q'} 3^{q''} \Phi_e(E_e) dE_e dt_e\cdot \frac{A_d}{r_e^2} \approx \tilde\Phi_d(E_d) dE_d dt_d\,, \quad(\text{for }E_e>E_\text{th}^{(e)}>E_{\nu_\alpha}^{(\nu)}) \,.
\ee
Substituting Eqs.~\eqref{eq:insta_90} and \eqref{eq:insta_92}, one gets
\be 
    dE_d \tilde\Phi_d(E_d) \approx \frac{dt_e}{dt_d} \frac{A_d}{r_e^2} \qty(\frac{E_\text{th}^{(e)}}{E_{\nu_\alpha}^{(\nu)}}) \;\qty[dE_e \Phi_e(E_e)\qty(\frac{E_e}{E_\text{th}^{(e)}})^{1/(1+r)}] \,, \quad(\text{for }E_e>E_\text{th}^{(e)}>E_{\nu_\alpha}^{(\nu)}) \,.
    \label{eq:dEdPhid_1}
\ee
We see that we cannot directly solve for $\tilde\Phi_d(E_d)$, without obtaining again (as in the continuous energy loss model) the derivative $\partial E_e/\partial E_d$, so we will keep the relation between emission and detection in terms of number of neutrinos instead of flux.

Let us consider now neutrinos emitted with energy $E_\text{th}^{(e)}>E_e>E_{\nu_\alpha}^{(\nu)}$, so they can only produce neutrino-antineutrino pairs. Let us also assume that in average they produce $q$ neutrino-antineutrino pairs until falling below the effective threshold, 
\be 
    E_e\; \Big(\avg{x}^{(\nu)}\Big)^q < E_{\nu_\alpha}^{(\nu)}\,.
\ee
We can approximate again $q$ by the number of decays necessary to get exactly $E_{\nu_\alpha}^{(\nu)}$,
\be 
    E_e\;(1/3)^q \approx E_{\nu_\alpha}^{(\nu)} \srarrow  3^q \approx \qty(\frac{E_e}{E_{\nu_\alpha}^{(\nu)}}) \,.
    \label{eq:insta_96}
\ee

Then, following again the instantaneous approximation, we can say that if $\Phi_e(E_e) dE_e dt_e$ neutrinos are emitted with energies between both thresholds, they will instantaneously decay in $3^{q} \Phi_e(E_e) dE_e dt_e$ neutrinos of energy $E_{\nu_\alpha}^{(\nu)}$. From that number of neutrinos, a fraction, given by the solid angle subtended by the detector from the source ($A_d/r_e^2$), will be detected, i.e.,
\be 
    3^{q} \Phi_e(E_e) dE_e dt_e\cdot \frac{A_d}{r_e^2} \approx \tilde\Phi_d(E_d) dE_d dt_d\,, \quad(\text{for }E_\text{th}^{(e)}>E_e>E_{\nu_\alpha}^{(\nu)}) \,.
\ee
Substituting Eq.~\eqref{eq:insta_96} in the previous equation we get
\be 
    dE_d \tilde\Phi_d(E_d) \approx \frac{dt_e}{dt_d} \frac{A_d}{r_e^2} \;\qty[dE_e \Phi_e(E_e) \qty(\frac{E_e}{E_{\nu_\alpha}^{(\nu)}})] \,, \quad(\text{for }E_\text{th}^{(e)}>E_e>E_{\nu_\alpha}^{(\nu)}) \,,
    \label{eq:dEdPhid_2}
\ee
where again we leave the expression in terms of the number of neutrinos, instead of flux.

Finally, we can study the case of neutrinos emitted below both thresholds, i.e., $E_\text{th}^{(e)}>E_{\nu_\alpha}^{(\nu)}>E_e$. In this case there is no multiplication of the number of neutrinos, so
\be 
    \Phi_e(E_e) dE_e dt_e\cdot \frac{A_d}{r_e^2} = \tilde\Phi_d(E_d) dE_d dt_d\,, \quad(\text{for }E_\text{th}^{(e)}>E_{\nu_\alpha}^{(\nu)}>E_e) \,.
\ee
In this case, as we know that $E_e=(1+z_e)E_d$, we can solve explicitly for $\tilde\Phi_d(E_d)$; however, in order to combine this expression with the results obtained for the other ranges of energies, we will keep the relation in terms of number of neutrinos,
\be 
    dE_d \Phi_d(E_d) \approx \frac{dt_e}{dt_d} \frac{A_d}{r_e^2} \;\Bigg[dE_e \Phi_e(E_e)\Bigg] \,, \quad(\text{for }E_\text{th}^{(e)}>E_{\nu_\alpha}^{(\nu)}>E_e) \,.
    \label{eq:dEdPhid_3}
\ee

Let us now consider a volumetric distribution of sources $\rho(\theta,\phi,r_e)$, all with same model of emission, and a certain bin of detection energy $[E_d^A,E_d^B]$, with $E_d^B>E_d^A$. Integrating to the relevant volume of the universe and to the energy range of the bin, one obtains
\be
    N_d(E_d^A,E_d^B) \coloneqq \iiint d\theta d\phi dr_e r_e^2 \sin\theta \rho(\theta,\phi,r_e) \int_{E_d^A}^{E_d^B} dE_d \tilde\Phi_d(E_d) \,.
\ee
If one considers an homogeneous and isotropic source distribution, we can define a radial source distribution $\rho(r_e)$ as
\be 
    \iint d\theta d\phi \sin\theta \rho(\theta,\phi,r_e) \eqqcolon 4\pi \rho(r_e) \,.
    \tag{See Eq.\eqref{eq:radial_rho}}
\ee
Then, the number of neutrinos inside the bin is simplified to
\be
    N_d(E_d^A,E_d^B) = 4\pi \int dr_e r_e^2 \rho(r_e) \int_{E_d^A}^{E_d^B} dE_d \tilde\Phi_d(E_d) \,.
\ee

We know the relation between $dE_d \tilde\Phi_d(E_d)$ and $dE_e \Phi_e(E_e)$, as a function of the emission energy, Eqs.~\eqref{eq:dEdPhid_1}, \eqref{eq:dEdPhid_2}, and \eqref{eq:dEdPhid_3}. Then, assuming that the bin of detected energies, $[E_d^A,E_d^B]$, corresponds to a certain bin of emission energies, $[E_e^A,E_e^B]$, we can study separately the cases in which the bin $[E_e^A,E_e^B]$ falls inside the three different ranges of energy (energies below, in-between, or above the two thresholds)\footnote{In fact, there are two additional possibilities which correspond to the cases in which the lower and upper bounds of the bin fall in different ranges. However, this can only happen for two specific bins, and their contribution will go to zero when we take the zero bin-width limit.}.

For the case of $E_\text{th}^{(e)}>E_{\nu_\alpha}^{(\nu)}>E_e\in[E_e^A,E_e^B]$ we can use Eq.~\eqref{eq:dEdPhid_3} to write
\be
    N_d(E_d^A,E_d^B) = 4\pi \int dr_e r_e^2 \rho(r_e) \frac{dt_e}{dt_d} \frac{A_d}{r_e^2} \int_{E_e^A}^{E_e^B} dE_e \Phi_e(E_e) \,, \quad(\text{for }E_\text{th}^{(e)}>E_{\nu_\alpha}^{(\nu)}>E_e) \,.
\ee
As the range of emission energies is below both thresholds, we know that the relation between the energies of emission and detection are classical, that is to say $E_d=E_e/(1+z_e)$, so
\be
    N_d(E_d^A,E_d^B) = 4\pi A_d \int dz_e \frac{\rho(z_e)}{(1+z_e)H(z_e)}\int_{(1+z_e)E_d^A}^{(1+z_e)E_e^B} dE_e \Phi_e(E_e)\,, \quad(\text{for }E_\text{th}^{(e)}>E_{\nu_\alpha}^{(\nu)}>E_e) \,,
\ee
where in the last step we have made a change of variables from the comoving radial distance $r_e$ to the redshift $z_e$. For that we have used that $dr_e=dz_e/H(z_e)$ and $dt_d=(1+z_e)dt_e$.

For the case of $E_\text{th}^{(e)}>E_e\in[E_e^A,E_e^B]>E_{\nu_\alpha}^{(\nu)}$ we can use Eq.~\eqref{eq:dEdPhid_2} to write
\be
    N_d(E_d^A,E_d^B) = 4\pi \int dr_e r_e^2 \rho(r_e) \frac{dt_e}{dt_d} \frac{A_d}{r_e^2} \int_{E_e^A}^{E_e^B} dE_e \Phi_e(E_e) \qty(\frac{E_e}{E_{\nu_\alpha}^{(\nu)}}) \,, \quad(\text{for }E_\text{th}^{(e)}>E_e>E_{\nu_\alpha}^{(\nu)}) \,.
\ee
As the range of emission energies is above the effective threshold of the neutrino-antineutrino pair emission, following the instantaneous cascade model, all the neutrinos, independently of the energy of emission, will be detected with energy $E_d=E_{\nu_\alpha}^{(\nu)}/(1+z_e)$. From this, we can conclude that all the range of energies (from $E_{\nu_\alpha}^{(\nu)}$ to $E_\text{th}^{(e)}$) will contribute to the same detection energy, $E_d=E_{\nu_\alpha}^{(\nu)}/(1+z_e)$. On the other hand, to get this energy of detection inside the bin $[E_d^A,E_d^B]$ one should impose
\be 
    E_d^A < E_d < E_d^B \srarrow E_{\nu_\alpha}^{(\nu)}/E_d^A-1 > z_e > E_{\nu_\alpha}^{(\nu)}/E_d^B-1 \,,
\ee
so we can write
\bea
    N_d(E_d^A,E_d^B) = 4\pi A_d \int_{E_{\nu_\alpha}^{(\nu)}/E_d^A-1}^{E_{\nu_\alpha}^{(\nu)}/E_d^B-1} dz_e \frac{\rho(z_e)}{(1+z_e)H(z_e)}\int_{E_{\nu_\alpha}^{(\nu)}}^{E_\text{th}^{(e)}} dE_e \Phi_e(E_e) &\qty(\frac{E_e}{E_{\nu_\alpha}^{(\nu)}})\,, \\ &\quad(\text{for }E_\text{th}^{(e)}>E_e>E_{\nu_\alpha}^{(\nu)}) \,,
\eea
where again we have made a change of variables from the comoving radial distance $r_e$ to the redshift $z_e$.

For the case $E_e\in[E_e^A,E_e^B]>E_\text{th}^{(e)}>E_{\nu_\alpha}^{(\nu)}$ the development is completely analogous to the previous case, but using now Eq.~\eqref{eq:dEdPhid_1}. The result obtained is
\bea
    N_d(E_d^A,E_d^B) = 4\pi A_d \int_{E_{\nu_\alpha}^{(\nu)}/E_d^A-1}^{E_{\nu_\alpha}^{(\nu)}/E_d^B-1} dz_e \frac{\rho(z_e)}{(1+z_e)H(z_e)} \qty(\frac{E_\text{th}^{(e)}}{E_{\nu_\alpha}^{(\nu)}}) \int_{E_\text{th}^{(e)}}^{E_\text{max}} dE_e & \Phi_e(E_e)\qty(\frac{E_e}{E_\text{th}^{(e)}})^{1/(1+r)}\,, \\ &\quad(\text{for }E_e>E_\text{th}^{(e)}>E_{\nu_\alpha}^{(\nu)}) \,.
\eea

Combining the results of the three cases for a generic bin $[E_d^A,E_d^B]$, we can write
\bea
    N_d(E_d^A,E_d^B) = 4\pi A_d \int &dz_e \frac{\rho(z_e)}{(1+z_e)H(z_e)}\int_{(1+z_e)E_d^A}^{(1+z_e)E_e^B} dE_e \Phi_e(E_e) \theta(E_{\nu_\alpha}^{(\nu)}-E_e) \\
    + 4\pi A_d \int_{E_{\nu_\alpha}^{(\nu)}/E_d^A-1}^{E_{\nu_\alpha}^{(\nu)}/E_d^B-1} &dz_e \frac{\rho(z_e)}{(1+z_e)H(z_e)} \;N_\text{eff} \,,
\eea
where $\theta(x_0-x)$ is the Heaviside step function and  $N_\text{eff}$ is just a shortcut for
\be 
    N_\text{eff} \coloneqq \int_{E_{\nu_\alpha}^{(\nu)}}^{E_\text{th}^{(e)}} dE_e \Phi_e(E_e) \qty(\frac{E_e}{E_{\nu_\alpha}^{(\nu)}}) + \qty(\frac{E_\text{th}^{(e)}}{E_{\nu_\alpha}^{(\nu)}}) \int_{E_\text{th}^{(e)}}^{E_\text{max}} dE_e \Phi_e(E_e)\qty(\frac{E_e}{E_\text{th}^{(e)}})^{1/(1+r)} \,.
\ee

Finally, we can now take the zero bin-width limit in order to recover the flux of detected neutrinos from the number of detected neutrinos inside a bin,
\be 
    \Phi_d(E_d) \equiv \lim_{\Delta\rightarrow 0} \;\frac{1}{\Delta} \; N_d(E_d,E_d+\Delta) \,,
\ee
which leads to the result
\be
    \Phi_d(E_d) = 4\pi A_d \int_{z_\text{min}}^{E_{\nu_\alpha}^{(\nu)}/E_d-1} dz_e \frac{\rho(z_e)}{H(z_e)} \Phi_e((1+z_e)E_d)
    + 4\pi A_d \; \frac{\rho(E_{\nu_\alpha}^{(\nu)}/E_d-1)}{H(E_{\nu_\alpha}^{(\nu)}/E_d-1)} \;\frac{N_\text{eff}}{E_d} \,.
    \label{eq:insta_flux}
\ee

\begin{figure}[p]
    \centering
    \includegraphics[width=0.8\textwidth]{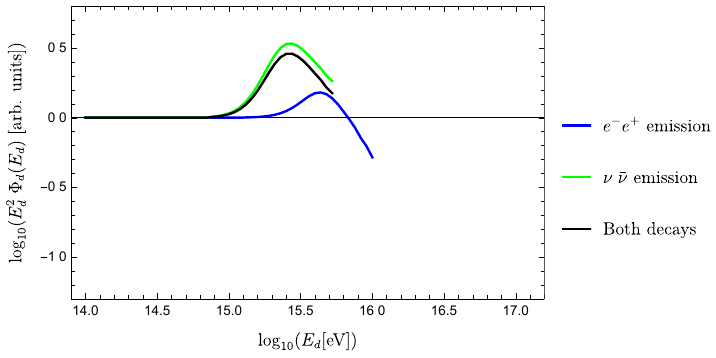}
    \caption{Neutrino flux at Earth using an analytical approximation of the instantaneous cascade model for $n=2$ and $\Lambda\approx 1.13\E{-2} M_P$ (such that $E_\text{th}^{(e)}=10^{16}$ eV). The only \abb{VPE} (blue) and only \abb{NSpl} (green) scenarios are also presented.}
    \label{fig:flux_instantaneous}
\end{figure}

In order to compute the flux we need to specify every function that appears in Eq.~\eqref{eq:insta_flux}. We can use $\rho(z_e)$ proportional to the \abb{SFR} distribution, as done in~\cite{Stecker:2014xja,Stecker:2014oxa,Stecker:2017gdy,Stecker:2022tzd}. The emission spectrum $\Phi_e(E_e)$ can be taken as an inverse power law from 100 TeV to 100 PeV, that is to say $\Phi_e(E_e)=(E_0/E_e)^2$, with $E_0$ a constant related with the luminosity of the source. The results for the computation of the flux are shown in Fig.~\ref{fig:flux_instantaneous}.

In contrast to the previous model, we have been able to include both effects, \abb{VPE} and \abb{NSpl}, in the computation of the flux. However, we can repeat all the development of this section considering only \abb{VPE} or  \abb{NSpl}. For the case in which one only considers the electron-positron pair emission one finds
\be
    \Phi_d(E_d) = 4\pi A_d \int_{z_\text{min}}^{E_\text{th}^{(e)}/E_d-1} dz_e \frac{\rho(z_e)}{H(z_e)} \Phi_e((1+z_e)E_d)
    + 4\pi A_d \; \frac{\rho(E_\text{th}^{(e)}/E_d-1)}{H(E_\text{th}^{(e)}/E_d-1)} \;\frac{N_\text{eff}^{(e)}}{E_d} \,,
    \label{eq:insta_flux_VPE}
\ee
with
\be 
    N_\text{eff}^{(e)} \coloneqq \int_{E_\text{th}^{(e)}}^{E_\text{max}} dE_e \Phi_e(E_e) \,,
\ee
and there is no multiplication of the number of neutrinos. If instead we consider the case with only \abb{NSpl} one can find that the flux is now
\be
    \Phi_d(E_d) = 4\pi A_d \int_{z_\text{min}}^{E_{\nu_\alpha}^{(\nu)}/E_d-1} dz_e \frac{\rho(z_e)}{H(z_e)} \Phi_e((1+z_e)E_d)
    + 4\pi A_d \; \frac{\rho(E_{\nu_\alpha}^{(\nu)}/E_d-1)}{H(E_{\nu_\alpha}^{(\nu)}/E_d-1)} \;\frac{N_\text{eff}^{(\nu)}}{E_d} \,,
    \label{eq:insta_flux_NSpl}
\ee
with
\be 
    N_\text{eff}^{(\nu)} \coloneqq \int_{E_{\nu_\alpha}^{(\nu)}}^{E_\text{max}} dE_e \Phi_e(E_e) \qty(\frac{E_e}{E_{\nu_\alpha}^{(\nu)}}) \,.
\ee
The computed flux for these scenarios are also shown in Fig.~\ref{fig:flux_instantaneous}.

We see how this model has problems in predicting the flux very close to the cutoff, because instead of showing a smooth suppression of the flux we get instead a discontinuity that goes from a non-zero number of neutrinos below $E_\text{cut}$ to zero after it. We can, however, still extract useful information from the results. 

Firstly, one can check the similarities between the green and black curves in Fig.~\ref{fig:flux_instantaneous}, showing how, for this value of $\Lambda$ (see Fig.~\ref{fig:threshold_comp} and the corresponding discussion), the flux is dominated by the \abb{NSpl} in both shape and cutoff. Secondly, as expected, under the instantaneous approximation, in which all the neutrinos above the threshold disintegrate, a scenario which only allows \abb{NSpl} (green curve) produces the larger enhancement of the flux. If instead one additionally includes now the electron-positron pair emission, there is a competition between both effects and each time the neutrino undergoes through \abb{VPE}, no multiplication happens.

\begin{figure}[p]
    \centering
    \includegraphics[width=0.95\textwidth]{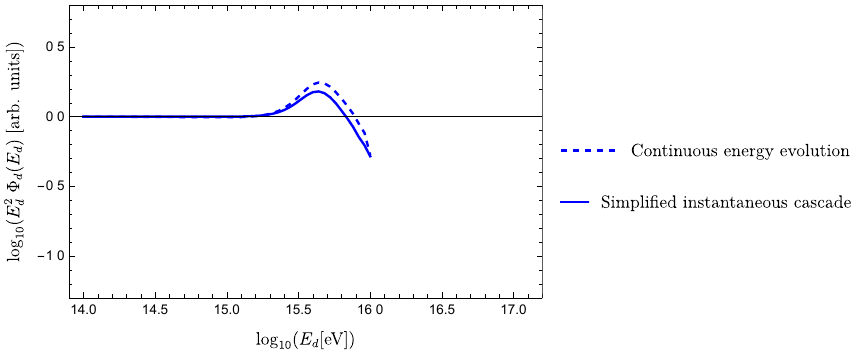}
    \caption{Comparison between the detected neutrino flux computed using two different models for $n=2$ and $\Lambda\approx 1.13\E{-2} M_P$ (such that $E_\text{th}^{(e)}=10^{16}$ eV).}
    \label{fig:flux_comp}
\end{figure}

We can finally compare the continuous energy evolution model (Appendix~\ref{sec:flux_continuous}) with this simplified version of the instantaneous cascade model, for the case of the electron-positron pair emission only. The results are shown in Fig.~\ref{fig:flux_comp}.
We see how the computed flux in the Appendix~\ref{sec:flux_continuous} produce an overestimation of the number of detected neutrinos in the bump and close to the cutoff with respect to the simplified instantaneous cascade. 

\clearpage
\section{Implementation of the instantaneous cascade model in SimProp}
\label{sec:simprop}

In order to characterize a \abb{LI} violating scenario we need at least two quantities, the scale of new physics $\Lambda$ (in units of the Plank mass) and the order of correction $n$. If any of these parameters are specified, the \abb{LIV} subroutine is activated. In case the other parameter is not specified, some default values are provided.
\begin{minted}[breaklines]{cpp}
double eLIV = 1.;
int nLIV = 2;
\end{minted}

When the program generates a neutrino, it calls a subroutine to propagate it. Once this subroutine is called, one should first characterize the particle that is being propagated. To do that, we ask for its energy, redshift, and flavour. The flavour can be positive or negative, in order to distinguish particles and antiparticles, but we also store the absolute value for later convenience (electron=1, muon=2 and tau=3).
\begin{minted}[breaklines]{cpp}
double E = input->GetEprod();
double z = input->GetZprod();
int flav = input->GetFlavor();
int abs_flav = sqrt(flav*flav);
\end{minted}

With this, both the \abb{LIV} scenario and the primary particle, are completely characterized. Then, one can now compute the relevant quantities for the instantaneous cascade, i.e., the kinematical and dynamical thresholds.
\begin{minted}[breaklines]{cpp}
double ErNSpl = pow(3.*c0*cfNSpl[abs_flav]*cnNSpl[nLIV],-1./(5.+3.*(double)nLIV)) / pow(pow(H0eV,1/(3.*nLIV))*eLIV*Mp,-3.*(double)nLIV/(5.+3.*(double)nLIV));
double ErVPE = pow(c0*cfVPE[abs_flav]*cnVPE[nLIV],-1./(5.+3.*(double)nLIV)) / pow(pow(H0eV,1/(3.*nLIV))*eLIV*Mp,-3.*(double)nLIV/(5.+3.*(double)nLIV));
double EthVPE = pow(2.*me*me*pow(eLIV*Mp,(double)nLIV),1./(2.+(double)nLIV));
\end{minted}
We have used the fact that the thresholds can be written in terms of the constants $c_0$, $c_{\nu_f}^{(A)}$, and $c_n^{(A)}$ of the decay width (see Sec.~\ref{sec:comparison}). These constants are stored as vectors; for $c_{\nu_f}^{(A)}$ the index corresponds to the flavour (without sign), and for $c_n^{(A)}$ the index corresponds to the value of $n$.
\begin{minted}[breaklines]{cpp}
double cfNSpl[4] = {0.00, 1.00, 1.00, 1.00};
double cfVPE[4] = {0.00, 1.68, 0.13, 0.13};
double cnNSpl[3] = {0.000, 0.024, 0.024};
double cnVPE[3] = {0.000, 0.144, 0.178};
\end{minted}

Following the instantaneous approximation, if the energy of the neutrino is below the threshold of a decay, then the probability of decay through that process is set to zero. If not, the decay width should be computed. As we are only interested in comparisons (ratios) between probabilities, we do not need to compute the whole decay widths, only the factors which are not common
(see Sec.~\ref{sec:comparison}).
\begin{minted}[breaklines]{cpp}
double EdVPE = (EthVPE>ErVPE) ? EthVPE : ErVPE;

double rate_NSpl = (E<ErNSpl) ? 0. : 3.0*cfNSpl[abs_flav]*cnNSpl[nLIV];
double rate_VPE  = (E<EdVPE) ? 0. : cfVPE[abs_flav]*cnVPE[nLIV];
double rate_total = rate_VPE + rate_NSpl;
\end{minted}

If the energy is below both thresholds, the neutrino will propagate trivially to Earth. We call this process ``Earth''. If instead the energy is above one of the thresholds, following the instantaneous approximation, the neutrino will always decay, so we just need to randomly choose through which process, using the probabilities we have just computed. 
\begin{minted}[breaklines]{cpp}
if (rate_total == 0) process = eEarth;
else{
    double u = gRandom->Uniform(0,1);
    if (u <= rate_VPE/rate_total) process = eVPE;
    else process = eNSpl;
}
\end{minted}

Once the process which the neutrino will undergo is set, we need to define which are the outputs of each process. If the particle is sent to Earth, the output will be the same initial neutrino, but with a redshifted energy, $E/(1+z)$, and with a new redshift equal to zero (i.e., the particle is now at Earth).
\begin{minted}[breaklines]{cpp}
switch (process){
    case eEarth: {
        input->SetEint(E/(1.+z));
        input->SetZint(0.);
        input->SetIntMult(0);
        break;
    }
    case eNSpl: {...}
    case eVPE: {...}
}
\end{minted}

If instead the neutrino undergoes \abb{VPE} or \abb{NSpl}, we will find three particles in the final state, whose energies should be randomly sampled from the energy distributions discussed in Sec.~\ref{sec:comparison}. In order to make this random sampling we have used the libraries included in the software CERN ROOT. We have defined a TF2 function (two variable function) for each probability distribution. These two-variable probability distributions are obtained applying the Dirac delta in Eqs.~\eqref{eq:energy_dist_VPE_electron} and \eqref{eq:energy_dist_VPE_muon}. Additionally we have made a change of variable from $(x',x_-)$ to $(x',y)$ with ${y\coloneqq(x_-)/(1-x_-)}$, because in this way, now the two variables go from zero to one.
\begin{minted}[breaklines]{cpp}
TF2 *NSplProbDistr = new TF2("NSplProbDistr","(1.-x) * (pow(1.+[0],3.)/(4.*[1])) * pow(1.-x,3.) * pow(1.-(1.-x)*y,3.) * pow(x+(1-x)*y,3.*[0]-1.)",0,1,0,1);
TF2 *VPEProbDistr = new TF2("VPEProbDistr","(1.-x) * (1./[1]) * pow(1.-pow(x,[0]+1.),3.) * ( [3]*pow(1.-(1.-x)*y,2.) + [2]*pow(x+(1.-x)*y,2.) )",0,1,0,1);
\end{minted}

When the neutrino produces a neutrino-antineutrino pair, we will find two neutrinos and an antineutrino in the final state, whose energies should be randomly sampled from the energy distribution given by Eq.~\eqref{eq:energy_dist_NSpl}.
\begin{minted}[breaklines]{cpp}
case eNSpl: {
    ...
    NSplProbDistr->SetParameters((double)nLIV,cnNSpl[nLIV]);
    NSplProbDistr->GetRandom2(x,y);
    E1 = E * x;
    E2 = E * (1.-x)*y;
    E3 = E - E1 - E2;
    ...
    output.push_back(Particle(0,0,nBr,E1,z,particle1));
    output.push_back(Particle(0,0,nBr,E2,z,particle2));
    output.push_back(Particle(0,0,nBr,E3,z,particle3));
    break;
}
\end{minted}
Let us note that the energy distribution is flavour independent.

If instead the neutrino produces an electron-positron pair, we will find a neutrino, an electron and a positron in the final state. Their energies must be randomly sampled from the energy distribution of the electron-positron decay, but it is different whether the initial neutrino is electronic (see Eqs.~\eqref{eq:energy_dist_VPE_electron} and \eqref{eq:energy_dist_VPE_electron_cos}) or muon/tau (see Eqs.~\eqref{eq:energy_dist_VPE_muon} and \eqref{eq:energy_dist_VPE_muon_cos}). So we need to set the flavour dependent parameters of the probability distribution before the sampling.
\begin{minted}[breaklines]{cpp}
double cos2theta, sin2theta;
switch(abs_flav){
    case 1:
        cos2theta = 0.97;
        sin2theta = 0.03;
        break;
    case 2:
    case 3:
        cos2theta = 0.61;
        sin2theta = 0.39;
        break;
}
\end{minted}
Once these parameters are set, the energy sampling is completely equivalent to the neutrino-antineutrino decay case.
\begin{minted}[breaklines]{cpp}
case eVPE: {
    ...
    VPEProbDistr->SetParameters((double)nLIV,cnVPE[nLIV],
    cos2theta,sin2theta);
    VPEProbDistr->GetRandom2(x,y);
    E1 = E * x;
    E2 = E * (1.-x)*y;
    E3 = E - E1 - E2;
    ...
    output.push_back(Particle(0,0,nBr,E1,z,particle1));
    output.push_back(Particle(0,charge2,nBr,E2,z,eElectron));
    output.push_back(Particle(0,charge3,nBr,E3,z,eElectron));
    break;
}
\end{minted}

Up to this point we have shown the basic ingredients one have to included in SimProp to implement the instantaneous cascade model. However, one can still add additional ingredients to gain some extra functionalities. For instance, we have defined a parameter \mintinline{cpp}{int fLIV = 0}, which can take integer values as \mintinline{cpp}{fLIV = 1} or \mintinline{cpp}{fLIV = 2}, in order to consider the \abb{VPE} only or \abb{NSpl} only scenarios. This can be accomplished just by setting the decay probabilities to zero for all the effects except the one in which we are interested.

Additionally, we can allow extra values for the parameter \mintinline{cpp}{fLIV}, in order to compute the three previously mentioned scenarios (\abb{VPE}+\abb{NSpl}, only \abb{VPE}, and only \abb{NSpl}), without randomly sampling the energies of the product particles, but using instead the mean values computed in Sec.~\ref{sec:comparison}. This allow us to gain computational speed in exchange for some accuracy of the computation. A comparison between the expected flux using or not this additional approximation is shown in Sec.~\ref{sec:monte_carlo}.

\end{subappendices}
\printbibliography[heading=subbibintoc,title={References}]
\end{refsection}

\begin{refsection}
\chapter{Cosmic Rays and cosmogenic neutrinos}
\graphicspath{{Chapter4/Figs/}}
\label{chap:cosmicrays}

Cosmic rays are positive charged particles, mostly consisting of atomic nuclei, from the lightest (protons and helium/$\alpha$-particles) to the heaviest elements (up to iron, $A=56$)~\cite{Gaisser:2016}. Due to the fact that they are charged, they are deflected by the magnetic fields during their propagation in the outer space, making challenging to identify their origin from their arrival directions. Additionally, they interact electromagnetically with the photon backgrounds introduced in Sec.~\ref{sec:cosmology}. In this section, we are interested in the most energetic part of the cosmic rays spectrum, for which the relevant backgrounds are the \abb{CMB} and \abb{EBL}.

\begin{figure}[htb]
    \centering
    \includegraphics[width=0.7\textwidth]{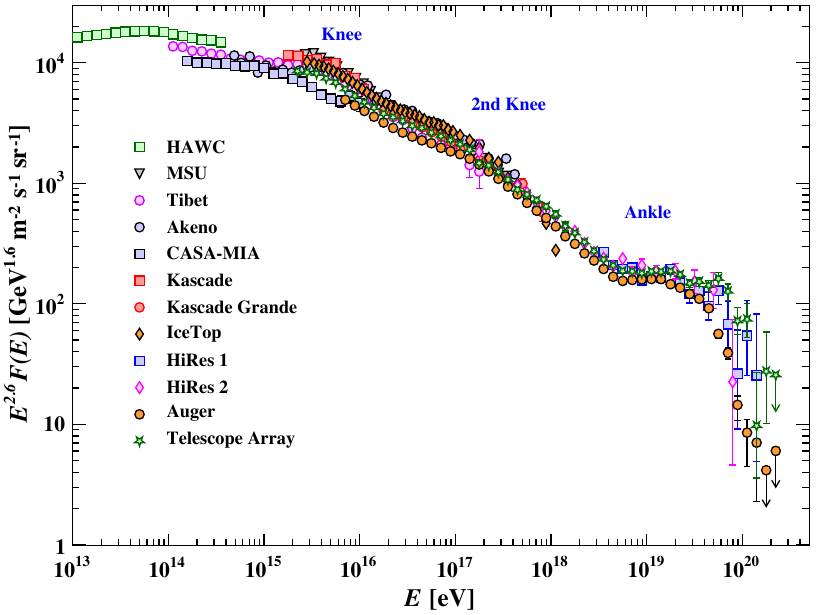}
    \caption{Cosmic rays all-particle spectrum measured by the different experiments. From~\cite{ParticleDataGroup:2022}.}
    \label{fig:cr_spec}
\end{figure}

The high-energy part of the detected cosmic rays spectrum at Earth contains four general features. There is a first \textit{knee} above $10^{15}\unit{eV}$ and a second one around $10^{17}\unit{eV}$, then there is an \textit{ankle} around $10^{19}\unit{eV}$, and finally a cutoff above $10^{20}\unit{eV}$. These characteristics are shown in Fig.~\ref{fig:cr_spec}. As a result of this behaviour, the best fit for the detected spectrum is a broken inverse power law, with an index $\alpha\sim 2.7$ below the first knee, $\alpha\sim 3.1$ after it, with an slightly change to $\alpha\sim 3.3$ after the second knee, and a recovering of $\alpha\sim 2.7$ above the ankle, up to the cutoff~\cite{Thoudam:2016syr}.

Several models have been proposed to explain this peculiar shape of the cosmic ray flux. Nowadays, the most accepted hypothesis is that the cosmic ray spectrum is composed by, at least, two very different contributions: a galactic one, up to the second knee, and an extragalactic one, above the ankle. The sources of the galactic contribution are thought to be supernova remnants or energetic binary systems, which could accelerate different species of nuclei up to a maximum energy proportional to their charge. This way, the shape of the flux up to $10^{17}\unit{eV}$ is thought to be composed by the superposition of different nucleus spectra. The maximum of the hydrogen and helium spectra would correspond to the first knee, and the successive cutoffs of heavier nuclei would end at the second knee with the iron cutoff (see~\cite{Sciascio:2022vkb,Abu-Zayyad:2018btv,Thoudam:2016syr}).

The sources of the extragalactic component above the ankle are thought to be very high-energy accelerators such as \abb{AGN} and \abb{GRB}. The spectrum starts from a predominantly dominated proton composition, and starts getting heavier as the energy increases (see~\cite{Condorelli:2023wsm,PierreAuger:2022atd}). This last part of the spectrum, above $10^{18}\unit{eV}$, is usually known as \abbdef{UHECR}{Ultra high-energy Cosmic Rays}, and it extends up to the cutoff. This cutoff can be explained from propagation effects, as the interactions with the \abb{CMB} (\abbdef{GZK}{Greinsen-Zatsepin-Kuzmin} cutoff~\cite{Greisen:1966jv,Zatsepin:1966}), in combination with a lack of sources able to accelerate the cosmic rays up to those energies.

The \abb{UHECR} interact significantly with the photon backgrounds. In the part of the flux dominated by protons, proton-photon interactions as the pion photoproduction ($p+\gamma\rightarrow \pi^0 + p$ and $p+\gamma\rightarrow \pi^+ + n$) and the Bethe-Heitler pair production ($p+\gamma\rightarrow p+e^-+e^+$) are the most relevant effects of energy loss. In the case of heavy nuclei, the dominant effect comes from the photodisintegration, in which a heavy nucleus is excited by a photon and then emits one or more protons, neutrons or alpha particles~\cite{Boncioli:2022ojf}.

The threshold energy of the protons for the photoproduction of pions (for head-on collisions) can be found to be
\be 
    E_\text{th}^{(\pi)} = \frac{m_\pi^2+2m_\pi m_p}{4\epsilon} \,,
\ee
with $\epsilon$ the energy of the background photon. Given that the \abb{CMB} energies are distributed around a central value of $\approx 10^{-4} \unit{eV}$, we obtain that only protons above $\approx 10^{20} \unit{eV}$ can produce pions when interacting with the \abb{CMB} photons. The cosmic rays above this threshold energy will lose about a 20\% of their energy due to pion production per interaction~\cite{Kachelriess:2008ze,Gaisser:2016}, leading to the \abb{GZK} cutoff. If one instead considers the photons of the \abb{EBL}, which has a first peak around $10^{-2}\unit{eV}$, the threshold energy will decrease around 2 orders of magnitude with respect to the \abb{CMB}. Let us note that even if the intensity of the \abb{EBL} is less than that of the \abb{CMB}, it could still be an important source of secondary particles.

The same procedure can be followed for the electron-positron pair production, whose threshold energy can be found to be 
\be 
    E_\text{th}^{(e)} = \frac{4m_e^2+8m_e m_p}{4\epsilon} \,,
\ee
which for photons of the \abb{CMB} is $\approx 10^{18}\unit{eV}$, and around 2 orders of magnitude less for photons of the \abb{EBL}. Let us note, however, that the electron-positron pair production is a much less important mechanism of energy loss for protons, as in each interaction the primary particle only loses around 0.1\% of its energy~\cite{Kachelriess:2008ze,Gaisser:2016}.

\section{Cosmogenic neutrinos}

Let us note that the produced particles from the proton-photon interactions are injected again in the cosmic ray flux; however, the hadrons are unstable and will decay before propagating large distances. For instance, the neutral pion can decay in a pair of photons ($\pi^0 \rightarrow \gamma + \gamma$); however, in this section we are more interested in the charged pions, which can decay into neutrinos.

\subsection{Mechanisms of neutrino production}

The positive charged pion will quickly decay into an antimuon and a muon neutrino\footnotemark ($\pi^+\rightarrow \mu^++\nu_\mu$). However, in turn, the antimuon will decay into a positron, an electron neutrino and a muon antineutrino ($\mu^+\rightarrow e^+ + \nu_e + \bar\nu_\mu$). We see how a muon neutrino, a muon antineutrino and an electron neutrino, of similar energies, can be obtained from the same primary proton.   
\footnotetext{Let us remember that the electron decay is helicity-suppressed and the tau one is not possible due to the large energy necessary to produce a tau particle.}
\be
\begin{array}{ l l l l l l l l l l l }
\bm{p} & + & \gamma  & \rightarrow  & \pi ^{+} & + & n &  &  &  & \\
 &  &  &  & \pi ^{+} & \rightarrow  & \mu ^{+} &  &  & + & \bm{\nu_\mu}\\
 &  &  &  &  &  & \mu ^{+} & \rightarrow  & e^{+} & + & \bm{\nu _e +\bar\nu_\mu} \,.
\end{array}
\ee

One should also take into account that when a positive charged pion is produced, ($p+\gamma\rightarrow \pi^+ + n$) a companion neutron also appears, which can undergo beta decay ($n\rightarrow p + e^- + \bar\nu_e$), producing an additional electronic antineutrino contribution of different energy,
\be
\begin{array}{ l l l l l l l l l l l l l }
\bm{p} & + & \gamma  & \rightarrow  & \pi ^{+} & + & n &  &  &  &  &  & \\
 &  &  &  &  &  & n & \rightarrow  & p & + & e^{-} & + & \bm{\bar\nu_e} \,.
\end{array}
\label{eq:beta_decay}
\ee
The neutron beta decay can also be produced by the neutrons that come from the photodisintegration of heavy nuclei, or from unstable isotopes decaying into their stable form.

Additionally, in more complex and subdominant processes (like multipion production), negative charged pions can be produced, replicating the chain of reactions of the $\pi^+$, but changing every particle by its antiparticle and vice versa. This produces an additional contribution of muon neutrinos and antineutrinos, plus a new electronic neutrino contribution.
\be
\begin{array}{ l l l l l l l l l l l }
\bm{p} & + & \gamma & \rightarrow  & \pi ^{-} & + & \cdots &  &  &  & \\
 &  &  &  & \pi ^{-} & \rightarrow  & \mu ^{-} &  &  & + & \bm{\bar\nu_\mu}\\
 &  &  &  &  &  & \mu ^{-} & \rightarrow  & e^{-} & + & \bm{\bar\nu _e +\nu_\mu} \,.
\end{array}
\ee
However, these subdominant processes are not taken into account in SimProp~\cite{Aloisio:2017iyh}, so they will not be considered in the following discussion.

In the case that the cosmic ray flux is dominated by protons, one expects a strong production of cosmogenic neutrinos through the mechanisms discussed above. However, as we discussed before, if one consider the cosmic ray flux at energies above the ankle, the flux gets heavier as the energy increase. In that case, the threshold of the proton-photon interaction increases, since the protons are now bounded inside the nuclei. The heavier the element, the more difficult for a photon of the background to interact with the protons of the nucleus, and as a consequence, the production of secondary neutrinos decrease. In this study, for simplicity, and as starting point for future work, we will consider an initial cosmic ray flux of pure protons.

\subsection{Cosmogenic neutrino flux using SimProp}

\begin{figure}[tbp]
        \centering
        \includegraphics[width=0.6\textwidth]{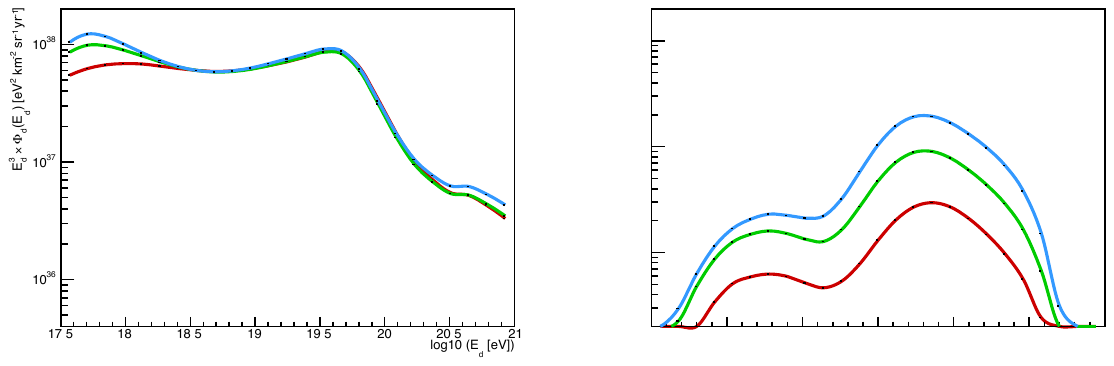}
        \caption{Cosmic rays flux at Earth for a uniform (red), \abb{SFR} (green) and \abb{AGN} (blue) source distribution with an spectral index $\gamma=2.6$, 2.5 and 2.4, respectively.}
        \label{fig:flux_cr}
\end{figure}

In order to simulate the cosmogenic neutrino production from the cosmic rays interactions, we need to fix the astrophysical scenario for the production and propagation of the cosmic rays. We are interested on very high-energy cosmic rays from extragalactic origin, so we are going to consider  energies from $10^{17}$ to $10^{21}$ eV and, for simplicity, redshifts within the range $[0,1]$. For the emission spectrum we will use the usual assumption taken in the literature~\cite{Boncioli:2022ojf,Aloisio:2017eqv,Aloisio:2015ega,Aloisio:2012mi}: an inverse power law (characterized by an spectral index $\gamma$) suppressed by an exponential cutoff (controlled by an energy $E_\text{sup}=10^{22}\unit{eV}$). However, we should notice that there exists a degeneracy between the emission spectrum and the source distribution, i.e., there exist different combinations of values $\gamma$ and source distributions that lead to a similar detected cosmic ray flux compatible with observations. If one considers three possible source distributions (see~\cite{Aloisio:2015ega}), uniform distribution, proportional to the \abb{SFR}, and the characteristic \abb{AGN} evolution, there are three different values for the emission spectral index: $\gamma=2.6$, 2.5 and 2.4, respectively, which reproduce the detected cosmic rays spectrum at Earth (see Fig.~\ref{fig:flux_cr}).

\begin{figure}[p]
    \centering
    \begin{minipage}{0.49\textwidth}
        \centering
        \includegraphics[width=\textwidth]{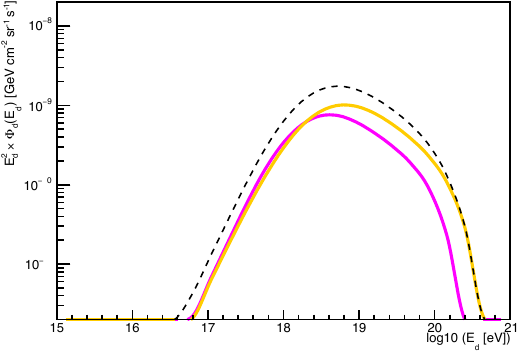}
        \caption{Cosmogenic muon neutrino (pink) and antineutrino (orange) flux at Earth considering \abb{CMB} only. Sum in dashed line.}
        \label{fig:flux_muon_neu_cmb}
    \end{minipage}%
    \hfill
    \begin{minipage}{0.49\textwidth}
        \centering
        \includegraphics[width=\textwidth]{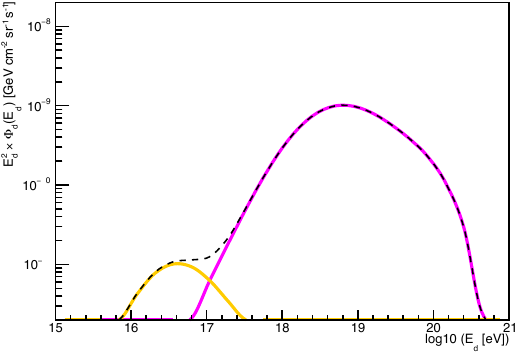}
        \caption{Cosmogenic electron neutrino (pink) and antineutrino (orange) flux at Earth considering \abb{CMB} only. Sum in dashed line.}
        \label{fig:flux_elec_neu_cmb}
    \end{minipage}
    \par\vspace{1em}
    \begin{minipage}{0.49\textwidth}
        \centering
        \includegraphics[width=\textwidth]{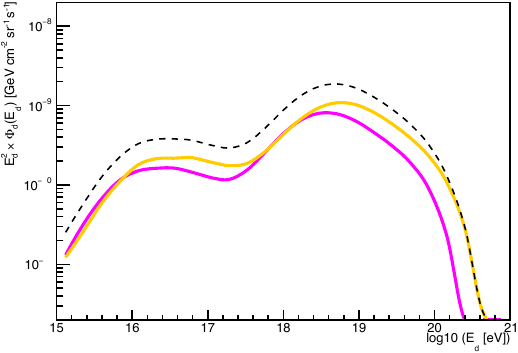}
        \caption{Cosmogenic muon neutrino (pink) and antineutrino (orange) flux at Earth considering \abb{CMB} and \abb{EBL}. Sum in dashed line.}
        \label{fig:flux_muon_neu}
    \end{minipage}%
    \hfill
    \begin{minipage}{0.49\textwidth}
        \centering
        \includegraphics[width=\textwidth]{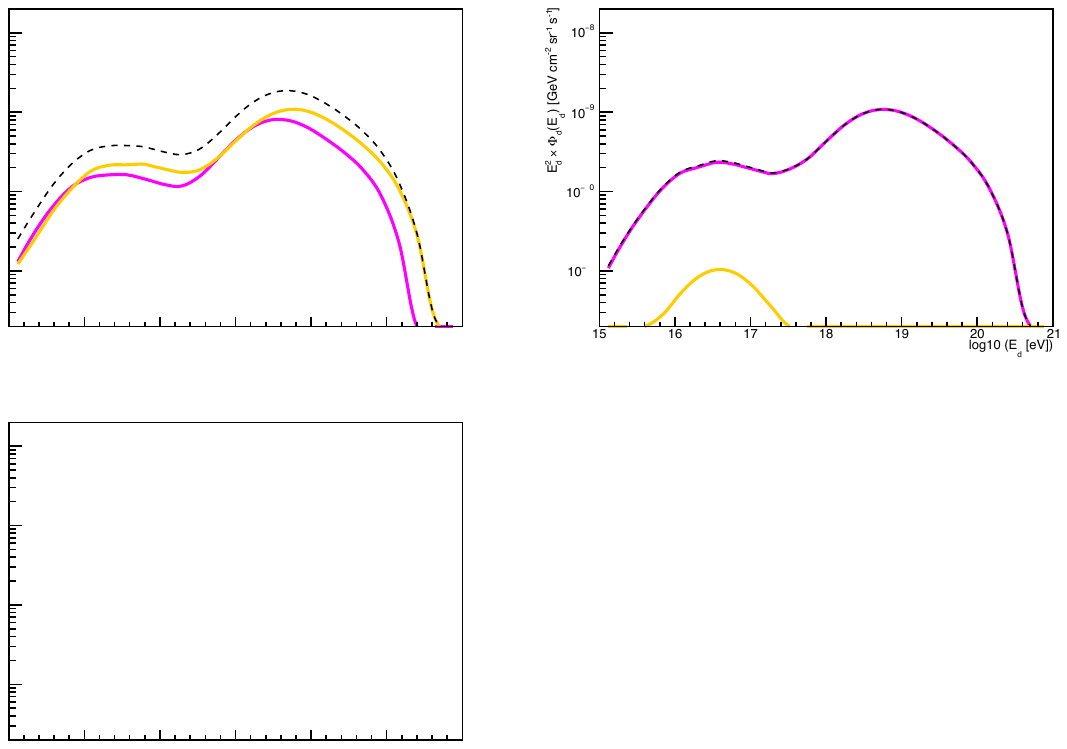}
        \caption{Cosmogenic electron neutrino (pink) and antineutrino (orange) flux at Earth considering \abb{CMB} and \abb{EBL}. Sum in dashed line.}
        \label{fig:flux_elec_neu}
    \end{minipage}
    \par\vspace{1em}
    \begin{minipage}{0.49\textwidth}
        \centering
        \includegraphics[width=\textwidth]{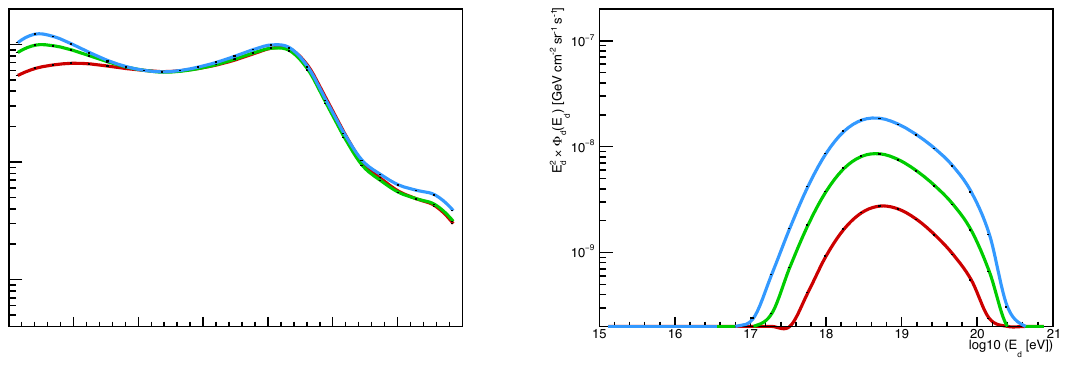}
        \caption{Cosmogenic neutrino flux at Earth (\abb{CMB} only) for a uniform (red), \abb{SFR} (green) and \abb{AGN} (blue) source distribution.}
        \label{fig:flux_all_neu_cmb}
    \end{minipage}%
    \hfill
    \begin{minipage}{0.49\textwidth}
        \centering
        \includegraphics[width=\textwidth]{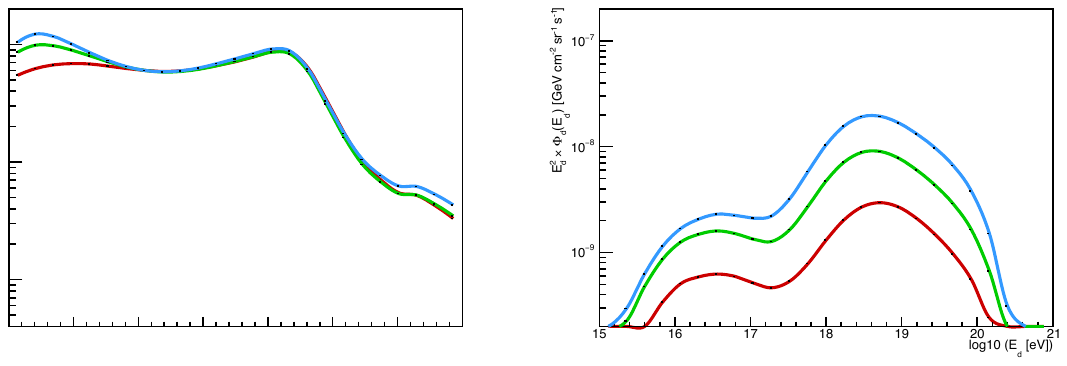}
        \caption{Cosmogenic neutrino flux at Earth (\abb{CMB} and \abb{EBL}) for a uniform (red), \abb{SFR} (green) and \abb{AGN} (blue) source distribution.}
        \label{fig:flux_all_neu}
    \end{minipage}
\end{figure}

Let us fix one of the astrophysical scenarios presented above (for instance, the uniform source evolution). The cosmic rays will be produced according to the selected model and they will interact in their propagation with the photon background through the processes discussed in the previous section. If the photon background is the \abb{CMB} ($\epsilon \approx 10^{-4}\unit{eV}$), we find that only protons above $10^{20}$ eV can produce pions. Assuming that in the pion photoproduction the resultant neutrinos/antineutrinos will inherit around a 5\%~\cite{Gaisser:2016} of the primary proton energy, we expect a muon neutrino, muon antineutrino, and electron neutrino flux contribution within approximately $10^{18.5}$ to $10^{19.5}$ eV. We can see this behaviour in Figs.~\ref{fig:flux_muon_neu_cmb} and \ref{fig:flux_elec_neu_cmb}.
Let us also notice that in Fig.~\ref{fig:flux_elec_neu_cmb} we can see an additional small peak of electron antineutrinos, corresponding to the contribution from the neutron beta decay, Eq.~\eqref{eq:beta_decay}. If one now considers also the interactions with the \abb{EBL}, the pion photoproduction threshold is decreased by 2 orders of magnitude. Then we expect a second peak in the muon neutrino, muon antineutrino, and electron neutrino flux at energies a couple of order of magnitudes below the ones produced by the \abb{CMB}. This behaviour is correctly reproduced in Figs.~\ref{fig:flux_muon_neu} and \ref{fig:flux_elec_neu}.

Let us note that, even if the three astrophysical models under consideration produce a similar cosmic ray flux\footnotemark, this is not necessarily the case for the cosmogenic neutrinos. The interactions of the protons with the \abb{CMB} lead to the \abb{GZK} cutoff, which limits the distance of the sources able to contribute to the \abb{UHECR} spectrum; however, neutrinos do not have this limitation. The expected neutrino spectrum can come from much farther distances, making them much more sensitive than the cosmic rays to the source distribution. A comparison between the cosmic neutrino flux predicted by each astrophysical model is shown in Figs.~\ref{fig:flux_all_neu_cmb} and \ref{fig:flux_all_neu}.
\footnotetext{Let us note, however, that we can see differences in the cosmic ray flux for each astrophysical model around $10^{18}\unit{eV}$, where the spectrum is more sensitive the contribution of distant sources.}

\section{LIV effects in cosmogenic neutrinos}

In this section we will use the same \abb{LIV} model for neutrinos we introduced in Sec.~\ref{sec:LIV_neu} and developed in Chapter~\ref{chap:neutrinos}. There, we saw that the strongest modification on the astrophysical neutrino flux is a cutoff preceded by an accumulation we called a \textit{bump}. When one applies the same scenario to cosmogenic neutrinos one would expect a similar behaviour. Playing with the scale of new physics one can make one of the cosmogenic neutrino peaks disappear\footnotemark, or even for some values of $\Lambda$ one can try to enhance the peaks matching them with the \abb{LIV} bump.
\footnotetext{A different \abb{LIV} model suppressing the photopion production would also produce a suppression of the highest part of the cosmogenic neutrino spectrum. Let us note, however, that in that case there would not be a bump before the cutoff~\cite{Scully:2010iv,Stecker:2004xm}.}

In order to know which values of the scale of new physics are relevant for our study we can adapt the relation we found between $\Lambda$ and $E_\text{cut}$ for astrophysical neutrinos (see Eqs.~\eqref{eq:cutoff_VPE} and \eqref{eq:cutoff_NSpl}, or Fig.~\ref{fig:Ecut}). Summarising the result of the previous chapter, the energy of the cutoff can be approximated by the quotient between the relevant energy threshold ($E_{\nu_\alpha}^{(\nu)}$ or $E_\text{th}^{(e)}$, depending on the value of $\Lambda$) and $(1+z_\text{min})$, with $z_\text{min}$ the redshift of the closest source. For the case of the cosmogenic neutrinos, instead of point-like neutrino sources one has a diffuse neutrino generation which can be produced during the propagation of cosmic rays. Thus, we can particularize Eqs.~\eqref{eq:cutoff_VPE} and \eqref{eq:cutoff_NSpl} for the case of cosmogenic neutrinos taking $z_\text{min}=0$, %as neutrinos can be emitted arbitrarily close to Earth:
\begin{align}
    E_\text{cut} &\approx E_{\nu_\alpha}^{(\nu)} =\Lambda^{3n/(5+3n)} \qty(\frac{96}{192\pi^3} \frac{G_F^2}{H_0}\,c_n^{(\nu)})^{-1/(5+3n)} &\text{for $E_{\nu_\alpha}^{(\nu)}>E_\text{th}^{(e)}$} \,,
    \label{eq:cutoff_VPE_cosmo} \\
    E_\text{cut} &\approx E_\text{th}^{(e)} = \Lambda^{n/(n+2)} \; (2m_e^2)^{1/(n+2)} &\text{for $E_{\nu_\alpha}^{(\nu)}<E_\text{th}^{(e)}$} \,.
    \label{eq:cutoff_NSpl_cosmo}
\end{align}
Using the previous equations we find that in order to make the cutoff (and the bump) to match the \abb{EBL} (low energy) and \abb{CMB} (high-energy) peaks, one needs a scale of new physics around $\Lambda \approx 2.19 M_P$ and $\Lambda\approx 1.13\E{4} M_P$, respectively, for $n=2$. If one considers the case $n=1$ instead, then $\Lambda \approx 5.90\E{11} M_P$ and $\Lambda\approx 1.57\E{17} M_P$ are necessary.

\subsection{Cosmogenic neutrino flux using SimPropLIV}

\begin{figure}[tbp]
    \centering
    \begin{minipage}{0.49\textwidth}
        \centering
        \includegraphics[width=\textwidth]{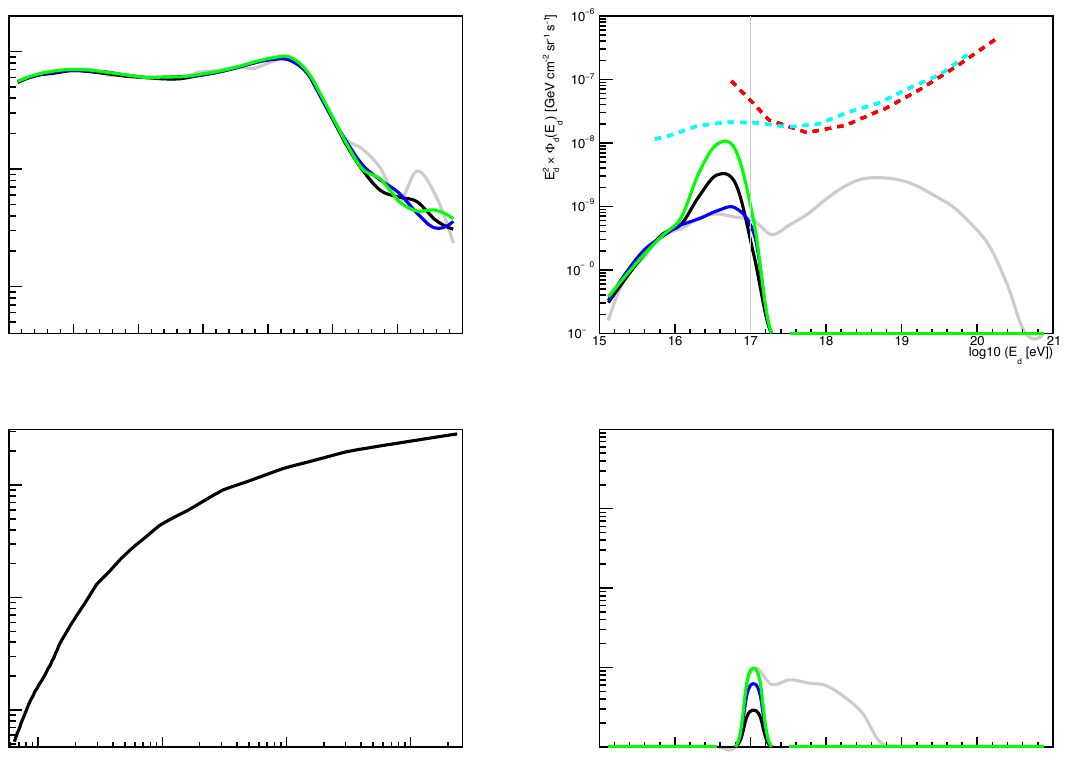}
        \caption{Cosmogenic neutrino flux at Earth (black) for $n=2$ and $\Lambda\approx 2.19 M_P$ ($E_\text{cut}=10^{17}\unit{eV}$). The only \abb{VPE} (blue) and only \abb{NSpl} (green) scenarios are also presented.}
        \label{fig:flux_low_peak_n2}
    \end{minipage}%
    \hfill
    \begin{minipage}{0.49\textwidth}
        \centering
        \includegraphics[width=\textwidth]{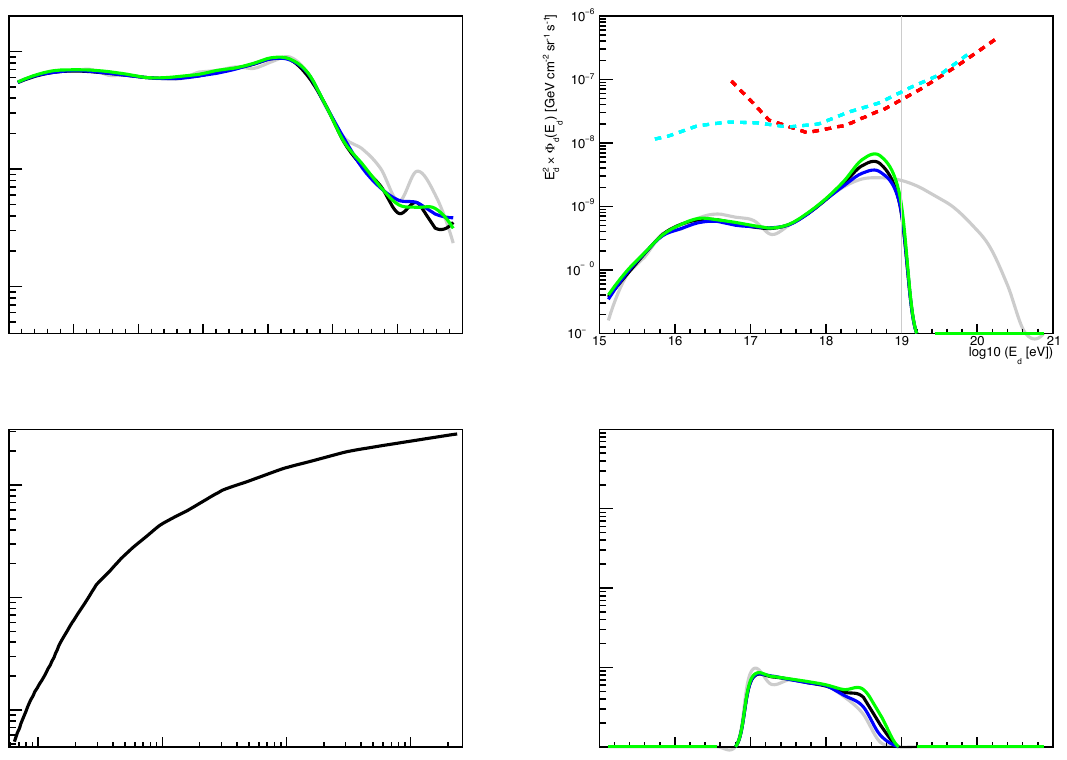}
        \caption{Cosmogenic neutrino flux at Earth (black) for $n=2$ and $\Lambda\approx 1.13\E{4} M_P$ ($E_\text{cut}=10^{19}\unit{eV}$). The only \abb{VPE} (blue) and only \abb{NSpl} (green) scenarios are also presented.}
        \label{fig:flux_high_peak_n2}
    \end{minipage}
\end{figure}

Let us fix the astrophysical scenario to a uniform distribution of sources (with $\gamma=2.6$) and $n=2$ for now. In Figs.~\ref{fig:flux_low_peak_n2} and \ref{fig:flux_high_peak_n2} we show the prediction for the cosmic neutrino flux for the previously mentioned values of $\Lambda$. We also show the 90\% CL upper limit for the detected flux by \abb{IC} (cyan) and \abb{PA} (red), from~\cite{IceCube:2018fhm} and \cite{PierreAuger:2019ens}, respectively.

In the same figures (\ref{fig:flux_low_peak_n2} and \ref{fig:flux_high_peak_n2}), we also show the only \abb{VPE} (in blue) and only \abb{NSpl} (in green) \abb{LIV} scenarios. Let us note that when we studied the astrophysical neutrinos, we considered values of $\Lambda$ less than the Plank scale. For these values of the scale of new physics, $E_\text{th}^{(e)}$ is always greater that $E_{\nu_{\mu,\tau}}^{(e)}$ (see Fig.~\ref{fig:threshold_comp}), and so the cutoff for the only \abb{VPE} scenario is controlled by $E_\text{th}^{(e)}$. As the cutoff for the only \abb{NSpl} scenario is controlled by $E_{\nu_{\alpha}}^{(\nu)}$, this implies a different cutoff and bump location for each effect (see Fig.~\ref{fig:flux_mean} or \ref{fig:flux_no_mean}). This same behaviour still holds in Fig.~\ref{fig:flux_low_peak_n2}, where we can see that the blue curve bump is slightly shifted with respect to the green one. However, when one uses larger values of the scale of new physics to affect the second peak, we finally reach the regime in which $E_{\nu_{\mu,\tau}}^{(e)}>E_\text{th}^{(e)}$, and as a consequence, the \abb{VPE} cutoff is now controlled by $E_{\nu_{\mu,\tau}}^{(e)}$. In Eq.~\eqref{eq:VPE_thresholds_comp}, we showed that both dynamical thresholds ($E_{\nu_{\mu,\tau}}^{(e)}$ and $E_{\nu_{\alpha}}^{(\nu)}$) are of the same order of magnitude. Then we now expect the cutoff of both scenarios to fall at approximately the same energy. This is the behaviour we find in Fig.~\ref{fig:flux_high_peak_n2}. From now on, only the \abb{VPE}+\abb{NSpl} case will be taken into account.

\begin{figure}[tbp]
        \centering
        \includegraphics[width=0.6\textwidth]{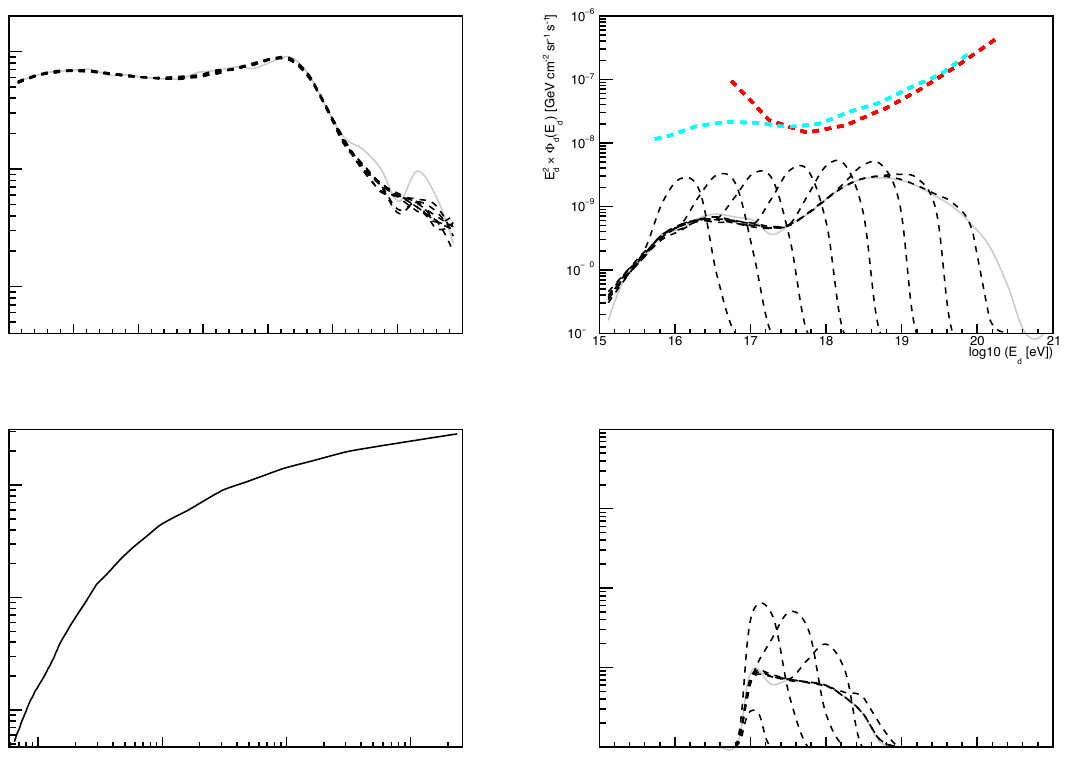}
        \caption{Cosmogenic neutrino flux at Earth for $n=2$ and increasing values of $\Lambda$ (such that $\log_{10}(E_\text{cut}\;[\text{eV}])$ goes from 16.5 to 20 in steps of 0.5).}
        \label{fig:flux_evol_n2}
\end{figure}

Let us come back to Fig.~\ref{fig:flux_low_peak_n2}. The inclusion of \abb{LIV} with a scale of new physics near the Plank scale (for $n=2$) produces the disappearance of the \abb{CMB} peak and an increase in the flux near the \abb{EBL} one. This scenario would be compatible with the detection of an excess of the cosmogenic neutrino flux within $10^{16}$ and $10^{17}$ eV, without the detection of the high-energy peak. If one focuses instead in Fig.~\ref{fig:flux_high_peak_n2}, we see that the increase in the flux which precedes the cutoff is much lower. In order to better understand this phenomenon, we show in Fig.~\ref{fig:flux_evol_n2} the evolution of the cutoff and the bump as we augment the scale of new physics.
We see that, as we increase the value of $\Lambda$, the bump produces less enhancement in the flux of neutrinos. This is due to the fact that raising the scale of new physics also increases the threshold energy of the neutrino decays. Cosmogenic neutrinos inherit a fraction of the primary cosmic rays, which have a strong suppression after $10^{20}\unit{eV}$; then, the produced neutrinos cannot have arbitrarily high energies. As $\Lambda$ increases, a less number of the produced neutrinos have enough energy to be multiplied through \abb{NSpl}. Additionally, now $E_{\nu_{\alpha}}^{(\nu)}>E_\text{th}^{(e)}$, that is to say, neutrinos stop decaying through \abb{NSpl} before they stop producing \abb{VPE}, which does not produce neutrino multiplication. 

In short, we find that the maximum change in the cosmogenic neutrino spectrum is produced when one considers values of $\Lambda\sim M_P$, i.e., when the bump coincides with the \abb{EBL} peak. However, let us note that the values of $\Lambda$ for which we find the maximum change does not necessarily correspond with the values that produce the largest absolute maximum of the spectrum. We will discuss this again later on.

\begin{figure}[tbp]
    \centering
    \begin{minipage}{0.49\textwidth}
        \centering
        \includegraphics[width=\textwidth]{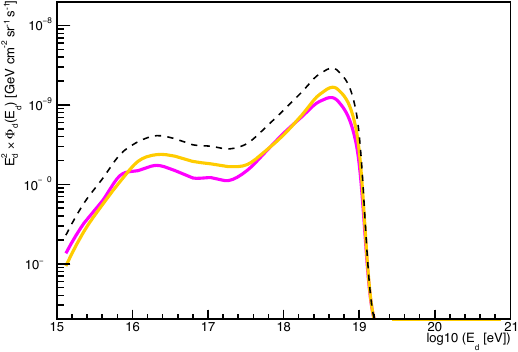}
        \caption{Cosmogenic muon neutrino (pink) and antineutrino (orange) flux at Earth for $n=2$ and $\Lambda\approx 1.13\E{4} M_P$ ($E_\text{cut}=10^{19}\unit{eV}$). Sum in dashed line.}
        \label{fig:flux_muon_LIV}
    \end{minipage}%
    \hfill
    \begin{minipage}{0.49\textwidth}
        \centering
        \includegraphics[width=\textwidth]{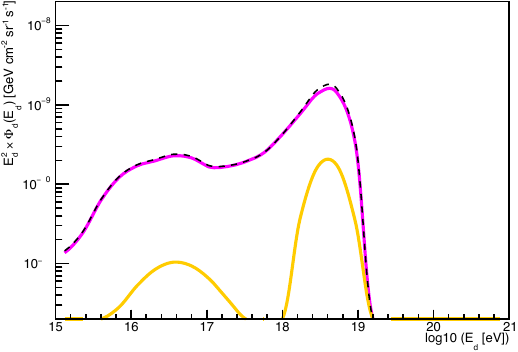}
        \caption{Cosmogenic electron neutrino (pink) and antineutrino (orange) flux at Earth for $n=2$ and $\Lambda\approx 1.13\E{4} M_P$ ($E_\text{cut}=10^{19}\unit{eV}$). Sum in dashed line.}
        \label{fig:flux_elec_LIV}
    \end{minipage}
    \par\vspace{1em}
    \begin{minipage}{0.49\textwidth}
        \centering
        \includegraphics[width=\textwidth]{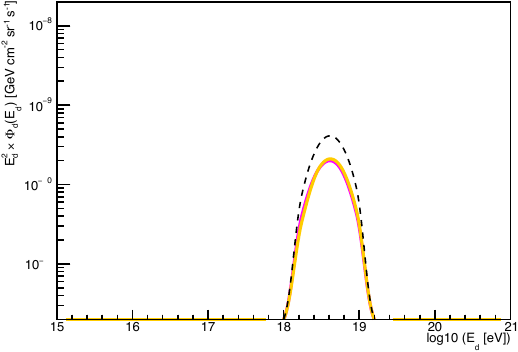}
        \caption{Cosmogenic tau neutrino (pink) and antineutrino (orange) flux at Earth for $n=2$ and $\Lambda\approx 1.13\E{4} M_P$ ($E_\text{cut}=10^{19}\unit{eV}$). Sum in dashed line.}
        \label{fig:flux_tau}
    \end{minipage}
\end{figure}

Another change in the cosmogenic neutrino flux due to the superluminal \abb{LIV} can be found in the flavour distribution of the particles and antiparticles. For instance, in Figs.~\ref{fig:flux_muon_LIV} and \ref{fig:flux_elec_LIV} we show the muon and electron contributions to the total flux in the case $n=2$ and $\Lambda\approx 1.13\E{4} M_P$ ($E_\text{cut}=10^{19}\unit{eV}$).
One can see in Fig.~\ref{fig:flux_elec_LIV} an additional electronic antineutrino contribution with respect to Fig.~\ref{fig:flux_elec_neu}, which comes from the \abb{NSpl}. However, let us remember that the \abb{NSpl} produces neutrinos of the three flavours with equal probability, so the most evident change in the flavour distribution comes from the appearance of tau neutrinos and antineutrinos (see Fig.~\ref{fig:flux_tau}), whose production was absent in the non-\abb{LIV} case\footnotemark. Nevertheless, in order to make any useful prediction of the flavour distribution at Earth, one must take into account the neutrino oscillations, which we are omitting in this study.
\footnotetext{Notice that even if in \abb{SR} no tau neutrinos are produced from the proton-photon interactions, one would still see tau neutrinos at Earth due to neutrino oscillations.}

\begin{figure}[tbp]
    \centering
    \begin{minipage}{0.49\textwidth}
        \centering
        \includegraphics[width=\textwidth]{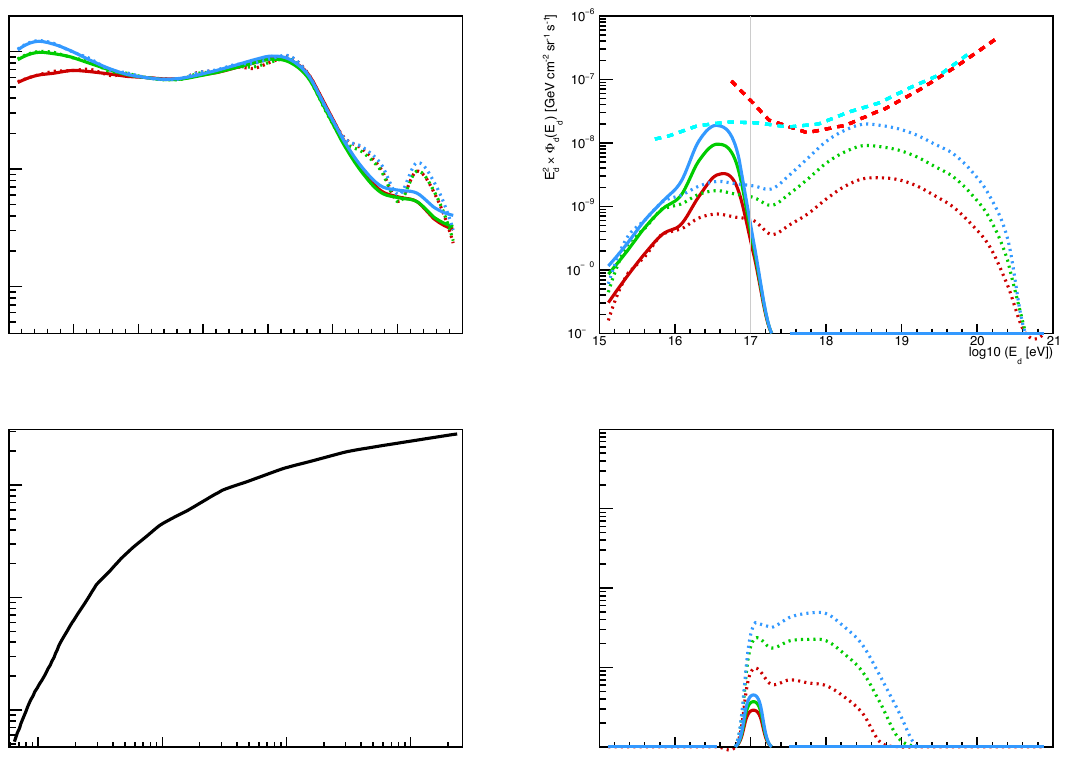}
        \caption{Cosmogenic neutrino flux at Earth for $n=2$ and $\Lambda\approx 2.19 M_P$ ($E_\text{cut}=10^{17}\unit{eV}$), for a uniform (red), \abb{SFR} (green), and \abb{AGN} (blue) source distribution. The \abb{SR} scenario for each case is shown in a dotted line.}
        \label{fig:flux_low_sources_n2}
    \end{minipage}%
    \hfill
    \begin{minipage}{0.49\textwidth}
        \centering
        \includegraphics[width=\textwidth]{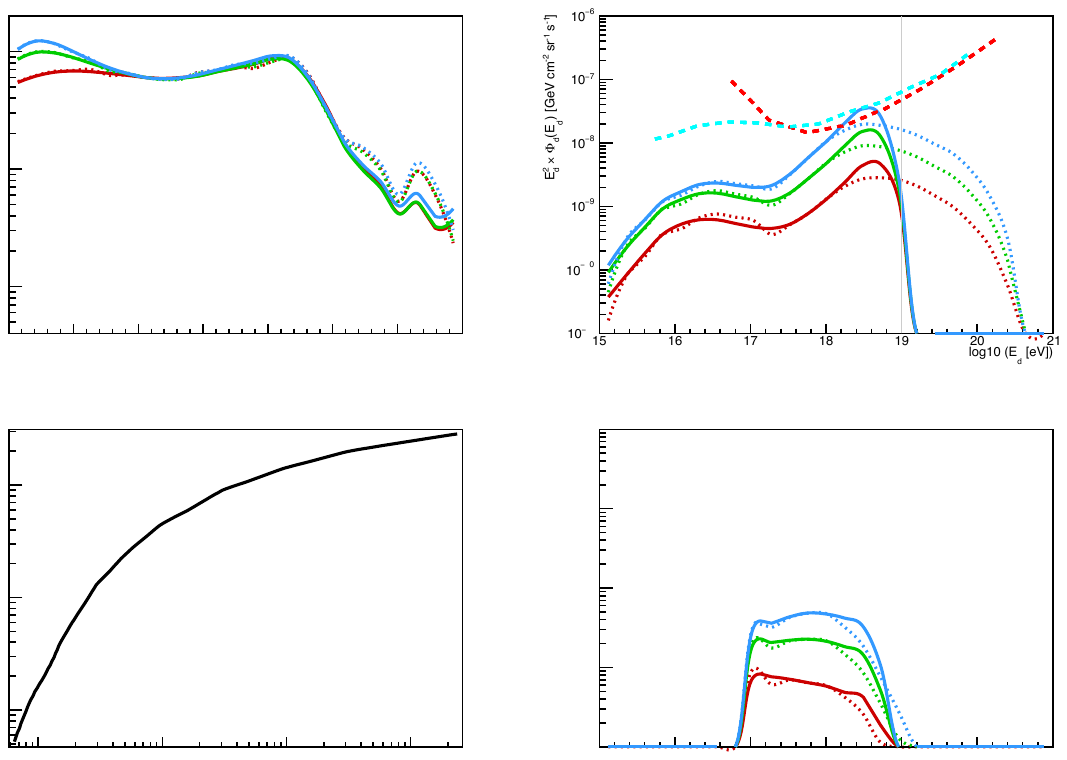}
        \caption{Cosmogenic neutrino flux at Earth for $n=2$ and $\Lambda\approx 1.13\E{4} M_P$ ($E_\text{cut}=10^{19}\unit{eV}$), for a uniform (red), \abb{SFR} (green), and \abb{AGN} (blue) source distribution. The \abb{SR} scenario for each case is shown in a dotted line.}
        \label{fig:flux_high_sources_n2}
    \end{minipage}
\end{figure}

In Fig.~\ref{fig:flux_evol_n2}, we see that for the chosen astrophysical scenario (uniform cosmic ray source distribution with $\gamma=2.6$), even with the addition of the new physics, we are still far from the sensitivity of current detectors (represented in Fig.~\ref{fig:flux_evol_n2} by the 90\% CL upper limits). We discussed before that the cosmogenic neutrino flux will be influenced by the chosen astrophysical scenario for the cosmic rays, so we have repeated the computation of the expected cosmogenic neutrino flux (showed in Figs~\ref{fig:flux_low_peak_n2} and \ref{fig:flux_high_peak_n2}) for the other two astrophysical scenarios mentioned before (\abb{SFR} and \abb{AGN}). The results are shown in Figs.~\ref{fig:flux_low_sources_n2} and \ref{fig:flux_high_sources_n2}.
 
\begin{figure}[tbp]
    \centering
    \begin{minipage}{0.49\textwidth}
        \centering
        \includegraphics[width=\textwidth]{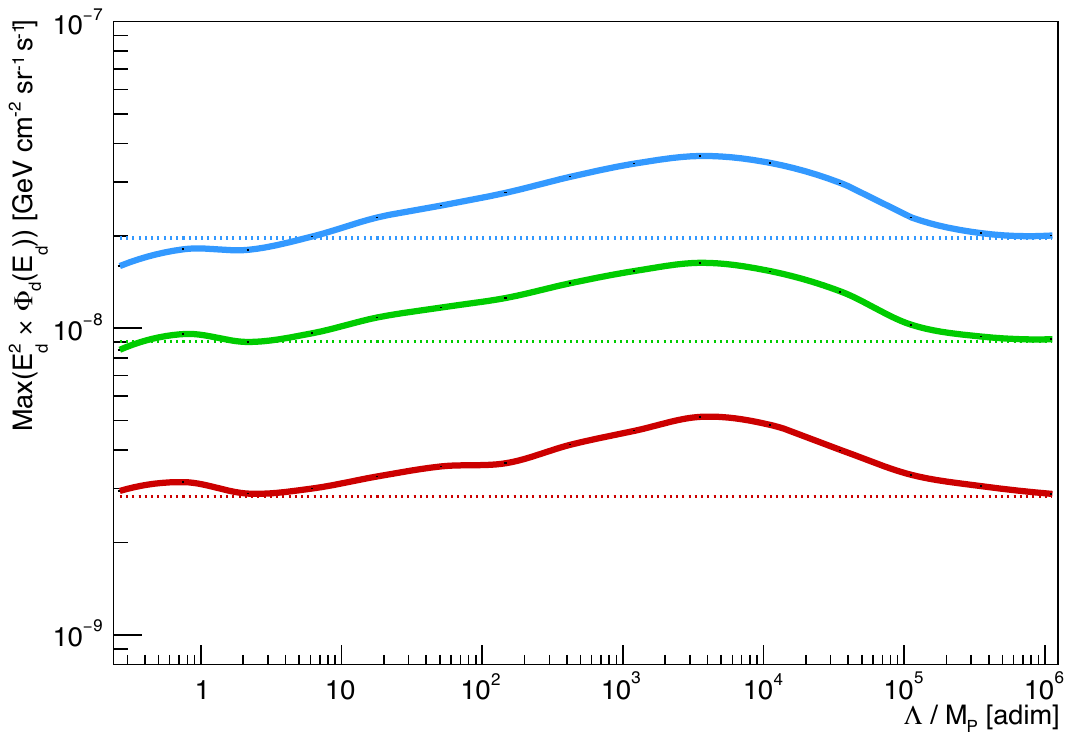}
        \caption{Maximum of $E^2\times\Phi_d$ as function of $\Lambda$, for a uniform (red), \abb{SFR} (green) and \abb{AGN} (blue) source distribution. The height of the \abb{CMB} maximum (\abb{SR} absolute maximum) is shown in a dotted line for each case.}
        \label{fig:valMax_n2}
    \end{minipage}%
    \hfill
    \begin{minipage}{0.49\textwidth}
        \centering
        \includegraphics[width=\textwidth]{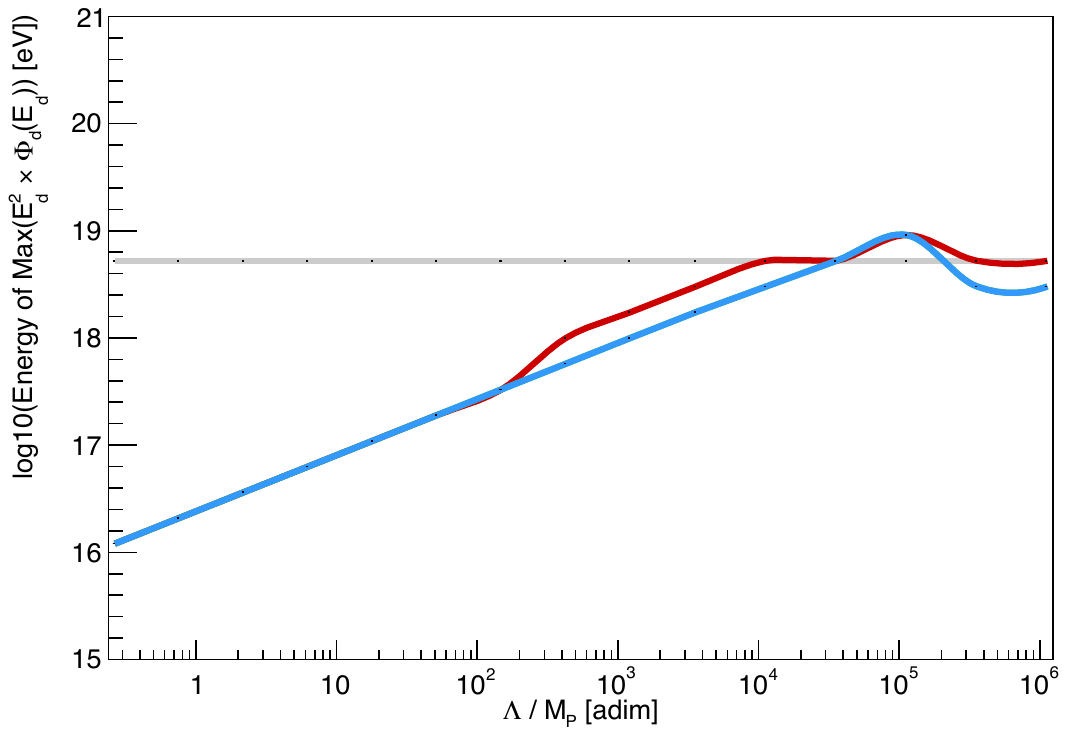}
        \caption{Energy at which $E^2\times\Phi_d$ is maximum as function of $\Lambda$, for a uniform (red), \abb{SFR} (green) and \abb{AGN} (blue) source distribution. The energy of the \abb{CMB} maximum (\abb{SR} absolute maximum) is shown in a gray line.}
        \label{fig:eneMax_n2}
    \end{minipage}
\end{figure}

We have repeated the same procedure for all the different values of $\Lambda$ shown in Fig.~\ref{fig:flux_evol_n2}, and we have identified the height and energy of the absolute maximum of the spectrum for each case. The evolution of the height of the maximum as we increase the scale of new physics is shown in Fig.~\ref{fig:valMax_n2}, and the corresponding energy in Fig.~\ref{fig:eneMax_n2}.
We can see that the cosmogenic neutrino flux is maximum (in the case $n=2$) for a scale of new physics within $10^{3}$--$10^{4}$ times the Plank mass (even though the change in the flux is larger at lower values). That is to say, absolute highest values of the neutrino spectrum are obtained when the \abb{LIV} bump overlaps with the \abb{CMB} peak.

However, the sensitivity of each experiment to this maximum will depend on the exposure of the detector. The exposure~\cite{PierreAuger:2019ens} can be defined as the instant aperture $A_p(E_d,t)$, a function that accounts for the probability of detecting an event of certain energy at a certain time, integrated to the time interval of detection,
\be
    \mathcal{E}(E_d) \coloneqq \int_{t_1}^{t_2} dt\; A_p(E_d,t) \approx A_p(E_d) \times \Delta t \,,
\ee
where in the last step we have considered a negligible time dependence of the instant aperture.

\begin{figure}[tbp]
    \begin{minipage}{0.49\textwidth}
        \centering
        \includegraphics[width=\textwidth]{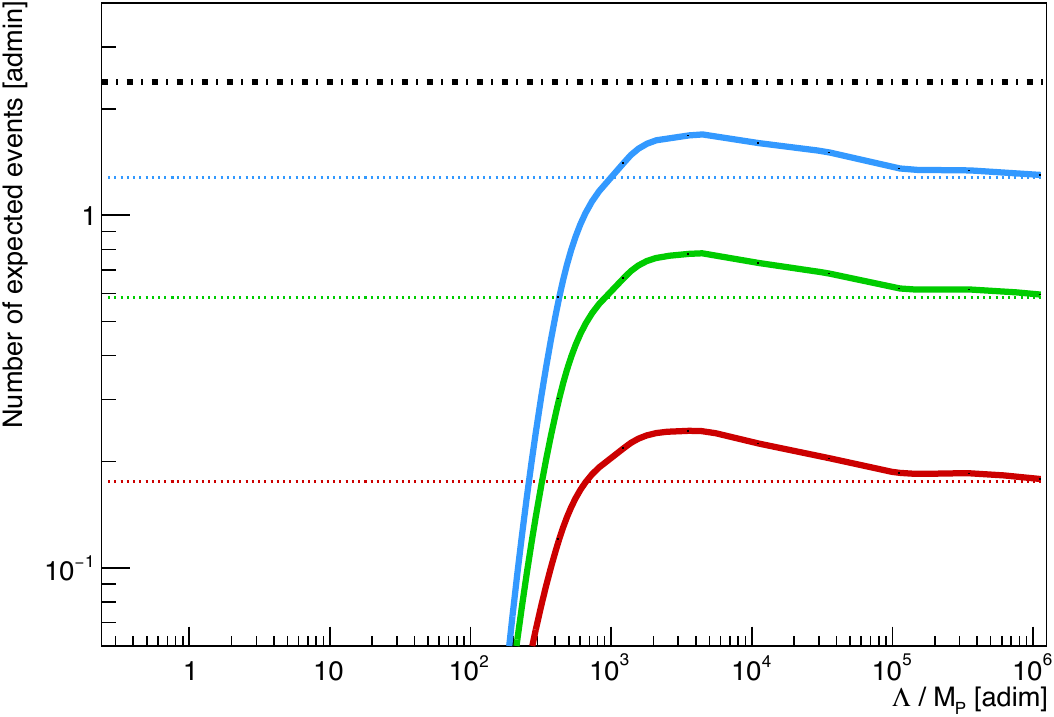}
        \caption{Number of expected events with energies between $10^{18}$ and $10^{19}$ eV as a function of $\Lambda$ and using the exposure of \abb{PA} from~\cite{PierreAuger:2019ens}. The statistical 90\% CL upper limit for absence of events ($N_d=2.39$)~\cite{Feldman:1997qc,PierreAuger:2019ens} is shown using a dot-dashed black line.}
        \label{fig:numEvents_n2_auger_high}
    \end{minipage}%
    \hfill
    \begin{minipage}{0.49\textwidth}
        \centering
        \includegraphics[width=\textwidth]{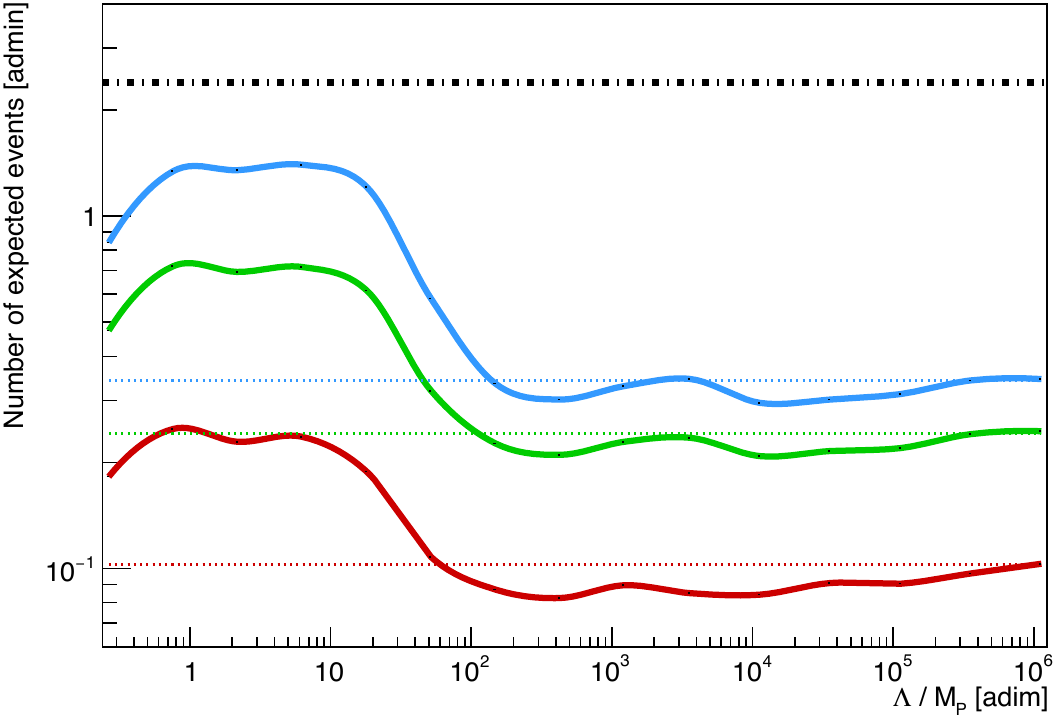}
        \caption{Number of expected events with energies between $10^{16}$ and $10^{17}$ eV as a function of $\Lambda$ and using the exposure of \abb{IC} from~\cite{IceCube:2015fuw}. The statistical 90\% CL upper limit for absence of events ($N_d=2.39$)~\cite{Feldman:1997qc,PierreAuger:2019ens} is shown using a dot-dashed black line.}
        \label{fig:numEvents_n2_icecube_low}
    \end{minipage}
\end{figure}

Looking at the 90\% CL upper limits of each experiment, we can get an idea of the expected sensitivity at each energy range. Then, we can identify that for the high-energy peak (\abb{CMB}), where the flux is maximum, we expect \abb{PA} to be more sensitive. In contrast, for the low energy peak (\abb{EBL}), where the change in the flux is maximum, \abb{IC} is the most sensitive one. In order to compare the sensitivity of both experiments, we have computed the number of expected events by \abb{PA} with energies between $10^{18}$ and $10^{19}$ eV (energies of the \abb{CMB} peak), and the number of expected events by \abb{IC} with energies between $10^{16}$ and $10^{17}$ eV (energies of the \abb{EBL} peak),
\be 
    N_d \coloneqq \int dE_d\;\mathcal{E}(E_d) \times \phi_d(E_d) \,.
\ee
For that, we have used the reported exposures $\mathcal{E}(E_d)$ from \abb{IC} and \abb{PA}, from~\cite{IceCube:2015fuw} and~\cite{PierreAuger:2019ens}, respectively, and the necessary time window to match the 90\% CL upper limits, i.e., up to 2018. The expected number of events, as a function of the scale $\Lambda$ are shown in Figs.~\ref{fig:numEvents_n2_auger_high} and \ref{fig:numEvents_n2_icecube_low}.

\begin{figure}[tbp]
        \centering
        \includegraphics[width=0.8\textwidth]{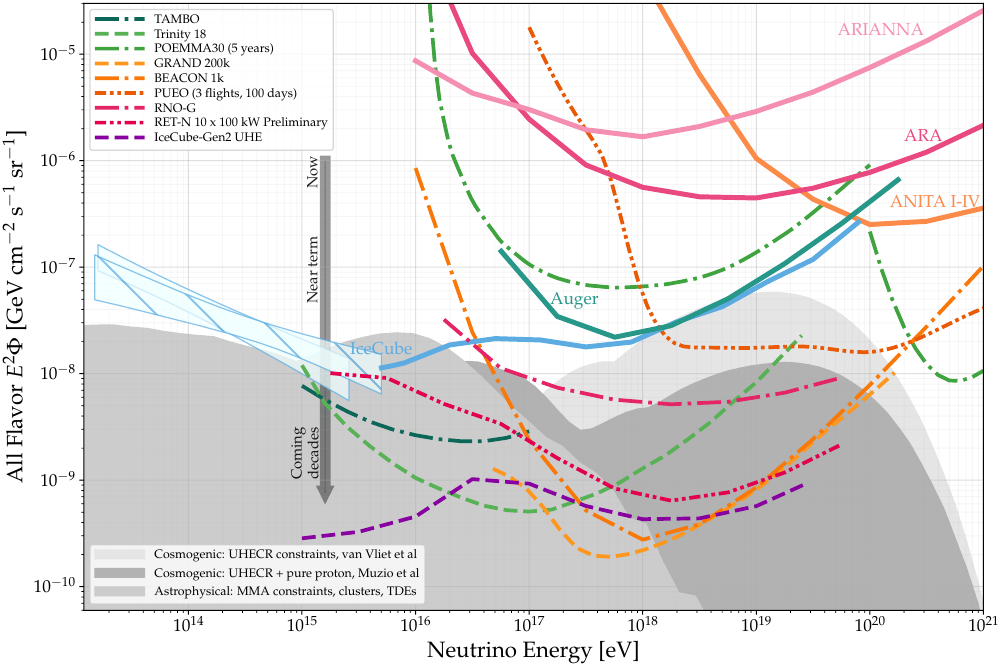}
        \caption{90\% CL upper limits for the detected flux for current (solid lines) and future (dashed and dot-dashed lines) experiments, from~\cite{Ackermann:2022rqc}. For the future experiments, a 10 years window time is considered, unless otherwise specified in the legend.}
        \label{fig:futureLimits}
\end{figure}

The non-observance of events in this energy range put a statistical 90\% CL upper limit of $N_d=2.39$ expected events (see~\cite{Feldman:1997qc,PierreAuger:2019ens} for more information) which can be used to reject some scenarios. We can check that the scenarios shown in Figs.~\ref{fig:numEvents_n2_auger_high} and \ref{fig:numEvents_n2_icecube_low} cannot be rejected yet. However, it shows that in the near future, the cosmogenic flux can be used as a powerful tool to reject and constrain some values of the model parameter space (including the parameters of new physics and the astrophysical parameters for the emission of cosmic rays), specially considering the new experiments in construction, which will provide much more sensitivity in these energy ranges (see Fig.~\ref{fig:futureLimits}). Let us also take into account that we assumed an pure proton cosmic ray composition and sources within $z\in[0,1]$. If one considers a heavier cosmic ray composition, we will expect less neutrino production. On the contrary, if one consider a larger volume of the universe, the expected number of neutrinos will increase.

\begin{figure}[tbp]
    \centering
    \begin{minipage}{0.49\textwidth}
        \centering
        \includegraphics[width=\textwidth]{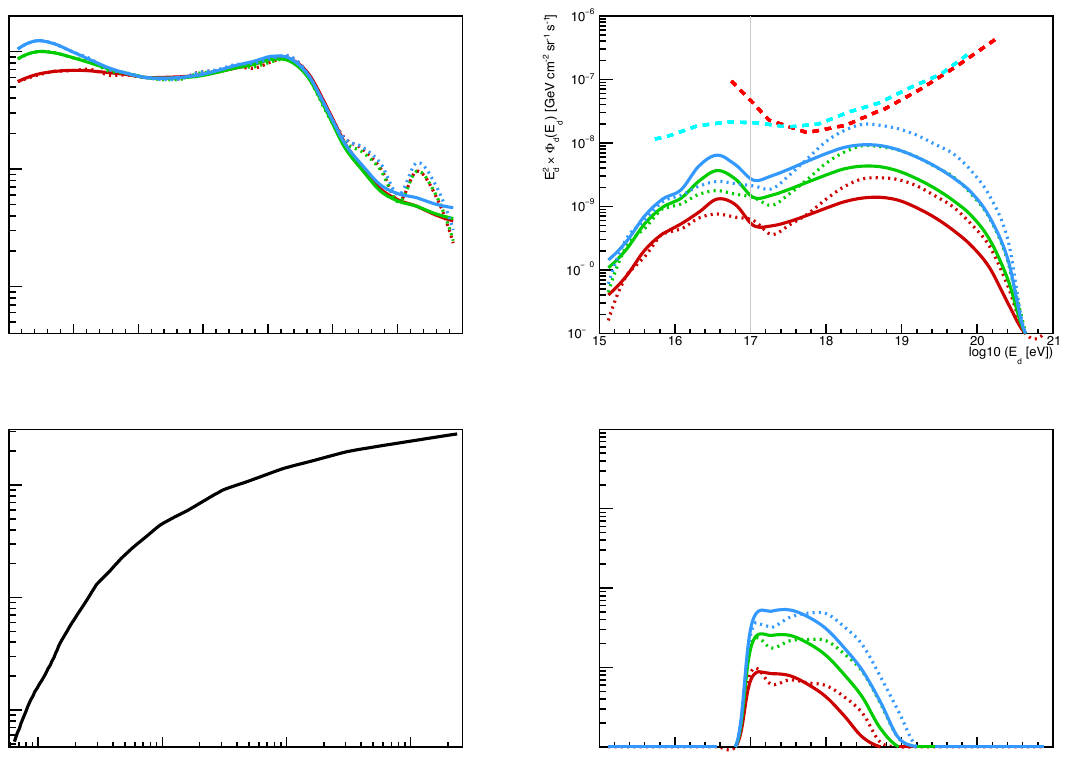}
        \caption{Cosmogenic neutrino flux at Earth for $n=1$ and $\Lambda\approx 5.90\E{11} M_P$ ($E_\text{cut}=10^{17}\unit{eV}$), for a uniform (red), \abb{SFR} (green), and \abb{AGN} (blue) source distribution. The \abb{SR} scenario for each case is shown in a dotted line.}
        \label{fig:flux_low_sources_n1}
    \end{minipage}%
    \hfill
    \begin{minipage}{0.49\textwidth}
        \centering
        \includegraphics[width=\textwidth]{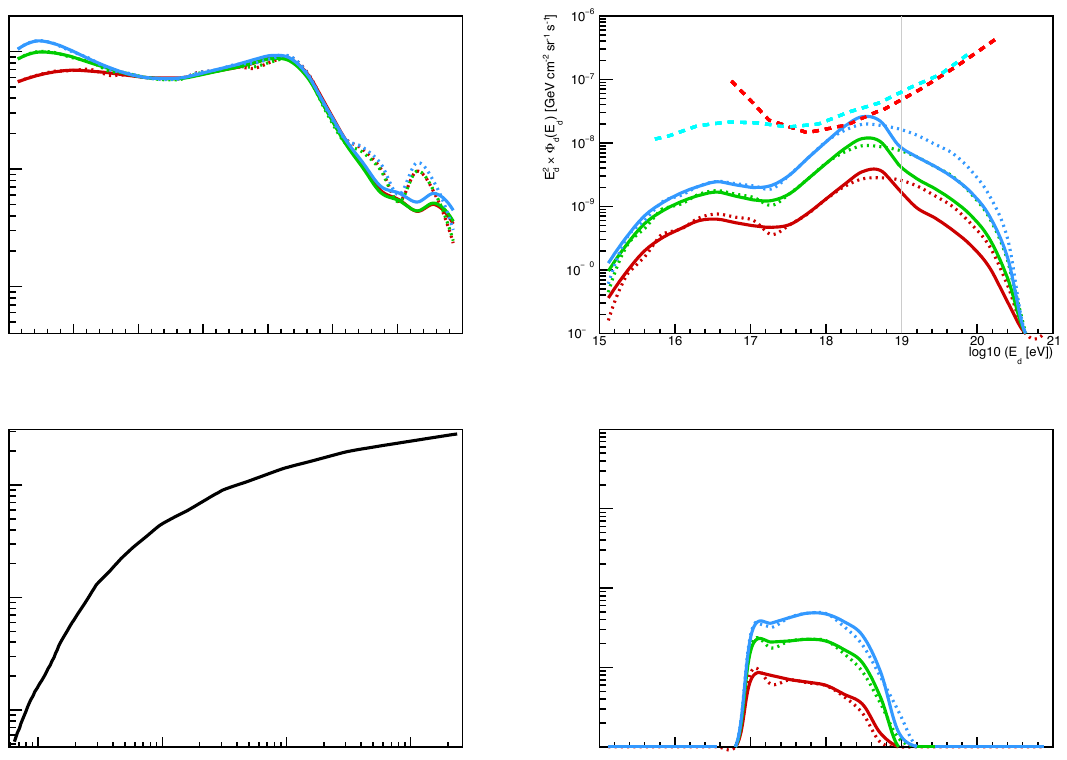}
        \caption{Cosmogenic neutrino flux at Earth for $n=1$ and $\Lambda\approx 1.57\E{17} M_P$ ($E_\text{cut}=10^{19}\unit{eV}$), for a uniform (red), \abb{SFR} (green), and \abb{AGN} (blue) source distribution. The \abb{SR} scenario for each case is shown in a dotted line.}
        \label{fig:flux_high_sources_n1}
    \end{minipage}
\end{figure}

Let us finally comment that, similarly to the results we obtained for the astrophysical neutrinos, the case $n=1$ produces much softer effects in the cosmogenic neutrino flux than the case $n=2$. This is again due to the fact that only neutrinos are superluminal and, in consequence, only the neutrino flux decreases after $E_\text{cut}$ and can undergo \abb{NSpl} enhancing the size of the bump. This behaviour can be seen in Figs.~\ref{fig:flux_low_sources_n1} and \ref{fig:flux_high_sources_n1}.
\begin{subappendices}
    \include{Chapter4/appendix4}
\end{subappendices}
\printbibliography[heading=subbibintoc,title={References}]
\end{refsection}

\begin{refsection}
\chapter{Gamma Rays}
\graphicspath{{Chapter5/Figs/}}
\label{chap:gammas}

\begin{figure}[htbp]
    \centering
    \includegraphics[width=0.9\textwidth]{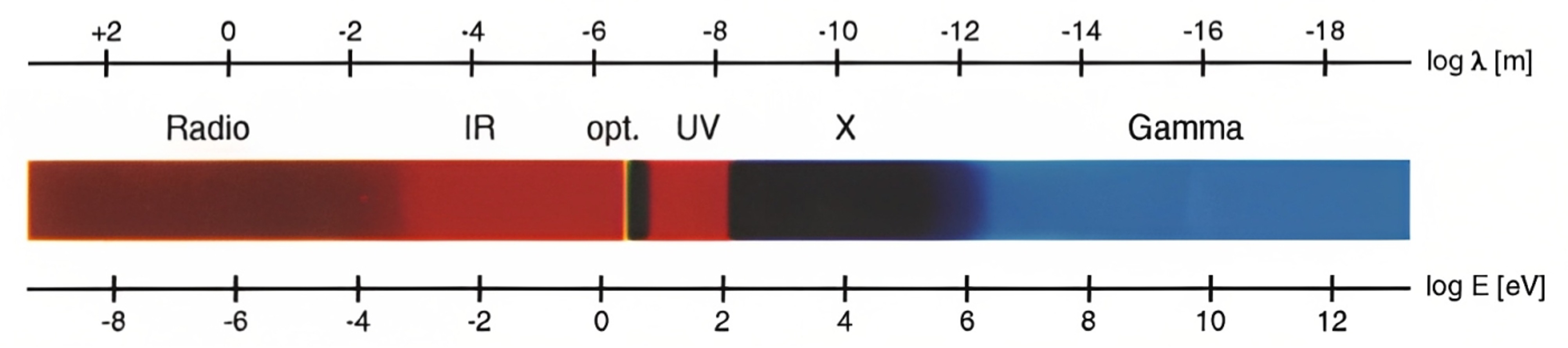}
    \caption{Energies and wavelengths of the electromagnetic spectrum. Adapted from~\cite{Schonfelder:2001}.}
    \label{fig:em_spec}
\end{figure}

Gamma rays are the most energetic part of the electromagnetic spectrum. They are so energetic that their wavelength is usually much shorter than the size of an atom, so they act more like a particle than a wave when they interact with matter. By convention, every photon with energy above $10^5\unit{eV}$, or equivalently, with a wavelength below of $10^{-11}\unit{m}$, is considered to be a photon gamma ray (see Fig.~\ref{fig:em_spec}).

Let us note that, in the same way the radio spectrum extends indefinitely to lower energies, the gamma ray spectrum also extends indefinitely for higher energies (see Fig.~\ref{fig:em_spec}). As a consequence, gamma rays cover a wide range of energies, which, in turn, can be divided in several regimes: low-energy range (below $10^8\unit{eV}$), high-energy range (up to $10^{11}\unit{eV}$), very high-energy range (up to $10^{14}\unit{eV}$), ultra high-energy range (up to $10^{17}\unit{eV}$) and extremely high-energy range (above $10^{17}\unit{eV}$). Gamma rays in the low- and high-energy ranges cannot penetrate the Earth's atmosphere without being absorbed and scattered, so their observation can only be performed from space satellite experiments; this is called ``spaceborn gamma ray astronomy''~\cite{Schonfelder:2001}. On the contrary, the observation of the very high- and extremely high-energy ranges is called ``ground-based gamma ray astronomy'', because gamma rays of these energies provoke strong electromagnetic cascades which can be detected by optical telescopes on ground~\cite{Schonfelder:2001}.

\begin{figure}[tbp]
    \centering
    \includegraphics[width=0.7\textwidth]{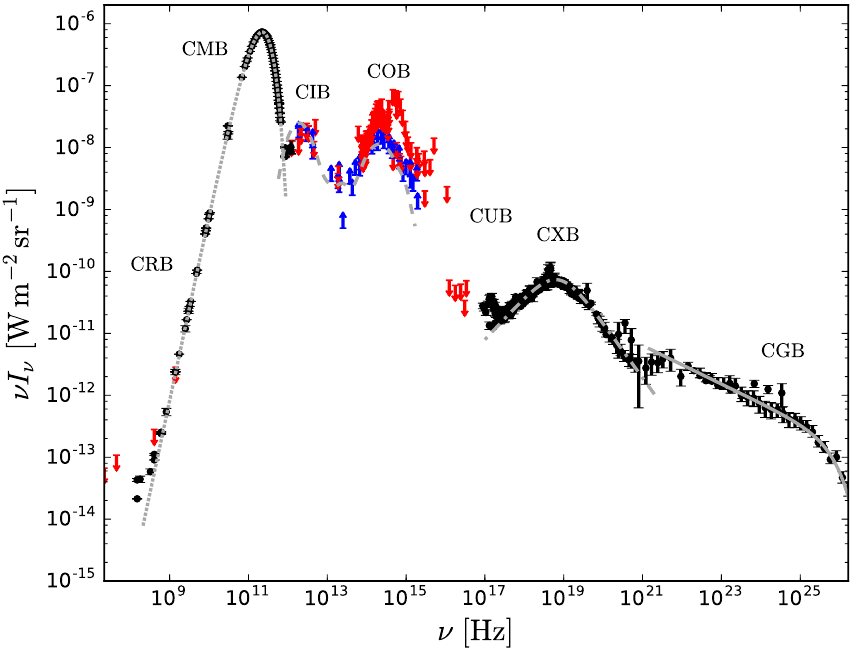}
    \caption{Complete cosmic background radiation. Gray lines for analytical models, black points for detections, and blue or red arrows for limits. From \cite{Hill:2018trh}.}
    \label{fig:debr}
\end{figure}

Due to the fact that gamma rays are neutral, their trajectories are not deflected by the galactic magnetic fields, so they travel in straight lines and their sources are relatively easy to determine. In fact, most of the more compact and energetic objects emit gamma rays, like neutron stars, massive black holes, supernova explosions, and even the cosmic rays when interacting with the photon background of the universe (through the production and decay of neutral pions, as we mentioned in Chapter~\ref{chap:cosmicrays}). Gamma rays can interact with photons of lower energies from the diffuse extragalactic background during their propagation, disappearing from the flux. As mentioned in Sec.~\ref{sec:cosmology}, the photon background is mostly dominated by the \abb{CMB} and \abb{EBL}. The latter, in turn, is composed by the peaks of the \abbdef{CID}{Cosmic Infrared Background} at $10^{-2}\unit{eV}$ ($10^{-4}\unit{m}$) and the \abbdef{COD}{Cosmic Optical Background} at $1\unit{eV}$ ($10^{-6}\unit{m}$) (see Fig.~\ref{fig:CMB_EBL}). There are additional photon backgrounds like \abbdef{CRB}{Cosmic Radio Background}, \abbdef{CXB}{Cosmic X rays Background}, or the mostly unexplored \abbdef{CUB}{Cosmic Ultraviolet Background} (see \cite{Hill:2018trh,Cooray:2016} for more information). In Fig.~\ref{fig:debr}, we show the spectrum of the extragalactic background light and we can see how the \abb{CMB} and \abb{EBL} are the most relevant components in terms of intensity; however, the different backgrounds become relevant depending on the energies of the particles they are interacting with. For the gamma ray energies we will study in this chapter, the \abb{CMB} will be the dominant background component.

The dominant process gamma rays undergo when they interact with background photons, is the annihilation through the emission of an electron-positron pair ($\gamma + \gamma_b \rightarrow e^- + e^+$). As the particles of the final state are massive, the gamma ray in the initial state requires a minimum energy to produce the electron-positron pair. Fixing the energy of the low-energy photon, the process has an energy threshold for the gamma rays, which (for head-on collisions) can be found to be
\be
    E_\text{th}^{(e)}=\frac{m_e^2}{\epsilon} \,.
\ee
For a background photon of the \abb{CMB} (around $10^{-4}$--$10^{-3}\unit{eV}$), the value is around $\approx 10^{15}\unit{eV}$. This implies a strong suppression of the flux above the PeV.

The study of deviations from this expected behaviour for the gamma ray flux can be used to test some models of new physics. Along this chapter we will mainly focus in how the opacity/transparency of the local universe changes when one considers a \abb{DSR} model. 

Together with the modification of the opacity/transparency, another possible observable consequence of a departure from \abb{SR} are time delays, i.e., a difference in the detection time of photons of different energies emitted simultaneously by a source. A discussion about the state of time delays in \abb{DSR} models is also held at the end of the chapter.

\section{Gamma ray transparency}

Let us assume a flux of gamma rays, $\Phi_e(E_e)$, emitted in the direction of the detector from a source located at a redshift $z$. At Earth, we will detect a flux $\Phi_d(E_d)$. If one disregards any interaction during the propagation, one would expect a total transparency of the universe to gamma rays,
\be 
    1=\frac{\Phi_d(E_d)}{\Phi_e((1+z_s)E_d)} \,,
\ee
that is to say, all the emitted gamma rays in the direction of the detector, are also detected. Let us note that we have to take into account that photons detected with energy $E_d$ have been emitted with energy $E_e=(1+z)E_d$.

However, we know that gamma rays interact with the low-energy photon background, producing electron-positron pairs and disappearing from the flux. We can characterize this decrement in the transparency by using the concept of optical depth,
\be
    \exp(-\tau(E_d,z))\eqqcolon\frac{\Phi_d(E_d)}{\Phi_e((1+z)E_d)} \,.
    \label{eq:opacity}
\ee

Let us say a gamma ray of energy $E$ is crossing an isotropic background of photons with spectral density $n(\epsilon)$ (number of background photons per unit of volume per unit of energy interval). The absorption probability per unit time, $d\tau/dt$, will be the product of the cross section of the process, $\sigma_{\gamma\gamma}(E,\epsilon,\theta)$, which will depend on the energy of both photons and the angle between them, and the flux of background photons seen by the gamma ray photon. This flux is the product of the number of photons per unit of volume per unit of energy interval in the direction of the gamma ray, $dn(\epsilon,\theta)$, and the difference of velocities between the two photons in that direction, $\Delta v(\theta)$.
\be 
    d\tau/dt=\sigma_{\gamma\gamma}(E,\epsilon,\theta) dn_b(\epsilon,\theta) \Delta v(\theta)\,.
\ee

On the one hand, the number of photons with energies within $\epsilon$ and $\epsilon +d\epsilon$ is just $n_b(\epsilon) d\epsilon$. On the other hand, for an isotropic background, the fraction of photons moving in a differential cone at angle $\theta$ is $(1/2)\sin\theta d\theta$. Then we can write $dn(\epsilon,\theta)=(1/2) n_b(\epsilon)\sin\theta d\epsilon d\theta$.

\begin{figure}[tbp]
    \centering
    \includegraphics[width=0.35\textwidth]{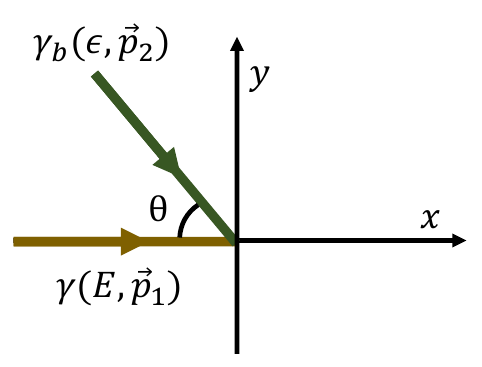}
    \caption{Interaction between a gamma ray and background photon.}
    \label{fig:gr_photon}
\end{figure}

Now, in a frame in which the four-momenta of the gamma ray and background photon are respectively $p=(E,p_1,0,0)$ and $p_b=(\epsilon,p_2 \cos \theta,-p_2 \sin \theta,0)$ (see Fig.~\ref{fig:gr_photon}), the difference of velocities along the direction of propagation of the gamma ray is simply $\Delta v(\theta)=(1-\cos\theta)$, where we have used that both photons travel at $c=1$.

Finally, let us compute the cross section. As discussed before, the process requires a minimum initial energy for the initial state. Fixing the energy of the gamma ray, we can compute the threshold energy for the background photon, which can be found to be
\be
    \epsilon_\text{th}(E,\theta)\coloneqq \frac{2m_e^2}{E(1-\cos{\theta})} \,.
\ee
The cross section $\sigma_{\gamma\gamma}(E,\epsilon,\theta)$ will be zero if for a fixed energy $E$ and angle $\theta$, the energy of the background photon is below $\epsilon_\text{th}(E,\theta)$. If the energy of the background photon is above the threshold, the cross section is given by the Breit-Wheeler cross section formula~\cite{Breit:1934zz,DeAngelis:2013jna}\footnotemark:
\footnotetext{Note that there is a typo in Eq.~(3) of~\cite{DeAngelis:2013jna}. A (3/4) factor is missing in the Breit-Wheeler cross-section formula.}
\be 
    \sigma_{\gamma\gamma}(\beta)\coloneqq \frac{\pi\alpha^2}{2m_e^2}W(\beta) \,,
    \label{eq:BW_crosssection}
\ee
with
\be
    W(\beta)\coloneqq(1-\beta^2)\qty[2\beta(\beta^2-2)+(3-\beta^4)\ln\frac{1+\beta}{1-\beta}] \\
    \text{and}\quad \beta(E,\epsilon,\theta)\coloneqq\sqrt{1-\frac{2m_e^2}{E\epsilon(1-\cos{\theta})}}  \,.
    \label{eq:W_beta}
\ee

\begin{figure}[tbp]
    \centering
    \begin{minipage}{0.49\textwidth}
        \centering
        \includegraphics[width=\textwidth]{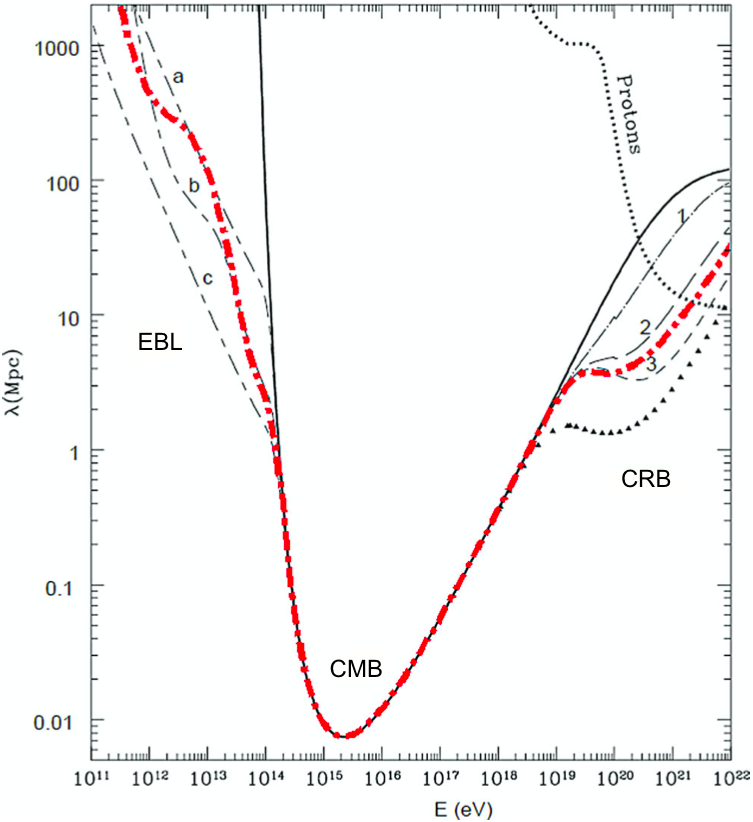}
        \caption{Mean free path of the process \mbox{($\gamma + \gamma_b \rightarrow e^- + e^+$)} considering the \abb{EBL}, \abb{CMB} and \abb{CRB}. Adapted from~\cite{DeAngelis:2013jna}.}
        \label{fig:lambda_DeAngelis}
    \end{minipage}%
    \hfill
    \begin{minipage}{0.49\textwidth}
        \centering
        \includegraphics[width=0.95\textwidth]{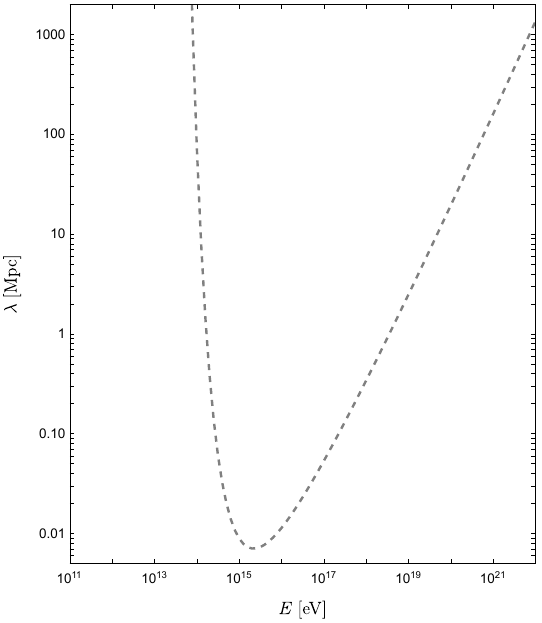}
        \caption{Mean free path of the process \mbox{($\gamma + \gamma_b \rightarrow e^- + e^+$)} considering the \abb{CMB}. Computed using Eq.~\eqref{eq:lambda_adim}.}
        \label{fig:lambda_SR}
    \end{minipage}
\end{figure}

The spectral density of the background $n(\epsilon)$ is the sum of the densities of all the relevant photon backgrounds; however, for gamma rays from $10^{14}$ to $10^{19}\unit{eV}$, the dominant background is the \abb{CMB}. If one wants to consider gamma rays of lower energies, one should include the \abb{EBL}, and, for higher energies, the \abb{CRB} (see Fig.~\ref{fig:lambda_DeAngelis}).

As discussed before, the spectral density of the \abb{CMB} can be computed analytically as the emitted spectrum of a black body at temperature $T(z)$,
\be 
    n(\epsilon,T(z))\coloneqq \frac{\epsilon^2/\pi^2}{\exp(\epsilon/(k_B T(z)))-1} \,,
\ee
where the value of the temperature at different epochs can be obtained from the current value ($T_0=2.73\unit{K}$) as $T(z)=(1+z)T_0$.

Gathering all the ingredients we can write the differential optical depth as
\be
    d\tau = \frac{1}{2}\sigma(E,\epsilon,\theta) n(\epsilon,T) (1-\cos{\theta})\sin{\theta} d\theta d\epsilon dl \,,
\ee
where we have used that for a photon, as $c=1$, $dt=dl$, being $dl$ a differential bit of the photon path. To obtain the total optical depth, we should now integrate the previous expression to all the angles, all the energies of the background photon above the threshold, and all along the gamma ray path, i.e.,
\be 
    \tau(E_d,z) = \frac{1}{2} \int_0^{z} dz'\, \frac{dl}{dz'} \int_{-1}^{1} d(\cos\theta)\; (1-\cos{\theta})  \int_{\epsilon_\text{th}(E(z),\theta)}^{\infty} d\epsilon \;\sigma(E(z),\epsilon,\theta) n(\epsilon,T(z)) \,,
    \label{eq:optical_depth}
\ee
with $E(z)=(1+z)E_d$ and $T(z)=(1+z)T_0$.

If we focus the study on gamma rays emitted within the local universe ($z\ll 1$), we can simplify the previous expression using that $E(z)\approx E_d$ and $T(z)\approx T_0$. In this way, the only redshift dependence is in the first integral, which is just the distance to the source ($D\equiv \int dz \,(dl/dz)$). Dividing Eq.~\eqref{eq:optical_depth} by $D$, one obtains the inverse of the mean free path,
\be 
    \frac{1}{\lambda(E_d)}\coloneqq \frac{\tau(E_d,z_s)}{D} = \frac{1}{2} \int_{-1}^{1} d(\cos\theta)\; (1-\cos{\theta})  \int_{\epsilon_\text{th}(E_d,\theta)}^{\infty} d\epsilon \;\sigma(E_d,\epsilon,\theta) n(\epsilon,T_0)  \,.
\ee

In order to solve the previous integral, one can realize that $\beta(E,\epsilon,\theta)$, Eq.~\eqref{eq:W_beta}, only depends on $E$, $\epsilon$, and $\theta$ in a very particular combination. From the definition of the invariant $s=2E\epsilon (1-\cos(\theta))$, we see that the functions $\beta$, $W$ and $\sigma$, only depend on the parameters of the system through $s$, that is to say,
\be 
    \beta(s)\coloneqq \sqrt{1-\frac{4m_e^2}{s}} \eqtext{and} \sigma(s)\coloneqq \frac{2\pi\alpha^2}{3m_e^2}W(\beta(s)) \,.
    \label{eq:beta_s}
\ee
This encourages us to make a change of variables from $(d\cos\theta,d\epsilon)$ to $(ds,d\epsilon)$. The Jacobian of the change of variables is $\abs{J}=1/(2E\epsilon)$, and the new limits of integration are $(4m_e^2<s<\infty)$ and $(s/(4E)<\epsilon<\infty)$, so we get 
\be 
    \frac{1}{\lambda(E_d)} = \frac{1}{8E_d^2} \int_{4m_e^2}^{\infty} ds \; s\; \sigma_{\gamma\gamma}(s) \int_{s/(4E_d)}^{\infty} d\epsilon \; \frac{n(\epsilon,T_0)}{\epsilon^2} \,.
    \label{eq:lambda_SR}
\ee

\begin{figure}[tbp]
    \centering
    \includegraphics[width=0.6\textwidth]{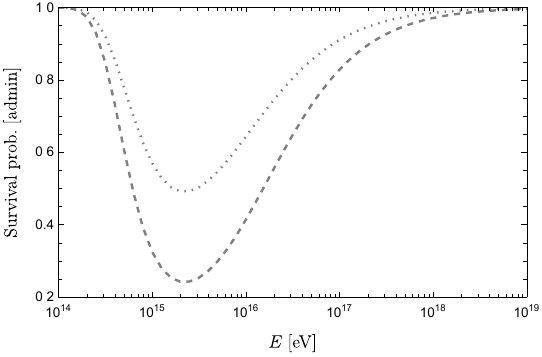}
    \caption{Gamma ray survival probability due to photon-photon interactions with the \abb{CMB} from a source at $D=5\unit{kpc}$ (dotted) and $D=10\unit{kpc}$ (dashed).}
    \label{fig:surv_SR}
\end{figure}

In order to make a numerical computation, it is better to work with adimensional quantities, so we can define
\be
    \bar s=\frac{s}{4m_e^2} \,,\quad \bar \epsilon = \frac{\epsilon}{k_B T_0}\,, \eqtext{and} \bar E_d=\frac{E_d}{m_e^2/(k_B T_0)} \,.
\ee
This way, the mean free path of the photon-photon pair emission can be written as
\be 
    \frac{1}{\lambda(E_d)} = \frac{\alpha^2(k_B T)^3}{m_e^2\pi} \frac{1}{\bar E_d^2} \int_{1}^{\infty} d\bar s \; \bar s\; \overline W(\bar s) \int_{\bar s/\bar E_d}^{\infty} d\bar \epsilon \; (e^{\bar \epsilon}-1)^{-1} \,,
    \label{eq:lambda_adim}
\ee
with $\overline W(\bar s)\coloneqq W(\bar \beta(\bar s))$ and $\bar \beta(\bar s)\coloneqq \sqrt{1-1/\bar s}$. The results of the computation of the mean free path are shown in Fig.~\ref{fig:lambda_SR}. Finally, let us recall that Eq.~\eqref{eq:optical_depth} defines the probability of survival of the gamma rays of a certain energy. For local sources we have
\be 
    \text{Prob}_{\gamma\rightarrow\gamma}(E_d,D) \approx \exp(-D/\lambda(E_d)) \,, \quad\text{for } z\ll1\,.
\ee
This probability will depend on the travelled distance, i.e., on the location of the source. One can see the obtained results for sources at $D=5\unit{kpc}$ and $D=10\unit{kpc}$ in Fig.~\ref{fig:surv_SR}.

\section{DSR effects in photons}
\label{sec:gamma_dsr}

\subsection{Gamma ray transparency in DSR}

As a simple example model of \abb{DSR} we are going to consider the one obtained in~\cite{Carmona:2019vsh}, derived by requiring the composition law to be isotropic, associative, non-commutative and linear on the four-momentum of each particle. The resulting composition law is called \abbdef{DCL1}{First-order Deformed Composition Law}, and can be written as
\be 
    (p\oplus q)_\mu = p_\mu + (1+p_0/\Lambda) q_\mu \,,
    \label{eq:DCL1_comp_law}
\ee
where the subscript $\mu$ corresponds to the four components of the four-momenta of the particles, and the subscript zero corresponds to the temporal coordinate ($\mu=0$). It is also easy to check that when one makes the scale of new physics go to infinity, one recovers the usual sum of \abb{SR}.

As discussed in Sec.~\ref{sec:DSR}, we can relate the composition law of momenta with the modified dispersion relation (in fact, at first order in $\Lambda$, this relation leads to the so-called golden rules~\cite{Carmona:2012un,Amelino-Camelia:2011gae}). Then we obtain that the energy-momentum relation on the \abb{DCL1} basis is given by
\be 
    \frac{p_0^2 - \vec{p}^2}{1+p_0/\Lambda} = m^2 \,.
    \label{eq:DCL1_disp}
\ee

In order to compute the modified mean free path and probability of survival of the gamma rays in this \abb{DSR} scenario, we need to compute first the modified cross section. Let us recall that in the previous section we saw that in \abb{SR} the cross section can be written in terms of the invariant $s$. Unfortunately, this quantity is not an invariant anymore, but using Eqs.~\eqref{eq:DCL1_comp_law} and \eqref{eq:DCL1_disp}, we can find a new invariant quantity, which can be written as
\be 
    \tilde s = \frac{(p\oplus q)^2}{1+(p\oplus q)_0/\Lambda} \,.
\ee
Particularizing the previous expression for a system of two photons with energies $E_1$ and $E_2$ we find that
\be 
    \tilde s = \frac{2 E_1 E_2 (1-\cos{\theta})}{1+E_2/\Lambda} \,.
\ee

Let us note that the modified invariant is not symmetric under a exchange of the two photons, so, in fact, we have two different invariants, 
\bea
    \tilde s_1 &= \frac{2 E \epsilon (1-\cos{\theta})}{1+\epsilon/\Lambda} \approx 2 E \epsilon (1-\cos{\theta}) \,, \\
    \tilde s_2 &= \frac{2 \epsilon E (1-\cos{\theta})}{1+E/\Lambda} \,,
\eea
where we have used that the energy of the background photon is much smaller than the scale of new physics. This statement does not hold for the gamma ray photon if its energy becomes comparable to the scale $\Lambda$. The existence of two different invariants for the same system implies two different channels for the process.

Comparing the modified invariants with respect to the one of \abb{SR}, $s=2 E \epsilon (1-\cos{\theta})$, we see that the first deformed invariant coincides with the usual definition. The second one, even though it is not equal, can be written in a similar way by making a change of variables:
\be 
    \tilde s_2 = 2 E' \epsilon (1-\cos{\theta})\,, \eqtext{where} E'\coloneqq\frac{E}{1+E/\Lambda} \,.
\ee

%Considering the expression of the relativistic invariant as a function of the energy of the gamma ray,
%\be
%    s(E_d)\coloneqq 2 E_d \epsilon (1-\cos{\theta}) \,,
%    \label{eq:invar_Ed}
%\ee
We see then that the modified invariants $\tilde s_1$ and $\tilde s_2$ have the same expression an in \abb{SR}, when applied at energies $E_d$ and ${E_d}'$, respectively. That is to say,
\bea
    \tilde s_1(E_d) &= s(E_d) = 2 E_d \epsilon (1-\cos{\theta}) \,, \quad \text{and} \\
    \tilde s_2(E_d) &= s({E_d}'(E_d))= 2 {E_d}'(E_d) \epsilon (1-\cos{\theta}) \,.
\eea
with ${E_d}'(E_d)=E_d/(1+E_d/\Lambda)$.

In \abb{SR} the invariant takes values from $4m_e^2$, corresponding to the particles at rest in the center of mass, to infinity; however, we could now have additional dependencies on $m_e$ and $\Lambda$\footnotemark. Taking into account that in the limit $\Lambda$ going to infinity we must recover the result of \abb{SR}, the new bottom limit can only depend on $m_e$ and $\Lambda$ through the quotient $(m_e/\Lambda)$. That is to say, \mbox{$\tilde{s}_\text{min}= 4m_e^2(1+O(m_e/\Lambda))$}, and so can be approximated as $\tilde{s}_\text{min}\approx 4m_e^2$.
\footnotetext{Only on $m_e$ and $\Lambda$, but not on $E_d$ or $\epsilon$, because the lower limit of $s$ must be also an invariant.}

Once the deformed relativistic invariant is set, we still need the proper generalization of the Breit-Wheeler cross section in \abb{DSR}, $\tilde \sigma_{\gamma\gamma}$. At present, we do not have a well established dynamical framework in \abb{DSR} which incorporates the relativistic deformed kinematics. However, if the cross section is still an invariant, it can only depend on invariant quantities such that $m_e$, $\tilde s$ and $\Lambda$. In \abb{SR}, the dependency of the cross section on $s$ is given by an adimensional function $W$, which only depends on the quotient $(m_e^2/s)$ (Eq.~\eqref{eq:beta_s}); then we want to find the new function $\widetilde W$ which gives the new dependency on the invariant quantities $m_e$, $\tilde s$ and $\Lambda$. Given that $\widetilde W$ is also adimensional, and that we expect to recover $W$ in the limit $\Lambda$ going to infinity, $\widetilde W$ can only depend on the quotients $(m_e^2/\Lambda^2)$, $(\tilde s/\Lambda^2)$, and $(m_e^2/\tilde s)$. Let us note that for gamma rays and values of $\Lambda$ around the PeV, and low energy photons of the \abb{CMB}, $(m_e^2/\tilde s) \gg (\tilde s/\Lambda^2),(m_e^2/\Lambda^2)$, then we can approximate $\widetilde W\approx \widetilde W (m_e^2/\tilde s)$. In the limit $\Lambda$ going to infinity, we know that $\tilde s\rarrow s$, so necessarily $\widetilde W \rarrow W$. Therefore, a reasonable approximation is that $\tilde \sigma_{\gamma\gamma}(\tilde s)\approx\sigma_{\gamma\gamma}(\tilde s)$ for the energies in consideration; that is to say, the modified cross section in \abb{DSR} can be approximated by the Breit-Wheeler formula applied at the modified invariant.

For the first channel, as the modified invariant coincides with the invariant of \abb{SR}, the cross section is not modified, so neither the mean free path nor the survival probability are. For the second channel, as the deformed invariant is just the one of \abb{SR} applied at a different energy, the same will happen for the cross section, mean free path, and survival probability. That is to say, for the first channel $\tilde \lambda_1(E_d)=\lambda(E_d)$, and for the second channel, $\tilde \lambda_2(E_d)=\lambda({E_d}'(E_d))$. 

\begin{figure}[tbp]
        \begin{center}
        \includegraphics[width=0.7\textwidth]{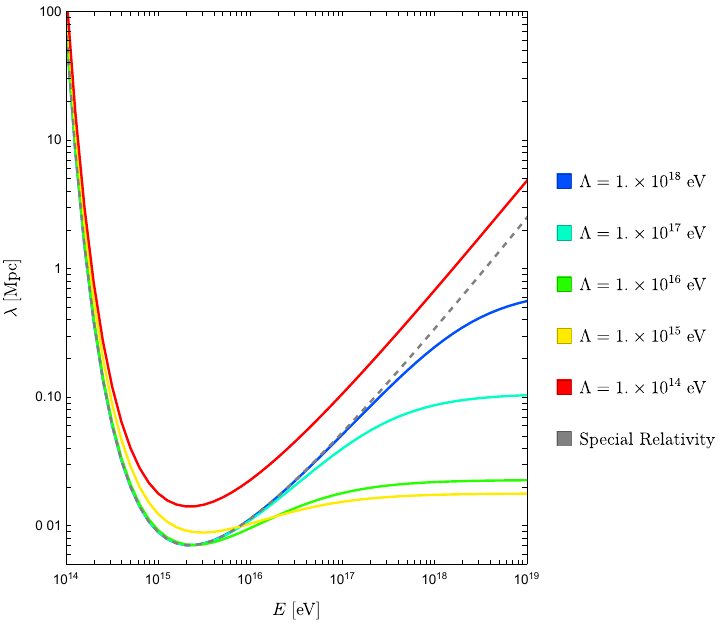}
        \caption{Modified mean free path of gamma rays within the range of energies in which the \abb{CMB} is the relevant background. The gray dashed line corresponds to the result of \abb{SR} and the colored solid curves represent the result for different values of the scale of new physics.}
        \label{fig:lambda_large}
        \end{center}
        \begin{center}
        \includegraphics[width=0.7\textwidth]{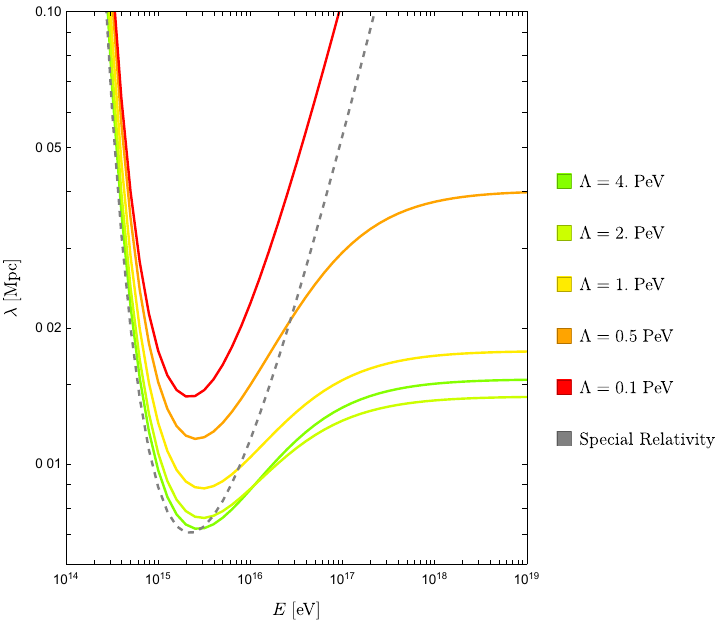}
        \caption{Modified mean free path of gamma rays within the range of energies in which the \abb{CMB} is the relevant background. The gray dashed line corresponds to the result of \abb{SR} and the colored solid curves represent the result for different values of the scale of new physics around the PeV.}
        \label{fig:lambda_short}
        \end{center}
\end{figure}

As we mentioned before, we lack a dynamical framework in \abb{DSR}, so we do not know what probability we should assign to each channel. An agnostic hypothesis is to consider both channels with equal probabilities. In that case, the mean free path can be computed as
\be
    \frac{1}{\tilde\lambda(E_d)}=\frac{1}{2} \qty(\frac{1}{\tilde\lambda_1(E_d)}+\frac{1}{\tilde\lambda_2(E_d)})=\frac{1}{2}\qty(\frac{1}{\lambda(E_d)}+\frac{1}{\lambda({E_d}'(E_d))}) \,,
\ee
with $\lambda(E_d)$ the mean free path of special relativity, given by Eq.~\ref{eq:lambda_adim}. We show the computation of the modified mean free path for different values of the scale of new physics in Fig.~\ref{fig:lambda_large}, and for values around the PeV in Fig.~\ref{fig:lambda_short}.

\begin{figure}[tbp]
    \centering
    \begin{minipage}{0.49\textwidth}
        \centering
        \includegraphics[width=\textwidth]{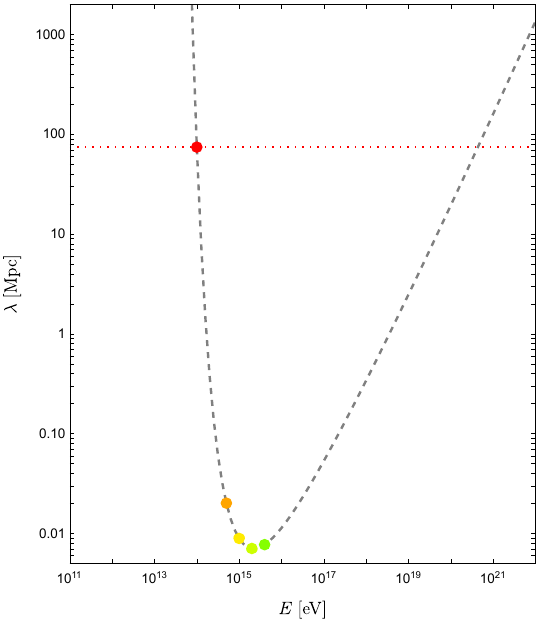}
        \caption{$\lambda(\Lambda)$ for the values of $\Lambda$ shown in Fig.~\ref{fig:lambda_short}. The horizontal line helps to identify the two energies with the same value of $\lambda$.}
        \label{fig:lambda_asym}
    \end{minipage}%
    \hfill
    \begin{minipage}{0.49\textwidth}
        \centering
        \includegraphics[width=1.01\textwidth]{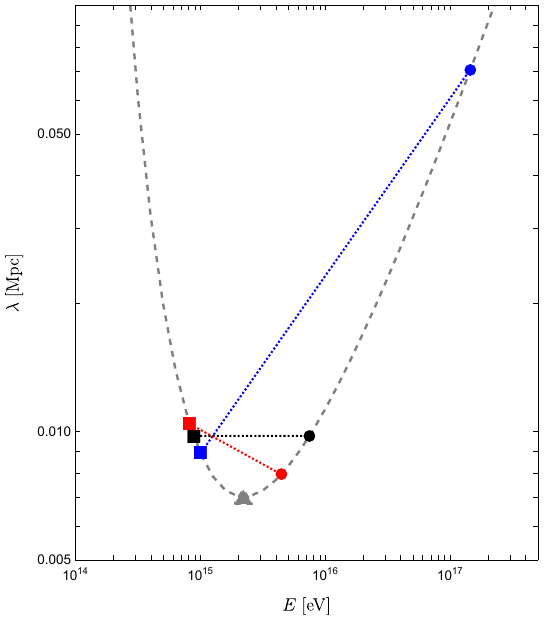}
        \caption{$E_0$ (triangle), $E^*$ (black circle), energies $E_1>E^*$ and $E_2<E^*$ (blue and red circles), and the corresponding ${E_i}'$ (squares).}
        \label{fig:lambda_Eu}
    \end{minipage}
\end{figure}

The first characteristic we can identify is that some curves tend to a constant value at high energies, i.e., when $E_d\gg\Lambda$. In order to understand better this behaviour we will focus on the values of $\Lambda$ of Fig.~\ref{fig:lambda_short}. Let us note that at energies much larger than the scale of new physics
\be
    \lim_{(E_d/\Lambda)\rightarrow\infty} E'(E_d) = \lim_{(E_d/\Lambda)\rightarrow\infty} \qty(\frac{E_d}{1+E_d/\Lambda}) = \Lambda \,.
\ee
So, for $E_d\gg\Lambda$, the modified mean free path tends to
\be 
    \lim_{(E_d/\Lambda)\rightarrow\infty} \qty(\frac{1}{\tilde\lambda(E_d)})= \frac{1}{2}\qty(\frac{1}{\lambda(E_d)}+\frac{1}{\lambda(\Lambda)}) \,.
    \label{eq:lambda_asym}
\ee
Then we see that the behaviour of $\tilde\lambda(E_d)$ at high energies depends on the comparison between $\lambda(E_d)$ and $\lambda(\Lambda)$ (where $\lambda$ is the mean free path of \abb{SR}). In order to understand better this comparison we show in Fig.~\ref{fig:lambda_asym} the values of $\lambda(\Lambda)$ for the different values of $\Lambda$ of Fig.~\ref{fig:lambda_short}.

For $\Lambda =0.1\unit{PeV}$, the value of $\lambda(\Lambda)$ (see red point in Fig.~\ref{fig:lambda_asym}), is not recovered until $E_d\approx 10^{21}\unit{eV}$. As a consequence, for all the energies much greater than $\Lambda$ shown in Fig.~\ref{fig:lambda_short}, $\lambda(E_d)\ll\lambda(\Lambda)$ and the limit of Eq.~\eqref{eq:lambda_asym} tends to $\tilde\lambda(E_d)\rightarrow 2\lambda(E_d)$. We can check this behaviour in the red curve of Fig.~\ref{fig:lambda_short}. If one continues computing the modified mean free path for energies above $10^{21}\unit{eV}$, the behaviour would be now the opposite, i.e., $\lambda(E_d)\gg\lambda(\Lambda)$. In this regime the modified mean free path tends to a constant value, $\tilde\lambda(E_d)\rightarrow 2\lambda(\Lambda)$. This is the behaviour we see in Fig.~\ref{fig:lambda_short} for all the other curves.

Let us also note that the constant value $2\lambda(\Lambda)$ follows the shape of the \abb{SR} mean free path, which has a minimum around $E_0\approx 2 \unit{PeV}$. As a consequence, the value of $2\lambda(\Lambda)$, as function of $\Lambda$, will decrease until $\Lambda\approx E_0$ and then will start increasing again (see yellow and green points in Fig.~\ref{fig:lambda_asym}). This is the reason why the curve of $\Lambda =4 \unit{PeV}$ ends at higher values than the curve of $\Lambda=2 \unit{PeV}$ in Fig.~\ref{fig:lambda_short}.

Another important characteristic of the modified mean free path are the crossing points with the values of \abb{SR}. These crossing points will separate the energies where the deformation of the kinematics will produce a larger mean free path (and so more transparency) from the energies at which it is smaller (and so more opacity) with respect to the case of \abb{SR}. The energy at which both mean free path curves intersect, let us call it $E^*(\Lambda)$, must satisfy
\be
  \frac{1}{\lambda(E^*)}\,=\,\frac{1}{\Tilde{\lambda}(E^*)} \srarrow \frac{1}{\lambda(E^*)}\,=\,\frac{1}{2}\,\left(\frac{1}{\lambda(E^*)}+\frac{1}{\lambda({E^*}')}\right)\,.
\ee
Simplifying, we get that the condition can be written as
\be
   \lambda(E^*)\,=\,\lambda({E^*}')\,,\quad \text{with}\,\quad {E^*}'\,=\,\frac{E^*}{1+ E^*/\Lambda}\,.
   \label{eq:corte}
\ee

For Eq.~\eqref{eq:corte} to be satisfied, $E^*$  and ${E^*}'$ must be to the right and left sides, respectively, of the minimum of $\lambda(E)$, which we denote as $E_0$. This is shown in Fig.~\ref{fig:lambda_Eu} (black circle and square, respectively). For a given $\Lambda$, in order to know if at a certain energy $E_d$ the universe is more or less transparent to gamma rays than in \abb{SR}, we have to compare $E_d$ and ${E}^*(\Lambda)$. The two possible outputs of the comparison are:
\begin{enumerate}
    
\item $E_d>{E}^*(\Lambda)$ (case of the energy $E_1$ in Fig.~\ref{fig:lambda_Eu} (blue circle)). Then,
\begin{equation}
    {E_1}'>{E^*}'(\Lambda) \srarrow \lambda(E_1)\,>\,\lambda(E'_1) \srarrow \Tilde{\lambda}(E_1)\,<\,\lambda(E_1)\,.
\end{equation}
Therefore, for energies greater than  $E^*(\Lambda)$, the modified mean free path is smaller than the one of \abb{SR}, and so the universe is less transparent for gamma rays.   
\item $E_d<{E}^*(\Lambda)$ (case of the energy $E_2$ in Fig.~\ref{fig:lambda_Eu} (red circle)). Then,
    \begin{equation}
        {E_2}'<{E^*}'(\Lambda) \srarrow \lambda(E_2)\,<\,\lambda(E'_2) \srarrow \Tilde{\lambda}(E_2)\,>\,\lambda(E_2).
    \end{equation}
Therefore, for energies smaller than $E^*(\Lambda)$, the modified mean free path is greater than the one of \abb{SR}, and so the universe is more transparent for gamma rays.  
    
\end{enumerate}

\begin{figure}[tbp]
    \centering
    \centering
    \includegraphics[width=0.8\textwidth]{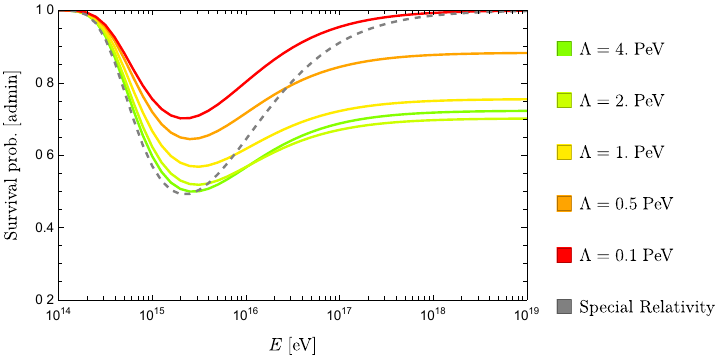}
    \caption{Modified gamma ray survival probability due to photon-photon interactions with the \abb{CMB} from a source at $D=5\unit{kpc}$.}
    \label{fig:surv_DSR_5}
\end{figure}

\begin{figure}[tbp]
    \centering
    \includegraphics[width=0.8\textwidth]{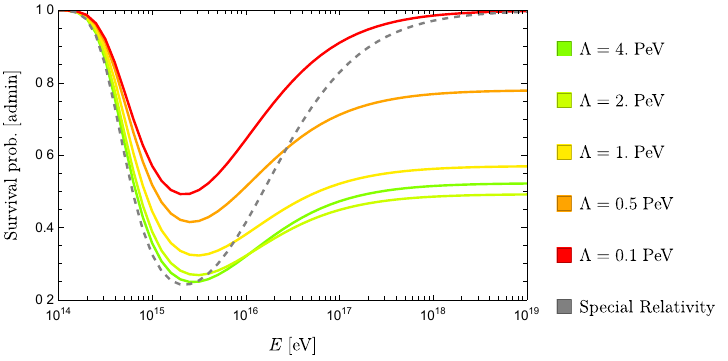}
    \caption{Modified gamma ray survival probability due to photon-photon interactions with the \abb{CMB} from a source at $D=10\unit{kpc}$.}
    \label{fig:surv_DSR_10}
\end{figure}

This behavior manifest in the mean free path shown in Fig.~\ref{fig:lambda_short}, where one can also check that the crossing point between the modified mean free path and the one of \abb{SR} appears before as we increase the scale of new physics $\Lambda$; but always for energies greater than the minimum of $\lambda(E)$. Also, as we did for the case of \abb{SR}, one can translate the behaviour of the mean free path to the probability of survival from a source at a certain distance. We show the results for sources at $D=5 \unit{kpc}$ and $D=10 \unit{kpc}$ in Figs.~\ref{fig:surv_DSR_5} and \ref{fig:surv_DSR_10}, respectively.

\begin{figure}[tbp]
    \centering
    \includegraphics[width=0.8\textwidth]{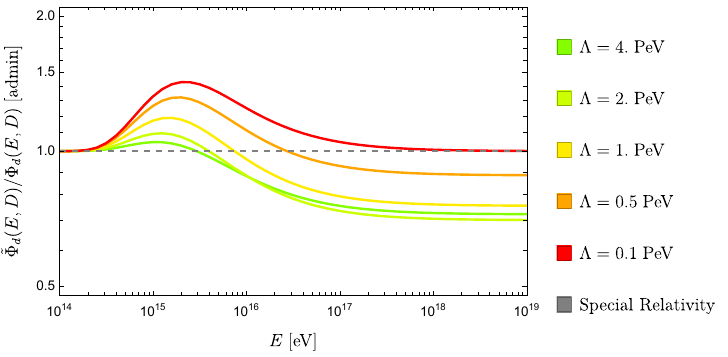}
    \caption{Ratio between the expected fluxes of gamma rays in the \abb{DSR} and \abb{SR} scenarios, considering a source at $D=10\unit{kpc}$.}
    \label{fig:flux_quo_5}
\end{figure}

\begin{figure}[tbp]
    \centering
    \includegraphics[width=0.8\textwidth]{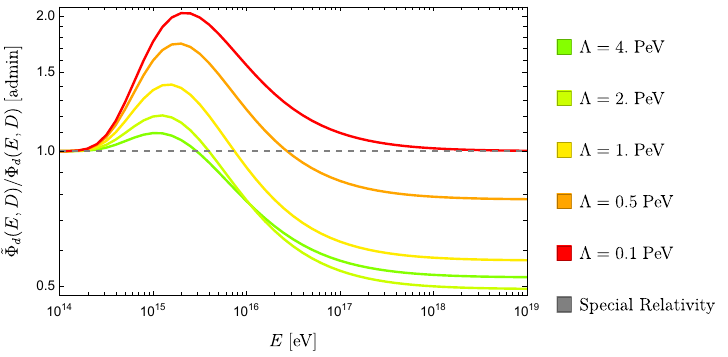}
    \caption{Ratio between the expected fluxes of gamma rays in the \abb{DSR} and \abb{SR} scenarios, considering a source at $D=10\unit{kpc}$.}
    \label{fig:flux_quo_10}
\end{figure}

The crossing points $E^*(\Lambda)$  will also distinguish, for each value of the scale $\Lambda$, where we expect a larger or smaller observed flux of gamma rays with respect to the case of \abb{SR}. We show in Figs.~\ref{fig:flux_quo_5} and \ref{fig:flux_quo_10} the ratio between the expected flux of gamma rays for the case of the \abb{DSR} ($\tilde\Phi_d(E,D)$) and \abb{SR} ($\Phi_d(E,D)$) scenarios, from a galactic source at distance $D=5$ and $10\unit{kpc}$, respectively, and for the values of $\Lambda$ showed in Fig.~\ref{fig:lambda_short}. We can take this ratio as a measure of the sensitivity to the deformed kinematics in the detection of high-energy photons from within our galaxy, and the different values of $D$ are examples of the role of the uncertainty in the location or identification of the sources.

We observe that in the \abb{DSR} scenario, respect to the case of \abb{SR}, there is an increase in the observed flux at energies close to the minimum of the mean free path, which goes down as the energy increase until the expected flux of both scenarios are the same at $E^*(\Lambda)$. Finally, the expected flux respect to \abb{SR} becomes smaller for energies above $E^*(\Lambda)$, tending to an asymptotic constant suppression for large energies. 

Experiments would then observe an anomalous high transparency at energies $E\lesssim E^*(\Lambda)$ and an anomalous low transparency at energies $E\gtrsim E^*(\Lambda)$, which would be more evident from sources at large distances. It is however important to notice that, even if larger distances would make the anomalies more manifest, the sources cannot be much farther than the mean free path of the photons in order to have an observable flux at the detector. 

\subsection{Photon time delays in DSR}
\label{sec:time_delays}

When one considers \abb{LIV} on photons, the kinematics is characterized by a modified energy-momentum relation from which a momentum-dependent velocity stems. This causes the existence of time delays between photons of different energies~\cite{MAGIC:2020egb,Pan:2020zbl,Levy:2021oec,Terzic:2021rlx,Martinez-Huerta:2020cut,Bolmont:2022yad,PerezdelosHeros:2022izj}. In contrast, the existence or absence of time delays for massless particles\footnotemark in \abb{DSR} is still a topic in study, due to the fact that one also has to take into account additional ingredients, such as relative locality, nontrivial translations, etc. (see~\cite{Amelino-Camelia:2011ebd,Amelino-Camelia:2013uya,Pfeifer:2018pty,Barcaroli:2016yrl,Barcaroli:2017gvg,Carmona:2017oit} for some studies in this subject). In the following, we will do a systematic analysis of the calculation of time delays on a flat spacetime (based on the idea presented in~\cite{frattulillo}), in order to throw some light over this question. We will not consider any particular basis, but instead we will state general properties that a model for time delays, consistent with relative locality, must fulfill.
\footnotetext{Let us note that, in contrast to \abb{LIV}, the effects of new physics in \abb{DSR} are universal. That is to say, if in \abb{DSR} models there are photon time delays, all the massless particles (or almost massless like the neutrino) will experience it.}

For simplicity, we will develop the study in $1+1$ dimensions; however, the generalization to $3+1$ dimensions is straightforward. Let us consider a canonical choice of coordinates for the $1+1$ dimensional phase space of one particle, i.e, we will chose some coordinates $(\Pi,\Omega)$ which represent the momentum and energy of the particle, and their conjugated variables $(x,t)$, as the canonical space-time coordinates. The canonical choice of coordinates of the phase space, $(x,t,\Pi,\Omega)$, is such that their Poisson brackets are
\be 
    \{t,\Omega\}=-1\,,\quad\quad \{x,\Pi\}=1\,,\quad\quad \{t,x\}=0\,, \\
    \{t,\Pi\}=0\,,\quad\quad \{x,\Omega\}=0\,,\quad\quad \{\Omega,\Pi\}=0\,.
    \label{eq:ps_brackets}
\ee

Once this is set, let us consider a completely general \abb{DSR} model at order $(1/\Lambda)$. This model will be characterized by the deformation of the generators of the Poincaré algebra. Let us write the most general deformation at order $(1/\Lambda)$ of the generator of the space-time translations and boosts, which we will call $E$, $P$, and $N$, respectively:
\be
    E=\Omega+\frac{a_1}{\Lambda}\Omega^2+\frac{a_2}{\Lambda}\Pi^2 \,,\quad\quad 
    P=\Pi+\frac{a_3}{\Lambda}\Omega\Pi\,,\\ N=x\Omega-t\Pi+\frac{a_4}{\Lambda}x\Omega^2+\frac{a_5}{\Lambda}x\Pi^2-\frac{a_6}{\Lambda} t \Omega \Pi \,.
\ee
In the previous equation, and from now on, we should understand the equalities only up to terms $(1/\Lambda)$. Let us note that $E\neq \Omega$ and $P\neq \Pi$, so the energy and momentum are not the generators of the one-particle space-time translations.

The deformed generators $(E,P,N)$ will satisfy a deformed algebra which will have a Casimir of the form
\be 
    C=\Omega^2-\Pi^2+\frac{\alpha_1}{\Lambda}\Omega^3+\frac{\alpha_2}{\Lambda}\Omega\Pi^2 \,,
\ee
for some $\alpha_1$ and $\alpha_2$, which can be determined as functions of the coefficients of the boost $N$, $(a_4,a_5,a_6)$, from the condition $\{N,C\}=0$,
\be 
    \{N,C\}=\frac{(3\alpha_1+2\alpha_2-2a_4+2a_6)}{\Lambda}\Omega^2\Pi+\frac{(\alpha_2-2a_5)}{\Lambda}\Pi^3=0 \,,
\ee
so then,
\be
    \alpha_1=\frac{2}{3}(a_4-2a_5-a_6) \,,\quad\quad \alpha_2=2a_5\,.
\ee

Setting $C=m^2$, we can find a relation between the energy $\Omega$ and the momentum $\Pi$. This is the modified energy-momentum relation or \abb{MDR}, which for massless particles is
\be
    \Omega=\Pi - \frac{(\alpha_1+\alpha_2)}{2 \,\Lambda} \,\Pi^2= \Pi-\frac{(a_4+a_5-a_6)}{3\Lambda}\Pi^2 \,.
    \label{eq:dispersion_relation}
\ee

One can also identify the deformed Poincaré algebra that the generators satisfy,
\be
    \{N,P\}=\Omega + \frac{(a_3+a_4)}{\Lambda}\Omega^2+\frac{(a_3+a_5)}{\Lambda}\Pi^2 \,\quad\quad
    \{N,E\}=\Pi + \frac{(2a_1+2a_2+a_6)}{\Lambda}\Omega\Pi \,,
\ee
which is more naturally written in terms of the generators themselves, 
\be
    \{N,P\}= E + \frac{w_1}{\Lambda}E^2+\frac{w_2}{\Lambda}P^2\,,\quad\quad 
    \{N,E\}= P + \frac{w_3}{\Lambda}EP\,,
\ee
where $(w_1, w_2, w_3)$ are the coefficients characterizing the deformed algebra (the basis of $\kappa$-Poincaré) and are related with the previous coefficients by
\be
    w_1 = a_4 - a_1 + a_3 \,, \qquad w_2 = a_5 - a_2 + a_3 \,, \qquad w_3 = a_6 + 2 a_1 + 2 a_2 - a_3\,.
\ee
Similarly, it is also useful to write the Casimir as a function of the generators $E$ and $P$,
\be
    C=E^2-P^2+\frac{2(w_1-2w_2-w_3)}{3\Lambda}E^3+\frac{2w_2}{\Lambda}EP^2 \,.
    \label{eq:casimir}
\ee
However, one should not confuse this previous expression, which is an operator written in terms of the deformed generators, with the \abb{MDR} shown in Eq.~\eqref{eq:dispersion_relation}, which is a function of the energy $\Omega$ and momentum $\Pi$ of the particle.

The Casimir, Eq.~\eqref{eq:casimir}, is the generator of translations along the worldline of a particle with a fixed energy and momentum, as a function of an arbitrary parameter $\tau$. Let us note that for a massless particle $\Omega$ and $\Pi$ are related by the \abb{MDR},  Eq.~\eqref{eq:dispersion_relation}, so specifying $\Pi$ is enough to completely characterize the worldline of the particle. Then we can write 
\be 
    \frac{dx}{d\tau}= \lambda(\tau) \{C,x\}\,,\quad\quad \frac{dt}{d\tau}=\lambda(\tau) \{C,t\} \,,
\ee
where $\lambda(\tau)$ is just an arbitrary function implementing the invariance under reparametrization of the worldline.

We can get rid of the arbitrary parameter $\tau$ defining the velocity of the particle
\be 
    v\coloneqq \frac{dx}{dt}=\frac{dx/d\tau}{dt/d\tau}=\frac{\{C,x\}}{\{C,t\}}=1-\frac{(\alpha_1+\alpha_2)}{3}\Pi =1-\frac{2(a_4+a_5-a_6)}{3}\Pi
    \label{eq:velocity} \,.
\ee
We see then that the velocity of the particle is modified exclusively by the coefficients $(a_4,a_5,a_6)$ from the boost generator, or equivalently, by $(\alpha_1,\alpha_2)$ from the \abb{MDR}.

In order to discuss time delays, we will consider the propagation of two photons, one of high energy,
i.e., which is affected by the effects of \abb{DSR} and whose velocity is given by Eq.~\eqref{eq:velocity}, and another one of low energy, which does not see any effect beyond \abb{LI} and so whose velocity is $v\approx 1$. We will define the time delay as the difference in the detection times of the two photons if their emission was simultaneous.

The existence of relative locality makes the analysis of time delays in \abb{DSR} very subtle. In contrast to \abb{SR} or \abb{LIV} scenarios, now one needs two observers in order to make a consistent analysis. Relative locality states that there is only one observer that is able to see as local each high-energy interaction, and it is the one that sees the interaction at its origin of coordinates (see Appendix~\ref{sec:relative_locality}). Therefore, since we have two high-energy interactions (the detection and emission of the high-energy photon), we need a pair of observers $\mathcal O$ and $\mathcal O'$, which see, respectively, each interaction as local. The effects of relative locality are proportional to the momenta involved in the interaction, so if we consider that the low-energy photon is emitted and detected in low-energy processes, one could expect that both its emission and detection are local for every observer, including $\mathcal O$ and $\mathcal O'$.
\begin{figure}[tbp]
    \centering
    \includegraphics[width=0.3\textwidth]{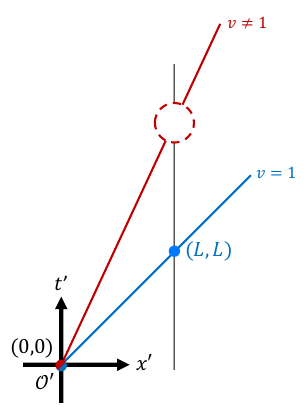}
    \hspace{2cm}
    \includegraphics[width=0.3\textwidth]{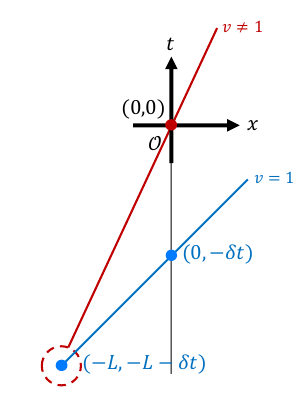}
    \caption{In \abb{DSR}, due to relative locality, the observer $\mathcal O'$ (left) sees the emission of the high-energy photon as local, but not the detection. The opposite behavior is shown for $\mathcal O$ (right), which sees the detection of the high-energy photon as local, but not the emission. The emission and detection of the low-energy photon are local for every observer.}
    \label{fig:obs_DSR}
\end{figure}

If the source and the detector are at relative rest, we expect that the coordinates that $\mathcal O$ and $\mathcal O'$ assign to the worldlines of each photon will be related by a translation of the form\footnotemark,
\footnotetext{One can write the finite transformation as the infinitesimal one because the Poisson brackets of $\{P,x\}$, $\{P,t\}$, $\{E,x\}$ and $\{E,t\}$ with $P$ and $E$ are zero.}
\be
    x'\,=\,x- \epsilon_1 \{P, x\} + \epsilon_0 \{E, x\} \,, \quad\quad
    t'\,=\,t - \epsilon_1\{P, t\} + \epsilon_0 \{E, t\} \,,
    \label{eq:translation}
\ee
with $(\epsilon_1,\epsilon_0)$ the parameters of the translation, which still needs to be determined.

The equation of the worldline for the high-energy photon seen from $\mathcal O$ and $\mathcal O'$ are $x=vt$ and $x'=vt'$, respectively. Then, using Eq.~\eqref{eq:translation}, one finds that the parameters of the translation are related by 
\be
    \epsilon_0 \qty[v\{E,t\} - \{E,x\}] \,=\, \epsilon_1 \qty[v\{P,t\} - \{P,x\}]\,.
\ee
Computing the Poisson brackets that appear in the previous expression,
\be
    \{E,t\}=1+\frac{2a_1}{\Lambda}\Omega\,,\quad \{E,x\}=-2\frac{a_2}{\Lambda}\Pi\,,\quad \{P,t\}=\frac{a_3}{\Lambda}\Pi\,,\quad \text{and}\quad \{P,x\}=-1-\frac{a_3}{\Lambda}\Omega\,,
\ee
and using the energy-momentum relation, Eq.~\eqref{eq:dispersion_relation}, one gets
\be
    \frac{\epsilon_0}{\epsilon_1} = 1+ \frac{2(a_4+a_5-a_6)}{3\Lambda}\,\Pi - \frac{2(a_1+a_2-a_3)}{\Lambda} \,\Pi \,.
    \label{eq:parameters}
\ee

Let us now consider the simultaneous emission of the high- and low-energy photons. We know that the coordinates that $\mathcal O'$ assigns to the high-energy photon at the emission are $(x',t')=(0,0)$. This is also true for the low-energy photon, because its emission is local and simultaneous to that of the high-energy photon. During the propagation of the low-energy photon, since its velocity is $v\approx 1$, the spatial and temporal coordinates coincide. As a consequence, the coordinates that $\mathcal O'$ will assign to the low-energy photon at the detection will be of the form $(L,L)$, where $L$ is what we call in \abb{SR} the ``distance to the detector from the source'' (see blue trajectory in Fig.~\ref{fig:obs_DSR} (left)).

Instead, from the point of view of $\mathcal O$, the coordinates of the high-energy photon at the detection are perfectly defined, $(x,t)=(0,0)$. Disregarding the size of the detector, the spatial coordinate of the low-energy photon at the detection must be the same, $x=0$; however, the temporal coordinate can be different, and this difference in the time of arrival between the two photons is just the definition of (minus) the time delay, $t=-\delta t$, (see Fig.~\ref{fig:obs_DSR} (right)).

The coordinates $(x',t')=(L,L)$ and $(x,t)=(0,-\delta t)$ correspond to the same point of the worldline of the low-energy photon (its detection), but seen from $\mathcal O'$ and $\mathcal O$, respectively. Therefore, they must be related by the translation given by Eq.~\eqref{eq:translation}, but particularized for the low-energy photon ($E\approx \Omega$ and $P\approx \Pi$),
\bea
    x'\,=\,x+ \epsilon_1 &\srarrow L\,=\,0+ \epsilon_1 \,, \\
    t'\,=\,t+ \epsilon_0 &\srarrow L\,=\,-\delta t+ \epsilon_0 \,.
\eea
This allows us to relate the parameters $(\epsilon_1,\epsilon_0)$ with $\delta t$,
\be
    \frac{\epsilon_0}{\epsilon_1} = 1+ \frac{\delta t}{L}\,,
    \label{eq:parameters_low}
\ee
and, using Eq.~\eqref{eq:parameters}, finally solve for the time delay,
\be 
    \delta t \,=\, L \qty[ \frac{2(a_4+a_5-a_6)}{3\Lambda}\,\Pi - \frac{2(a_1+a_2-a_3)}{\Lambda} \,\Pi]\,.
    \label{delta t}
\ee

From the previous equation we can see that the time delay has two contributions, one from the momentum-dependent velocity, which depends only on the coefficients $(a_4,a_5,a_6)$ from the boost, and one from the non-trivial translations, which depends only on the coefficients $(a_1,a_2,a_3)$ from the generators of the translations. One will not have observable consequences of \abb{DSR} at the level of time delays if the deformation of the Poincaré generators is such that
\be 
    \frac{1}{3}(a_4+a_5-a_6)=(a_1+a_2-a_3)\,.
\ee

We can also check that the value of the time delay is independent of the initial choice of canonical coordinates. Let us consider a generic coordinate transformation from $(t,x,\Omega,\Pi)$ to some other canonical coordinates $(\overline t,\overline x,\overline \Omega,\overline \Pi)$, related with the first ones by three parameters $(\delta_1,\delta_2,\delta_3)$ in the following way
\be
    \overline \Omega=\Omega +\frac{\delta_1}{\Lambda}\Omega^2+ \frac{\delta_2}{\Lambda}\Pi^2\,,\quad \overline \Pi=\Pi +\frac{\delta_3}{\Lambda}\Omega\Pi \,,\\
    \overline t=t -\frac{2\delta_1}{\Lambda}t\Omega+ \frac{\delta_3}{\Lambda}x\Pi\,,\quad \overline x=x +\frac{2\delta_2}{\Lambda}t\Pi- \frac{\delta_3}{\Lambda}x\Omega\,.
    \label{eq:change_variable}
\ee
Then the generators of the translations and boosts will have a new expansion in this choice of coordinates,
\be
    E=\overline\Omega+\frac{\overline a_1}{\Lambda}\overline\Omega^2+\frac{\overline a_2}{\Lambda}\overline\Pi^2\,,\quad 
    P=\overline\Pi+\frac{\overline a_3}{\Lambda}\overline\Omega\,\overline\Pi\,,\quad\text{and}\\
    N=\overline x\overline\Omega-\overline t\overline\Pi+\frac{\overline a_4}{\Lambda} \overline x\overline\Omega^2+\frac{\overline a_5}{\Lambda}\overline x\overline\Pi^2-\frac{\overline a_6}{\Lambda}\overline t\bar\Omega,\overline\Pi \,,
\ee
where
\be 
    \overline a_1=a_1-\delta_1\,,\quad \overline a_2=a_2-\delta_2\,,\quad \overline a_3=a_3-\delta_3\,, \\
    \overline a_4=a_4-\delta_1+\delta_3\,,\quad \overline a_5=a_5-\delta_2+\delta_3\,,\quad \overline a_6=a_6+2\delta_1+2\delta_2-\delta_3 \,.
\ee
Then, it is straightforward to check that
\bea
    \delta \overline t \,&=\, L\;\frac{2(\overline a_4+\overline a_5-\overline a_6-3\overline a_1-3\overline a_2+3\overline a_3)}{3\Lambda}\,\overline \Pi \\ \,&=\, L\;\frac{2(a_4+a_5-a_6-3a_1-3 a_2+3 a_3)}{3\Lambda}\,\Pi \,=\, \delta t \,.
\eea
From this result one concludes that the value of the time delay, i.e., Eq~\eqref{delta t}, is independent of the choice of energy-momentum variables. This is a property we would like for any observable.

We can use this freedom of choice of the energy-momentum variables to simplify the computation of the time delay. For instance, from Eq.~\eqref{eq:velocity}, one can check that the condition to have a momentum-independent velocity for massless particles is $(\overline a_4 + \overline a_5 - \overline a_6) = 0$. Therefore, when performing a change of phase-space variables such that $\delta_1 + \delta_2 - \delta_3 = (1/3) (a_4 + a_5 - a_6)$, the time delay will have a contribution only from the non-trivial translations. On the other hand, from Eq.~\eqref{delta t}, we see that the condition to have trivial translations is $(\overline a_1+ \overline a_2- \overline a_3)=0$. Then, when performing a change of phase-space variables such that $\delta_1 + \delta_2 - \delta_3 = (a_1 + a_2 - a_3)$, the time delay  will be exclusively due to the momentum dependence of the velocity.

Let us use a choice of energy-momentum variables such that they coincide with the generators of space-time translations (like in \abb{SR}),  $(\overline\Omega,\overline\Pi)=(E, P)$, and so
\be
\overline{t} \,=\, t - \frac{2 a_1}{\Lambda} \,t \,\Omega + \frac{a_3}{\Lambda} \,x \,\Pi\,, \quad\quad\quad
\overline{x} \,=\, x + \frac{2 a_2}{\Lambda} \,t \,\Pi - \frac{a_3}{\Lambda} \,x \,\Omega\,.
\ee
In this choice of coordinates, the space-time translations are not deformed, so the time delay stems only from the momentum-dependent velocity,
\be 
    \delta \overline t=L\qty[\frac{1}{\overline v}-1] \,.
\ee
But the velocity can be read directly from the Casimir in terms of $E$ and $P$,
\be
\overline v = \frac{d\overline{x}}{d\overline{t}} \,=\, \frac{d\overline\Omega}{d\overline\Pi} \,=\, \frac{dE}{dP}\bigg |_{C=0}.
\ee
This way, the time delay, Eq.~\eqref{delta t}, can be written in terms of the relation between $E$ and $P$, which does not depend on the choice of phase space variables,
\be
\delta t \,=\, \delta \overline t \,=\, L\qty[\qty(\frac{dE}{dP}\bigg |_{C=0})^{-1}-1] \,=\, L \, \frac{2(w_1+w_2-w_3)}{3 \Lambda} \,P\,.
\ee
Then, instead of writing the condition of absence of time delays as a constraint on the coefficients of the expansion of the generators in a certain choice of the phase space,  we can write the time delay in terms of the algebra of $(E,P,N)$,
\be
(w_1 + w_2 - w_3) \,=\, 0\,,
\label{no-time-delay-2}
\ee
which is independent of the choice of phase-space variables.

Let us note that a very different situation arises if one, instead of making a change of phase-space coordinates, now changes the deformed generators of space-time translations, i.e., if one changes the basis of $\kappa$-Poincaré. This implies that we change $(E,P,N) \to (\overline{E},\overline{P},N)$, with
\be
\overline{E} \,=\, E + \frac{\Delta_1}{\Lambda} \,E^2 + \frac{\Delta_2}{\Lambda} \,P^2\,, 
\quad\quad\quad
\overline{P} \,=\, P + \frac{\Delta_3}{\Lambda} \,E \,P\,.
\ee
Then we have a new deformed algebra given by
\be
\{N, \overline{P}\} \,=\, \overline{E} + \frac{\overline{w}_1}{\Lambda} \,\overline{E}^2 + \frac{\overline{w}_2}{\Lambda} \,\overline{P}^2\,, 
\quad\quad\quad
\{N, \overline{E}\} \,=\, \overline{P} + \frac{\overline{w}_3}{\Lambda} \,\overline{E} \,\overline{P}\,,
\ee
where
\be
\overline{w}_1 \,=\, w_1 - \Delta_1 + \Delta_3 \,, \quad\quad
\overline{w}_2 \,=\, w_2 - \Delta_2 + \Delta_3 \,, \quad\quad
\overline{w}_3 \,=\, w_3 + 2 \Delta_1 + 2 \Delta_2 - \Delta_3 \,,
\ee
and, as a consequence, the result for time delays is also modified,
\bea
\delta \overline{t} \,=\, \delta t - L \,\frac{2(\Delta_1+\Delta_2-\Delta_3)}{\Lambda} \,\Pi\,.
\eea
Therefore, different bases of $\kappa$-Poincaré correspond to different physical models going beyond \abb{SR}, with different testable predictions for observables like the time delays.

If we get the condition of absence of time delays, Eq.~\eqref{no-time-delay-2}, and we apply it to the generic form of the Casimir of Eq.~\eqref{eq:casimir}, we find that
\bea
    C &=E^2-P^2+\frac{2(w_1-2w_2-w_3)}{3\Lambda}E^3+\frac{2w_2}{\Lambda}EP^2 \\
    &= E^2 - P^2 - \frac{2w_2}{\Lambda} E^3 + \frac{2w_2}{\Lambda} E P^2 \\
    &= \qty(E^2 - P^2) \qty(1- \frac{2w_2}{\Lambda} E) = 0 \,.
    \label{eq:casimir_notimedelay}
\eea
So we conclude that any \abb{DSR} model whose Casimir can be written in the form of Eq.~\eqref{eq:casimir_notimedelay} at first order in $\Lambda$, will not produce any time delays for massless particles. Some examples are the classical basis~\cite{Borowiec2010},
\be 
    C=E^2-P^2=0 \,,
    \label{eq:cas_classical}
\ee
the \abb{DCL1} basis~\cite{Carmona:2019vsh},
\be 
    C=\frac{E^2-P^2}{\qty(1-E/\Lambda)}=0 \,,
    \label{eq:cas_DCL1}
\ee
and the Magueijo-Smolin basis~\cite{Magueijo:2001cr},
\be 
    C=\frac{E^2-P^2}{\qty(1-E/\Lambda)^2}=0 \,.
    \label{eq:cas_MS}
\ee

With this, we have shown how to obtain the value of the photon time delay from the algebra of the deformed Poincaré generators; however, as we mentioned in Sec.~\ref{sec:DSR}, we can define the \abb{DSR} model in an equivalent way using the corresponding composition law of momenta. In Appendix~\ref{sec:timedelay_composition}, we see how to relate the ingredients developed in this section with the composition law of momenta, at linear order in $(1/\Lambda)$.

Let us also note that in order to derive the time delay we ensured the consistency with the relative locality of each of the two high-energy interactions (the emission and detection of the high-energy photon). As a consequence, we have been able to solve the time delay, even if we do not know which coordinates the observer $\mathcal O'$ (local to the emission) assigns to the high-energy photon at the detection (see red bubble in Fig.~\ref{fig:obs_DSR} (left)) and vice versa, the coordinates that $\mathcal O$ (local to the detection) assigns to the high-energy photon at the emission (see red bubble in Fig.~\ref{fig:obs_DSR} (right)). However, in the computation of the time delay, we found the translation that connects the coordinates of the worldlines of the photons seen from each observer (Eq.~\eqref{eq:translation}). If we combine this information with the action formulation of relative locality we showed in Appendix~\ref{sec:relative_locality}, we are able to compute the coordinates of all the incoming and outgoing particles at each interaction, seen for a nonlocal observer. For more details see Appendix~\ref{sec:timedelay_locality}. Let us also note that it is remarkable that the obtained formula for time delays in \abb{DSR}, while being compatible with the relative locality framework, depends only on the four-momentum of the high-energy photon and is independent of the details of its emission and detection.

Finally, let us also recall that in Sec.~\ref{sec:DSR} we mentioned that some authors in the literature, in order to avoid dealing with relative locality, have proposed that the physical spacetime, where the trajectories of the particles have to be described, is not the canonical spacetime, but a noncommutative spacetime where relative locality is absent~\cite{Carmona:2017oit,Carmona:2019oph,Amelino-Camelia:2011ebd,Mignemi:2016ilu}. In Appendix~\ref{sec:timedelay_noncommutative} we show how the time delay does not change if one chooses to describe the propagation of the photons using these noncommutative coordinates.
\begin{subappendices}
    \section{Time delays and the modified composition law}
\label{sec:timedelay_composition}

We have seen that there can exist models of \abb{DSR} without time delays, for example the ones showed in Eqs.~\eqref{eq:cas_classical}, \eqref{eq:cas_DCL1} and \eqref{eq:cas_MS}. This is because time delays are not an essential ingredient of \abb{DSR}, but a consequence, in the one-particle sector, of a more fundamental ingredient, the deformed algebra of the Poincaré generators. However, we discussed in Sec.~\ref{sec:DSR} that this deformed algebra can be obtained by providing the modified composition law. In this appendix we will see how to relate the ingredients used in the computation of time delays with the modified composition law of momenta.

In the same way we defined the space-time translation generators, $E$ and $P$, we can also define the generators of translations in energy-momentum space, let us call them $X$ and $T$. Similarly to what we did for $E$ and $P$, let us consider the most general expansion, linear in the space-time coordinates, of the generators $X$ and $T$,
\be 
    X=x+\frac{b_1}{\Lambda}x\Omega+\frac{b_2}{\Lambda}t\Pi \,,\qquad
    T=t+\frac{b_3}{\Lambda}x\Pi+\frac{b_4}{\Lambda}t\Omega \,,
    \label{eq:XT}
\ee
where $(b_1,b_2,b_3,b_4)$ are coefficients to be determined. 

If we assume that the generators $X$ and $T$, together with the boost generator $N$, are the generators of a Lie group of transformations acting in momentum space, they have to fulfill the Lie algebra,
\be
    \{T,X\}=\frac{1}{\Lambda}X\,,\qquad
    \{N,X\}=T\,,\qquad
    \{N,T\}=X+\frac{1}{\Lambda}N \,.
    \label{eq:lie}
\ee
This allows us to relate the coefficients ($b_1$, $b_2$, $b_3$, $b_4$), which define ($X$, $T$), with the coefficients ($a_4$, $a_5$, $a_6$), which define $N$, leading to
\bea
    &b_1=\frac{2a_4+2a_5-a_6}{3}-1 \,,\qquad
    b_2=\frac{-4a_4+2a_5+2a_6}{3}+1 \,,\\
    &b_3=\frac{-2a_4-2a_5+a_6}{3} \,,\qquad
    b_4=\frac{-2a_4+4a_5-2a_6}{3} \,.
    \label{eq:b1b2b3b4}
\eea

Now, we can use these generators to define a translation in momentum space of the form
\be
    \Omega'=\Omega+\pi\{X,\Omega\}-\omega\{T,\Omega\}\,,\qquad
    \Pi'=\Pi+\pi\{X,\Pi\}-\omega\{T,\Pi\}\,,
    \label{eq:mom_tras}
\ee
which, starting from the point $(\Pi, \Omega)$ in momentum space, performs a translation of parameters $(\pi, \omega)$ to the point $(\Pi', \Omega')$. Let us note that the parameters $(\pi, \omega)$ of the translation can be interpreted like another point of the same energy-momentum space. That is to say, we can define the momentum variables $q=(q_1,q_0)=(\Pi, \Omega)$ and $p=(p_1,p_0)=(\pi, \omega)$, and the translation generated by $(X,T)$ represents an operation that takes two momentum variables $p$ and $q$ and produce a new momentum variable that we can call $(p\oplus q)=(\Pi', \Omega')$. This is just a composition law of momentum variables.

A generic composition law of momenta can be written as
\bea
(p\oplus q)_1 \,&=\, p_1 + q_1 + \frac{\gamma_1}{\Lambda} \,p_0 \,q_1 + \frac{\gamma_2}{\Lambda} \,p_1\,q_0 \,\eqqcolon\, \Pi'\,,\\
(p\oplus q)_0 \,&=\, p_0 + q_0 + \frac{\beta_1}{\Lambda} \,p_0\,q_0 + \frac{\beta_2}{\Lambda} \,p_1\,q_1 \,\eqqcolon\, \Omega' \,.
\label{eq:composition}
\eea
Comparing Eqs.~\eqref{eq:mom_tras} and \eqref{eq:composition}, it is straightforward to show the one-to-one correspondence between the coefficients  $(\beta_1,\beta_2,\gamma_1,\gamma_2)$ and ($b_1$, $b_2$, $b_3$, $b_4$),
\be
\gamma_1 \,=\, - b_3  \,,\qquad
\gamma_2 \,=\, b_1  \,,\qquad
\beta_1 \,=\, b_4  \,,\qquad
\beta_2 \,=\, - b_2  \,.
\label{eq:gb}
\ee

This way, a deformed composition of momenta with coefficients $(\beta_1,\beta_2,\gamma_1,\gamma_2)$ is in correspondence with an energy-momentum translation of coefficients ($b_1$, $b_2$, $b_3$, $b_4$), which in turn  determines the coefficients ($a_4$, $a_5$, $a_6$) of the deformed boost generator using Eq.~\eqref{eq:b1b2b3b4}. Finally, the boost coefficients ($a_4$, $a_5$, $a_6$) also determine the modified energy-momentum relation, Eq.~\eqref{eq:dispersion_relation}. These correspondences can be summarized in the following identities,
\bea
\alpha_1 \,&=\, \frac{2a_4-4a_5-2a_6}{3} \,=\, - b_4 \,=\, - \beta_1\,, \\
\alpha_2 \,&=\, 2 a_5 \,=\, b_1 + b_2 - b_3 \,=\, \gamma_1 + \gamma_2 - \beta_2\,,
\eea
which are nothing more than the golden rules of a relativistic kinematics, derived from different arguments (from the compatibility of the deformed composition of momenta with the implementation of Lorentz transformations in a two-particle system) in other works~\cite{Carmona:2012un,Amelino-Camelia:2011gae}.

\clearpage
\section{Time delays and relative locality}
\label{sec:timedelay_locality}

In Sec.~\ref{sec:time_delays} we introduced two observers, $\mathcal{O}'$ and $\mathcal{O}$, which were necessary to construct a model for time delays consistent with relative locality. The first one, $\mathcal{O}'$, was defined as the observer which sees as local (and so at its origin of coordinates) the emission of the high-energy photon. Similarly, $\mathcal{O}$ was defined as the one which sees as local its detection.

As a direct consequence of its definition, we know that $\mathcal{O}'$ assigns coordinates $(x',t')=(0,0)$ to the start of the worldlines of all the outgoing particles participating in the interaction and to the end of the worldlines of the incoming ones. Similarly, observer $\mathcal{O}$ will assign coordinates $(x,t)=(0,0)$ to the starts or ends of the worldlines of the particles participating in the detection. However, when one asks $\mathcal{O}'$ about the detection, or $\mathcal{O}$ about the emission, as they see the interactions as non-local, the worldlines of the particles do not meet and the starting and ending points have different coordinates for each particle. In fact, we saw in Appendix~\ref{sec:relative_locality} that the values of the coordinates can be obtained using
\be
    t_i(0)\,=\,-z_1\frac{\partial \mathcal{P}}{\partial \Omega_i} + z_0\frac{\partial \mathcal{E}}{\partial \Omega_i}\,, \qquad
    x_i(0)\,=\, z_1\frac{\partial \mathcal{P}}{\partial \Pi_i} - z_0\frac{\partial \mathcal{E}}{\partial \Pi_i}\,, 
    \tag{See Eq.~\eqref{eq:RLpoint}}
\ee
where $\mathcal{P}$ and $\mathcal{E}$ are the total momentum and energy of the interaction, and $(z_1,z_0)$ are the pair of Lagrange multipliers that the observer assigns to the interaction.

It is clear that observer $\mathcal{O}$ assigns $(z_1,z_0)=(0,0)$ to the interaction corresponding to the detection of the high-energy photon, so that all the trajectories start or end at $(x,t)=(0,0)$. If we knew the values of $(z_1,z_0)$ that $\mathcal{O}$ assigns to the emission, using Eq.~\eqref{eq:RLpoint} and the total energy and momentum of the interaction, we would be able to compute the coordinates that $\mathcal{O}$ assigns to the start or ends of the worldlines of all the particles participating in the emission.

Let us remember that in Sec.~\ref{sec:time_delays}, during the development of the model of time delays, we found the one-particle space-time translation that connects the coordinates of worldlines of the photons assigned by $\mathcal{O}$ and $\mathcal{O}'$,
\be
    x'\,=\,x- \epsilon_1 \{P, x\} + \epsilon_0 \{E, x\} \,, \quad\quad
    t'\,=\,t - \epsilon_1\{P, t\} + \epsilon_0 \{E, t\} \,,
    \tag{See Eq.~\eqref{eq:translation}}
\ee
with $\epsilon_1=L$ and $\epsilon_0=L+\delta t$. Then, we can use this relation to find the coordinates $(x,t)$ that the observer $\mathcal{O}$ assigns to the start of the worldline of the high-energy photon at the emission, which seen from $\mathcal{O}'$ are just $(x',t')=(0,0)$,
\bea
 x\,&=\, \epsilon_1 \{P,x\} -\epsilon_0 \{E,x\} \,=\, -L \left[ 1+ \frac{(-2a_2+a_3)}{\Lambda}\Pi \right] \,, \\
 t\,&=\, \epsilon_1 \{P,t\} -\epsilon_0 \{E,t\} \,=\, -L \left[ 1 + \frac{(-2a_2+a_3)}{\Lambda} \Pi + \frac{2(a_4+a_5-a_6)}{3\Lambda}\Pi \right]\,. 
\eea

Now that we know the coordinates that $\mathcal{O}$ assigns to the starting point of the trajectory of one of the particles participating in the emission, we can substitute this value in Eq.~\eqref{eq:RLpoint} and solve for the values of $(z_1,z_0)$ that observer $\mathcal{O}$ assigns to the emission,
\bea
z_1 \,&=\, - L \,\left[1 + \frac{(-2 a_2 + a_3)}{\Lambda} \Pi + \left(1 + \frac{\partial\mathcal{E}}{\partial\Pi} - \frac{\partial\mathcal{P}}{\partial\Pi}\right)\right]\,, \\
 z_0 \,&=\, - L \,\left[1 + \frac{(-2 a_2 + a_3)}{\Lambda} \Pi + \frac{2(a_4+a_5-a_6)}{\Lambda} \Pi + \left(1 - \frac{\partial\mathcal{E}}{\partial\Omega} + \frac{\partial\mathcal{P}}{\partial\Omega}\right)\right]\,.  
\eea

Finally, since we know the value of the Lagrange multipliers that observer $\mathcal{O}$ assigns to the interaction corresponding to emission of the high-energy photon, we can compute the starting or ending points (respectively) of all the outgoing and incoming particles participating in the interaction using Eq.~\eqref{eq:RLpoint}. Additionally, this method can also be used to determine the parameters $(z'_1,z'_0)$ that the observer $\mathcal{O}'$ assigns to the interaction corresponding to the detection.

\clearpage
\section{Time delays in noncommutative spacetime}
\label{sec:timedelay_noncommutative}

In Appendix~\ref{sec:timedelay_composition} we introduced the generators of the translations in energy-momentum space, $(X,T)$, which close a Lie algebra with the generator of the boosts, $N$,
\be
    \{T,X\}=\frac{1}{\Lambda}X\,,\qquad
    \{N,X\}=T\,,\qquad
    \{N,T\}=X+\frac{1}{\Lambda}N \,.
    \tag{See Eq.~\eqref{eq:lie}}
\ee
One can make a model in which the physical space-time coordinates are $(X,T)$ instead of the canonical ones used in Sec.~\ref{sec:time_delays}\footnotemark. This kind of choice leads to a noncommutative spacetime, because as we can see in Eq.~\eqref{eq:lie}, the Poisson brackets of the spatial and temporal coordinates are not trivial.
\footnotetext{Note that the $(X,T)$ coordinates are not canonical; that is to say, it does not exists any change of variables of the form of Eq.~\eqref{eq:change_variable} that goes from $(x,t)$ to $(X,T)$.}

If now $(X,T)$ are the coordinates in which the trajectories of the particles must be described, we can wonder what is the expression of the time delay we found in Sec.~\ref{sec:time_delays} in these new coordinates. In order to do that, we need a relation between $(X,T)$ and $(x,t)$, which is given by the expansion of the generators in the canonical coordinates, 
\be 
    X=x+\frac{b_1}{\Lambda}x\Omega+\frac{b_2}{\Lambda}t\Pi \,,\qquad
    T=t+\frac{b_3}{\Lambda}x\Pi+\frac{b_4}{\Lambda}t\Omega \,.
    \tag{See Eq.~\eqref{eq:XT}}
\ee
From the previous equation we can find
\be
    \frac{dX}{dt} \,=\,\frac{dx}{dt}+\frac{b_1}{\Lambda}\frac{dx}{dt}\Omega+\frac{b_2}{\Lambda}\Pi\,, \quad\quad\quad
    \frac{dT}{dt} \,=\,1+\frac{b_3}{\Lambda}\frac{dx}{dt}\Pi+\frac{b_4}{\Lambda}\Omega\,,
\ee
and merging both equations, and expanding until order $1/\Lambda$, we get that the velocity in the noncommutative spacetime is given by
\begin{align}
    \frac{dX}{dT} &\,=\, \frac{dx}{dt}+ \frac{(b_1+b_2-b_3-b_4)}{\Lambda}\Pi  \notag \\ &\,=\, \left(1 - \frac{2\,(a_4+a_5+a_6)}{3\, \Lambda}\Pi \right)+  \frac{2\,(a_4+a_5+a_6)}{3\, \Lambda}\Pi  \,=\, 1  \,,
\end{align}
where in the next-to-last step we have used Eqs.~\eqref{eq:velocity} and \eqref{eq:b1b2b3b4} to write the canonical velocity and $(b_1,b_2,b_3,b_4)$ in terms of the coefficients of the boost. Then, we get that the velocity of a massless particle in the noncommutative spacetime is simply $(dX/dT)=1$.

Let us now consider again the two observers $\mathcal O$ and $\mathcal O'$, defined to see as local the detection and emission of the high-energy photon, respectively. From Eq.~\eqref{eq:XT}, one can check that a point with coordinates $(x,t)=(0,0)$ has coordinates $(X,T)=(0,0)$. Therefore, the detection of the high-energy photon will take place at $(X,T)=(0,0)$ for observer $\mathcal O$, and the emission at $(X',T')=(0,0)$ for observer $\mathcal O'$. Imposing that the observers are at relative rest, the coordinates that each observer assigns to the worldline of the photons must be related by a space-time translation of the form
\be
    X' \,=\, X + \epsilon_0 \{E, X\} - \epsilon_1 \{P, X\} \,, \quad\quad
    T' \,=\, T + \epsilon_0 \{E, T\} - \epsilon_1 \{P, T\} \,,
\ee
with $(\epsilon_1,\epsilon_0)$ the parameters of the translation. Using the fact that $(dX/dT)=1$, the new worldline of the high-energy photon seen from $\mathcal O$ is $X=T$, and $X'=T'$ from $\mathcal O'$, from what we can derive that the parameters of the translation are now related by 
\be
    \epsilon_0 \left[\{E,T\} - \{E,X\}\right] \,=\, \epsilon_1 \left[\{P,T\} - \{P,X\}\right]\,.
\ee
Computing the Poisson brackets that appear in previous expression,
\begin{align}
    &\{E,T\} \,=\, 1 + \frac{(b_4 + 2 a_1)}{\Lambda} \Omega\,, \quad\quad
    \{E,X\} \,=\, \frac{(b_2 - 2 a_2)}{\Lambda} \Pi\,, \nonumber \\
    &\{P,T\} \,=\, \frac{(-b_3 + a_3)}{\Lambda} \Pi\,, \quad\quad
    \{P,X\} \,=\, -1 - \frac{(b_1+a_3)}{\Lambda} \Omega\,,
\end{align}
and using once more the energy-momentum relation, Eq.~\eqref{eq:dispersion_relation}, one finds
\bea
    \frac{\epsilon_0}{\epsilon_1} \,&=\, 1 + \frac{(b_1+b_2-b_3-b_4)}{\Lambda} \Pi - \frac{2(a_1+a_2-a_3)}{\Lambda} \Pi\\ \,&=\, 1 + \frac{2(a_4+a_5+a_6)}{3 \Lambda} \Pi - \frac{2(a_1+a_2-a_3)}{\Lambda} \Pi \,. 
    \label{eq:parameters_noncom}
\eea
Then, the relation between the parameters of the translation between observers $\mathcal{O}$ and $\mathcal{O'}$ are the same as the one obtained using canonical coordinates. Since for the trajectory of the low-energy photon, which allows to define the time delay from Eq.~\eqref{eq:parameters_noncom}, there is no distinction between commutative and noncommutative coordinates, the result for the time delay $\delta T$ coincides with the result for $\delta t$, Eq.~\eqref{delta t}, obtained using the commutive (i.e. canonical) spacetime.
\end{subappendices}
\printbibliography[heading=subbibintoc,title={References}]
\end{refsection}

\begin{refsection}
\chapter{Summary of the conclusions}
\graphicspath{{Chapter6/Figs/}}
\label{chap:conclusions}

As discussed in Sec.~\ref{sec:scope} of Chapter~\ref{chap:introduction}, the main objective of this dissertation has been to explore possible phenomenological windows of physics beyond Lorentz Invariance, i.e., Lorentz Invariance Violation and Doubly Special Relativity, using the most favorable scenario we have at out disposal, the (ultra) very high-energy observations of astroparticles from distant sources. We decided to focus the study of new physics on the propagation effects, due to the fact that they grow with the travelled distance, providing the necessary amplification to make them observable. However, we should take in mind that the effects at the production can also be relevant in the study of the modified fluxes at Earth. We focused on the study of Lorentz Invariance Violation in neutrinos (due to the fact that the effects are stronger than in the case of Doubly Special Relativity) and the study of Doubly Special Relativity in gamma rays (because, in contrast to Lorentz Invariance Violation, it is mostly unexplored). In this section we will collect the most relevant results found during the course of this dissertation. We will also discuss their limitations and possible future extensions.

In Chapter~\ref{chap:LIV-DSR}, we explained the main characteristics of the framework behind Lorentz Invariance Violation and Doubly Special Relativity. Thanks to the work of Alan Kostelecký and many others, there exists a precise mathematical description to study the scenarios of Lorentz Invariance Violation inside the framework of Effective Field Theories. In contrast, the idea of Doubly Special Relativity, proposed by Giovanni Amelino-Camelia, is still in an early phase of study, and, in consequence, many open questions remain, like the \textit{soccer ball problem}, the \textit{spectator problem}, the physical role of a change of energy-momentum variables, or the existence of time delays and their compatibility with relative locality, which still need to be solved in order to consider Doubly Special Relativity a consistent theory.

In the Appendix~\ref{sec:new_perspective} of Chapter~\ref{chap:LIV-DSR}, we propose a novel perspective of Doubly Special Relativity, which can avoid these paradoxes and problems. We can summarize the change of paradigm in the fact that one can resign from the idea of considering the effects of Doubly Special Relativity as effects produced, at a fundamental level, by the interaction of the particles with a quantum spacetime, but consider instead that Doubly Special Relativity is a manifestation of a departure from locality of interactions. This implies the need to go beyond a local Quantum Field Theory description of the interactions. However, the relation between a fundamental Quantum Gravity theory and this new perspective, as its footprint, is debatable and needs further investigation.

In the same chapter, we also discuss about how the loss of the Lorentz Invariance complicates the computation of decay widths and cross sections. We propose the use of the collinearity of the high-energy interactions to simplify the computations when applied to very high-energy astroparticles. In Sec.~\ref{sec:collinear} we show an example applying the aforementioned approximation to the decay of a neutrino in three particles in a superluminal Lorentz Invariance Violation scenario.

The previous result is used in Sec.~\ref{sec:neu_VPE} and \ref{sec:neu_NSpl} of Chapter~\ref{chap:neutrinos} to compute explicitly, for the first time, the decay widths and energy distributions of electron-positron and neutrino-antineutrino pair emissions in the propagation of superluminal neutrinos. We have also analysed and compared the obtained decay widths, and computed the mean energy fractions of the particles of the final state. 

Using the results mentioned in the previous paragraph, in Secs.~\ref{sec:neu_prop_cont} and \ref{sec:neu_prop_ins} of Chapter~\ref{chap:neutrinos} we developed a model for the propagation of neutrinos in the superluminal Lorentz Invariance Violation scenario. The proposed models are based on, and in consequence limited by, the instantaneity of the decay. This approximation splits the energy of the neutrino in ``weak'' and ``strong'' regimes for each decay, in which one can consider the decay as negligible or instantaneous, respectively. The existence of a kinematical threshold for the electron-positron pair emission, ensures the strong regime for every neutrino with enough energy to disintegrate (for values of $\Lambda<M_P$). However, this is not necessarily true for the neutrino-antineutrino pair emission when the energy of the neutrino is close to where both regimes meet. The limitations of this approximation need further investigation.

As a consequence of the two previously mentioned decays, the neutrino flux at Earth is modified by a very strong suppression after a certain energy $E_\text{cut}$. The use of the instantaneous approximation leads to an analytical and straightforward approximate relation between the parameters of new physics and the energy of the cutoff. We used the most energetic neutrino detection, the neutrino of the Glashow resonance recently reported by IceCube, to update the constraints on the scale of new physics for a first and second order dominant modification of the neutrino dispersion relation.

In the last section of of Chapter~\ref{chap:neutrinos}, Sec.~\ref{sec:monte_carlo}, after several attempts of developing a flux model, we decided to implement the aforementioned decays of new physics in SimProp, a Monte Carlo software which simulates particle propagation in a standard scenario. As a result, we obtained a prediction for the detected neutrino spectrum at Earth (assuming certain astrophysical conditions) which can be confronted with real data. We show, as an example, a comparison between the Monte Carlo simulation and the High Energy Starting Events from IceCube, assuming the astrophysical conditions proposed by Stecker et al. in a previous work. This procedure can be done using future data and different values of the parameters of new physics and the astrophysical conditions, in order to constraint the possible scenarios of Lorentz Invariance Violation.

In the next chapter, Chapter~\ref{chap:cosmicrays}, we extend this analysis to cosmogenic neutrinos, those which are produced in the propagation of cosmic rays due to the interaction with the Cosmic Microwave Background and the Extragalactic Background Light. Current experiments have not yet detected neutrino events in this energy range; however, from the absence of events, one can still put an upper limit for the detected flux, which can be compared with the Monte Carlo simulation predictions. We computed, as an example of the procedure, the expected number of events by IceCube and Pierre Auger. In the close future, the upcoming experiments will have much more sensitivity in this energy range, and they can be used to put constraints in both the parameters of new physics and the astrophysical conditions for the cosmic rays.

In Chapter~\ref{chap:gammas} we change the paradigm in both the messenger and the physics. In this chapter we study the effects of a Doubly Special Relativity scenario in gamma rays. We can distinguish different phenomenological windows for the photons: one refers to effects in the gamma ray flux and the other to anomalies in the times of flight. In Sec.~\ref{sec:gamma_dsr}, we study how the transparency of the universe is modified due to the effects of Doubly Special Relativity in the range of energies where the Cosmic Microwave Background is the most relevant source of low energy photons. In order to do that, we focus in the local universe and in a specific realization of Doubly Special Relativity, the DCL1 model. As a result, we find that the transparency of the universe increases and then decreases asymptotically to a constant value. This peculiar behaviour is clearly distinguishable from the case of Lorentz Invariance Violation, so observations could potentially be used to reject some models of new physics. In the future, we would like to improve the analysis including also the energy ranges in which the Extragalactic Background Light dominates, as well as considering farther sources, which would require the use of the opacity instead of the mean free path. One can also repeat the analysis using a different model of Doubly Special Relativity; specifically, the one proposed in Appendix.~\ref{sec:new_perspective} for the new perspective of Doubly Special Relativity.

In the last section of Chapter~\ref{chap:gammas}, Sec.~\ref{sec:time_delays}, we discuss the other phenomenological window, photon time delays. In Lorentz Invariance Violation scenarios, a modified dispersion relation for photons necessarily implies an energy dependent velocity of propagation and time delays between photons of different energies; however, the existence of additional ingredients in Doubly Special Relativity, such as relative locality, nontrivial translations, etc., makes the discussion much more complex, up to the point that the existence of time delays in Doubly Special Relativity scenarios is still an open question up to now. In Sec.~\ref{sec:time_delays}, we developed a general analysis of the computation of time delays in a flat spacetime in Doubly Special Relativity, paying particular attention to the consistency with relative locality and the role of the choice of energy-momentum variables. As a result, we find that there exist models of Doubly Special Relativity without time delays, and we provide a formula to obtain the time delay directly from the algebra of the Poincaré generators defining the model. We also realize that the change of the choice of energy-momentum variables does not affect the result, in contrast to a change of the generators, which corresponds to a change of the model. As a future work, we would like to extend this analysis, including an expanding universe. This would allow to confront the model with real data, and to put constraints on the scale of new physics.

We can conclude from all the previously mentioned results that there are several windows of possible footprints of a quantum gravity theory in the study of observations of very high-energy astroparticles. This kind of studies will be very relevant in the near future, since we expect the construction of new and more sensitive astroparticle observatories in the next decade. In this dissertation we made some first steps in different lines of research of new physics using very high-energy astroparticles. We hope this work can continue being developed and extended in the future, especially considering that we are at the dawn of the multimessenger era.

\setcounter{chapter}{5}
\renewcommand{\chaptername}{Capítulo}
\chapter{Resumen de las conclusiones}
\graphicspath{{Chapter6/Figs/}}

Como se discutió en la Sec.~\ref{sec:scope} del Capítulo~\ref{chap:introduction}, el objetivo principal de esta tesis es explorar posibles ventanas fenomenológicas de la física más allá de la Invariancia Lorentz, es decir, la Violación de la Invariancia Lorentz y la Relatividad Doblemente Especial, utilizando el escenario más favorable posible que tenemos a nuestra disposición, las observaciones de astropartículas de muy (ultra) alta energía procedentes de fuentes lejanas. Decidimos centrar el estudio de la física en los efectos en la propagación, debido a que crecen con la distancia recorrida, consiguiendo la amplificación necesaria para hacerlos observables. Sin embargo, debemos tener en cuenta que los efectos en la producción también pueden ser relevantes en el estudio de anomalías en flujos en la Tierra. Nos centramos en el estudio de la Violación de la Invariancia Lorentz en neutrinos (debido a que los efectos son más fuertes que en el caso de la Relatividad Doblemente Especial) y en el estudio de la Relatividad Doblemente Especial en rayos gamma (debido a que, en contraste con la Violación de la Invariancia Lorentz, ha sido mucho menos explorada). En esta sección recogeremos los resultados más relevantes encontrados a lo largo de esta tesis. También discutiremos sus limitaciones y futuras posibles extensiones.

En el Capítulo~\ref{chap:LIV-DSR}, explicamos las principales características del marco que subyace a la Violación de la Invariancia Lorentz y la Relatividad Doblemente Especial. Gracias al trabajo de Alan Kostelecký y muchos otros, disponemos de una descripción matemática precisa para estudiar escenarios de Violación de la Invariancia Lorentz dentro del marco de Teorías de Campos Efectivas. Por el contrario, la idea de la Relatividad Doblemente Especial, propuesta por Giovanni Amelino-Camelia, se encuentra todavía en una fase temprana de estudio, y, en consecuencia, existen muchas preguntas abiertas, como el \textit{problema del balón de fútbol}, el \textit{problema del espectador}, el papel físico del cambio de las variables energía-momento, o la existencia de diferencias de tiempos vuelo y su compatibilidad con la localidad relativa, que todavía necesitan ser resueltas para poder considerar la Relatividad Doblemente Especial como una teoría consistente.

En el Apéndice~\ref{sec:new_perspective} del Capítulo~\ref{chap:LIV-DSR}, proponemos una nueva perspectiva de la Relatividad Doblemente Especial, que puede evitar estas paradojas y problemas. Podemos resumir el cambio de paradigma, en el hecho de que se puede renunciar a la idea de considerar los efectos de la Relatividad Doblemente Especial como efectos producidos, a nivel fundamental, por la interacción de las partículas con un espacio-tiempo cuántico, para considerar en cambio, es una manifestación de una pérdida de la localidad de las interacciones. Esto implica la necesidad de ir más allá de una descripción de las interacciones mediante una Teoría Cuántica de Campos local. Sin embargo, la relación entre una teoría fundamental de Gravedad Cuántica y esta nueva perspectiva, como una huella de la misma, es discutible y necesita más investigación.

En el mismo capítulo, también discutimos cómo la pérdida de la Invariancia Lorentz complica el cálculo de las anchuras de desintegración y de las secciones eficaces. Proponemos el uso de la colinealidad de las interacciones de alta energía para simplificar los cálculos cuando se aplican a astropartículas de muy alta energía. En la Sec.~\ref{sec:collinear} mostramos un ejemplo aplicando dicha aproximación a la desintegración de un neutrino en tres partículas en un escenario superlumínico de Violación de la Invariancia Lorentz.

El resultado anterior se utiliza en las Secciones ~\ref{sec:neu_VPE} y \ref{sec:neu_NSpl} del Capítulo~\ref{chap:neutrinos} para calcular explícitamente, por primera vez, las anchuras de desintegración y las distribuciones de energía de la emisión de pares neutrino electrón-positrón y neutrino-antineutrino. También analizamos y comparamos las anchuras de desintegración obtenidas, y calculamos el valor medio de fracciones de energía de las partículas del estado final. 

Utilizando los resultados del párrafo anterior, en las Secciones ~\ref{sec:neu_prop_cont} y \ref{sec:neu_prop_ins} del Capítulo~\ref{chap:neutrinos}, desarrollamos un modelo para la propagación de neutrinos en el escenario superlumínico de Violación de la Invariancia Lorentz. Los modelos desarrollados se basan en, y en consecuencia están limitados por, la instantaneidad de la desintegración. Esta aproximación divide la energía del neutrino en los regímenes ``débil'' y ``fuerte'' para cada desintegración, en los que se puede considerar que la desintegración es despreciable o instantánea, respectivamente. La existencia de un umbral cinemático para la emisión del par electrón-positrón, asegura que los neutrinos con suficiente energía para desintegrarse están el régimen fuerte (para valores de $\Lambda<M_P$). Sin embargo, esto no es necesariamente cierto para la emisión de pares neutrino-antineutrino cuando la energía del neutrino está cerca de donde ambos regímenes se encuentran. Las limitaciones de esta aproximación requieren más investigación.

Como consecuencia de las dos desintegraciones mencionadas anteriormente, el flujo de neutrinos en la Tierra se ve modificado por una supresión muy fuerte a partir de una cierta energía $E_\text{cut}$. El uso de la aproximación instantánea conduce a una aproximada relación analítica y bastante directa entre los parámetros de la nueva física y la energía del corte. Utilizamos la detección más energética de un neutrino, el neutrino de la resonancia de Glashow recientemente reportado por IceCube, para actualizar las cotas sobre la escala de la nueva física para una corrección dominante de primer y segundo orden.

En la última sección del Capítulo~\ref{chap:neutrinos}, Sec.~\ref{sec:monte_carlo}, después de varios intentos de desarrollar un modelo para el flujo, decidimos implementar las desintegraciones de nueva física previamente mencionadas en SimProp, un software de Montecarlo que simula la propagación de partículas en un escenario estándar. Como resultado, obtenemos una predicción para el espectro de neutrinos detectado en la Tierra (suponiendo ciertas condiciones astrofísicas) que puede ser confrontada con datos reales. Mostramos, a modo de ejemplo, una comparación entre la simulación Montecarlo y los Eventos de Alta Energía de IceCube, asumiendo las condiciones astrofísicas propuestas por Stecker et al. en un trabajo previo. Este procedimiento puede realizarse con futuros datos y para diferentes valores de los parámetros de nueva física y de las condiciones astrofísicas, permitiendo así restringir los posibles escenarios de Violación de la Invariancia Lorentz.

En el siguiente capítulo, Capítulo~\ref{chap:cosmicrays}, extendemos este análisis a los neutrinos cosmogénicos, aquellos producidos en la propagación de los rayos cósmicos debido a su interacción con el Fondo Cósmico de Microondas y la Luz Extragaláctica de Fondo. Los experimentos actuales aún no han detectado ningún evento de neutrinos en este rango de energías; sin embargo, a partir de la ausencia de eventos, uno todavía puede poner un límite superior al flujo detectado que se puede comparar con las predicciones de la simulación de Montecarlo. Hemos calculado, como ejemplo del procedimiento, el número esperado de eventos por IceCube y Pierre Auger. En un futuro próximo, los nuevos experimentos tendrán mucha más sensibilidad en este rango de energías, y podrán ser utilizados para poner cotas tanto en los parámetros de la nueva física como en las condiciones astrofísicas de los rayos cósmicos.

En el capítulo~\ref{chap:gammas} cambiamos de paradigma tanto en el mensajero como en la física. En este capítulo estudiamos los efectos de un escenario de Relatividad Doblemente Especial en los rayos gamma. Podemos distinguir diferentes ventanas fenomenológicas para los fotones: uno son los efectos en el flujo de rayos gamma y el otro se refiere a las anomalías en los tiempos de vuelo. En la Sec.~\ref{sec:gamma_dsr}, estudiamos cómo se modifica la transparencia del universo debido a los efectos de la Relatividad Doblemente Especial en el rango de energías donde el Fondo Cósmico de Microondas es la fuente más relevante de fotones de baja energía. Para ello, nos centramos en el universo local y en un modelo específico de Relatividad Doblemente Especial, el modelo DCL1. Como resultado, encontramos que la transparencia del universo aumenta y luego disminuye asintóticamente hasta un valor constante. Este peculiar comportamiento se distingue claramente del caso de Violación de la Invariancia Lorentz, por lo que las observaciones se pueden utilizar potencialmente para rechazar algunos modelos de la nueva física. En el futuro, nos gustaría mejorar el análisis incluyendo también rangos de energía en los que domina la Luz de Fondo Extragaláctica, así como considerando fuentes más lejanas, lo cual requeriría el uso de la opacidad en lugar del recorrido libre medio. También se puede repetir el análisis utilizando un modelo diferente de Relatividad Doblemente Especial; específicamente, el  propuesto en el Apéndice.~\ref{sec:new_perspective} para la nueva perspectiva de la Relatividad Doblemente Especial.

En la última sección del mismo capítulo, Sec.~\ref{sec:time_delays} del Capítulo~\ref{chap:gammas}, discutimos la otra ventana fenomenológica, los retrasos en tiempos de vuelo del fotón. En escenarios de Violación de la Invariancia Lorentz, una relación de dispersión modificada para fotones implica necesariamente una velocidad de propagación dependiente de la energía y por tanto la existencia de retrasos en tiempos de vuelo entre fotones de diferentes energías; sin embargo, la existencia de ingredientes adicionales en Relatividad Doblemente Especial, tales como la localidad relativa, traslaciones no triviales, etc., hacen la discusión mucho más compleja, hasta el punto de que la existencia de retrasos en tiempos de vuelo en escenarios de Relatividad Doblemente Especial sigue siendo una cuestión abierta hasta el día de hoy. En la Sec.~\ref{sec:time_delays}, desarrollamos un análisis general del cálculo de retrasos en tiempos de vuelo en un espacio-tiempo plano en Relatividad Doblemente Especial, prestando especial atención a la consistencia con la localidad relativa y al papel de la elección de las variables de energía-momento. Como resultado, encontramos que existen modelos de Relatividad Doblemente Especial sin retrasos en tiempos de vuelo, y proporcionamos una fórmula para obtener dicho retraso directamente a partir del álgebra de los generadores de Poincaré que definen el modelo. También nos damos cuenta de que el cambio de la elección de las variables de energía-momento no afecta al resultado, en contraste con un cambio de los generadores, que corresponden a un cambio de modelo. Como trabajo futuro, nos gustaría ampliar este análisis, incluyendo un universo en expansión. Esto permitiría confrontar el modelo con datos reales, y poner restricciones a la escala de la nueva física.

Podemos concluir de todos los resultados anteriores que existen varias ventanas a posibles huellas de una teoría de gravedad cuántica en el estudio de las observaciones de astropartículas de muy alta energía. Este tipo de estudios serán muy relevantes en un futuro próximo, ya que se espera la construcción de nuevos y más sensibles observatorios de astropartículas en la próxima década. En esta tesis hemos dado algunos primeros pasos en diferentes líneas de investigación de la nueva física utilizando astropartículas de muy alta energía. Esperamos que este trabajo pueda seguir desarrollándose y ampliándose en el futuro, sobre todo teniendo en cuenta que nos encontramos en los albores de la era multimensajero.

%\begin{subappendices}
%    \include{Chapter6/appendix6}
%\end{subappendices}
\printbibliography[heading=subbibintoc,title={References}]
\end{refsection}

%\begin{subappendices}
%    \include{Chapter1/appendix1}
%\end{subappendices}

% ********************************** Back Matter *******************************
% Backmatter should be commented out, if you are using appendices after References

%\backmatter

% ********************************** Bibliography ******************************
%\begin{spacing}{0.9}

% To use the conventional natbib style referencing
% Bibliography style previews: http://nodonn.tipido.net/bibstyle.php
% Reference styles: http://sites.stat.psu.edu/~surajit/present/bib.htm
%\nocite{*}
%\bibliographystyle{apsrev4-1}
%\bibliographystyle{apalike}
%\bibliographystyle{unsrt} % Use for unsorted references  
%\bibliographystyle{plainnat} % use this to have URLs listed in References
%\cleardoublepage
%\bibliography{References/Tesis,References/QuGraPheno} % Path to your References.bib file

% If you would like to use BibLaTeX for your references, pass `custombib' as
% an option in the document class. The location of 'reference.bib' should be
% specified in the preamble.tex file in the custombib section.
% Comment out the lines related to natbib above and uncomment the following line.

%\printbibliography[heading=bibintoc, title={References}]

%\end{spacing}

% ********************************** Appendices ********************************

%\begin{appendices} % Using appendices environment for more functionality

%\include{}

%\end{appendices}

% *************************************** Index ********************************
\printthesisindex % If index is present

\cleardoublepage
\includepdf[pages=-]{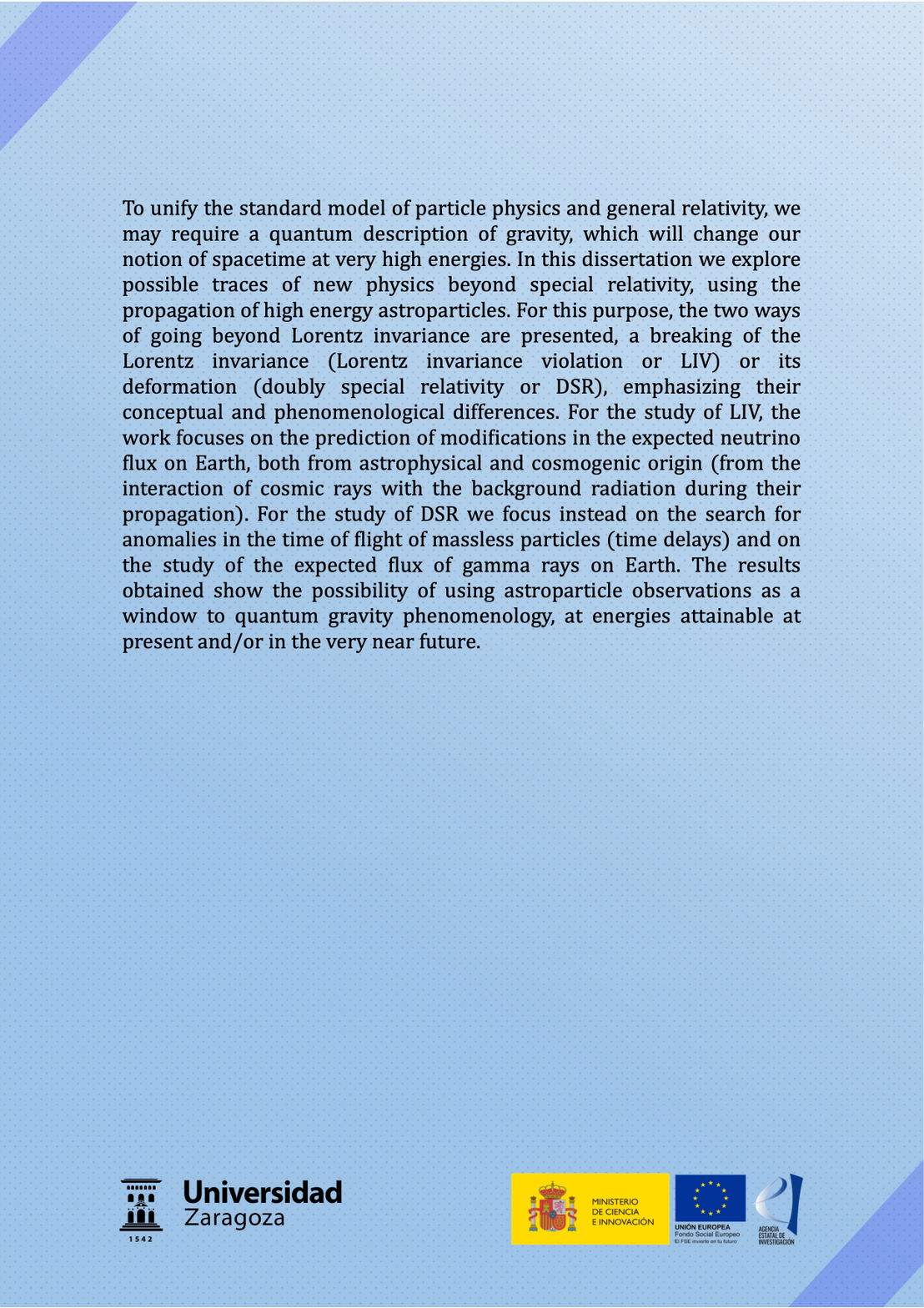}

\end{document}